\pdfoutput=1
\documentclass[11pt]{article}

\usepackage[english]{babel}
\usepackage{xcolor}
\usepackage{graphicx}

\usepackage[colorlinks,allcolors=blue]{hyperref}

\usepackage{cite}
\usepackage{subfigure}
\usepackage{amsmath}

\usepackage{placeins}
\usepackage{booktabs}

\usepackage[top=15mm,bottom=12mm,left=30mm,right=30mm,foot=12mm,includefoot]{geo
metry}

\allowdisplaybreaks[3]

\numberwithin{equation}{section}

\newcommand{\sect}[1]{section~#1}
\newcommand{\app}[1]{appendix~#1}

\newcommand{\fig}[1]{figure~#1}
\newcommand{\figs}[1]{figures~#1}

\newcommand{\eqn}[1]{equation~#1}
\newcommand{\eqs}[1]{equations~#1}

\newcommand{\as}{a_s}

\newcommand{\half}{{\textstyle\frac{1}{2}}}

\newcommand{\gev}{\operatorname{GeV}}

\newcommand{\fm}{\operatorname{fm}}

\newcommand{\ms}{\mskip 1.5mu}
\newcommand{\bs}{\mskip -1.5mu}

\newcommand{\tr}{\operatorname{tr}\ms{}}

\newcommand{\tvec}[1]{\boldsymbol{#1}}

\newcommand{\ub}{\bar{u}}
\newcommand{\zb}{\bar{z}}

\newcommand{\msbar}{$\overline{\text{MS}}$ }

\newcommand{\msbars}{\scalebox{0.6}{$\overline{\mathrm{MS}}$}}

\newcommand{\us}{\text{us}}

\newcommand{\bus}{B \ms \text{us}}

\newcommand{\conv}[1]{\underset{#1}{\otimes}}

\newcommand{\pr}[2]{{}^{#1}\bs #2}      
\newcommand{\prb}[2]{{}^{#1}\! #2}      
\newcommand{\prn}[2]{{}^{#1} #2}        

\newcommand{\Rp}{R^{\ms \prime}}
\newcommand{\Rpp}{R^{\ms \prime\prime}}
\newcommand{\Rbar}{\overline{R}}
\newcommand{\Rpbar}{\overline{R}{}^{\ms \prime}}

\newcommand{\RR}{R_1\bs R_2}

\newcommand{\tenbar}{\scalebox{0.7}{$\overline{10}$}}
\newcommand{\ten}{\scalebox{0.7}{$10\phantom{\overline{1}}\hspace{-1ex}$}}
\newcommand{\tenbarten}{\tenbar\, \ten}
\newcommand{\tentenbar}{\ten\ms \tenbar}

\newcommand{\tentenbartext}{\scalebox{0.95}{$10\, \overline{10}$}}
\newcommand{\tenbartentext}{\scalebox{0.95}{$\overline{10}\, 10$}}

\graphicspath{{graphs/}, {Plots_gg_CT14/}, {Plots_misc/}}

\newcommand{\rev}[1]{#1}

\begin{document}

\begin{flushright}
DESY 21-015 \\
MAN/HEP/2021/002 \\
\href{https://arxiv.org/abs/2105.08425}{arXiv:2105.08425 [hep-ph]}
\end{flushright}

\begin{center}
\vspace{4\baselineskip}
\textbf{\Large Two-loop splitting in double parton distributions:\\[0.2em]
the colour non-singlet case} \\
\vspace{3\baselineskip}
M.~Diehl$^{\ms 1}$, J.~R.~Gaunt$^{\ms 2}$ and P.~Pl{\"o}{\ss}l$^{\ms 1}$
\end{center}

\vspace{2\baselineskip}

${}^{1}$ Deutsches Elektronen-Synchrotron DESY, Notkestr.~85, 22607 Hamburg, Germany \\
${}^{2}$ Department of Physics and Astronomy, University of Manchester, Manchester, M13 9PL, \\ \phantom{${}^{2}$} United Kingdom \\
\vspace{3\baselineskip}

\parbox{0.9\textwidth}{
}

At small inter-parton distances, double parton distributions receive their dominant contribution from the splitting of a single parton.  We compute this mechanism at next-to-leading order in perturbation theory for all colour configurations of the observed parton pair.  Rapidity divergences are handled either by using spacelike Wilson lines or by applying the $\delta$~regulator.  We investigate the behaviour of the two-loop contributions in different kinematic limits, and we illustrate their impact in different channels.

\vfill

\newpage

\tableofcontents

\begin{center}
\rule{0.6\textwidth}{0.3pt}
\end{center}


\section{Introduction}
\label{sec:intro}

To interpret measurements at the Large Hadron Collider in its high-luminosity phase, it is of great importance to understand the strong-interaction part of proton-proton collisions as much as possible.  This provides a strong motivation for the study of double parton scattering (DPS), which is a mechanism in which two pairs of partons initiate two separate hard-scattering processes in a single collision.  Due to an increased parton flux for decreasing parton momentum fractions, the mechanism becomes more important for higher collision energy.  It is therefore relevant also to the planning of future hadron colliders at the energy frontier.

The theoretical investigation of DPS started long ago \cite{Landshoff:1978fq,Kirschner:1979im,Politzer:1980me,Paver:1982yp,Shelest:1982dg,Mekhfi:1983az,Sjostrand:1986ep}, and in the last decade several approaches to a systematic description in QCD have been put forward \cite{Blok:2010ge,Diehl:2011tt,Gaunt:2011xd,Ryskin:2011kk, Blok:2011bu,Diehl:2011yj,Manohar:2012jr, Manohar:2012pe,Ryskin:2012qx,Gaunt:2012dd,Blok:2013bpa, Diehl:2017kgu,Diehl:2018wfy}.  Following early experimental studies \cite{Akesson:1986iv,Alitti:1991rd}, a variety of DPS processes have been analysed at the Tevatron and the LHC, see \cite{Abe:1997xk,Abazov:2015nnn,Aaij:2016bqq,Aaboud:2018tiq,Sirunyan:2019zox} and references therein.  There is also a large body of phenomenological work on such processes.  An example is the recent study \cite{Fedkevych:2020cmd} of four-jet production, which also contains a wealth of further references.  For a detailed account of theoretical and experimental aspects of DPS, we refer to the monograph~\cite{Bartalini:2017jkk}.

Whilst DPS is often suppressed compared with single parton scattering (SPS), it is enhanced in certain kinematic regions, in particular when the products of the two hard-scattering processes have small transverse momenta \cite{Diehl:2011tt, Diehl:2011yj, Blok:2010ge, Blok:2011bu} or when they are far apart in rapidity \cite{Kom:2011bd, Kom:2011nu, Aaij:2016bqq}.  There are also channels in which SPS is suppressed compared with DPS by coupling constants.  A prominent example is the production of like-sign gauge boson pairs $W^+ W^+$ or $W^- W^-$ \cite{Kulesza:1999zh,Gaunt:2010pi,Sirunyan:2019zox,Ceccopieri:2017oqe, Cotogno:2018mfv,Cotogno:2020iio}, which also provides a background to the search for new physics with like-sign lepton pairs \cite{Khachatryan:2016kod, Sirunyan:2018yun}.

To compute DPS cross sections, one needs double parton distributions (DPDs), which generalise the concept of parton distribution functions (PDFs) to the case of two partons.  To determine DPDs from experimental data is considerably more complicated than the determination of PDFs, and our knowledge of DPDs remains quite fragmentary.  There is a large number of calculations of DPDs in quark models \cite{Chang:2012nw, Rinaldi:2013vpa, Broniowski:2013xba,  Rinaldi:2014ddl, Broniowski:2016trx, Kasemets:2016nio, Rinaldi:2016jvu, Rinaldi:2016mlk, Rinaldi:2018zng, Courtoy:2019cxq, Broniowski:2019rmu}, and information about the Mellin moments of DPDs can be obtained in lattice QCD \cite{Zimmermann:2019quf, Bali:2020mij}.  One may also try to construct an ansatz for DPDs that fulfils non-trivial theory constraints \cite{Gaunt:2009re,Golec-Biernat:2014bva,Golec-Biernat:2015aza,Diehl:2020xyg}.

A regime in which DPDs can actually be computed is when the transverse distance $y$ between the two partons becomes small.  The two observed partons are then produced from a single parton in a perturbative splitting process, so that the DPD is given by the convolution of a splitting kernel with a PDF.  Several phenomenological studies suggest that this splitting mechanism can have a substantial impart on DPS observables \cite{Blok:2013bpa, Blok:2015rka, Blok:2015afa, Ryskin:2012qx, Snigirev:2014eua, Golec-Biernat:2014nsa, Gaunt:2014rua, Diehl:2017kgu}.  The study \cite{Cabouat:2019gtm} finds visible effects even for like-sign $W$ pairs, where the splitting contribution starts at order $\alpha_s^2$ rather than $\alpha_s^{}$.

The perturbative splitting mechanism in DPS is intimately connected with a  double counting problem between DPS and SPS in physical cross sections \cite{Cacciari:2009dp, Diehl:2011yj}.  A systematic framework to solve this problem has been presented in \cite{Diehl:2017kgu}.  Broadly speaking, the perturbative splitting mechanism in DPS is cut off for small inter-parton distances $y$, where its contribution is better described as a higher-order corrections to SPS since the splitting occurs at the same scale as the hard collision.  Double counting is removed by a subtraction term, which can easily be constructed using the perturbative splitting formula for DPDs.

The quantum numbers of two partons in a hadron can be correlated in various ways \cite{Mekhfi:1983az,Diehl:2011yj,Manohar:2012jr}.  Among these, the correlation between their colour state is least well studied, and most investigations are restricted to the case in which the parton colour is uncorrelated.  It was realised early on that colour correlations between the partons are suppressed in DPS cross sections by Sudakov factors \cite{Mekhfi:1988kj}.  A simple numerical estimate in \cite{Manohar:2012jr} found this suppression to be quite strong for  hard scattering process at the electroweak scale, whereas it becomes marginal for scales around $10 \gev$.  Colour correlations in DPS could thus have a visible effect in the production of additional jets with moderate $p_T$ or of mesons with open or hidden charm.  In the context of perturbative splitting, colour correlations between the partons appear naturally, and how strongly they are suppressed by Sudakov logarithms was not addressed in \cite{Manohar:2012jr}.

To leading order (LO) in $\alpha_s$, the splitting kernels for DPDs have been computed in \cite{Diehl:2011yj}, and it was found that colour correlations between the two partons are generically large.  In many situations, LO results receive substantial corrections from higher orders.  This is typically true for hard-scattering cross sections in hadron-hadron collisions, which motivates an analysis of DPS at next-to-leading order (NLO).  Since such an analysis requires all perturbative ingredients to be available at NLO, we computed the  splitting kernels for DPDs at this order in \cite{Diehl:2019rdh}, restricting ourselves to unpolarised partons and uncorrelated parton colours.  The purpose of the present paper is to extend that computation to the colour correlated case.  Among other aspects, this will enable us to investigate the colour structure of small $x$ logarithms.  In a different setting, colour effects in small $x$ dynamics for double hard scattering were investigated in \cite{Bartels:2011qi}, using the BFKL formalism.

The computational techniques employed in \cite{Diehl:2019rdh} can readily be applied to the colour correlated case.  However, a major new element is the appearance of so-called rapidity divergences, which are closely related with the Sudakov logarithms mentioned above.  Such rapidity divergences are familiar from factorisation for processes with small measured transverse momenta, and it is interesting that they appear in DPS even if transverse momenta are integrated over.  A variety of regulators to handle these divergences can be found in the literature \cite{Becher:2010tm,Becher:2011dz, Collins:2011zzd, GarciaEchevarria:2011rb,Echevarria:2012js, Echevarria:2015byo,Echevarria:2016scs, Chiu:2011qc,Chiu:2012ir, Li:2016axz, Ebert:2018gsn}.  In this work, we use two of them, which provides a strong cross check of our results.  As one alternative, we take spacelike Wilson lines as proposed by Collins \cite{Collins:2011zzd}.  This regulator has been used in the factorisation proof for Drell-Yan production in \cite{Collins:2011zzd} and in our work on factorisation for DPS \cite{Diehl:2011yj,Diehl:2015bca, Diehl:2017kgu,Buffing:2017mqm,Diehl:2018wfy}.  It may be regarded as a descendant of the formalism in the original work of Collins, Soper, and Sterman \cite{Collins:1981uk,Collins:1984kg}, where an axial gauge was used.  Our second alternative is the so-called $\delta$ regulator of Echevarria et al.\ in the form described in \cite{Echevarria:2015byo,Echevarria:2016scs}.  This regulator has been used in several two-loop calculations for transverse-momentum dependent (TMD) factorisation \cite{Echevarria:2015byo,Echevarria:2016scs} and also for the two-loop soft factor in DPS \cite{Vladimirov:2016qkd}.  To the best of our knowledge, our work presents the first use of the Collins regulator in a two-loop setting.

This paper is organised as follows.  In \sect{\ref{sec:theory}}, we recall the basics of DPDs with non-trivial colour dependence and then work out the formalism necessary for treating ultraviolet and rapidity divergences in the calculation of perturbative splitting DPDs.  Details of the two-loop calculation are given in \sect{\ref{sec:two-loops}}.  Analytical results for the two-loop kernels are presented in \sect{\ref{sec:results}}, and some numerical illustrations of two-loop effects are shown in \sect{\ref{sec:numerics}}.  We conclude in  \sect{\ref{sec:conclude}}.  A number of formulae and technical explanations can be found in appendices~\ref{sec:projectors} to \ref{sec:distrib}.

\section{General theory}
\label{sec:theory}

We begin this section by recalling the basic properties of DPDs with a non-trivial dependence on the parton colour in the formalism set up in \cite{Diehl:2011yj, Buffing:2017mqm}.  We then describe the interplay between ultraviolet renormalisation and the treatment of rapidity divergences, and we show how these operations are to be carried out at the level of hard splitting kernels for DPDs at small inter-parton distance $y$.  The main result of this section is a set of formulae that specify how to compute the two-loop splitting kernels for renormalised DPDs from bare kernels and the soft factor for double parton scattering.


\subsection{Overview}

\begin{figure}[ht]
\begin{center}
\includegraphics[width=0.8\textwidth]{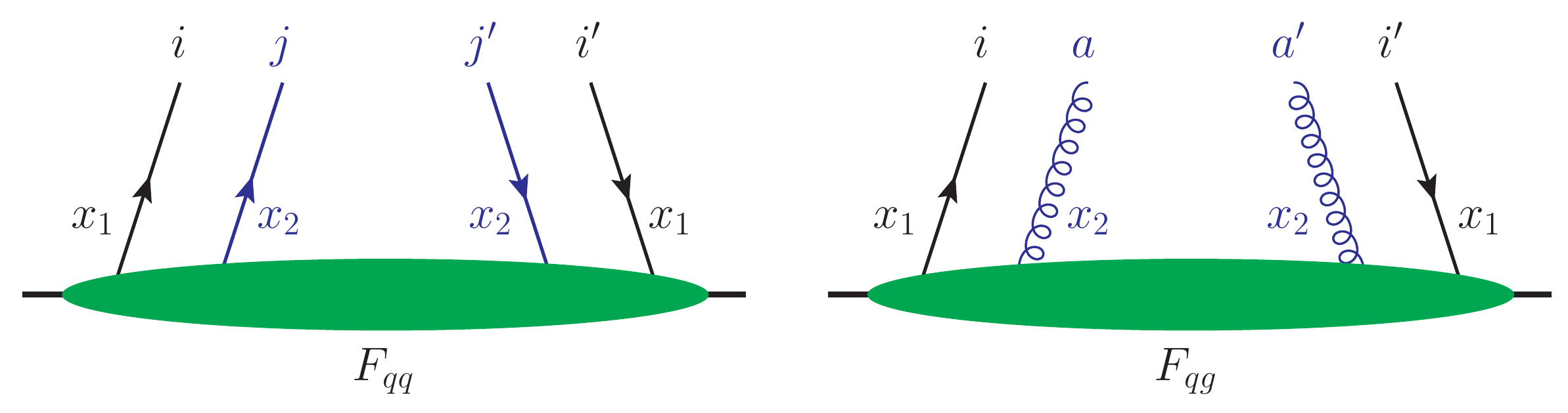}
\end{center}
\caption{\label{fig:distrib} Assignment of colour labels for a two-quark distribution (left) and a quark-gluon distribution (right).}
\end{figure}

We are interested in collinear, i.e.\ transverse-momentum integrated DPDs, which we denote by $\pr{\RR}{F}_{a_1 a_2}(x_1,x_2, y,\mu,\zeta_p)$.  These distributions depend on a number of variables.
\begin{itemize}
\item $x_1$ and $x_2$ are the longitudinal momentum fractions of partons $a_1$ and $a_2$ in a fast moving proton.  In general, the labels $a_1$ and $a_2$ also specify the polarisation of the two partons, but throughout this work we consider the case where both partons are unpolarised.  With proper adjustments, the results of the present section readily generalise to the polarised case.
\item $y$ is the spatial distance between the two partons in the plane transverse to the proton momentum.  We write $\tvec{y}$ for the distance vector and $y$ for its length $|\tvec{y}|$.
\item $\mu$ is the renormalisation scale.  As shown in \cite{Buffing:2017mqm}, it is possible to define separate scales $\mu_1$ and $\mu_2$ associated with the two partons.  Throughout the present work, we take these scales to be equal.  This is natural when computing the contribution to DPDs in which the two partons originate from a single hard splitting process.  From this, one can obtain distributions with different $\mu_1$ and $\mu_2$ by using the appropriate evolution equations.
\item $\zeta_p$ is a parameter associated with the regularisation of rapidity divergences.  It is boost invariant and has dimension mass squared.  Its precise definition depends on the rapidity regulator used and will be given below.  A corresponding dependence is familiar from transverse-momentum dependent parton distributions (TMDs).  It is remarkable that we encounter the same type of dependence for collinear distributions.
\end{itemize}
We now discuss the colour dependence.  The operator matrix elements associated with DPDs have four open colour indices, two for the partons $a_1$ and $a_2$ in the process amplitude and two for the same partons in the conjugate process amplitude.  Examples are shown in \fig{\ref{fig:distrib}}.  We organise the colour dependence by coupling the two indices for $a_1$ to an irreducible representation $R_1$ and those for $a_2$ to an irreducible representation $R_2$ such that one has an overall colour singlet.  The relevant representations depend on the parton type.  In all cases, both partons can be in the colour singlet ($R_1 = R_2 = 1$).  In addition, we have the following possibilities.
\begin{itemize}
\item For $F_{q q}$, the two quark pairs can each be coupled to an octet ($R_1 = R_2 =8$).
\item For $F_{g q}$, the two gluons can be coupled to an asymmetric or a symmetric octet ($R_1 = A$ or $R_1 = S$).  The two quarks are coupled to an octet ($R_2 = 8$) in both cases.  Analogous combinations exist for $F_{q g}$.
\item For $F_{g g}$, each of the two gluons can be coupled to an antisymmetric or a symmetric octet (in all four combinations), or to the 27 representation ($R_1 = R_2 = 27$).  Furthermore, one gluon can be coupled to a decuplet and the other to an anti-decuplet ($R_1 = 10$ and $R_2 = \overline{10}$ or vice versa).
\end{itemize}
We have the same representations when quarks are replaced by antiquarks.  Of course, the octet representations of SU(3) are readily generalised to the adjoint representations of SU($N$).  The coupling of two adjoint indices in SU($N$) with $N \neq 3$ involves the generalisations of $R = 10$, $\overline{10}$, $27$, and an additional irreducible representation \cite{Cvitanovic:2008zz,Keppeler:2012ih}.  Throughout this work, we give results for singlet and octet representations for general number of colours~$N$.  Results for the decuplet and antidecuplet are given for general $N$ or for $N=3$, whereas for $R = 27$ we always set $N = 3$.

We call the colour space basis for DPDs just described the ``$t$ channel basis''.  An alternative, which we call ``$s$ channel basis'', is to couple the indices of $a_1$ and $a_2$ to irreducible representations in the amplitude and in the conjugate amplitude \cite{Mekhfi:1983az}.  The conversion between the two colour bases can be found in \cite{Kasemets:2014yna}.  We note that the  combinations of $R_1$ and $R_2$ considered in \cite{Mekhfi:1983az,Diehl:2011yj,Kasemets:2014yna} and in the original version of \cite{Buffing:2017mqm} are incomplete; this mistake is corrected in the erratum of \cite{Buffing:2017mqm}.

We now recall some basic properties of the DPDs, which were derived in \cite{Buffing:2017mqm}.  Their $\zeta_p$ dependence is described by a Collins-Soper equation
\begin{align}
\label{CS-F}
\frac{\partial}{\partial \log\zeta_p} \,
   \log \prn{\RR}{F}_{a_1 a_2}(x_1,x_2, y,\mu,\zeta_p)
&= \prb{R_1}{J}(y, \mu) \big/ 2 \,,
\end{align}
which is analogous to the one derived by Collins and Soper for TMDs \cite{Collins:1981uk}.  We write $J$ instead of $K$ for the kernel relevant to collinear DPDs in order to avoid confusion with the TMD case.  The kernel $\prb{R}{J}$ depends only on the multiplicity of the representation $R$, so that $\prb{R_1}{J} = \prb{R_2}{J}$ for all possible colour combinations in a DPD.  Moreover, it is independent of the type of partons $a_1$ and~$a_2$.
In the singlet channel, one has the exact relation
\begin{align}
\pr{1}{J}(y, \mu) &= 0 \,,
\end{align}
so that colour singlet DPDs have no rapidity dependence.  Quite remarkably, the octet kernel $\pr{8}{J}(y, \mu)$ is equal to the Collins-Soper kernel for the rapidity evolution of gluon TMDs \cite{Vladimirov:2016qkd}.
The renormalisation scale dependence of $J$ is given by
\begin{align}
\frac{\partial}{\partial \log \mu^2}\, \prb{R}{J}(y, \mu)
  &= {}- \prn{R}{\gamma}_J(\mu)
\end{align}
with an anomalous dimension that has a perturbative expansion
\begin{align}
\label{J-expansion}
\prn{R}{\gamma}_J(\mu)
  &= \sum_{n=0}^\infty \as^{n + 1}(\mu)\, \prn{R}{\gamma}_J^{(n)} \,.
\end{align}
Here and in the following we write
\begin{align}
\label{as-def}
\as(\mu) &= \frac{\alpha_s(\mu)}{2\pi} \,.
\end{align}
The one-loop coefficients in \eqref{J-expansion} read
\begin{align}
\label{gammaJ-LO}
\prn{8}{\gamma}_J^{(0)} = \prn{A}{\gamma}_J^{(0)} = \prn{S}{\gamma}_J^{(0)}
   &= 2 C_A \,,
&
\prn{10}{\gamma}_J^{(0)} = \prn{\tenbar}{\gamma}_J^{(0)}
& \underset{N=3}{=} 12 \,,
&
\prn{27}{\gamma}_J^{(0)} & \underset{N=3}{=} 16
\end{align}
with $C_A = N$.

The scale dependence of the DPDs is given by the double DGLAP equations \cite{Buffing:2017mqm}
\begin{align}
\label{DGLAP-F}
\frac{\partial}{\partial \log\mu^2}\,
   \prn{\RR}{F}_{a_1 a_2}(x_1,x_2, y,\mu,\zeta_p)
&= \sum_{b_1, \Rp_1}
   \prn{R_1^{} \Rpbar_1}{P}_{a_1 b_1}(\mu, x_1^2\ms \zeta_p) \conv{1}
   \prn{\Rp_1 R_2^{}}{F}_{b_1 a_2}(y,\mu,\zeta_p)
\nonumber \\
& + \sum_{b_2, \Rp_2}
   \prn{R_2^{} \Rpbar_2}{P}_{a_2 b_2}(\mu, x_2^2\ms \zeta_p) \conv{2}
   \prn{R_1^{} \Rp_2}{F}_{a_1 b_2}(y,\mu,\zeta_p) \,,
\end{align}
where $\Rbar$ denotes the conjugate of the representation $R$.  Note that conjugation is only relevant for the decuplet and anti-decuplet, whilst $\Rbar = R$ for the singlet, all octets, and $R=27$.  The convolutions are defined as
\begin{align}
\label{conv-1-def}
\Bigl[ P \conv{1} F \ms\Bigr] (x_1, x_2)
&= \int\limits_{x_1}^{1} \frac{dz}{z}\; P(z)\,
  F\biggl( \frac{x_1}{z}, x_2 \biggr) \,,
\nonumber \\
\Bigl[ P \conv{2} F \ms\Bigr] (x_1, x_2)
&= \int\limits_{x_2}^{1} \frac{dz}{z}\; P(z)\,
  F\biggl( x_1, \frac{x_2}{z} \biggr) \,.
\end{align}
The effective integration range in these convolutions is smaller than indicated, because DPDs $F(x_1, x_2)$ vanish for $x_1 + x_1 > 1$ as a consequence of momentum conservation.  The same support property holds for DPD splitting kernels and renormalisation factors.

Note that the factors $x_1^2$ and $x_2^2$ that multiply $\zeta_p$ in the arguments of the evolution kernels in \eqref{DGLAP-F} are fixed by the arguments of the DPD on the l.h.s., i.e., they are not part of the convolution integrals on the r.h.s.  The reason for this rescaling of $\zeta_p$ is explained in \sect{\ref{sec:renormalisation}}.

The rapidity dependence of the DGLAP kernels is given by
\begin{align}
\label{DGLAP-zeta}
\prb{R \Rp}{P}_{ab}(z, \mu, \zeta)
&= \frac{\prn{R}{\gamma}_J(\mu)}{4}\,
   \delta_{R \Rpbar}\, \delta_{a b} \, \delta(1-z) \log \frac{\mu^2}{\zeta}
   + \prb{R \Rp}{\hat{P}}_{a b}(z, \mu) \,,
\end{align}
where we abbreviate $\prb{R \Rp}{\hat{P}}(z, \mu) = \prb{R \Rp}{P}(z, \mu, \mu^2)$.   The evolution kernels have a perturbative expansion
\begin{align}
\label{DGLAP-expansion}
\prb{R \Rp}{P}_{ab}(z, \mu, \zeta)
&= \sum_{n=0}^\infty \as^{n + 1}(\mu)\;
   \prb{R \Rp}{P}_{ab}^{(n)}(z, \mu^2/\zeta) \,,
\end{align}
and the colour singlet kernels $\pr{11}{P}(z, \mu)$ are the same as the DGLAP kernels for ordinary PDFs.

Notice that Collins-Soper and DGLAP evolution of collinear DPDs is diagonal in the $t$ channel colour basis, as seen in \eqref{CS-F} and \eqref{DGLAP-F}.  This does not hold in the $s$ channel basis, where different representations mix under evolution.


\paragraph{Perturbative splitting.}
In the limit of small $y$, the DPDs are given by the perturbative splitting form
\begin{align}
\label{split-master}
& \pr{\RR}{F}_{a_1 a_2}(x_1,x_2,y; \mu,\zeta_p)
\nonumber \\
&\qquad\qquad
   = \frac{1}{\pi\ms y^2}\, \sum_{a_0}\,
  \int\limits_{x_1+x_2}^1 \!\! \frac{d z}{z^2} \;
  \prn{\RR}{V}_{a_1 a_2, a_0}\biggl( \frac{x_1}{z}, \frac{x_2}{z}, \as(\mu),
     L, \log \frac{\mu^2}{x_1 x_2\ms \zeta_p} \biggr) \,
  f_{a_0}(z,\mu) \,,
\end{align}
where we abbreviate
\begin{align}
\label{L-b0-def}
L &= \log\frac{y^2 \mu^2}{b_0^2} \,,
&
b_0 &= 2 e^{-\gamma} \,,
\end{align}
with $\gamma$ being the Euler-Mascheroni constant.  Corrections to \eqref{split-master} are of order $y^2 \Lambda^2$, where $\Lambda$ is a hadronic scale.  They arise from twist-four contributions in the operator product expansion, as explained in \sect{3.3} of \cite{Diehl:2017kgu}.

We define a convolution product between a two-variable and a one-variable function as
\begin{align}
\label{conv-12-def}
\Bigl[ V \conv{12} f \ms\Bigr] (x_1,x_2)
&= \int\limits_{x}^{1} \frac{dz}{z^2}\;
   V\biggl( \frac{x_1}{z}, \frac{x_2}{z} \biggr) \, f(z)
= \frac{1}{x} \int\limits_{x}^{1} dz\;
   V( u z, \ub z ) \, f\biggl( \frac{x}{z} \biggr)
\end{align}
with abbreviations
\begin{align}
\label{x-u-def}
x &= x_1 + x_2 \,,
&
u &= \frac{x_1}{x_1+x_2} \,,
&
\ub &= 1 - u \,.
\end{align}
When denoting the kernel in the rightmost expression of \eqref{conv-12-def} by $V(z_1, z_2)$, we also have the relations
\begin{align}
\label{z-u-def}
z &= z_1 + z_2 \,,
&
u &= \frac{z_1}{z_1 + z_2} \,,
\end{align}
so that $u$ can be expressed either in terms of the momentum fractions of the full convolution integral or the momentum fractions of its integrand.
Note that the factors $x_1 x_2$ multiplying $\zeta_p$ on the r.h.s.\ of \eqref{split-master} are fixed by the arguments of the DPD on the l.h.s., in analogy to our earlier discussion for evolution kernels.

Convolutions w.r.t.\ different variables are associative, i.e.\ we have
\begin{align}
[ A \conv{1} D ] \conv{12} B
  &= A \conv{1} [ D \conv{12} B ] \,,
&
[ A \conv{1} B \conv{2} D ] \conv{12} C
  &= A \conv{1} B \conv{2} [ D \conv{12} C ] \,,
\end{align}
where $A$, $B$ and $C$ depend on one momentum fraction, whilst $D$ depends on two momentum fractions and a rapidity parameter.  We can hence write these convolutions without brackets.  Here we have defined the combined convolution
\begin{align}
\label{conv-comb-def}
A \conv{1} B \conv{2} D &= A \conv{1} [ B \conv{2} D ] \,.
\end{align}
We also have
\begin{align}
[D \conv{12} A ] \conv{12} B
  &= D \conv{12} [ A \otimes B ]  \,,
\end{align}
where $A \otimes B$ is the usual Mellin convolution for functions of a single momentum fraction.  We note in passing that for rescaled functions, the convolution integrals take the same form as for the usual Mellin convolution,
\begin{align}
x_1\ms x_2\, \Bigl[ V \conv{12} f \Bigr](x_1, x_2)
&= \int\limits_{x}^{1} \frac{dz}{z}\;
   \widetilde{V}\biggl( \frac{x_1}{z}, \frac{x_2}{z} \biggr) \, \tilde{f}(z)
 = \int\limits_{x}^{1} \frac{dz}{z}\;
   \widetilde{V}( u z, \ub z ) \, \tilde{f}\biggl( \frac{x}{z} \biggr) \,,
\end{align}
where $\widetilde{V}(z_1, z_2) = z_1\ms z_2\, V(z_1, z_2)$ and $\tilde{f}(z) = z f(z)$.

We extend our notation for convolutions to include appropriate summation over indices denoting the parton type (quarks, antiquarks, gluons).  For functions $A_{ab}(x), B_{ab}(x)$ and $f_a(x)$ of a single momentum fraction, we write as usual
\begin{align}
\label{parton-indices-1}
[A \otimes B]_{ac} &= \sum_b A_{a b} \otimes B_{b c} \,,
&
[A \otimes f]_{a} &= \sum_b A_{a b} \otimes f_b \,.
\end{align}
We do \emph{not} use the summation convention for parton indices, i.e.\ sums are indicated explicitly when indices are given.  For functions $V_{a_1 a_2, a_0}(x_1,x_2)$ and $F_{a_1 a_2}(x_1,x_2)$ of two momentum fractions, we define
\begin{align}
\label{parton-indices-2}
[A \conv{1} V]_{a_1 a_2, a_0}
   &= \sum_b A_{a_1 b} \conv{1} V_{b\ms a_2, a_0} \,,
&
[A \conv{1} F]_{a_1 a_2}
   &= \sum_b A_{a_1 b} \conv{1} F_{b\ms a_2} \,,
\nonumber \\
[A \conv{2} V]_{a_1 a_2, a_0}
   &= \sum_b A_{a_2 b} \conv{2} V_{a_1 b, a_0} \,,
&
[A \conv{2} F]_{a_1 a_2}
   &= \sum_b A_{a_2 b} \conv{2} F_{a_1 b} \,,
\nonumber \\
[V \conv{12} A]_{a_1 a_2, a_0}
  &= \sum_b V_{a_1 a_2, b} \conv{12} A_{b\ms a_0} \,,
&
[V \conv{12} f]_{a_1 a_2}
  &= \sum_b V_{a_1 a_2, b} \conv{12} f_{b} \,.
\end{align}
For the combination of convolutions in the first and second momentum fraction, we then have
\begin{align}
\label{parton-indices-3}
[A \conv{1} B \conv{2} V]_{a_1 a_2, a_0}
&= \sum_{b_1, b_2} A_{a_1 b_1}
   \conv{1} B_{a_2 b_2} \conv{2} V_{b_1 b_2, a_0} \,.
\end{align}


\subsection{Bare distributions and soft factors}

As specified in \cite{Buffing:2017mqm}, DPDs are constructed from hadronic matrix elements, which we call ``unsubtracted'' following the nomenclature of Collins \cite{Collins:2011zzd}, and from the soft factor for DPS.  In analogy to the modern definition of TMDs in \cite{Collins:2011zzd}, this construction implements the cancellation of rapidity divergences.  It also avoids the explicit appearance of the soft factor in the factorisation formula for the cross section.  This is important for minimising the amount of non-perturbative functions, given that the cross section includes the region of large $y$, where the soft factor cannot be computed in perturbation theory.

To treat rapidity divergences, we use two schemes in our work.  One is the $\delta$ regulator in the form described in \cite{Echevarria:2015byo,Echevarria:2016scs}, and the other is the regulator of Collins using spacelike Wilson lines \cite{Collins:2011zzd}.  With compatible definitions of the rapidity parameter $\zeta_p$ in the two schemes, we must obtain identical results for the final two-loop splitting kernels, which constitutes a strong cross check of our procedure and computation.

We note that several other schemes have been used in the literature, such as the analytic regulator of \cite{Becher:2010tm, Becher:2011dz}, the CMU $\eta$ regulator of \cite{Chiu:2011qc, Chiu:2012ir}, the exponential regulator of \cite{Li:2016axz}, and the ``pure rapidity regulator'' of \cite{Ebert:2018gsn}.
There are also two earlier variants of the $\delta$ regulator, described in \cite{GarciaEchevarria:2011rb} and \cite{Echevarria:2012js}, and compared with
the Collins regulator in \cite{Collins:2012uy} and \cite{Echevarria:2012js},
respectively.  A detailed discussion of TMDs in the above schemes is given in \app{B} of \cite{Ebert:2019okf}.

The construction described in the following is laid out in \cite{Buffing:2017mqm} for the Collins regulator and adapted to the $\delta$ regulator as described in app.~B of that reference.  We regulate ultra\-violet divergences by working in $4 - 2\epsilon$ space-time dimensions.
To start, we define bare (i.e.\ unrenormalised) and unsubtracted distributions with open colour indices,
\begin{align}
\label{dpd-def}
F^{r_1^{}\bs r_1' \, r_2^{} r_2'}_{\bus,\ms a_{1} a_{2}}(x_1,x_2,\tvec{y})
 & = (x_1\ms p^+)^{-n_1}\, (x_2\ms p^+)^{-n_2}\; 2 p^+ \int d y^-\,
        \frac{d z^-_1}{2\pi}\, \frac{d z^-_2}{2\pi}\;
          e^{i\ms ( x_1^{} z_1^- + x_2^{} z_2^-)\ms p_{}^+}
\nonumber \\[0.2em]
 & \quad \times
    \langle\ms p \ms|\,
      \mathcal{O}^{r_1^{}\bs r_1'}_{a_1}(y,z_1)\,
      \mathcal{O}^{r_2^{} r_2'}_{a_2}(0,z_2)
    \,|\ms p \ms\rangle \bigl|_{y^+ = 0} \;,
\end{align}
where $n_i = 0$ for quarks and antiquarks, and $n_i = 1$ for gluons.  We use light-cone coordinates $v^\pm = (v^0 \pm v^3) /\sqrt{2}$ for any four-vector $v^\mu$ and write its transverse part in boldface, $\tvec{v} = (v^1, v^2)$.  It is understood that the transverse proton momentum $\tvec{p}$ is zero and that the proton polarisation is averaged over.  The colour indices $r_i^{}$ and $r_i'$ ($i=1,2$) are in the fundamental or the adjoint representation as appropriate.  For unpolarised quarks and gluons, the twist-two operators in \eqref{dpd-def} read
\begin{align}
\label{op-defs}
\mathcal{O}^{i i'}_{q}(y, z) &=
   \bar{q}^{}_{j'}(\xi_-) \,
      \bigl[\ms W^\dagger(\xi_-, v_L) \ms\bigr]{}^{}_{j' i'} \;
   \frac{\gamma^+}{2}\,
   \bigl[\ms W^{}(\xi_+, v_L) \ms\bigr]{}^{}_{i j}\, q^{}_{j}(\xi_+) \,,
\nonumber \\
\mathcal{O}_{g}^{a a'}(y, z) &=
   \bigl[\ms G^{+ k}(\xi_-) \ms\bigr]^{b'}\,
      \bigl[\ms W^\dagger(\xi_-, v_L) \ms\bigr]^{b' a'} \,
   \bigl[\ms W(\xi_+, v_L) \ms\bigr]^{a b}\,
      \bigl[\ms G^{+ k}(\xi_+) \ms\bigr]^{b} \,,
\nonumber \\[0.5em]
& \quad\, \text{with~~} \xi_\pm = y \pm z/2 \,,
\quad
z^+ = 0 \,,
\quad
\tvec{z} = \tvec{0} \,.
\end{align}
Repeated colour indices are summed over.  $W(\xi, v)$ is a Wilson line along the path parameterised by $\xi - s v$ with $s$ going from $0$ to $\infty$.  Details on the direction $v$ of the path are given below.  The twist-two operator for antiquarks is given by $\mathcal{O}^{i i'}_{\bar{q}}(y, z) = {}- \mathcal{O}^{i i'}_q(y, - z)$.  All operators (including Wilson lines) are constructed from bare fields.
DPDs for definite colour representations of the two partons are now defined by
\begin{align}
\label{F-R-projection}
\prn{\RR}{F}_{\bus}^{} &= M_{\RR}^{r_1^{}\bs r_1' \, r_2^{} r_2'} \;
                     F_{\bus}^{r_1^{}\bs r_1' \, r_2^{} r_2'}
\end{align}
with the colour space matrices $M_{\RR}$ given in \app{\ref{sec:projectors}}.

In the splitting formula for bare DPDs, we also need the definition of bare PDFs, which reads
\begin{align}
\label{pdf-def}
f_{B \bs, a}(x) &= (x\ms p^+)^{-n}  \int \frac{d z^-}{2\pi}\,
    e^{i\ms x^{} z^- p_{}^+} \langle\ms p \ms|\,
    \mathcal{O}^{r r}_{a}(0,z) \,|\ms p \ms\rangle \,.
\end{align}
Here the operators \eqref{op-defs} appear in the colour singlet representation, i.e.\ with their colour indices set equal and summed over.

Bare unsubtracted DPDs in momentum space are defined by a Fourier transform w.r.t.\ the inter-parton distance,
\begin{align}
\label{dpd-mom-def}
F_{\bus, a_{1} a_{2}}(x_1,x_2,\tvec{\Delta})
&= \int d^{2-2\epsilon} \tvec{y}\;
  e^{i \tvec{y} \tvec{\Delta}}\, F_{B, a_{1} a_{2}}(x_1,x_2,\tvec{y}) \,.
\end{align}
These are the primary quantities we will compute at two loops in momentum space, as described in \sect{\ref{sec:two-loops}}.

The bare soft factor ${S}_{B, a_1 a_2}$ for DPS is defined as the vacuum matrix element of  Wilson line operators along two different directions $v_L$ and $v_R$, as specified in \sect{3.2} of \cite{Buffing:2017mqm}.  It is a matrix in colour space with $8$ indices that can be turned into a matrix $\prn{R_1^{} R_2^{}, \Rp_1 \Rp_2}{S}_{a_1 a_2}$ in representation space by multiplication with appropriate projectors.  A non-trivial result for collinear DPS factorisation (but not for its TMD counterpart) is that this matrix is independent of the parton types $a_1$ and $a_2$, that it has nonzero entries only for $R_1^{} = \Rpbar_{1}$ and $R_2^{} = \Rpbar_2$, and that these entries depend only on the common multiplicity of the four representations.  One can thus denote them by $\pr{R}{S}$, with $R$ being any one of the four representations.

The soft factor depends on a variable $\ell$ related with the rapidity dependence, and it satisfies a Collins-Soper equation
\begin{align}
\label{SB-CS}
\frac{\partial}{\partial \ell} \, \log \pr{R}{S}_B(y, \ell)
  &= \prb{R}{J}_B(y)
\end{align}
for $|\ell| \gg 1$.   The kernel $J_B(y)$ is the unrenormalised analogue of $J(y, \mu)$ in \eqref{CS-F}.
One easily derives a composition law
\begin{align}
\label{S-decompose}
\pr{R}{S}_B(y, \ell_R + \ell_L) &=
  \sqrt{\pr{R}{S}_B(y, 2\ms \ell_R)} \,
  \sqrt{\pr{R}{S}_B(y, 2\ms \ell_L)}
\end{align}
for the region where $|\ell_R|, |\ell_L|$ and $|\ell_R + \ell_L|$ are large, using the differential equation \eqref{SB-CS} and the initial condition $\ell_R = \ell_L$.  This composition law is used to absorb part of the soft factor into the DPD for the right-moving proton and the other part into the DPD for the left-moving one.

We now specify $\ell_R$ and $\ell_L$, as well as the rapidity parameters $\zeta$ and $\bar{\zeta}$, which are respectively associated with the right- and the left-moving proton.  We write $x_{1,2}$ ($\bar{x}_{1,2}$) for the parton momentum fractions of the DPD for the right (left) moving proton in the collision, with the large momentum component of the proton being $p^+$ ($\bar{p}^-$).
\begin{itemize}
\item In the scheme of Collins, one takes Wilson lines with finite rapidity, where the rapidity of a Wilson line along $v$ is defined as $Y = \half \log |v^+ / v^-|$.  The variables $\ell_R$ and $\ell_L$ are given by
\begin{align}
\ell_R + \ell_L &= Y_R - Y_L \,,
&
\ell_R &= Y_R - Y_C \,,
&
\ell_L &= Y_C - Y_L \,,
\end{align}
where $Y_R$ and $Y_L$ are the rapidities of the Wilson lines in the soft factor and unsubtracted DPDs, and $Y_C$ is an additional rapidity introduced for defining subtracted DPDs.  For an explanation of the rationale behind this, we refer to \sect{10.11.1} in \cite{Collins:2011zzd} and \sect{2.1} in \cite{Gaunt:2018eix}.
Removing the regulator corresponds to taking the lightlike limit of Wilson lines, i.e.\ $Y_R \to \infty$ and $Y_L \to - \infty$, whilst keeping $Y_C$ fixed.

Rapidity parameters for DPS are defined as
\begin{align}
	\label{zeta-def}
\zeta_p &= 2 (p^+)^2\, e^{-2 Y_C} \,,
&
\zeta_{\bar{p}} &= 2 (\bar{p}^-)^2\, e^{2 Y_C}
\end{align}
and satisfy the normalisation condition
\begin{align}
   \label{zeta-product}
\zeta_p\ms \zeta_{\bar{p}} &= (2 p^+ \bar{p}^{\,-})^2 = s^2 \,,
\end{align}
where $s$ is the squared centre-of-mass energy of the proton-proton collision. Note that the definition \eqref{zeta-def} refers to the momenta of the colliding protons.  This differs from the convention in the modern TMD literature, where $\zeta$ refers to the momentum of the extracted parton \cite{Collins:2011zzd}.  Such a definition would be awkward in the case of DPDs, where the parton momenta often appear in convolution integrals.

Denoting by $v_R$ ($v_L$) the direction of the Wilson line with positive (negative) rapidity, we get
\begin{align}
\ell_R &= \frac{1}{2} \log\frac{v_R^+}{|v_R^-|}
   + \frac{1}{2} \log\frac{2 (\bar{p}^-)^2}{\zeta_{\bar{p}}} \,,
&
\ell_L &= \frac{1}{2} \log\frac{2 (p^+)^2}{\zeta_p}
   - \frac{1}{2} \log\frac{|v_L^+|}{v_L^-} \,.
\end{align}
It is understood that $v_R^+$ and $v_L^-$ are positive, whilst $v_L^+$ and $v_R^-$ are negative, such that the Wilson lines are spacelike.  Concentrating now on the right-moving proton, we define the boost invariant quantity
\begin{align}
\label{rho-def-collins}
\rho &= \frac{2 (x_1 p^+)\ms (x_2 \ms p^+)\, v_L^-}{|v_L^+|} \,,
\end{align}
so that
\begin{align}
\label{l-var-collins}
2 \ell_L &= \log\frac{\rho}{x_1 x_2\ms \zeta_p} \,.
\end{align}

\item With the $\delta$ regulator, Wilson lines are taken along lightlike paths.  Rapidity divergences are regulated by exponential damping factors $\exp(\delta^- z^+)$ and $\exp(\delta^+ z^-)$ in the Wilson lines pointing in the plus or minus directions, where $z^+$ and $z^-$ go from $0$ to $-\infty$.  The regulating parameters $\delta^+$ and $\delta^-$ have dimension of mass and transform like plus or minus components under boosts along the collision axis.  Both parameters are sent to zero when the regulator is removed.

The matrix element defining the bare soft factor depends on the product $\delta^+ \delta^-$ because of boost invariance, and for dimensional reasons the logarithms associated with rapidity divergences depend on $y^2 \, \delta^+ \delta^-$.  We write
\begin{align}
\label{l-variables-def}
\ell_R + \ell_L &= \log\frac{b_0^2}{2 \delta^+ \delta^-\ms y^2} \,,
\nonumber \\
\ell_R &= \log \frac{\bar{p}^-}{\delta^-}
   + \frac{1}{2} \log \frac{b_0^2}{y^2 \ms \zeta_{\bar{p}}} \,,
&
\ell_L &= \log \frac{p^+}{\delta^+}
   + \frac{1}{2} \log \frac{b_0^2}{y^2 \ms \zeta_p} \,,
\end{align}
where $\zeta$ and $\bar{\zeta}$ again satisfy \eqref{zeta-product}.  Introducing the boost invariant variable
\begin{align}
\label{rho-def-delta}
\rho &= \frac{(x_1 p^+)\ms (x_2 \ms p^+)}{(\delta^+)^2}
\end{align}
we have
\begin{align}
\label{l-var-delta}
2 \ell_L &= \log \rho + \log\frac{\mu^2}{x_1 x_2\ms \zeta_p} - L \,.
\end{align}
Notice that $\rho$ is dimensionless here, whereas it has dimension of mass squared for the Collins regulator.
\end{itemize}
In both schemes, the bare unsubtracted DPD $F_{\bus}$ for a right-moving proton depends on the plus-momentum fractions $x_1, x_2$, on $y$, and on the rapidity regulator variable for Wilson lines along $v_L$.  Because of boost invariance, the dependence on the regulator variable is via the combination $\rho$.  For the Collins regulator, there are two boost invariant variables $p^+ \ms v_L^-$ and $v_L^+ \ms v_L^-$, which must appear in the combination \eqref{rho-def-collins} because the Wilson lines are invariant under a rescaling of $v_L$.

The subtracted (but still unrenormalised) DPD $F_B$ is defined as \cite{Buffing:2017mqm}
\begin{align}
\label{DPD-subt-def}
\prn{\RR}{F}_{B}(x_1,x_2,y, \zeta_p) &= \lim\limits_{\rho \to \infty}
  \frac{\prn{\RR}{F}_{\bus}(x_1,x_2,y, \rho)}{
  \sqrt{\rule{0pt}{1.95ex} \pr{R_1}{S}_B(y, 2\ms \ell_L)}} \,,
\end{align}
where $\ell_L$ depends implicitly on $\rho$ and $\zeta_p$ as specified in \eqref{l-var-collins} or \eqref{l-var-delta}.  An analogous definition is made for the left moving proton.  We recall that bare quantities are independent of $\mu$ if expressed in terms of the bare coupling $\alpha_0$.  It is understood that all factors in \eqref{DPD-subt-def} are to be taken in $4 - 2\epsilon$ dimensions, until renormalisation is performed at a later stage.


\subsection{Bare distributions in the short-distance limit}

In the limit of large $\Delta$ or small $y$, bare unsubtracted DPDs can be expressed as a convolution of splitting kernels with bare PDFs.  We write
\begin{align}
\label{WV-defs}
\prn{\RR}{F}_{\bus}(\Delta, \rho)
   &= \prn{\RR}{W}_{\bus}(\Delta, \rho) \conv{12} f_B \,,
\nonumber \\
\prn{\RR}{F}_{\bus}(y, \rho)
   &= \frac{\Gamma(1-\epsilon)}{(\pi y^2)^{1-\epsilon}} \;
      \prn{\RR}{V}_{\bus}(y, \rho) \conv{12} f_B
\end{align}
with kernels being related by
\begin{align}
\label{FT-WV}
\frac{\Gamma(1-\epsilon)}{(\pi y^2)^{1-\epsilon}} \;
      \prn{\RR}{V}_{\bus}(y, \rho)
&= \int \frac{d^{2-2\epsilon} \tvec{\Delta}}{(2 \pi)^{2-2\epsilon}} \,
    e^{-i \tvec{\Delta} \tvec{y}} \,\, \prn{\RR}{W}_{\bus}(\Delta, \rho)
\end{align}
in accordance with \eqref{dpd-mom-def}.  The kernels have perturbative expansions
\begin{align}
\label{Vus-alpha}
\prn{\RR}{W}_{\bus} &= \sum_{n=1}^\infty
    \biggl( \frac{\alpha_0}{2\pi} \biggr)^n\, \prn{\RR}{W}_{0}^{(n)}
  = \sum_{n=1}^\infty a_s^n\, \biggl( \frac{\mu}{\Delta} \biggr)^{2\epsilon n}
     Z_\alpha^n\; \prn{\RR}{W}_{\bus}^{(n)} \,,
\nonumber \\
\prn{\RR}{V}_{\bus} &= \sum_{n=1}^\infty
    \biggl( \frac{\alpha_0}{2\pi} \biggr)^n\, \prn{\RR}{V}_{0}^{(n)}
  = \sum_{n=1}^\infty a_s^n\, \biggl( \frac{y \mu}{b_0} \biggr)^{2\epsilon n}
     Z_\alpha^n\; \prn{\RR}{V}_{\bus}^{(n)}
\end{align}
with
\begin{align}
\label{WVB-def}
\prn{\RR}{W}_{\bus}^{(n)} &=
   \Delta^{2\epsilon n}\, S_{\epsilon}^{-n}\; \prn{\RR}{W}_0^{(n)} \,,
&
\prn{\RR}{V}_{\bus}^{(n)} &=
   (b_0 / y)^{2\epsilon n}\, S_{\epsilon}^{-n}\; \prn{\RR}{V}_0^{(n)} \,.
\end{align}
The bare and renormalised couplings are related by $\alpha_0 = \mu^{2\epsilon}\, \alpha_s\, Z_\alpha /S_\epsilon$, where $Z_\alpha$ is the renormalisation constant for the running coupling.  We give our results for two definitions of the \msbar scheme parameter,
\begin{align}
   \label{MSbar-param}
S_\epsilon &=
\begin{cases}
(4\pi)^{\epsilon} \big/ \Gamma(1-\epsilon)
   & \text{(Collins in \cite{Collins:2011zzd})}
\\
(4\pi e^{-\gamma})^{\epsilon}
   & \text{(standard choice)}
\end{cases}
\end{align}

The construction \eqref{DPD-subt-def} is multiplicative in $y$ space and hence involves a convolution in $\Delta$ space.  We therefore exclusively work in $y$ space when constructing subtracted DPDs.  For the associated splitting kernel we then have
\begin{align}
\label{FB-split}
\prn{\RR}{F}_{B}(y, \zeta_p)
&= \frac{\Gamma(1-\epsilon)}{(\pi y^2)^{1-\epsilon}} \;
  \prn{\RR}{V}_{B}(y, x_1 x_2\ms \zeta_p) \conv{12} f_B \,,
\nonumber \\
\prn{\RR}{V}_{B}(y, x_1 x_2\ms \zeta_p) &= \lim\limits_{\rho \to \infty}
  \frac{\prn{\RR}{V}_{\bus}(y, \rho)}{
  \sqrt{\rule{0pt}{1.95ex} \pr{R_1}{S}_B(y, 2\ms \ell_L)}} \,.
\end{align}
Notice that $V_{B}$ depends on the combination $x_1 x_2\ms \zeta_p$, which according to \eqref{zeta-def} and \eqref{l-variables-def} depends the plus momenta $x_1 p^+$ and $x_2 \ms p^+$ of the partons in the DPD, as does $\rho$ defined in \eqref{rho-def-collins} or \eqref{rho-def-delta}.  In fact, the kernels $V_{B}$ and $V_{\bus}$ cannot depend on the proton momentum but only on quantities involving parton kinematics.  We emphasise that $x_1 x_2\ms \zeta_p$ and $\rho$ remain fixed when the kernels are convoluted with a PDF.

We expand the bare soft factor ${S}_B(y, 2\ms \ell_L)$ and the bare splitting kernel $V_B(y,\zeta)$ in analogy to $V_{\bus}(y,\rho)$ in \eqref{Vus-alpha},
\begin{align}
\label{VB-alpha}
\pr{R}{S}_{B} &= \, 1 \, + \,
   \sum_{n=1}^\infty a_s^n\, \biggl( \frac{y \mu}{b_0} \biggr)^{2\epsilon n}
     Z_\alpha^n\; \pr{R}{S}_{B}^{(n)} \,,
&
\prn{\RR}{V}_{B} &=
   \sum_{n=1}^\infty a_s^n\, \biggl( \frac{y \mu}{b_0} \biggr)^{2\epsilon n}
     Z_\alpha^n\; \prn{\RR}{V}_{B}^{(n)} \,,
\end{align}
and then obtain
\begin{align}
\label{bare-V}
\prn{\RR}{V}_{B}^{(1)} &= \prn{\RR}{V}_{\bus}^{(1)} \,,
\nonumber \\[0.1em]
\prn{\RR}{V}_{B}^{(2)} &= \lim\limits_{\rho \to \infty} \, \biggl\{
  \prn{\RR}{V}_{\bus}^{(2)}(\rho)
  - \frac{1}{2}\, \pr{R_1}{S}_B^{(1)}(2 \ell_L)\,
    \prn{\RR}{V}_{B}^{(1)} \biggr\} \,.
\end{align}
The $\rho$ dependence of $V_{\bus}$ first appears at NLO.  From the Collins-Soper equation \eqref{SB-CS} and the relations \eqref{l-var-collins} and \eqref{l-var-delta}, it follows that $S_B^{(n)}(2 \ell_L)$ is a polynomial in $\log \rho$ of degree $n$.  The cancellation of rapidity divergences in \eqref{bare-V} then requires $V_{\bus}^{(2)}$ to be a linear function of $\log \rho$.

The relation between splitting kernels in $\Delta$ and $y$ space is different for the two regulator schemes and will be discussed next.

\paragraph{$\delta$ regulator:}
The kernels ${W}_{\bus}^{(n)}$ and ${V}_{\bus}^{(n)}$ depend on $\delta^+$ via the dimensionless ratio $\rho$.  Furthermore, they depend on $\epsilon$ and on the parton momentum fractions $z_1$ and $z_2$.  The kernel ${W}_{\bus}^{(n)}$ does \emph{not} depend on $\Delta$, because it is dimensionless and there is no other quantity with mass dimension to form a boost invariant and dimensionless ratio.
With the Fourier integrals given in \app{\ref{sec:FT}}, one can perform the Fourier integral in \eqref{FT-WV} and obtains
\begin{align}
\label{VW-relation}
\prn{\RR}{V}_{\bus}^{(n)}
  &= n \epsilon\ms T_{\epsilon, n}^{}  \prn{\RR}{W}_{\bus}^{(n)}
\end{align}
with
\begin{align}
\label{T-def}
T_{\epsilon,n}
 &= \frac{\Gamma(1-\epsilon-\epsilon n)}{\Gamma(1+\epsilon n)\,
    \Gamma(1-\epsilon)}\, e^{- 2 n \gamma \epsilon}
  = 1 + n \ms \zeta_2 \ms \epsilon^2 + \mathcal{O}(\epsilon^3) \,,
\end{align}
where $\zeta_2 = \pi^2 /6$.

As noted below \eqref{bare-V}, $V^{(2)}_{\bus}$ is a linear function of $\log \rho$, which we write as
\begin{align}
\label{V-log-delta}
\prn{\RR}{V}^{(2)}_{\bus}(\rho) &=
  \prn{\RR}{V}^{(2,0)}_{\bus}
  + \log \rho \; \prn{\RR}{V}^{(2,1)}_{\bus}
\end{align}
with coefficients ${V}^{(2,m)}_{\bus}$ depending only on $\epsilon$, $z_1$, and $z_2$.  By virtue of \eqref{VW-relation}, an analogous decomposition holds for ${W}_{\bus}^{(n)}$.

\paragraph{Collins regulator:}
The coefficients ${W}_{\bus}^{(n)}$ in the perturbative expansion \eqref{Vus-alpha} depend on the rapidity regulator via the dimensionless and boost invariant ratio $\rho /\Delta^2$.
Since $V_{\bus}^{(2)}$ is a linear function of $\log \rho$, the same must be true for $W_{\bus}^{(2)}$.  We write the latter as
\begin{align}
\prn{\RR}{W}_{\bus}^{(2)}(\rho /\Delta^2) &=
  \prn{\RR}{W}_{\bus}^{(2,0)}
  + \log\frac{\rho}{\Delta^2}\, \prn{\RR}{W}_{\bus}^{(2,1)} \,.
\end{align}
Fourier transform to $y$ space then gives
\begin{align}
\label{V-log-collins}
\prn{\RR}{V}_{\bus}^{(2)}(\rho \ms y^2) &= \prn{\RR}{V}_{\bus}^{(2,0)} +
     \biggl( \log\frac{\rho\ms y^2}{b_0^2}
        + \frac{1}{2\epsilon}
        - 2 \gamma - \psi(1 + 2\epsilon) - \psi(1 - 3\epsilon)
        \biggl) \prn{\RR}{V}_{\bus}^{(2,1)}
\nonumber \\
&= \prn{\RR}{V}_{\bus}^{(2,0)} +
     \biggl( \log\frac{\rho}{\mu^2} + L + \frac{1}{2\epsilon}
        + \epsilon \zeta_2 + \mathcal{O}(\epsilon^2) \biggl)
        \prn{\RR}{V}_{\bus}^{(2,1)}
\end{align}
with the integrals in \app{\ref{sec:FT}}, where
\begin{align}
\label{VW-rel-log}
\prn{\RR}{V}_{\bus}^{(2,m)}
  &= 2 \epsilon\ms T_{\epsilon, 2}^{} \ms \prn{\RR}{W}_{\bus}^{(2,m)}
\end{align}
in analogy to \eqref{VW-relation}.  Again, the coefficients ${V}^{(2,m)}_{\bus}$ depend only on $\epsilon$, $z_1$, and $z_2$.
Notice that the highest pole in $\epsilon$ has the same order for ${V}_{\bus}^{(2)}$ and ${W}_{\bus}^{(2)}$, in contrast to the case of the $\delta$ regulator.


\subsection{One-loop soft factor for collinear DPS}

\begin{figure}
\begin{center}
\includegraphics[width=0.95\textwidth]{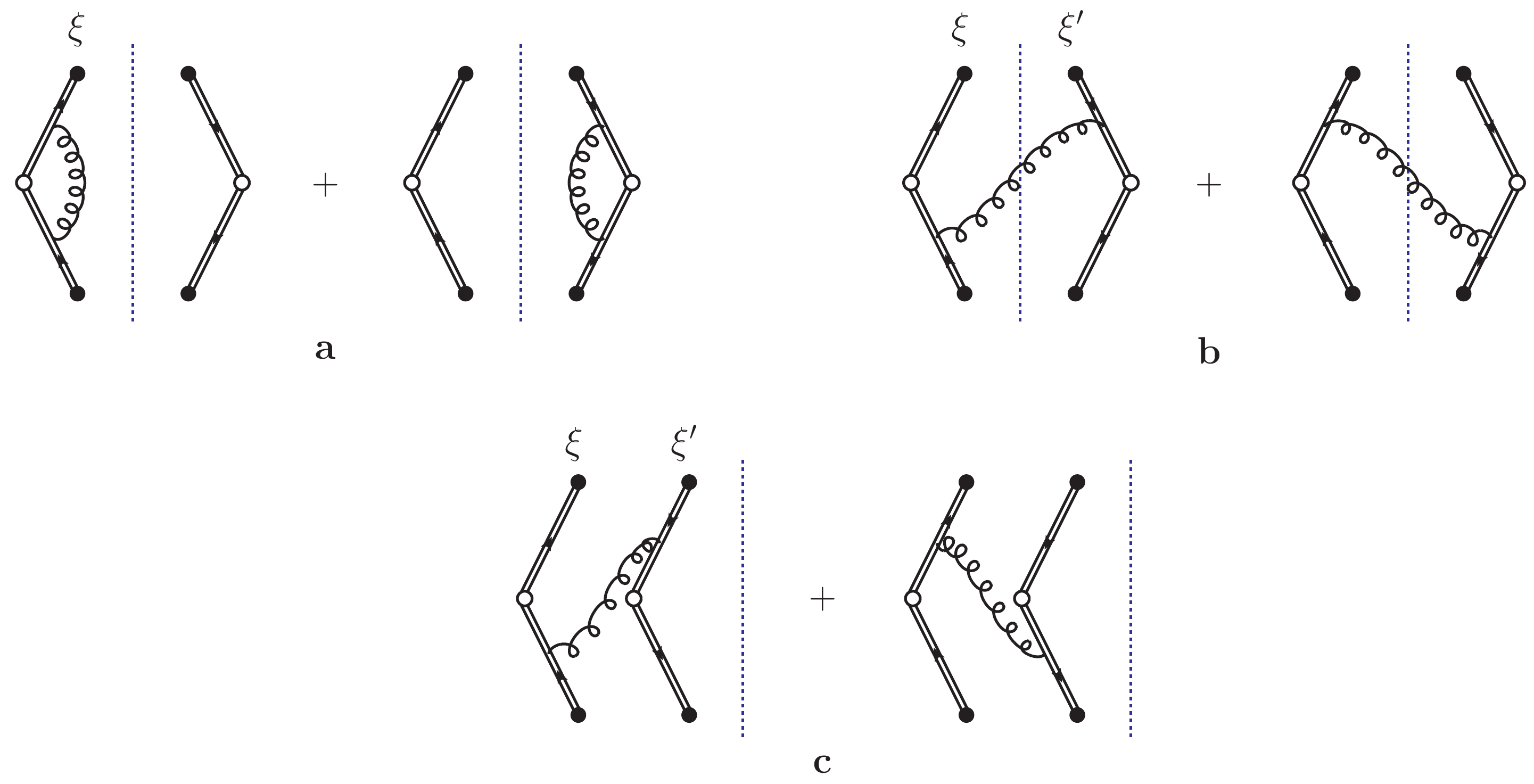}
\end{center}
\caption{\label{fig:soft-basic} Basic graphs for the soft factor at one loop.  Full graphs are obtained by adding further disconnected eikonal lines as in \fig{25} of \protect\cite{Diehl:2011yj}.  The upper eikonal lines have direction $v_R$ and the lower ones have direction $v_L$.  The position vectors $\xi$ and $\xi'$ on either side of the final state cut are purely transverse.  They can be equal or different.}
\end{figure}

At small $y$, the soft factor can be computed in perturbation theory.  At order $\as$, the graphs for the bare soft factor have one gluon exchanged between two Wilson lines with different rapidity.  The corresponding Wilson lines are  joined to each other as in \fig{\ref{fig:soft-basic}}a, or they are separated as in \figs{\ref{fig:soft-basic}}b and c.  Graphs a and b are familiar from the soft factor for single-parton TMDs, whereas graph c is specific to DPS.  The complete set of graphs is given in \fig{25} of \cite{Diehl:2011yj}.

An important result is that graphs b and c give the same result for equal transverse distance between the two Wilson lines.  The parts of the graphs that are independent of the colour structure read
\begin{align}
\label{soft-C}
U_b(\tvec{\xi} - \tvec{\xi}') &=
2 \, \frac{\mu^{2\epsilon} g^2}{S_\epsilon}
\int \frac{d^{4 - 2\epsilon} \ell}{(2\pi)^{3 - 2\epsilon}} \,
\theta(\ell^+) \, \delta(2 \ell^+ \ell^- - \tvec{\ell}^2) \;
e^{i ( \tvec{\xi}' - \tvec{\xi}) \tvec{\ell}} \,
\frac{i}{- \ell^+ v_R^- - \ell^- v_R^+ + i\epsilon} \,
\bigl(- i v_R^\alpha \ms\bigr) \,
\nonumber \\
& \quad \times
(- g_{\alpha\beta}) \,
\bigl(i v_L^\beta \ms\bigr) \,
\frac{-i}{\ell^+ v_L^- + \ell^- v_L^+ + i\epsilon} \,
\nonumber \\
U_c(\tvec{\xi} - \tvec{\xi}') &=
2 \, \frac{\mu^{2\epsilon} g^2}{S_\epsilon}
\int \frac{d^{4 - 2\epsilon} \ell}{(2\pi)^{4 - 2\epsilon}} \;
e^{i ( \tvec{\xi}' - \tvec{\xi}) \tvec{\ell}} \,
\frac{i}{- \ell^+ v_R^- - \ell^- v_R^+ + i\epsilon} \,
\bigl(- i v_R^\alpha \ms\bigr) \,
\nonumber \\
& \quad \times
\frac{- i g_{\alpha\beta}}{2 \ell^+ \ell^- - \tvec{\ell}^2 + i\epsilon} \,
\bigl(-i v_L^\beta \ms\bigr) \,
\frac{i}{\ell^+ v_L^- + \ell^- v_L^+ + i\epsilon} \,
\end{align}
with the Collins regulator.  For details on the Feynman rules used, we refer to \sect{3} of \cite{Diehl:2011yj} and to \app{D} of \cite{Buffing:2017mqm}. A factor 2 in \eqref{soft-C} takes into account the second graph in \figs{\ref{fig:soft-basic}}b and c.
With $v_L^-$ and $v_R^+$ being positive and $v_L^+$ and $v_R^-$ negative, one finds that in $U_c$ all poles in $\ell^-$ are on the same side if $\ell^+ < 0$, so that one obtains zero in this case.  For $\ell^+ > 0$, one can evaluate the $\ell^-$ integral by picking up the residue of the gluon propagator pole and thus obtains $U_c(\tvec{y}) = U_b(\tvec{y})$.
For the $\delta$ regulator, one needs to replace
\begin{align}
\label{C-to-delta-1}
v_R^+ &\to 1 \,,
&
v_R^- &\to 0 \,,
&
v_L^- &\to 1 \,,
&
v_R^+ &\to 0
\end{align}
and
\begin{align}
\label{C-to-delta-2}
- \ell^+ v_R^- - \ell^- v_R^+ + i\epsilon
   &\to - \ell^- + i\delta^- \,,
&
\ell^+ v_L^- + \ell^- v_L^+ + i\epsilon &\to \ell^+ + i\delta^+
\end{align}
in \eqref{soft-C}.  One can then repeat the argument just given for the Collins regulator.

We note that the analyticity properties of the regulated eikonal propagators are essential to obtain the above (this includes the fact of using spacelike and not timelike vectors $v_L$ and $v_R$ for the Collins regulator).  Other regulators, such as the $\eta$ regulator \cite{Chiu:2011qc, Chiu:2012ir}, do not readily allow for the contour deformations in the integration over $\ell^-$, so that further considerations are required in such cases.

The graphs in \fig{\ref{fig:soft-basic}}a are evaluated along the same lines, and one readily finds $U_a = - U_b(\tvec{0})$ for both regulators.

Following the analysis in \cite{Collins:2011zzd}, we exclude graphs with a gluon exchanged between eikonal lines of equal rapidity.  Such graphs are zero for lightlike Wilson lines and tree-level gluon propagators, but not for Wilson lines with finite rapidity.  The corresponding Wilson line self-interactions do not appear in the derivation of factorisation, and their appearance in the soft factor defined as a vacuum matrix element of Wilson lines may be regarded as an artefact.  In fact, the corresponding loop integrals are not even well defined for spacelike Wilson lines, because they contain pinched poles (see \app{B} in \cite{Ebert:2019okf}).  A more systematic treatment of this issue would be desirable but is beyond the scope of this work.

Let us now evaluate the loop integrals \eqref{soft-C} for the two regulators.


\paragraph{Collins regulator:}  For the Collins regulator the result of performing the $\ell^-$ and $\ell^+$ integrations can be obtained from \eqs{(3.41)} and (3.44) in \cite{Diehl:2011yj} by setting the gluon mass used in that work to zero.  Taking the limit $Y_R - Y_L \gg 1$, we obtain
\begin{align}
U(\tvec{y}) &= U_{a} + U_{b}(\tvec{y})  = U_{a} + U_{c}(\tvec{y})
= \frac{4 \alpha_s\ms \mu^{2\epsilon}}{S_\epsilon} \,
   ( Y_R - Y_L )
   \int \frac{d^{2 - 2\epsilon} \tvec{\ell}}{(2\pi)^{2 - 2\epsilon}}
   \bigl( e^{-i \tvec{\ell} \tvec{y}} - 1 \bigr) \,
   \frac{1}{\tvec{\ell}^2} \,,
\end{align}
where we have added graphs that always appear in combination in the DPS soft factor.  We note that the infrared divergences of the individual graphs cancel in this combination.
Carrying out the Fourier transformation (see \app{\ref{sec:FT}}), we obtain
\begin{align}
\label{soft-U-result}
U(\tvec{y})  &= - 2 \as\, \biggl( \frac{y \mu}{b_0} \biggr)^{2\epsilon} \,
  \frac{1}{\epsilon}\, \frac{1}{R_\epsilon} \, (Y_R - Y_L) \,.
\end{align}
Here we have defined\footnote{Note that $R_\epsilon$ in this work differs from $R_\epsilon$ in \eqn{(53)} of \protect\cite{Diehl:2019rdh}.}
\begin{align}
\label{R-ratio-def}
R_\epsilon &=
   \frac{S_\epsilon}{(4\pi e^{-2 \gamma})^{\epsilon}\, \Gamma(1-\epsilon)}
= 1 - (\zeta_2 + c_{\msbars})\, \epsilon^2 + \mathcal{O}(\epsilon^3) \,,
\end{align}
where the expression \eqref{MSbar-param} of $S_\epsilon$ gives
\begin{align}
   \label{MSbar-c}
c_{\msbars} &=
\begin{cases}
\phantom{-} 0    & \text{for the Collins definition}
\\
- \zeta_2 / 2    & \text{for the standard choice}
\end{cases}
\end{align}
of the \msbar scheme.

The TMD soft factor for quarks (gluons) is obtained from \eqref{soft-U-result} by multiplying with $C_F$ ($C_A$), and the soft factor $\pr{R}{S}$ for collinear DPDs is obtained by multiplying with colour factors $c_R$ that can be deduced from \sect{7.2.1} of \cite{Buffing:2017mqm}.  They are $c_8 = C_A$, $c_{10} = 6$, and $c_{27} = 8$, where the latter two values are specific to SU(3).  Details on how the different graphs contribute to the DPS soft factor are given in \sect{3.3.2} of \cite{Diehl:2011yj}.

We can then write the one-loop coefficient of the soft factor needed in \eqref{DPD-subt-def} as
\begin{align}
\label{SB-collins}
\pr{R}{S}_{B}^{(1)}(2\ms \ell_L) &= {}- \prn{R}{\gamma}_J^{(0)}\;
  \frac{1}{R_\epsilon}\,
  \frac{1}{\epsilon} \log\frac{\rho}{x_1 x_2\ms \zeta_p}
\end{align}
with $\prn{R}{\gamma}_J^{(0)}$ given in \eqref{gammaJ-LO}.
Replacing $\prn{R}{\gamma}_J^{(0)}$ with $2 C_F$, we  can compare with the expression of the TMD soft factor for quarks in \cite{Ebert:2019okf} and find that our result is consistent with \eqn{(B.11)} in that work.


\paragraph{$\delta$ regulator:}
Starting from \eqref{soft-C} with the replacements \eqref{C-to-delta-1} and \eqref{C-to-delta-2}, and performing the integrals over $\ell^-$ and $\ell^+$, one obtains
\begin{align}
U(\tvec{y})  &= U_{a} + U_{b}(\tvec{y})  = U_{a} + U_{c}(\tvec{y})
\nonumber \\
&= \frac{4 \alpha_s\ms \mu^{2\epsilon}}{S_\epsilon} \int
   \frac{d^{2 - 2\epsilon} \tvec{\ell}}{(2\pi)^{2 - 2\epsilon}}
   \bigl( e^{-i \tvec{\ell} \tvec{y}} - 1 \bigr) \,
   \frac{1}{\tvec{\ell}^2 - 2 \delta^+ \delta^-} \,
   \log \frac{\tvec{\ell}^2}{2 \delta^+ \delta^-} \,.
\end{align}
For $\epsilon > 0$ one can replace $1 / (\tvec{\ell}^2 - 2 \delta^+ \delta^-)$ with $1 / \tvec{\ell}^2$ because the difference between the two corresponding integrals goes to zero for $\delta^+ \delta^- \to 0$.  The Fourier integral then gives
\begin{align}
U(\tvec{y})  &= - 2 \as\, \biggl( \frac{y \mu}{b_0} \biggr)^{2\epsilon} \,
  \frac{1}{\epsilon}\, \frac{1}{R_\epsilon} \,
  \biggl[ \log\frac{b_0^2}{2 \delta^+ \delta^- \ms y^2}
      + \frac{1}{\epsilon} + \gamma + \psi(1 - \epsilon) \biggr] \,.
\end{align}
Using \eqref{l-var-delta} and expanding in $\epsilon$, we get
\begin{align}
\label{SB-delta}
\pr{R}{S}_{B}^{(1)}(2\ms \ell_L) &=
{}- \prn{R}{\gamma}_J^{(0)}\, \frac{1}{R_\epsilon}
  \Biggl[ \frac{1}{\epsilon}\,
    \Bigl( \log \rho + \log\frac{\mu^2}{x_1 x_2\ms \zeta_p} - L \Bigr)
    + \frac{1}{\epsilon^2} - \zeta_2 + \mathcal{O}(\epsilon)
  \Biggr] \,.
\end{align}
Replacing $\prn{R}{\gamma}_J^{(0)}$ with $2 C_F$ and setting $x_1 = x_2 = x$, we find that our result is consistent with \eqn{(B.27)} in \cite{Ebert:2019okf}.  To see this, we use their \eqn{(B.2)} and take into account that they define light-cone coordinates as $v^\pm = v^0 \mp v^3$.


\subsection{Renormalised DPDs and splitting kernels}
\label{sec:renormalisation}

The renormalisation of $\prn{\RR}{F}_B$ has been briefly described in \sect{5.2} of \cite{Buffing:2017mqm}.  It proceeds by convolution with a renormalisation factor $\prb{R \Rp}{Z}$ for each of the two partons, in close similarity to what is done in the colour singlet case \cite{Diehl:2011yj, Diehl:2018kgr, Diehl:2019rdh}.  A new aspect for colour non-singlet channels is that DPDs and renormalisation factors depend on the rapidity parameter $\zeta\ms$:
\begin{align}
\label{renorm-F}
& \prn{\RR}{F}(x_1,x_2, y,\mu,\zeta_p)
 = \sum_{\Rp_1 \Rp_2} \prn{R_1^{} \Rpbar_1}{Z}(\mu, x_1^2\ms \zeta_p)
  \conv{1} \prn{R_2^{} \Rpbar_2}{Z}(\mu, x_2^2\ms \zeta_p)
  \conv{2} \prn{\Rp_1 \Rp_2}{F}_B(y,\mu,\zeta_p) \,.
\end{align}
Parton indices are suppressed here and should be reinstated according to \eqref{parton-indices-1}, \eqref{parton-indices-2}, and \eqref{parton-indices-3}.  This should be done before carrying out the sum over the colour representations $\Rp_1$ and $\Rp_2$, whose possible values differ between quarks and gluons.

The scale dependence of $\pr{R \Rp}{Z}$ is given by
\begin{align}
\label{P-def}
\frac{d}{d\log\mu^2}\, \prb{R \Rpp}{Z}(\mu, \zeta)
  &= \sum_{\Rp} \prb{R \Rpbar}{P}(\mu, \zeta) \otimes
      \prb{\Rp \Rpp}{Z}(\mu, \zeta) \,,
\end{align}
where $\pr{R \Rp}{P}$ denotes the splitting functions in the double DGLAP equation \eqref{DGLAP-F}.  Note that in the \msbar scheme the $Z$ factors depend on $\epsilon$, whereas the splitting functions $P$ do not (see e.g.\ \sect{7} of \cite{Diehl:2018kgr}).
With the perturbative expansions \eqref{J-expansion}, \eqref{DGLAP-expansion} and
\begin{align}
\prb{R \Rp}{Z}_{ab}(x, \mu, \zeta) &= \delta_{R \Rpbar}\,
   \delta_{ab}\, \delta(1-x)
  + \sum_{n=1}^\infty \as^n(\mu)\; \prb{R \Rp}{Z}_{ab}^{(n)}(x, \mu^2/\zeta)
\end{align}
we can easily construct the one-loop renormalisation factor in the \msbar scheme as
\begin{align}
\label{one-loop-Z}
\prb{R \Rp}{Z}^{(1)}_{ab}(x, \mu^2/\zeta) &=
  - \frac{\prn{R}{\gamma}_J^{(0)}}{4} \biggl[\ms
    \frac{1}{\epsilon^2} + \frac{1}{\epsilon}\ms \log \frac{\mu^2}{\zeta} \ms\biggr]
    \delta_{R \Rpbar}\, \delta_{ab}\, \delta(1-x)
    - \frac{1}{\epsilon}\ms \prb{R \Rp}{\hat{P}}_{ab}^{(0)}(x) \,.
\end{align}
This follows from the RGE derivative
\begin{align}
\label{beta-dimreg}
\frac{d}{d\log \mu^2} &= \frac{\partial}{\partial \log \mu^2}
  + \biggl[ \frac{\beta(\as)}{2\pi} - \epsilon \ms \as \biggr]\,
    \frac{\partial}{\partial \as}
\end{align}
to be used in \eqref{P-def}, together with the requirement that $\prb{R \Rp}{P}$ has the form \eqref{DGLAP-zeta} and is $\epsilon$ independent.

Inserting the splitting form \eqref{FB-split} for $\prn{\RR}{F}$ into \eqref{renorm-F}, we obtain
\begin{align}
\label{colour-split}
\prn{\RR}{F}(x_1,x_2, y,\mu,\zeta_p)
&= \frac{1}{\pi\ms y^2} \sum_{\Rp_1 \Rp_2}
   \Bigl[ \prn{R_1^{} \Rpbar_1}{Z}(\mu, x_1^2\ms \zeta_p)
   \conv{1} \pr{R_2^{} \Rpbar_2}{Z}(\mu, x_2^2\ms \zeta_p)
\nonumber \\
&\quad \hspace{4.5em}
   \conv{2} \prn{\Rp_1 \Rp_2}{V}_B(y,\mu,x_1 x_2\ms \zeta_p)
   \conv{12} \bigl( \pr{11}{Z} \ms\bigr)^{-1}(\mu) \Bigr] \conv{12} f(\mu)
\nonumber \\
&= \frac{1}{\pi\ms y^2} \, \prn{\RR}{V}(y,\mu,x_1 x_2\ms \zeta_p)
      \conv{12} f(\mu) \,,
\end{align}
where we have used $f_B = \bigl( \pr{11}{Z} \ms\bigr)^{-1} \conv{} f$, with $Z^{-1}$ defined such that
\begin{align}
[ Z^{-1} \otimes Z ]_{ab} &= \delta_{ab}\, \delta(1-x) \,,
\end{align}
which can be solved for $Z^{-1}$ order by order in $\as$.  We see that the factor $\pr{11}{Z}(\mu)$ renormalises both colour-singlet DPDs and ordinary PDFs.  It has no $\zeta$ dependence, because $\prn{1}{\gamma}_J = 0$.

We also need the one-loop coefficient of the renormalisation constant for the running coupling in \eqref{VB-alpha}, which reads
\begin{align}
\label{beta0-def}
Z_\alpha(\mu) &= 1 - \frac{\as(\mu)}{\epsilon}\, \frac{\beta_0}{2}
  + \mathcal{O}(\as^2) \,,
&
\beta_0 &= \frac{11}{3}\ms C_A - \frac{4}{3}\, T_F\ms n_F \,,
\end{align}
where $T_F = 1/2$, and $n_F$ is the number of active quark flavours.  The running coupling in $4$~dimensions satisfies
\begin{align}
\frac{d \as}{d\log \mu^2} &= \frac{\beta(\as)}{2\pi}
  = - \frac{\beta_0}{2}\, \as^2 + \mathcal{O}(\as^3)
\end{align}


\paragraph{Rescaling of the rapidity parameter.}
As is the case in the evolution equation \eqref{DGLAP-F}, the rescaling factors $x_1^2$ and $x_2^2$ in \eqref{colour-split} are fixed by the momentum fractions on the l.h.s.\ and are not part of the convolution integrals.  They appear because the renormalisation factor for parton $a_i$ can depend on the plus-momentum $x_i\ms p^+$ of that parton, but not on the plus-momentum of the other parton or the plus-momentum of the proton.  This becomes evident when renormalisation is formulated at the level of twist-two operators and not of their matrix elements \cite{Buffing:2017mqm}.  The definition of $\zeta_p$ in \eqref{zeta-def} or \eqref{l-variables-def} refers to the squared proton plus-momentum $p^+$, so that the rescaled factors $x_1^2\ms \zeta_p$ and $x_2^2 \ms\zeta_p$ refer to the square of the appropriate parton plus-momentum.

Let us show how these rescaling factors combine to a dependence on $x_1 x_2\ms \zeta_p$ in the last line of \eqref{colour-split}.  Adapting the argument given for the renormalisation of two-parton TMDs in \sect{3.4} of \cite{Buffing:2017mqm}, one finds that the renormalization factor $\pr{R}{Z}$ satisfies a Collins-Soper equation
\begin{align}
\label{CS-Lambda}
\frac{\partial}{\partial \log\zeta}\, \log \prb{R \Rp}{Z}_{a b}(z,\mu,\zeta)
&= \frac{1}{2}\, \pr{R}{\Lambda}(\mu) \,,
\end{align}
where we make all arguments explicit, including the parton momentum fraction $z$.  Here $\Lambda(\mu)$ is the additive counterterm that renormalises the Collins-Soper kernel
\begin{align}
\pr{R}{J}(y,\mu) &= 2 \; \pr{R}{\Lambda}(\mu) + \pr{R}{J}_B(y) \,.
\end{align}
We note that the right-hand side becomes $\prn{R}{\Lambda}(\mu_1) + \prn{R}{\Lambda}(\mu_2) + \prn{R}{J}_B(y)$ if one takes different renormalisation scales for the two partons.  Using \eqref{CS-Lambda}, we have
\begin{align}
\prb{R \Rp}{Z}_{a b}(z,\mu, x_1^2\ms \zeta_p)
&= \exp\biggl[ \frac{1}{2}\, \pr{R}{\Lambda}(\mu) \log \frac{x_1}{x_2}
   \biggr] \, \prb{R \Rp}{Z}_{a b}(z,\mu, x_1 x_2\ms \zeta_p) \,.
\end{align}
Using this and its analogue for the factor with $x_2^2\ms \zeta_p$ in \eqref{colour-split}, we get
\begin{align}
& \prn{R_1^{} \Rpbar_1}{Z}(\mu, x_1^2\ms \zeta_p)
  \conv{1} \prn{R_2^{} \Rpbar_2}{Z}(\mu, x_2^2\ms \zeta_p)
  \conv{2} \ldots
\nonumber \\
& \qquad = \prn{R_1^{} \Rpbar_1}{Z}(\mu, x_1 x_2\ms \zeta_p)
   \conv{1} \prn{R_2^{} \Rpbar_2}{Z}(\mu, x_1 x_2\ms \zeta_p)
   \conv{2} \ldots \,.
\end{align}
Of course, this only happens if one takes equal renormalisation scales for the two partons.


\subsection{Two-loop splitting kernels for DPDs}
\label{sec:two-loop-kernels}

Expanding the renormalised splitting kernel in \eqref{colour-split} to second order, we obtain
\begin{align}
\label{V-expansion}
& \pr{\RR}{F}(x_1,x_2, y,\mu,\zeta_p)
\nonumber \\
&\qquad\qquad
   = \frac{1}{\pi y^2}\, \Bigl[ \as(\mu)\, \prn{\RR}{V}^{(1)}
     + \as^2(\mu)\, \prn{\RR}{V}^{(2)}(y, \mu, x_1 x_2\ms \zeta_p)
     + \mathcal{O}(\as^3) \ms
   \Bigr] \conv{12} f(\mu) \,,
\end{align}
where it is understood that ${V}^{(1)}$ and ${V}^{(2)}$ also depend on two parton momentum fractions, which are integrated over as specified in \eqref{conv-12-def}.
At one loop, the bare splitting kernels $V_B$ are finite at $\epsilon = 0$, so that their renormalised version in \eqref{V-expansion} is simply given by
\begin{align}
\label{one-loop-kernel}
\prn{\RR}{V}^{(1)}
  &= \lim\limits_{\epsilon \to 0} \prn{\RR}{V}_{B}^{(1)}(\epsilon)
   = \bigl[ \prn{\RR}{V}_{B}^{(1)} \bigr]_{0} \,.
\end{align}
Here we use the following notation for the expansion of an $\epsilon$ dependent quantity $Q(\epsilon)$ in a Laurent series around $\epsilon = 0$:
\begin{align}
Q(\epsilon) &= \sum_{k} \epsilon^{k} \ms \bigl[ Q \bigr]_{k} \,.
\end{align}
In particular, $[ Q ]_0$ gives the finite piece, and $[ Q ]_{-1}$ gives the residue of the $1/\epsilon$ pole.
Using \eqref{VB-alpha} together with \eqref{colour-split} and the one-loop counterterms in \eqref{one-loop-Z} and \eqref{beta0-def}, we find that the renormalised two-loop kernel has the form
\begin{align}
\label{V2-master}
\prn{\RR}{V}^{(2)} &= \lim_{\epsilon \to 0} \,
  \Bigl[ \, e^{2 \epsilon L} \; \prn{\RR}{V}_{B}^{(2)}(\epsilon)
   + \prn{\RR}{V}_{\text{ct\,1}}^{}(\epsilon)
   + \prn{\RR}{V}_{\text{ct\,2}}^{}(\epsilon) \, \Bigr]
\end{align}
with $\prn{\RR}{V}_{B}^{(2)}$ given in \eqref{bare-V} and the counterterm contributions given by
\begin{align}
\label{Vct-def}
\prn{\RR}{V}_{\text{ct\,1}}(\epsilon)
&= \bs {}- e^{\epsilon L} \, \frac{1}{\epsilon} \,
 \biggl( \sum_{\Rp_1} \prn{R_1^{} \Rpbar_1}{\hat{P}}^{(0)}
         \conv{1} \prn{\Rp_1 R_2^{}}{V}_{B}^{(1)}(\epsilon)
       + \sum_{\Rp_2} \prn{R_2^{} \Rpbar_2}{\hat{P}}^{(0)}
         \conv{2} \prn{R_1^{} \Rp_2}{V}_{B}^{(1)}(\epsilon)
\nonumber \\
& \hspace{5em}
       - \,\prn{\RR}{V}_{B}^{(1)} (\epsilon)\conv{12} \pr{11}{P}^{(0)}
       + \frac{\beta_0}{2}\, \prn{\RR}{V}_{B}^{(1)}(\epsilon) \biggr) ,
\nonumber \\
\prn{\RR}{V}_{\text{ct\,2}}(\epsilon)
&= \bs {}- e^{\epsilon L} \,
      \biggl( \frac{1}{\epsilon^2} + \frac{1}{\epsilon}\,
         \log\frac{\mu^2}{x_1 x_2\ms \zeta_p}
      \biggr) \,
      \frac{\prn{R_1}{\gamma}_J^{(0)}}{2}\;
         \prn{\RR}{V}_{B}^{(1)}(\epsilon) \,.
\end{align}
It is useful to rewrite the second counterterm as
\begin{align}
\label{rewrite-ct2}
\prn{\RR}{V}_{\text{ct\,2}}(\epsilon)
&= e^{2 \epsilon L} \,
  \Biggl[
    - \frac{1}{\epsilon^2}
    + \frac{1}{\epsilon} \, \biggl( L - \log\frac{\mu^2}{\zeta} \biggr)
    + L \log\frac{\mu^2}{\zeta}
    - \frac{L^2}{2} + \zeta_2 + c_{\msbars}
    + \mathcal{O}(\epsilon) \,
  \Biggr]
\nonumber \\[0.2em]
& \quad \times
  \frac{\prn{R_1}{\gamma}_J^{(0)}}{2}\,
  \frac{\prn{\RR}{V}_{B}^{(1)}(\epsilon)}{R_{\epsilon}} \,,
\end{align}
where we abbreviate
\begin{align}
\zeta = x_1 x_2\ms \zeta_p
\end{align}
for the remainder of this section.
Since ${V}_{B}^{(1)}(\epsilon)$ has no poles in $1/\epsilon$, the $\mathcal{O}(\epsilon)$ terms inside the square brackets of \eqref{rewrite-ct2} are not relevant.

We now derive an explicit form of the renormalised two-loop kernels for the two regulators.  For the sake of legibility, we omit colour representation labels in the following.


\paragraph{$\delta$ regulator:}  With the decomposition \eqref{V-log-delta} of the bare unsubtracted kernel and the explicit form \eqref{SB-delta} of the one-loop soft factor for the $\delta$ regulator, we obtain
\begin{align}
\label{VB2-start-delta}
V_B^{(2)}(\epsilon) &= V_{\bus}^{(2,0)}(\epsilon)
  + \lim_{\rho \to \infty} \, \Biggl\{
    \log \rho \, V_{\bus}^{(2,1)}(\epsilon)
    + \frac{1}{\epsilon}\, \log \rho \,
      \frac{\gamma_J^{(0)}}{2}\, \frac{V_{B}^{(1)}(\epsilon)}{R_{\epsilon}}
   \Biggr\}
\nonumber \\
& \quad
  + \biggl[ \frac{1}{\epsilon}\, \Bigl( \log\frac{\mu^2}{\zeta} - L \Bigr)
    + \frac{1}{\epsilon^2} - \zeta_2 + \mathcal{O}(\epsilon) \biggr] \,
    \frac{\gamma_J^{(0)}}{2}\, \frac{V_{B}^{(1)}(\epsilon)}{R_{\epsilon}} \,.
\end{align}
Rapidity divergences must cancel in the presence of the ultraviolet regulator, i.e.\ the terms with $\log \rho$ must cancel in $4 - 2\epsilon$ dimensions.  This leads to the consistency relation
\begin{align}
\label{rap-relation}
V_{\bus}^{(2,1)}(\epsilon)
  &= - \frac{1}{\epsilon}\, \frac{\gamma_J^{(0)}}{2}\,
       \frac{V_{B}^{(1)}(\epsilon)}{R_{\epsilon}} \,.
\end{align}
Because $V_{\bus}^{(2,1)} \propto S_\epsilon^{-2}$, $V_{B}^{(1)} \propto S_\epsilon^{-1}$, and $R_\epsilon \propto S_\epsilon$, this relation can indeed hold for different choices of the \msbar scheme parameter $S_\epsilon$.
Using \eqref{rewrite-ct2} and \eqref{rap-relation}, we obtain
\begin{align}
\label{first-comb-delta}
e^{2 \epsilon L} \, V_{B}^{(2)}(\epsilon) + V_{\text{ct\,2}}(\epsilon)
&= e^{2 \epsilon L} \, \Biggl[ V_{\bus}^{(2,0)}
      + \biggl( L \log\frac{\mu^2}{\zeta}
      - \frac{L^2}{2} + c_{\msbars} + \mathcal{O}(\epsilon)
   \biggr) \,
   \frac{\gamma_J^{(0)}}{2}\, \frac{V_{B}^{(1)}(\epsilon)}{R_{\epsilon}}
   \Biggr] \,.
\end{align}
It follows that $V_{\bus}^{(2,0)}$ can have only single poles in $\epsilon$, which must cancel against those in the counterterm $V_{\text{ct\,1}}$.  This implies the condition
\begin{align}
\label{single-pole-delta}
\Bigl[ V_{\bus}^{(2,0)} \Bigr]_{-1}
&= \hat{P}^{(0)} \conv{1} V^{(1)}
   + \hat{P}^{(0)} \conv{2} V^{(1)}
   - V^{(1)} \conv{12} P^{(0)}
   + \frac{\beta_0}{2}\, V^{(1)} \,,
\end{align}
where we used \eqref{one-loop-kernel}.
Combining the finite terms, we find
\begin{align}
\label{two-loop-final}
V^{(2)} &= V^{(2)}_{\text{fin}}
  - \biggl(
         \hat{P}^{(0)} \conv{1} \bigl[ V_{B}^{(1)} \bigr]_{1}
       + \hat{P}^{(0)} \conv{2} \bigl[ V_{B}^{(1)} \bigr]_{1}
       - \bigl[ V_{B}^{(1)} \bigr]_{1} \conv{12} P^{(0)}
       + \frac{\beta_0}{2}\, \bigl[ V_{B}^{(1)} \bigr]_{1}
    \biggr)
\nonumber \\
& \quad
  + \biggl( L \log\frac{\mu^2}{\zeta} - \frac{L^2}{2}
    + c_{\msbars} \biggr) \, \frac{\gamma_J^{(0)}}{2} \, V^{(1)}
\nonumber \\
& \quad
  + L\, \biggl(
           \hat{P}^{(0)} \conv{1} V^{(1)}
         + \hat{P}^{(0)} \conv{2} V^{(1)}
         - V^{(1)} \conv{12} P^{(0)}
         + \frac{\beta_0}{2}\, V^{(1)}
      \biggr)
\end{align}
with
\begin{align}
\label{V_fin-delta}
V^{(2)}_{\text{fin}} &= \Bigl[ V_{\bus}^{(2,0)} \Bigr]_{0} \,.
\end{align}
The difference between the two definitions of the \msbar scheme parameter $S_\epsilon$ in \eqref{MSbar-param} starts at $\mathcal{O}(\epsilon^2)$, as can be seen from \eqref{R-ratio-def} and \eqref{MSbar-c}.   Since $V_{\bus}^{(2,0)}$ has only single poles in $\epsilon$, we conclude that $V^{(2)}_{\text{fin}}$ is identical in the two schemes.
We note that the convolutions with $\hat{P}$ in \eqref{single-pole-delta} and \eqref{two-loop-final} involve a sum over representation labels when these are restored according to~\eqref{Vct-def}.

Let us check that the $\zeta_p$ dependence of $V^{(2)}$ is as required by the Collins-Soper equation \eqref{CS-F}.  According to \eqref{V-expansion} and \eqref{two-loop-final}, we have
\begin{align}
\frac{\partial}{\partial \log \zeta} \, V(y, \mu,\zeta) \conv{12} f(\mu)
&= - \as^2(\mu) \, L \,
     \frac{\gamma_J^{(0)}}{2} \, V^{(1)} \conv{12} f(\mu)
   + \mathcal{O}(\as^3)
\nonumber \\
&= \frac{1}{2} \, J(y, \mu) \,
   \Bigl[  V(y, \mu,\zeta) \conv{12} f(\mu) \Bigr] + \mathcal{O}(\as^3) \,,
\end{align}
where we used the short-distance expansion of $J(y)$ derived in \sect{7.2} of \cite{Buffing:2017mqm}:
\begin{align}
J(y, \mu) &= - \as\, \gamma_J^{(0)} \ms L + \mathcal{O}(\as^2) \,.
\end{align}
As a further cross check, we take the derivative of $V(y, \mu\,\zeta) \conv{} f(\mu)$ w.r.t.\ $\mu$.  Using \eqref{DGLAP-zeta}, we find that the double DGLAP equation \eqref{DGLAP-F} is satisfied to the required order in $\as$.


\paragraph{Collins regulator:}  The decomposition \eqref{V-log-collins} of the bare unsubtracted kernel and the explicit form \eqref{SB-collins} of the one-loop soft factor for the Collins regulator give
\begin{align}
V_B^{(2)}(\epsilon) = V_{\bus}^{(2,0)}(\epsilon)
    + \lim_{\rho \to \infty}
& \Biggl\{
    \biggl(
       \log\frac{\rho}{\mu^2} + L
       + \frac{1}{2\epsilon} + \epsilon \ms \zeta_2 + \mathcal{O}(\epsilon^2)
    \biggl) \, V_{\bus}^{(2,1)}(\epsilon)
\nonumber \\
& + \frac{1}{\epsilon}\, \log\frac{\rho}{\zeta}\,
    \frac{\gamma_J^{(0)}}{2}\, \frac{V_{B}^{(1)}(\epsilon)}{R_{\epsilon}}
  \Biggr\} \,.
\end{align}
The cancellation of rapidity divergences in $4 - 2\epsilon$ dimensions leads to the same relation \eqref{rap-relation} as for the $\delta$ regulator.  Using this relation, we obtain
\begin{align}
V_B^{(2)}(\epsilon) &= V_{\bus}^{(2,0)}(\epsilon)
  + \Biggl[
      - \frac{1}{2 \epsilon^2}
      + \frac{1}{\epsilon}\, \Bigl( \log\frac{\mu^2}{\zeta} - L \Bigr)
      - \zeta_2 + \mathcal{O}(\epsilon)
    \Biggr] \,
    \frac{\gamma_J^{(0)}}{2}\, \frac{V_{B}^{(1)}(\epsilon)}{R_{\epsilon}}
\end{align}
and hence
\begin{align}
\label{first-comb-Coll}
& e^{2 \epsilon L} \, V_{B}^{(2)}(\epsilon) + V_{\text{ct\,2}}(\epsilon)
\nonumber \\
& \qquad =
   e^{2 \epsilon L} \; \Biggl[ V_{\bus}^{(2,0)}(\epsilon)
      + \biggl( - \frac{3}{2 \epsilon^2} + L \log\frac{\mu^2}{\zeta}
      - \frac{L^2}{2} + c_{\msbars} + \mathcal{O}(\epsilon)
   \biggr) \,
   \frac{\gamma_J^{(0)}}{2}\, \frac{V_{B}^{(1)}(\epsilon)}{R_{\epsilon}}
   \Biggr] \,.
\end{align}
Since $V_{\text{ct\,1}}$ contains only single poles in $\epsilon$, we find that the coefficient of the double pole in $V_{B}^{(2,0)}$ must be
\begin{align}
\label{double-pole-collins}
\Bigl[ V_{\bus}^{(2,0)} \Bigr]_{-2}
  &= \frac{3}{2} \, \frac{\gamma_J^{(0)}}{2}\, V^{(1)} \,,
\end{align}
where we used \eqref{one-loop-kernel} and $R_\epsilon = 1 + \mathcal{O}(\epsilon^2)$.  The remaining single pole of \eqref{first-comb-Coll} must cancel against the one in $V_{\text{ct\,1}}$, which implies
\begin{align}
\label{single-pole-collins}
\Bigl[ V_{\bus}^{(2,0)} \Bigr]_{-1}
  - \frac{3}{2}\, \frac{\gamma_J^{(0)}}{2}\, \Bigl[ V_{B}^{(1)} \Bigr]_{1}
&= \hat{P}^{(0)} \conv{1} V^{(1)} + \hat{P}^{(0)} \conv{2} V^{(1)}
       - V^{(1)} \conv{12} P^{(0)} + \frac{\beta_0}{2}\, V^{(1)} \,.
\end{align}
Combining the finite terms in \eqref{first-comb-Coll} with those of $V_{\text{ct\,1}}$, we find that the renormalised two-loop kernel $V^{(2)}$ has the form \eqref{two-loop-final} with
\begin{align}
\label{V_fin-collins}
V^{(2)}_{\text{fin}} &=
  \Bigl[ V_{\bus}^{(2,0)} \Bigr]_{0}
  - \frac{3}{2} \, \frac{\gamma_J^{(0)}}{2}\,
    \Bigl[ R_{\epsilon}^{-1} \, V_{B}^{(1)} \Bigr]_2 \,.
\end{align}
Using \eqref{double-pole-collins}, we can rewrite
\begin{align}
\Bigl[ V_{\bus}^{(2,0)} \Bigr]_{0}
 &= \Bigl[ R_\epsilon^{2} \, V_{\bus}^{(2,0)} \Bigr]_0
    - 2 \bigl[ R_\epsilon \bigr]_{2} \, \Bigl[ V_{\bus}^{(2,0)} \Bigr]_{-2}
  = \Bigl[ R_\epsilon^{2} \, V_{\bus}^{(2,0)} \Bigr]_0
    - 3 \bigl[ R_\epsilon \bigr]_{2} \, \frac{\gamma_J^{(0)}}{2}\, V^{(1)} \,,
\nonumber \\
\Bigl[ R_{\epsilon}^{-1} \, V_{B}^{(1)} \Bigr]_2
 &= \Bigl[ R_{\epsilon} V_{B}^{(1)} \Bigr]_2
    - 2 \bigl[ R_\epsilon \bigr]_{2} \, V^{(1)} \,,
\end{align}
so that
\begin{align}
V^{(2)}_{\text{fin}} &=
  \Bigl[ R_\epsilon^{2} \, V_{\bus}^{(2,0)} \Bigr]_{0}
  - \frac{3}{2} \, \frac{\gamma_J^{(0)}}{2}\,
    \Bigl[ R_{\epsilon} \, V_{B}^{(1)} \Bigr]_2 \,.
\end{align}
The dependence on the \msbar scheme parameter $S_\epsilon$ cancels out in $R_\epsilon^{2} \, V_{\bus}^{(2,0)}$ and $R_{\epsilon} V_{B}^{(1)}$, so that $V^{(2)}_{\text{fin}}$ is identical for the standard and the Collins definition of the \msbar scheme.


\paragraph{Scheme dependence.}
We have seen that $V^{(2)}_{\text{fin}}$ is independent of the \msbar scheme implementation, both for the $\delta$ and for the Collins regulator.  The only scheme dependence of the two-loop kernels therefore comes from the term with $c_{\msbars}$ in \eqref{two-loop-final}.  This leads to a \msbar scheme dependence of the DPD at small $y$,
\begin{align}
\label{scheme-dep}
\prn{\RR}{F}^{(2)}_{a_1 a_2, a_0}
&= c_{\msbars}\, \frac{\as\ms \prn{R_1}{\gamma}_J^{(0)}}{2} \,
   \prn{\RR}{F}^{(1)}_{a_1 a_2, a_0} + \{ \text{scheme independent} \} \,,
\end{align}
where
\begin{align}
\label{F-expansion}
\prn{\RR}{F}^{(n)}_{a_1 a_2, a_0}
&= \frac{1}{\pi y^2}\; \as^n \;
   \sum_{b} \prn{\RR}{V}^{(n)}_{a_1 a_2, b} \ms \conv{12} \ms f_b
\end{align}
denotes the term of order $\as^n$ in the expansion \eqref{V-expansion}.
Of course, this dependence must cancel in a physical quantity.  To see how this happens, let us consider the double Drell-Yan process.  As is well-known, the unsubtracted hard scattering cross section $\pr{R \Rp}{\hat{\sigma}}_{\us}$ for $R = \Rp = 1$ has a single $1/\epsilon$ pole at $\mathcal{O}(\alpha_s)$, due to collinear divergences related to the incoming quark and antiquark.  Double poles $1/\epsilon^2$ cancel between real and virtual corrections.  This cancellation does not take place if $R \neq 1$, because the colour factors of real and virtual graphs differ in this case \cite{Mekhfi:1988kj}.  At order $\as$, the product $S_\epsilon^{-1} \ms \epsilon^{-2}$ leads to a scheme dependent term proportional to $c_{\msbars}$ in $\pr{R \Rp}{\hat{\sigma}}_{\us}$.  We can determine its coefficient by using that the subtraction of collinear and soft-collinear poles is tantamount to the convolution with an inverse renormalisation factor $( \prb{R \Rp}{Z} )^{-1}$ for each of the incoming partons.  This ensures the $\mu$ independence of the physical cross section.  According to \eqref{one-loop-Z} we have
\begin{align}
\bigl( \prb{R \Rp}{Z} \ms\bigr)^{-1}_{a b}
&= \delta_{R \Rpbar}\, \delta_{ab}\, \delta(1-x) \,
   \Bigl[ 1 + \as \ms \prn{R}{\gamma}_J^{(0)}
   \bigl/ (4 \epsilon^2) + \ldots \, \Bigr]
\end{align}
for the double pole at $\mathcal{O}(\as)$.  The scheme dependent part of the subtracted hard cross section at this order is therefore $- c_{\msbars}\, \as \prn{R_1}{\gamma}_J^{(0)} \bigl/ 2$ times the Born term $\hat{\sigma}^{(0)}$.  Expanding the double Drell-Yan cross section to NLO, we have twice this factor times the LO cross section from the scheme dependence of the hard cross sections.  This cancels the scheme dependence due to the NLO part \eqref{scheme-dep} of the splitting DPDs for each incoming proton.

\section{Two-loop calculation}
\label{sec:two-loops}

Most of the techniques described in \sect{4} of \cite{Diehl:2019rdh} for the calculation of the two-loop kernels with $R=1$ carry over to general representations $R$.  We hence begin this section by recalling just briefly the main steps of the calculation.  We then discuss the aspects specific for colour non-singlet channels, namely the colour factors and the treatment of the rapidity dependence.

To compute bare unsubtracted two-loop kernels, we apply the factorisation formula \eqref{WV-defs} to the DPD $F_{\bus,\ms a_1 a_2 / a_0}({\Delta}, \rho)$ of partons $a_1$ and $a_2$ in a parton $a_0$.  Expanding the formula in $\as$ and writing $f_{B\bs,\ms b /\bs a_0}$ for the bare PDF of parton $b$ in parton $a_0$  we obtain
\begin{align}
\label{FW-master}
F^{(2)}_{\bus,\ms a_1 a_2 / a_0}({\Delta}, \rho)
&= \sum_b \Bigl[
  W_{\bus,\ms a_1 a_2, b}^{(2)}({\Delta}, \rho)
      \conv{12} f^{(0)}_{B\bs,\ms b /\bs a_0}
+ W_{\bus,\ms a_1 a_2, b}^{(1)}({\Delta}, \rho)
      \conv{12} f^{(1)}_{B\bs,\ms b /\bs a_0}
\Bigr]
\nonumber \\
&= W_{\bus,\ms a_1 a_2, a_0}^{(2)}({\Delta}, \rho)
\end{align}
for the second order term.  The bare unsubtracted two-loop kernel is hence directly given by the two-loop graphs for the bare unsubtracted DPD of partons $a_1$ and $a_2$ in an on-shell parton $a_0$.  We compute with massless quarks and gluons, treating both ultraviolet and infrared divergences by dimensional regularisation.  In the second step of \eqref{FW-master} we used
\begin{align}
f^{(0)}_{B\bs,\ms b /\bs a_0}(x) &= \delta_{b\ms a_0}\, \delta(1-x) \,,
&
f^{(1)}_{B\bs,\ms b /\bs a_0} &= 0 \,,
\end{align}
where the second relation holds because the corresponding loop integrals do not depend on any mass scale.


\subsection{Channels, graphs, and colour factors}
\label{sec:graphs}

Depending on the order at which a splitting process $a_0 \to a_1 a_2$ first appears, we distinguish between
\begin{itemize}
\item LO channels $g \to g g$, $g \to q \bar{q}$, $q \to q g$, and
\item NLO channels $g \to q g$, $q \to g g$, as well as $q_j \to q_j q_k$, $q_j \to q_j \bar{q}_k$, $q_j \to q_k \bar{q}_k$, where $j$ and $k$ label quark flavours that may be equal or different.
\end{itemize}
More LO and NLO channels are obtained by interchanging partons $a_1$ and $a_2$ or by charge conjugation, where the kernels for charge conjugated channels are identical or opposite in sign (see \sect{\ref{sec:results}}).  At order $\as^2$, a dependence on the rapidity parameter $\zeta_p$ only appears in the LO channels.

\begin{figure}
  \begin{center}
    \subfigure[\label{fig:ggg-LD}]{\includegraphics[width=0.23\textwidth]{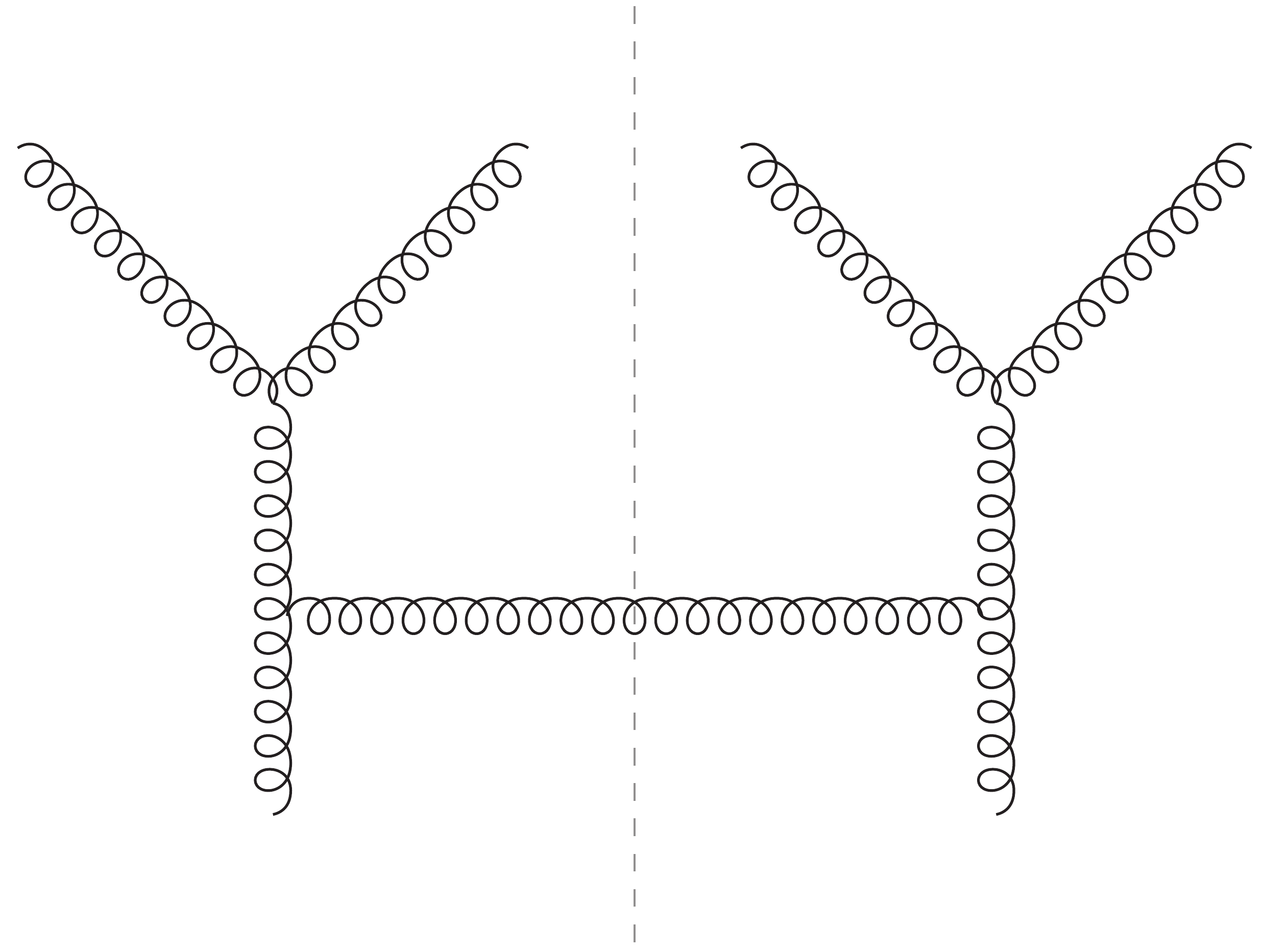}}
    \hspace{0.3em}
    \subfigure[\label{fig:ggg-UD}]{\includegraphics[width=0.23\textwidth]{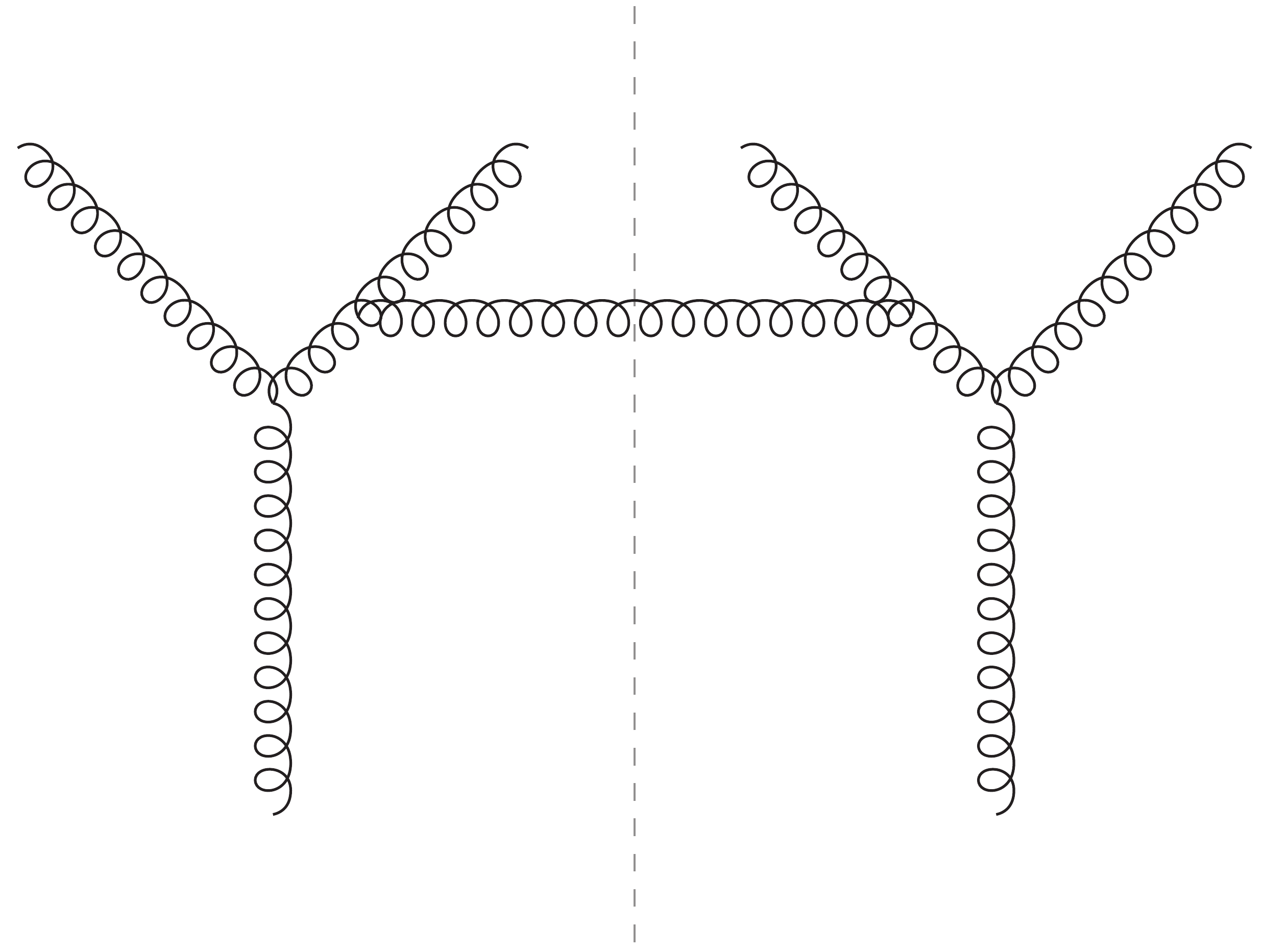}}
    \hspace{0.3em}
    \subfigure[UND \label{fig:ggg-UND}]{\includegraphics[width=0.23\textwidth]{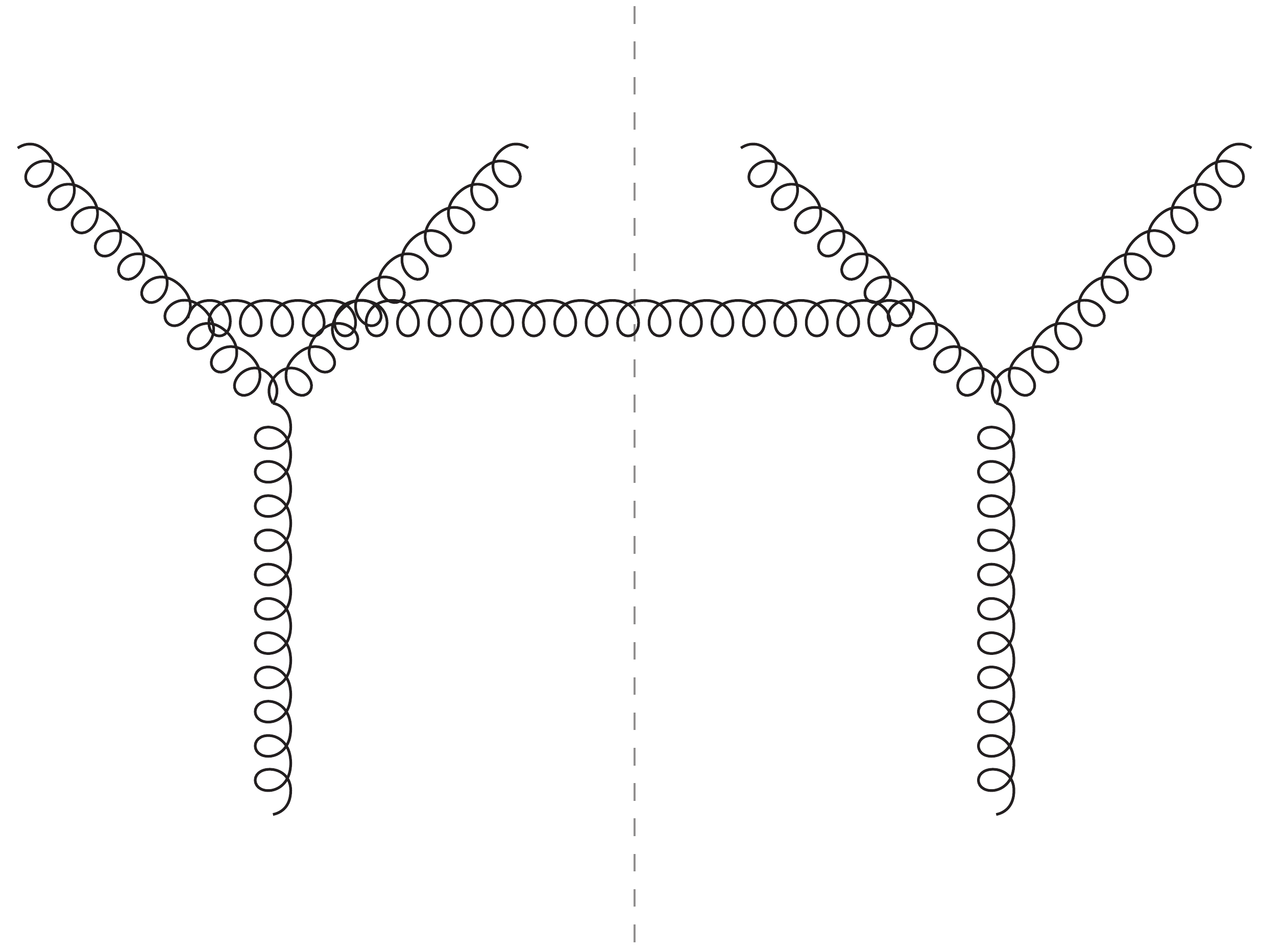}}
    \hspace{0.3em}
    \subfigure[\label{fig:ggg-T2B}]{\includegraphics[width=0.23\textwidth]{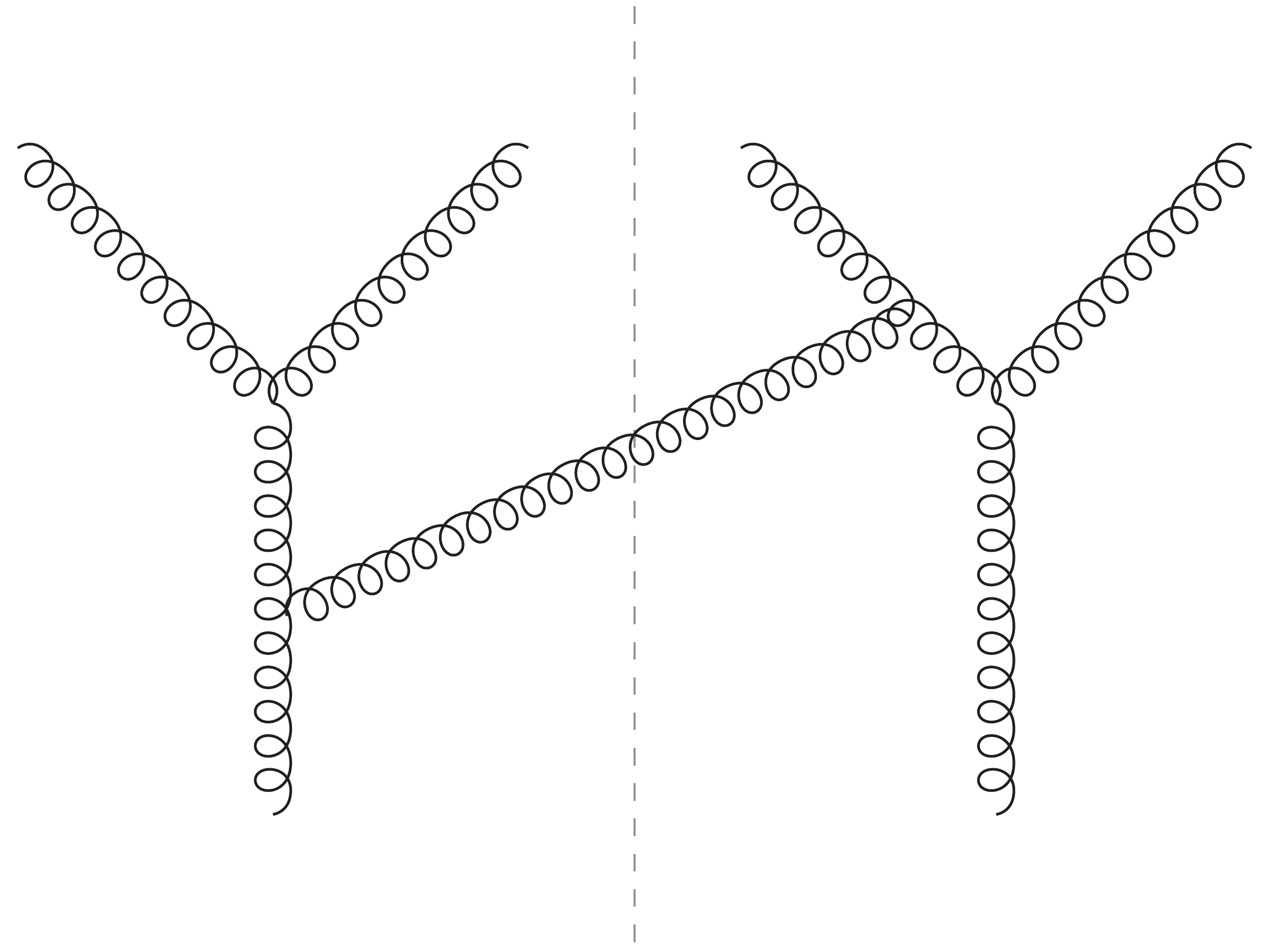}}
\\[2em]
    \subfigure[\label{fig:ggg-4gx2}]{\includegraphics[width=0.23\textwidth]{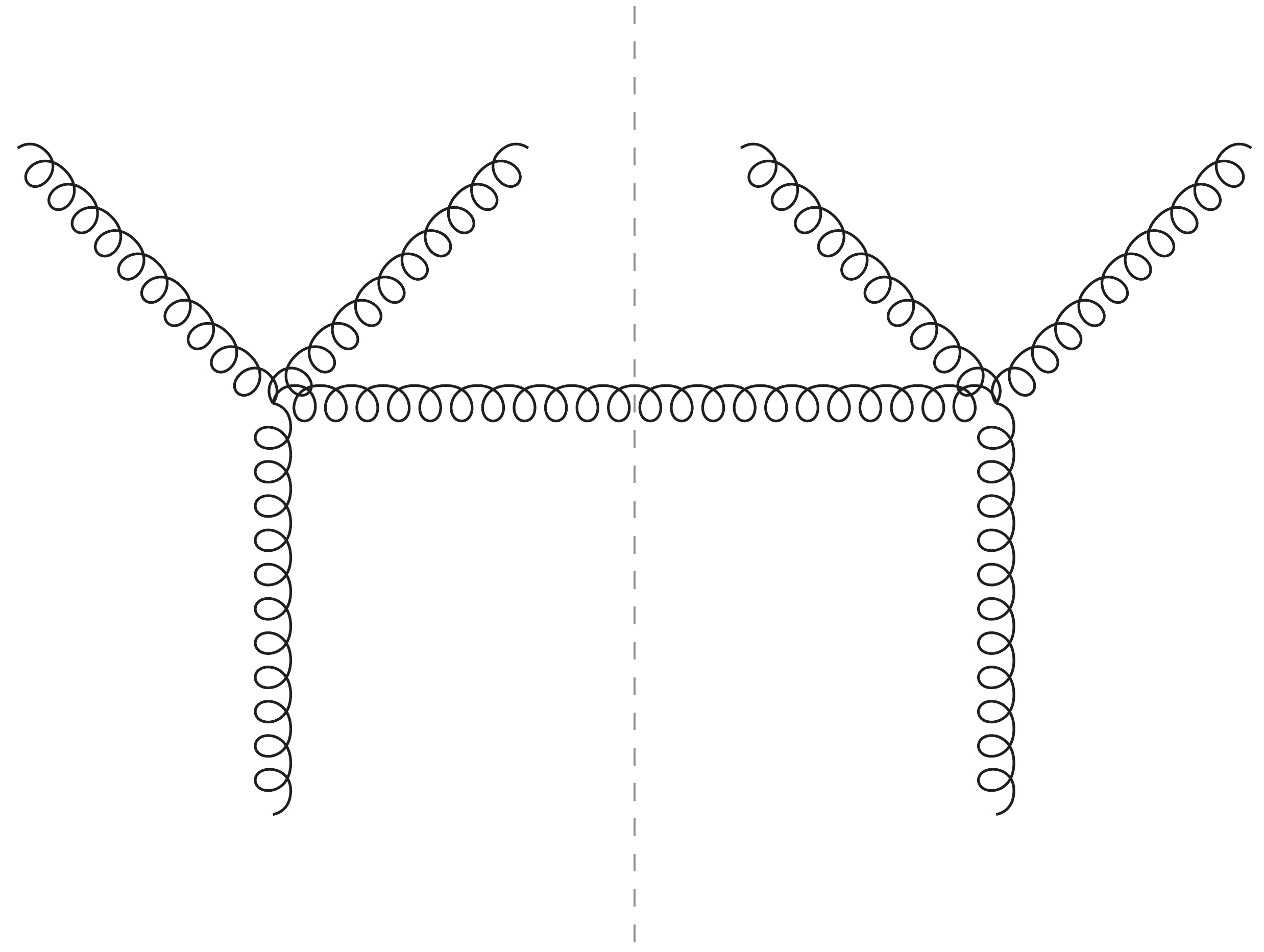}}
    \hspace{0.3em}
    \subfigure[\label{fig:ggg-4g2up}]{\includegraphics[width=0.23\textwidth]{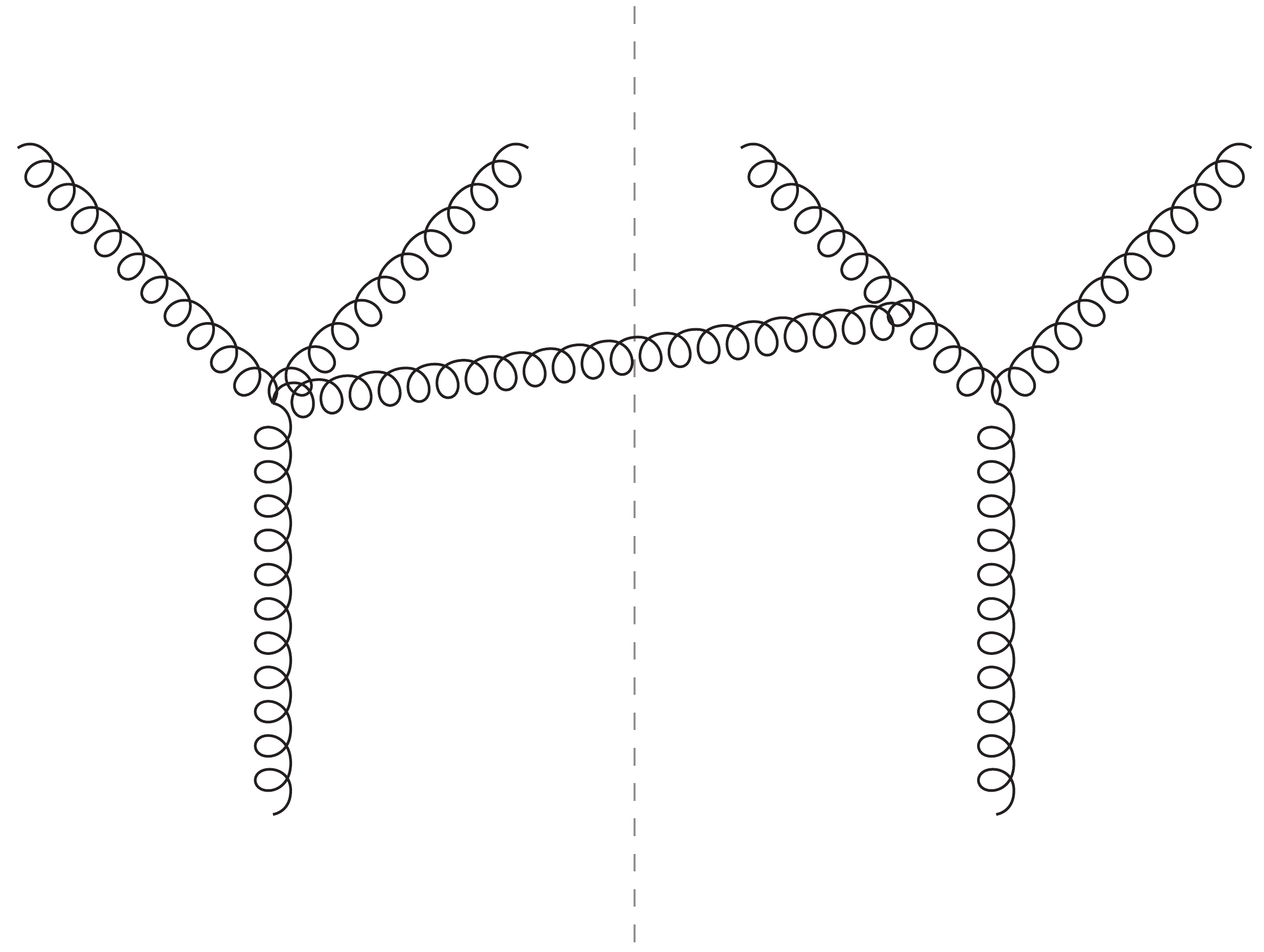}}
    \hspace{0.3em}
    \subfigure[\label{fig:ggg-4g2down}]{\includegraphics[width=0.23\textwidth]{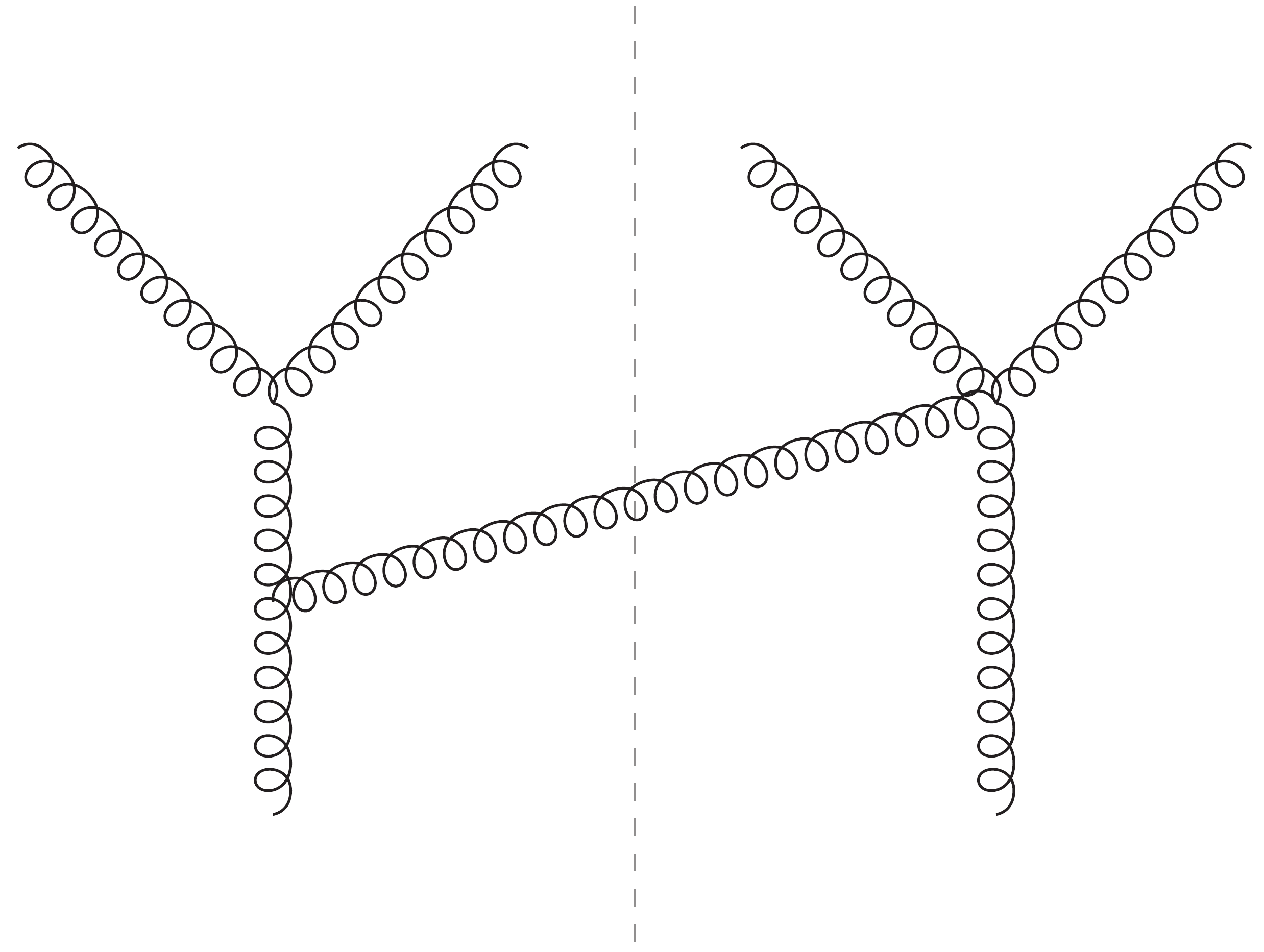}}
\\[2em]
    \subfigure[\label{fig:gqqbar-LD}]{\includegraphics[width=0.23\textwidth]{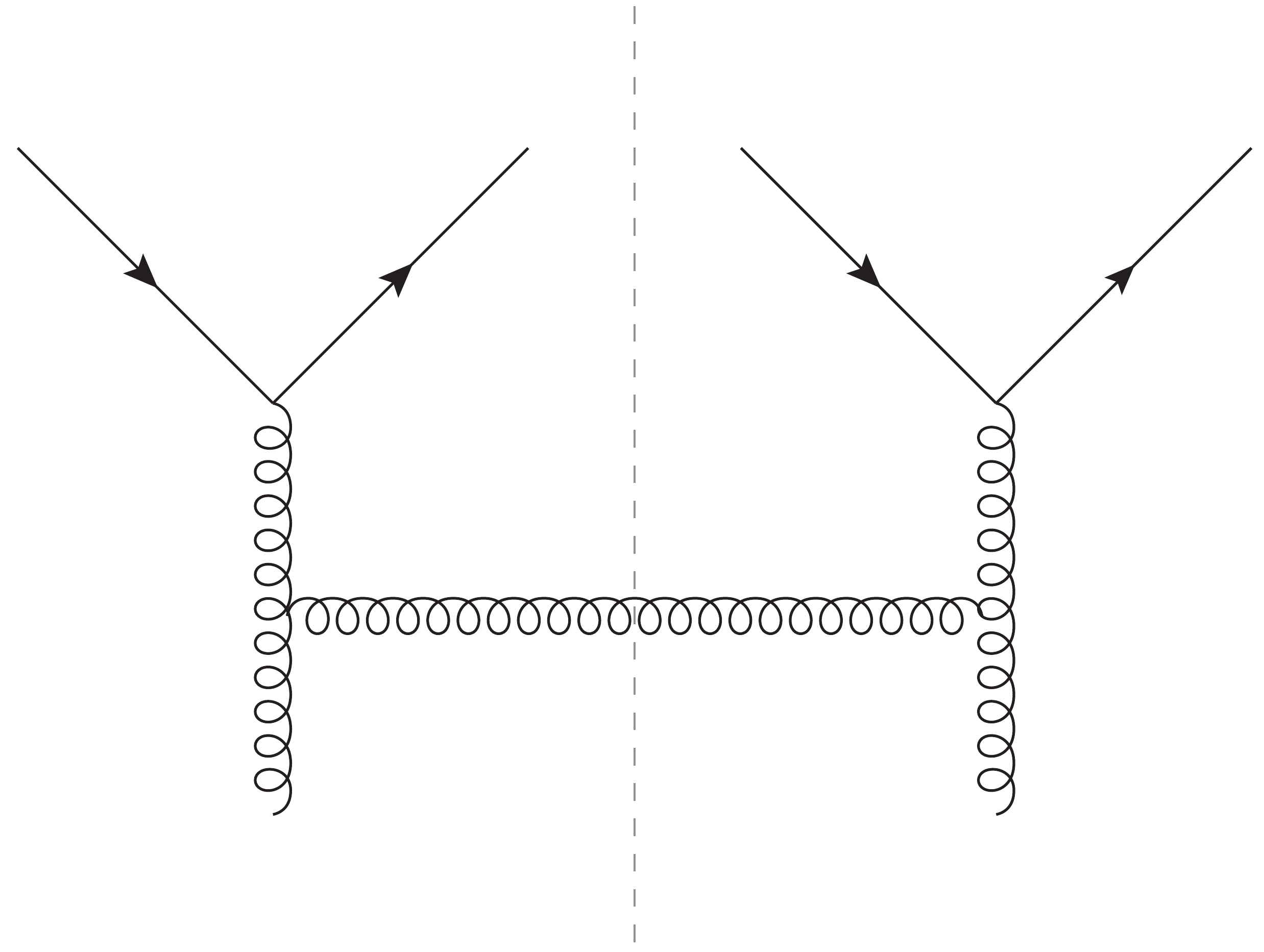}}
    \hspace{0.3em}
    \subfigure[]{\includegraphics[width=0.23\textwidth]{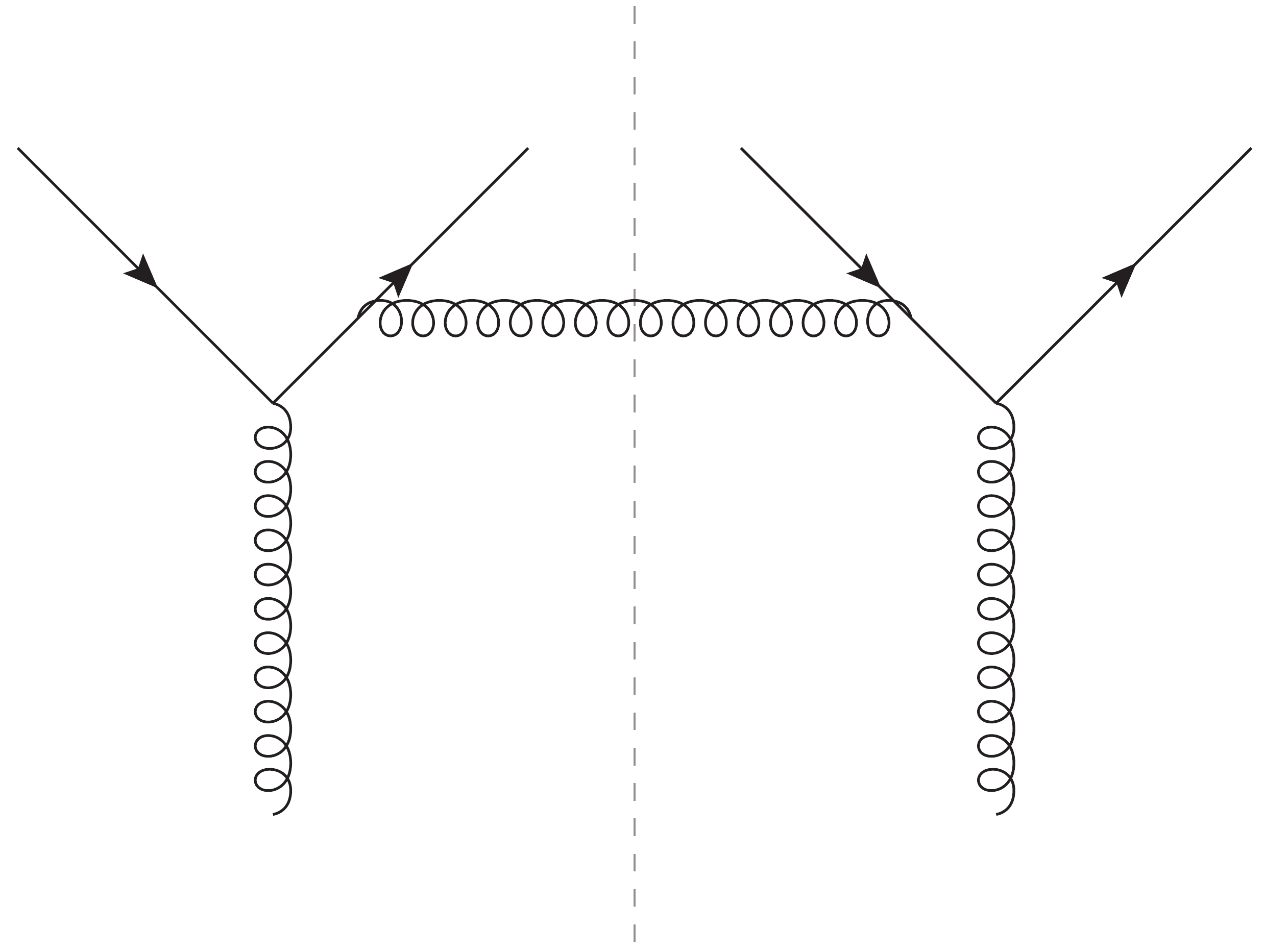}}
    \hspace{0.3em}
    \subfigure[UND \label{fig:gqqbar-UND}]{\includegraphics[width=0.23\textwidth]{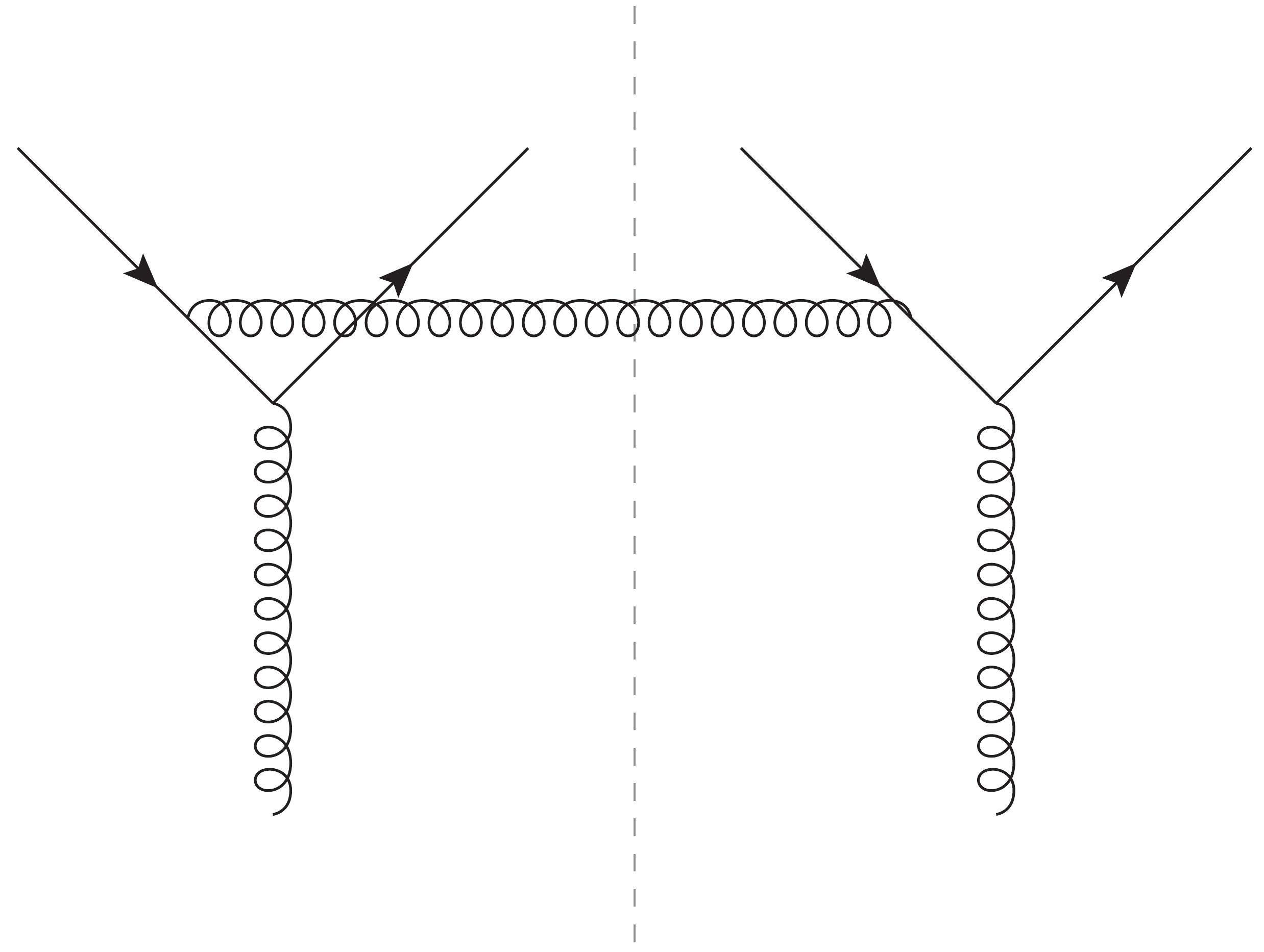}}
    \hspace{0.3em}
    \subfigure[]{\includegraphics[width=0.23\textwidth]{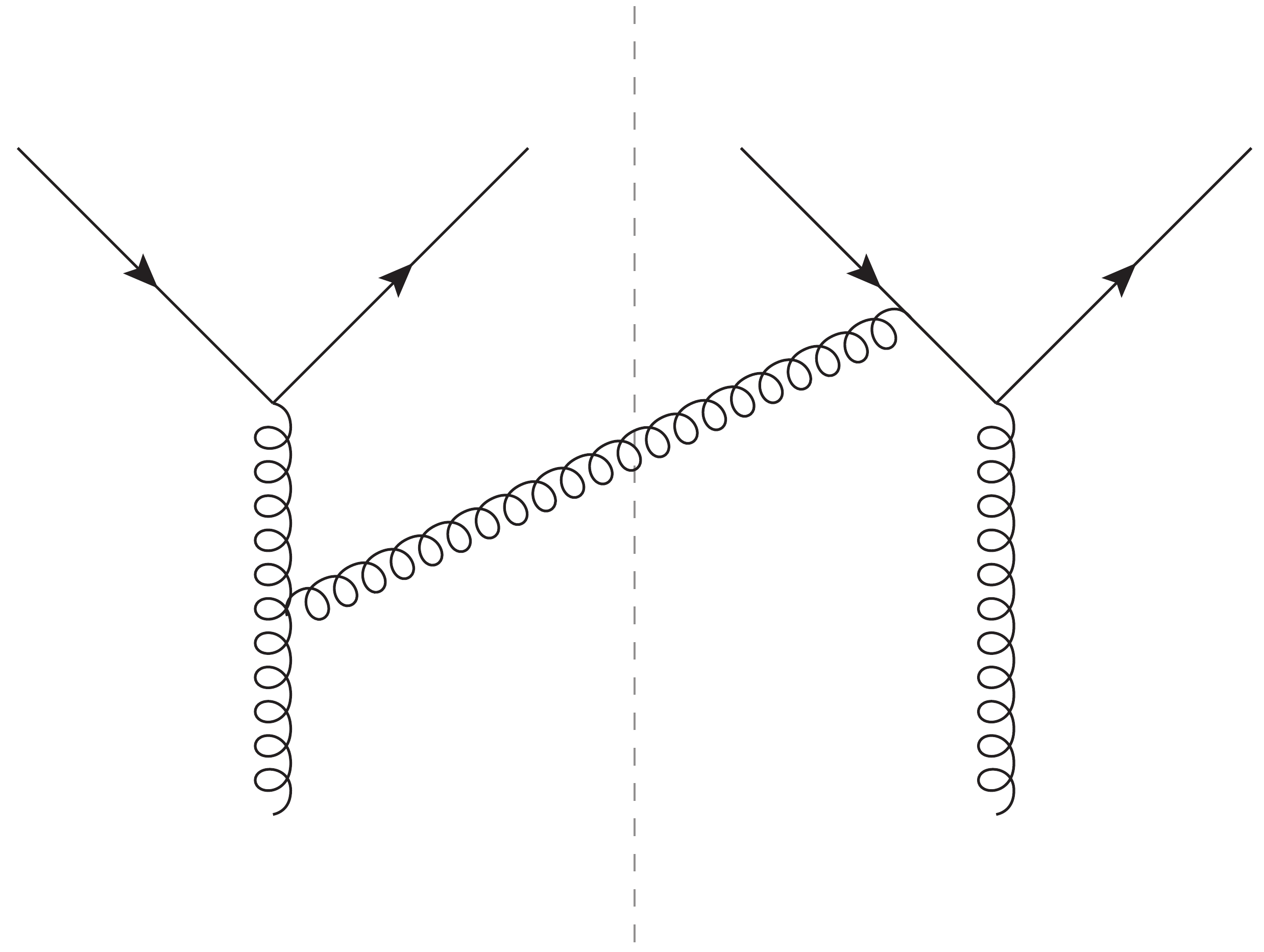}}
\\[2em]
    \subfigure[\label{fig:qgq-LD}]{\includegraphics[width=0.23\textwidth]{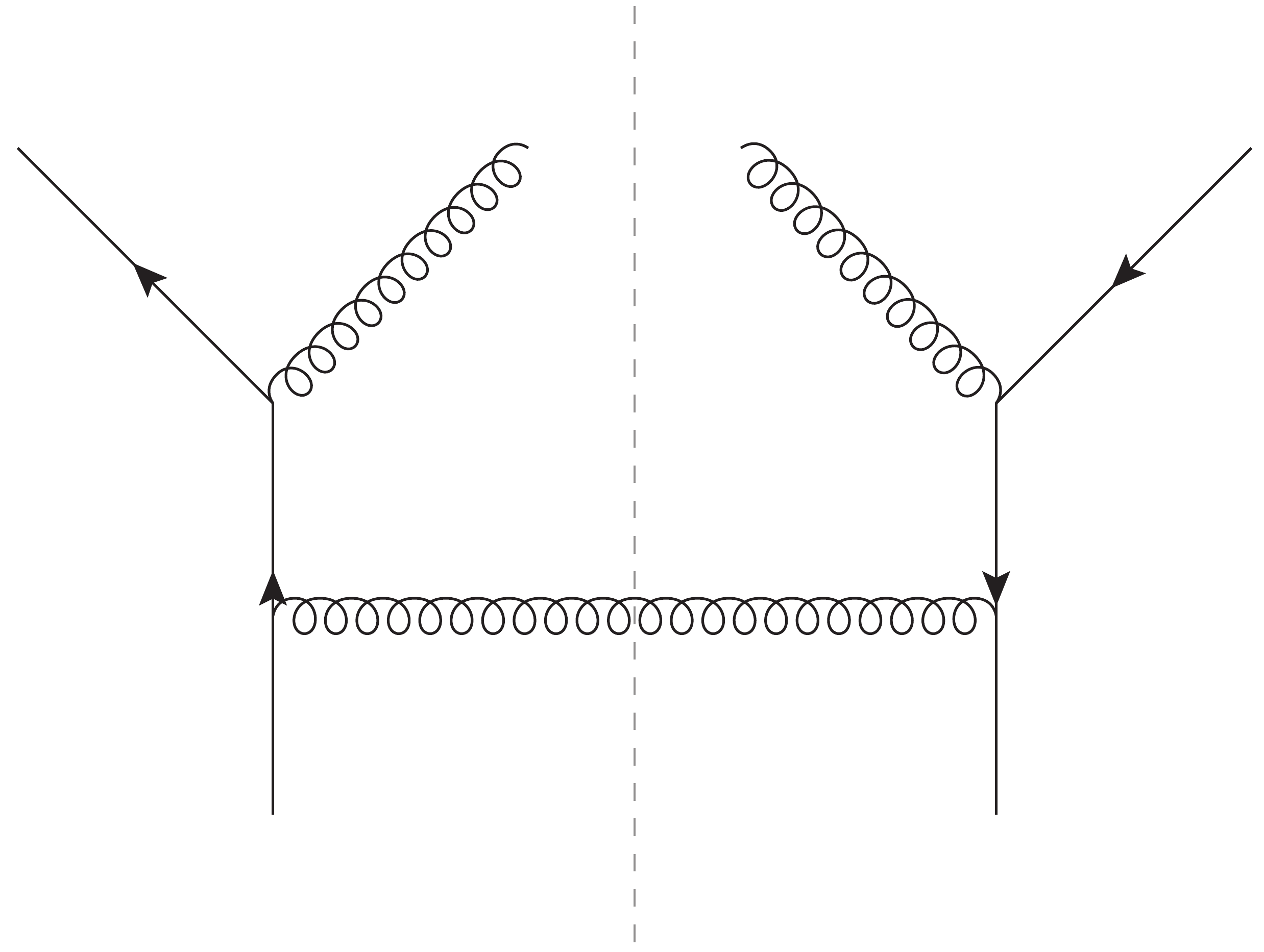}}
    \hspace{0.3em}
    \subfigure[]{\includegraphics[width=0.23\textwidth]{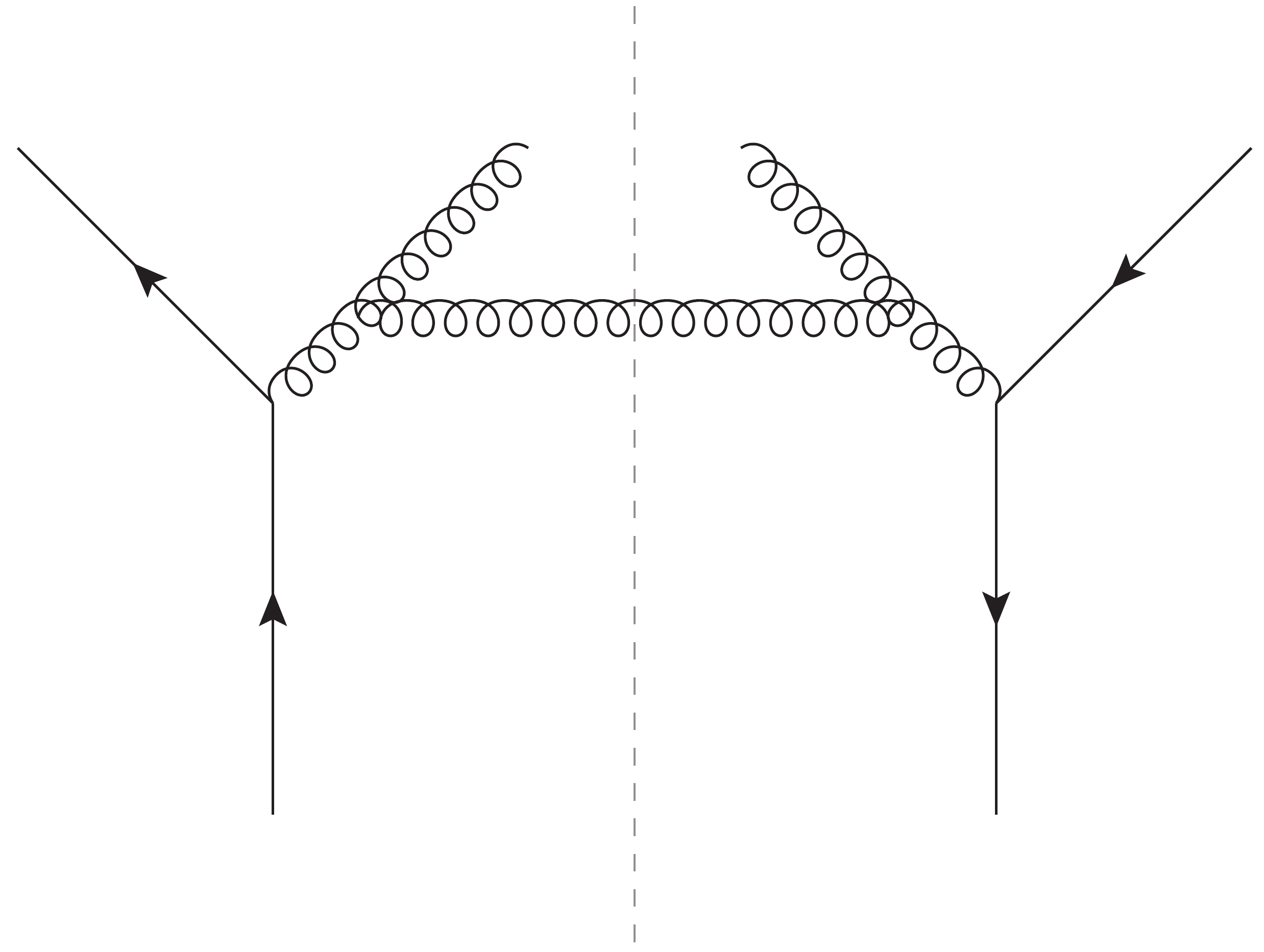}}
    \hspace{0.3em}
    \subfigure[]{\includegraphics[width=0.23\textwidth]{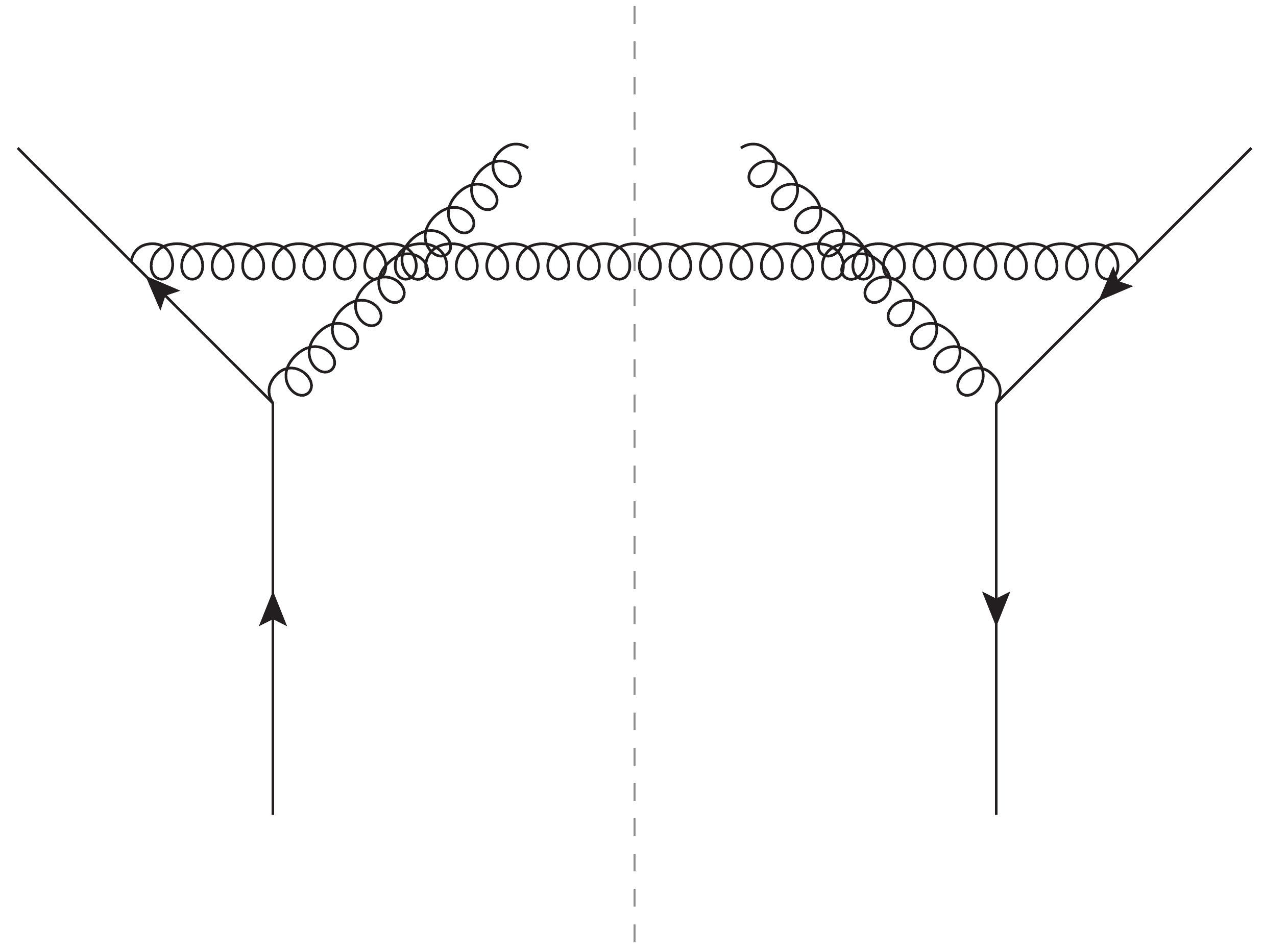}}
    \hspace{0.3em}
    \subfigure[UND \label{fig:qgq-UND}]{\includegraphics[width=0.23\textwidth]{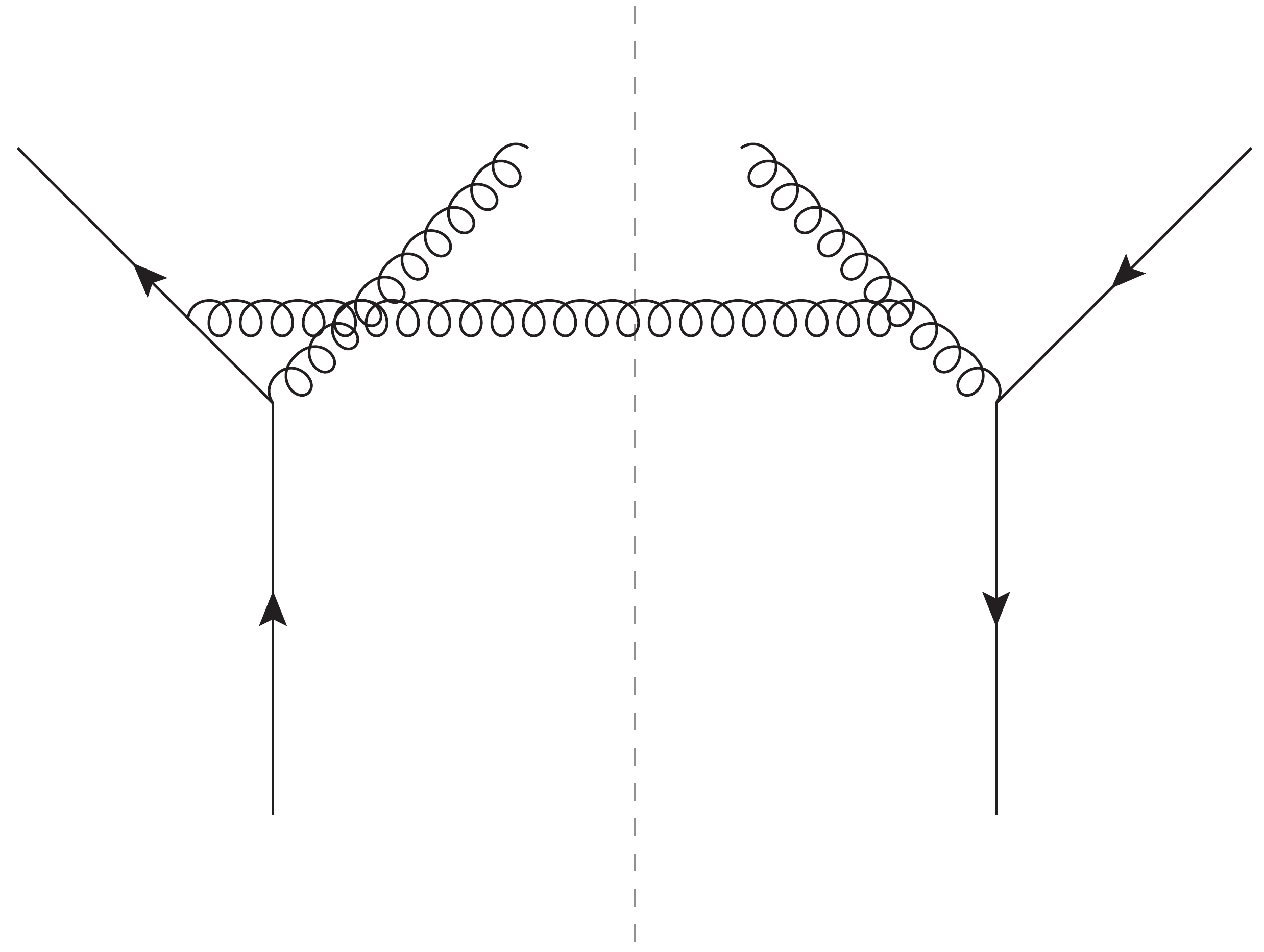}}
\\[2em]
    \subfigure[]{\includegraphics[width=0.23\textwidth]{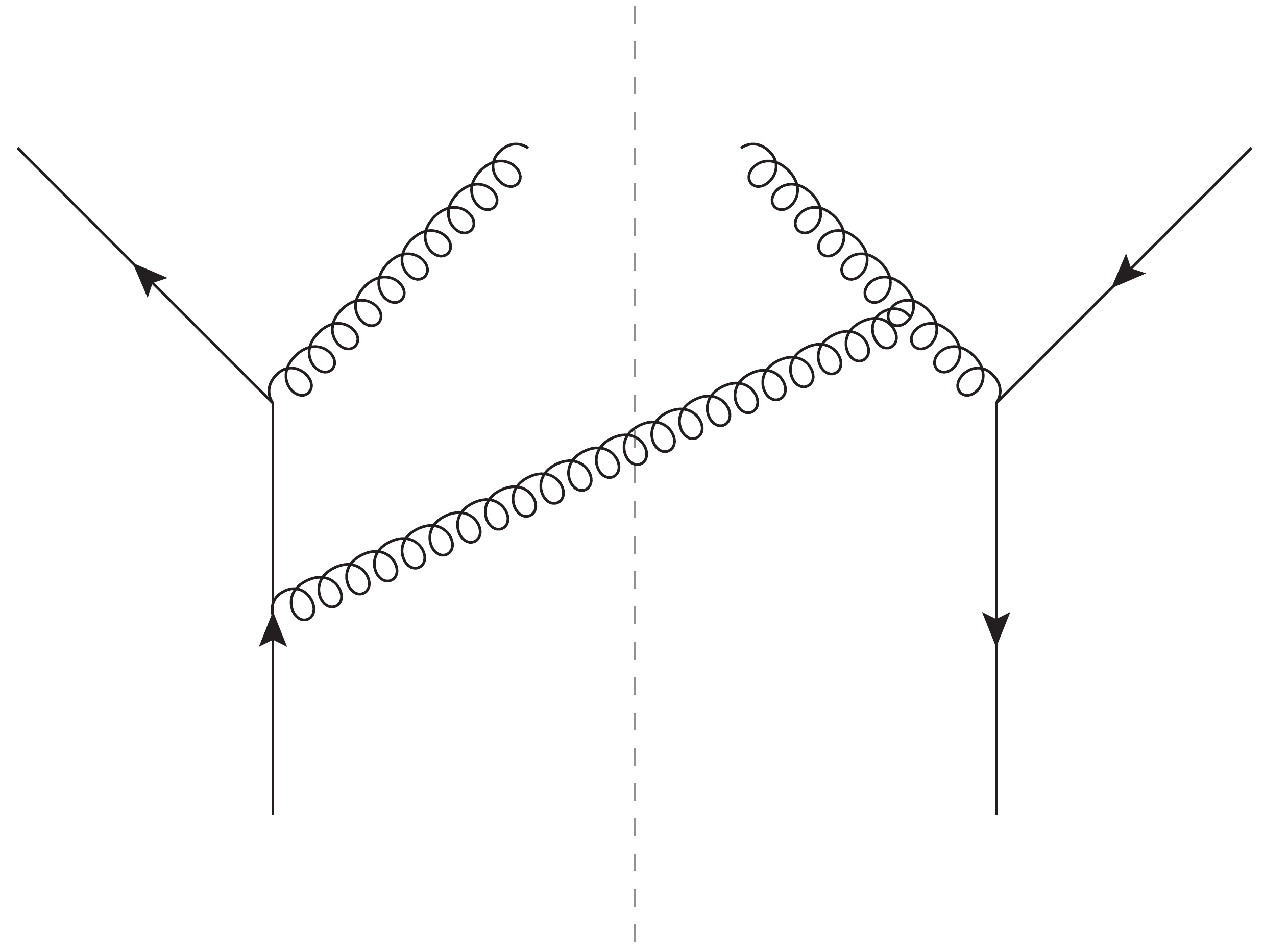}}
    \hspace{0.3em}
    \subfigure[]{\includegraphics[width=0.23\textwidth]{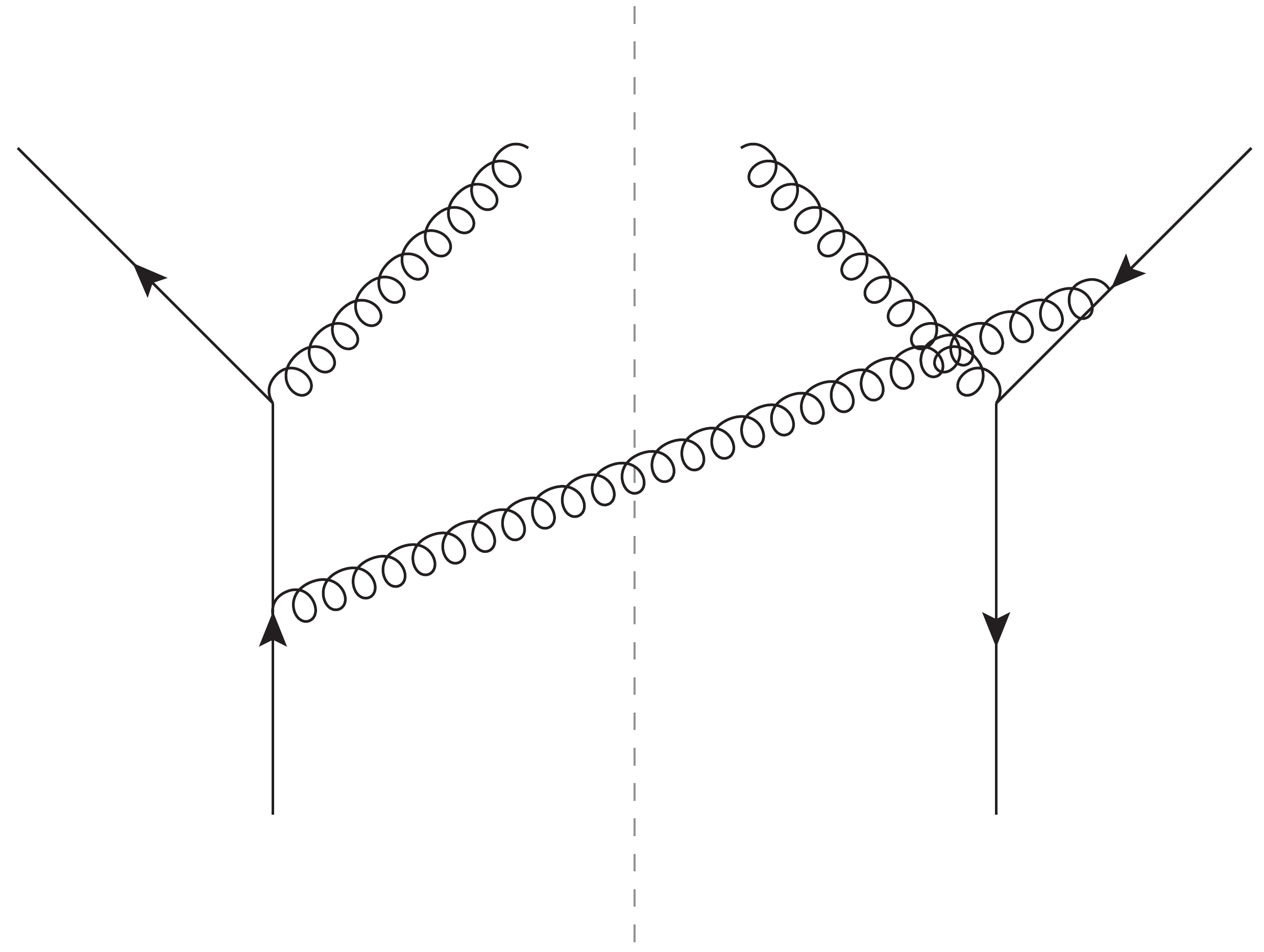}}
    \caption{\label{fig:real-LO} Real graphs for LO channels.  The parton lines at the bottom of the graph correspond to $a_0$, and those on top to $a_1$, $a_2$, $a_2$ and $a_1$ from left to right.  Additional graphs are obtained by interchanging $a_1 \leftrightarrow a_2$ and by reflection w.r.t.\ the final state cut.  Graphs with topology UND (upper non-diagonal) are marked here and further discussed in the text.}
  \end{center}
\end{figure}

\begin{figure}
  \begin{center}
    \subfigure[\label{fig:gqg-LD}]{\includegraphics[width=0.23\textwidth]{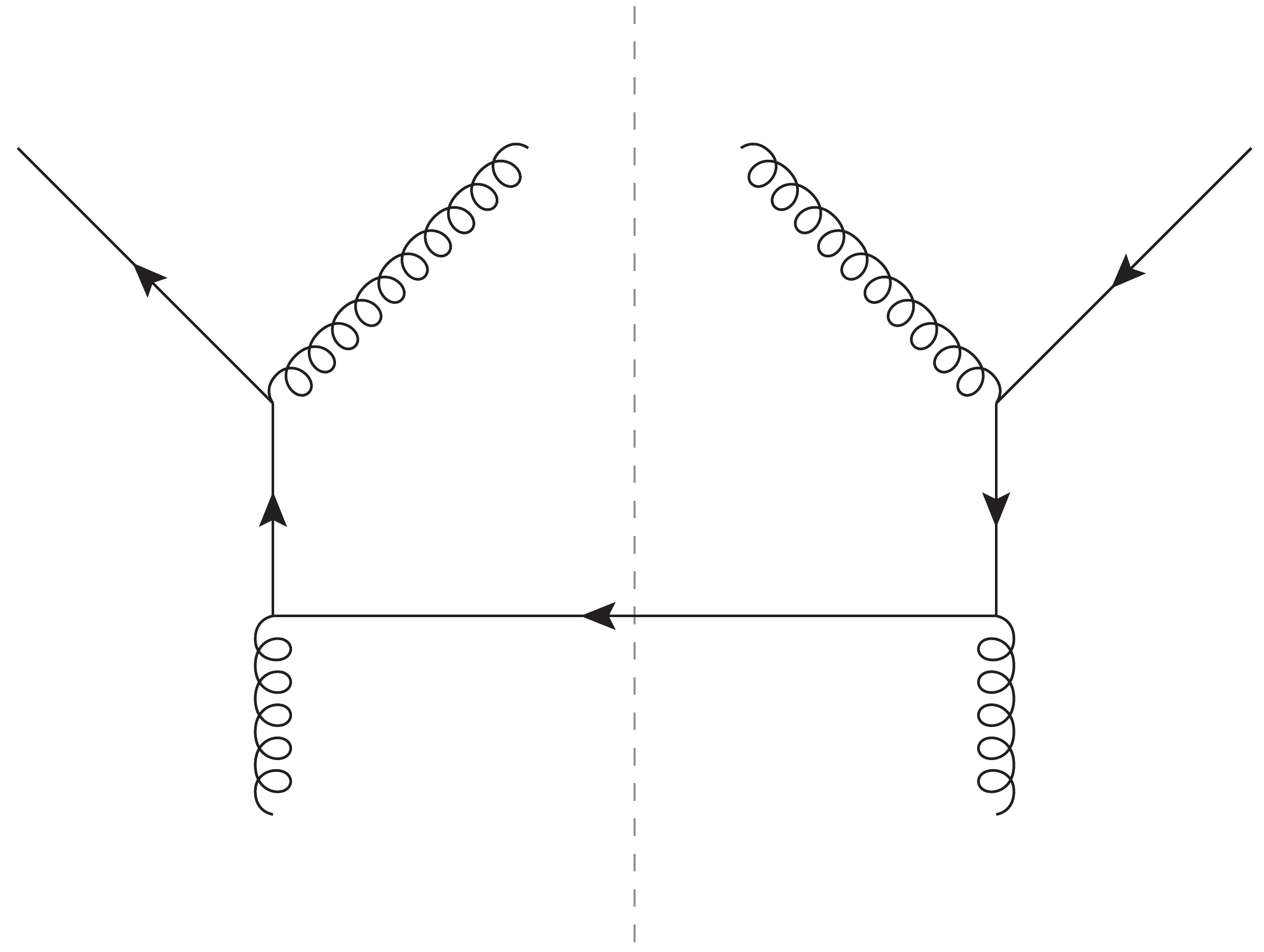}}
    \hspace{0.3em}
    \subfigure[\label{fig:gqg-UD}]{\includegraphics[width=0.23\textwidth]{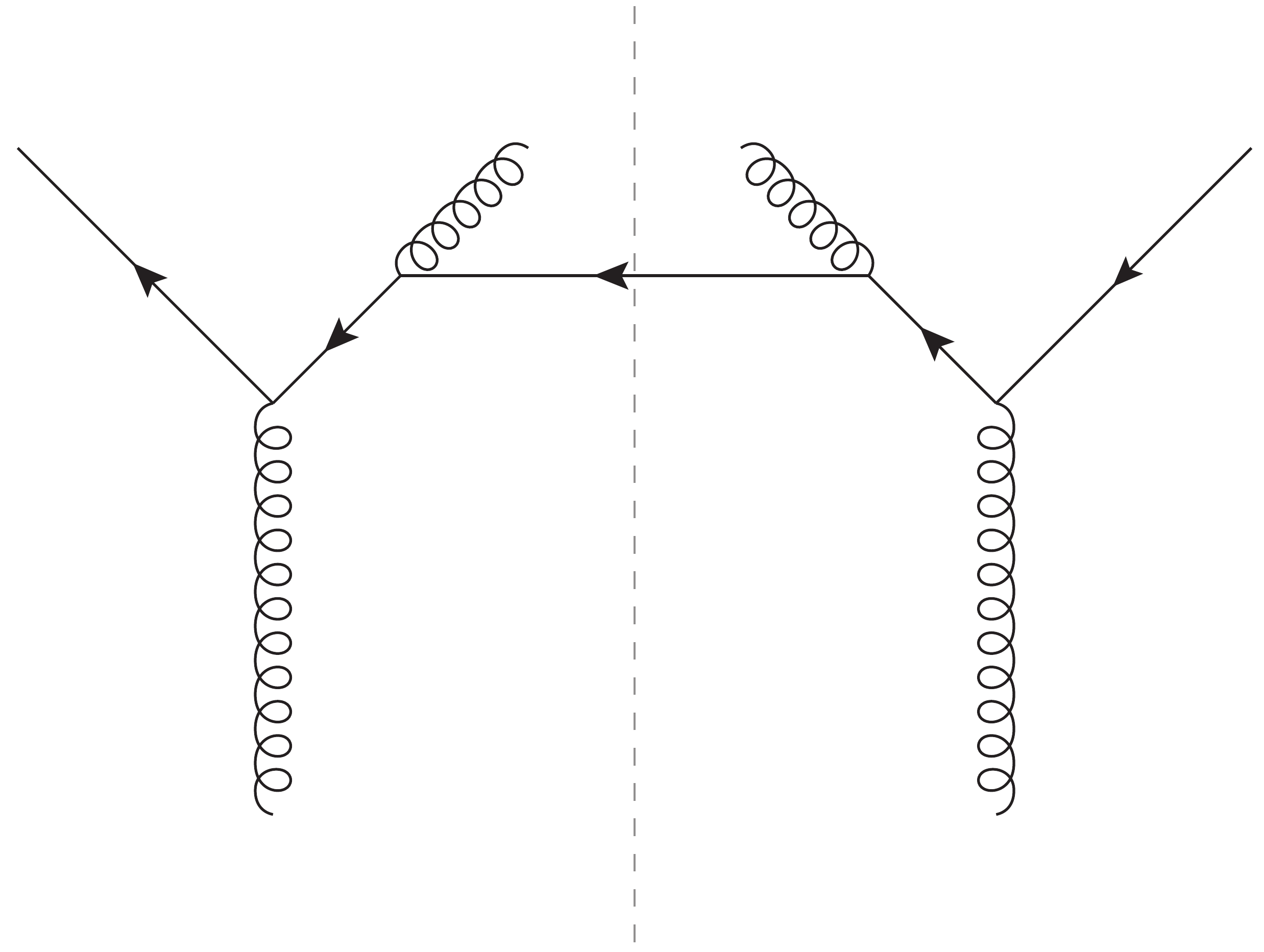}}
    \hspace{0.3em}
    \subfigure[\label{fig:gqg-UD-2}]{\includegraphics[width=0.23\textwidth]{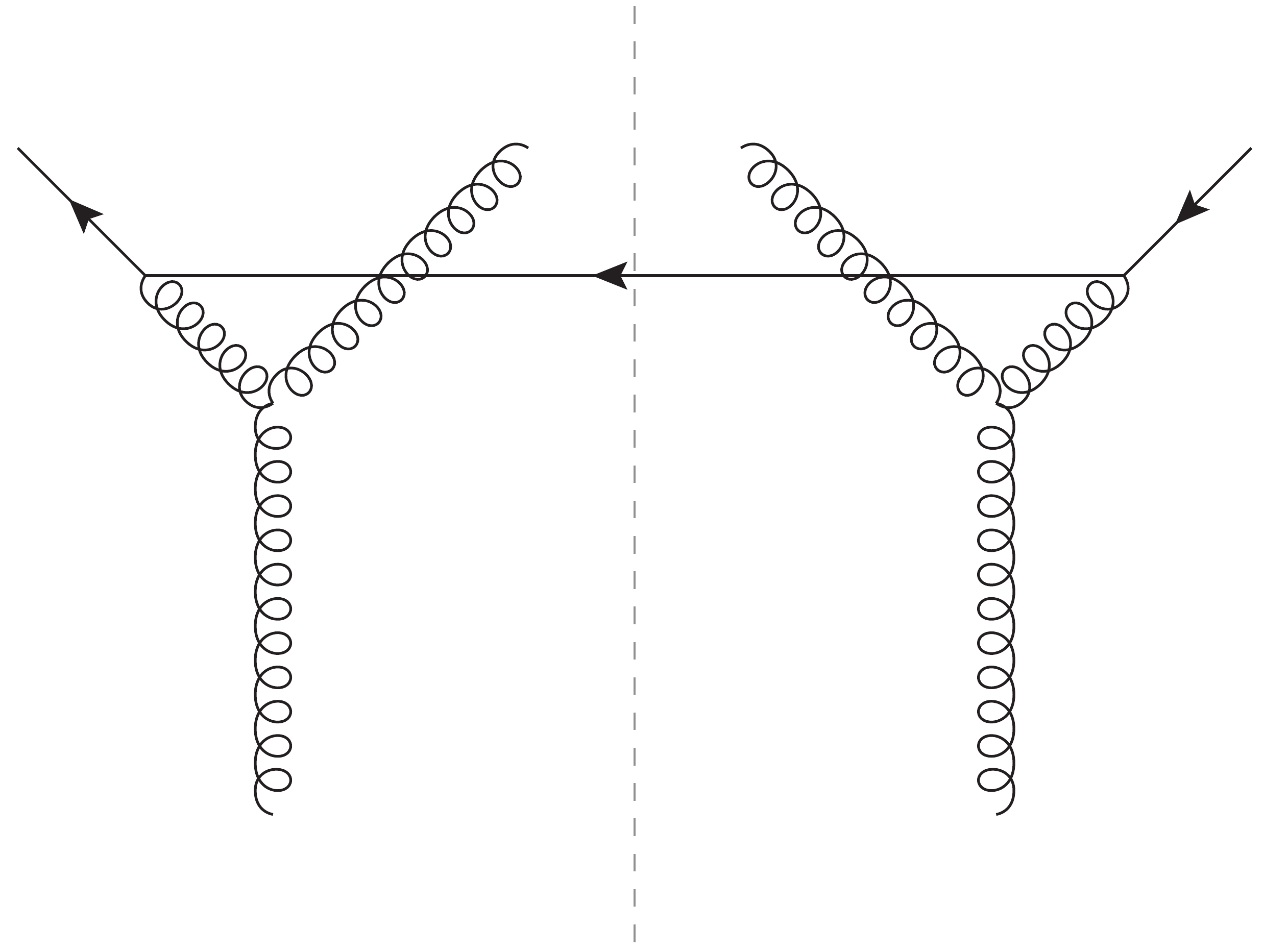}}
    \hspace{0.3em}
    \subfigure[\label{fig:gqg-UND}]{\includegraphics[width=0.23\textwidth]{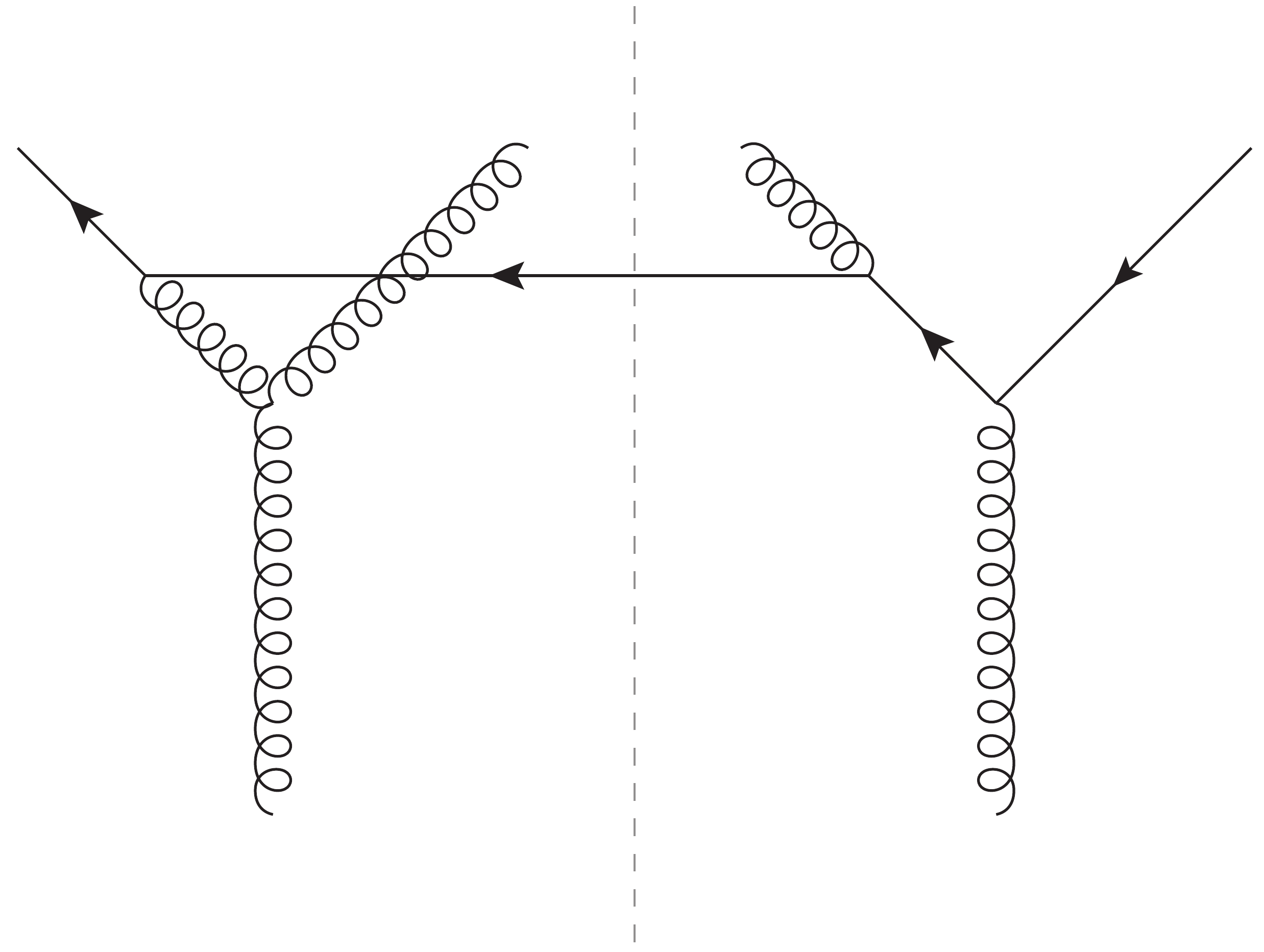}}
\\[2em]
    \subfigure[]{\includegraphics[width=0.23\textwidth]{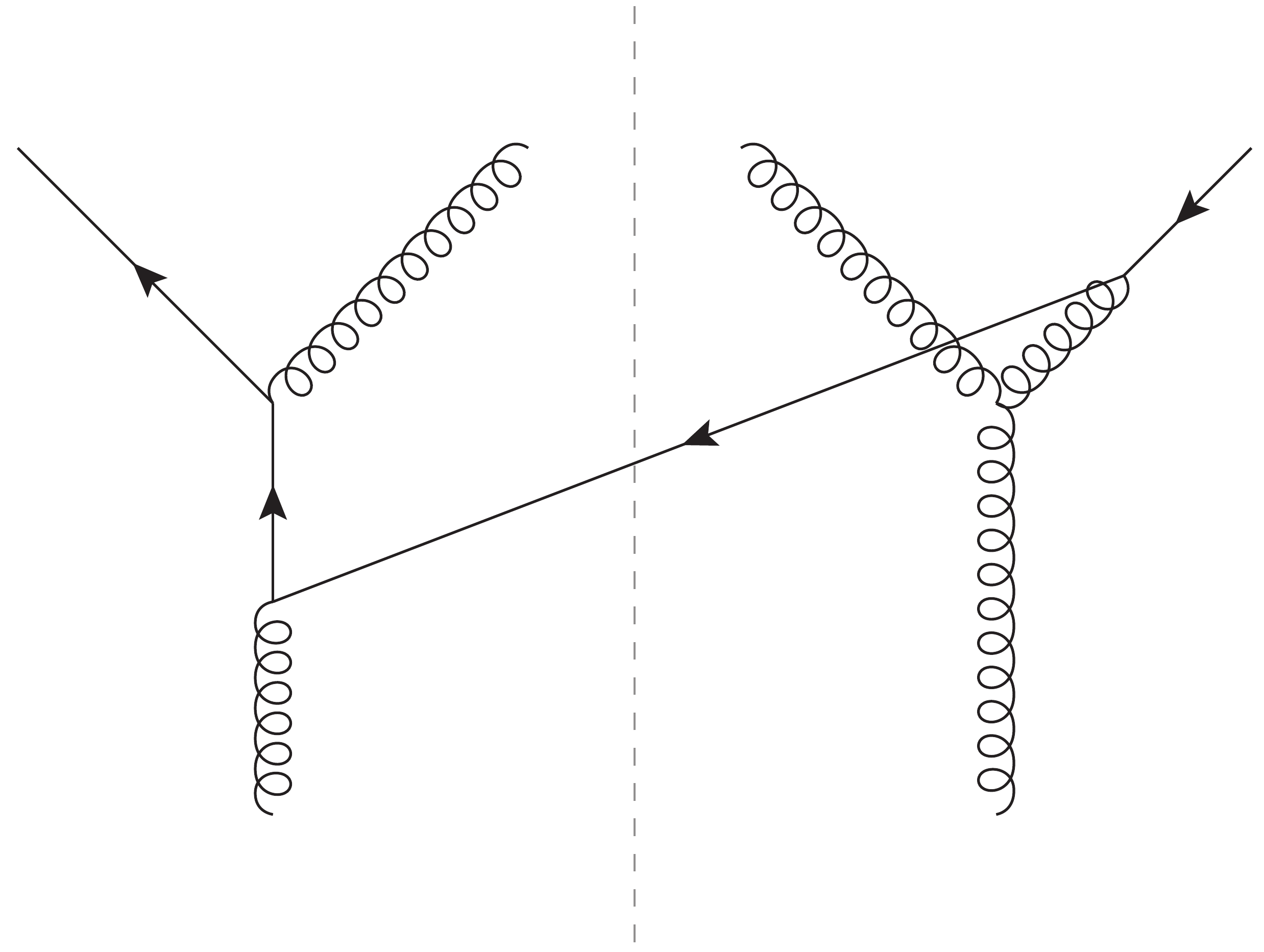}}
    \hspace{0.3em}
    \subfigure[]{\includegraphics[width=0.23\textwidth]{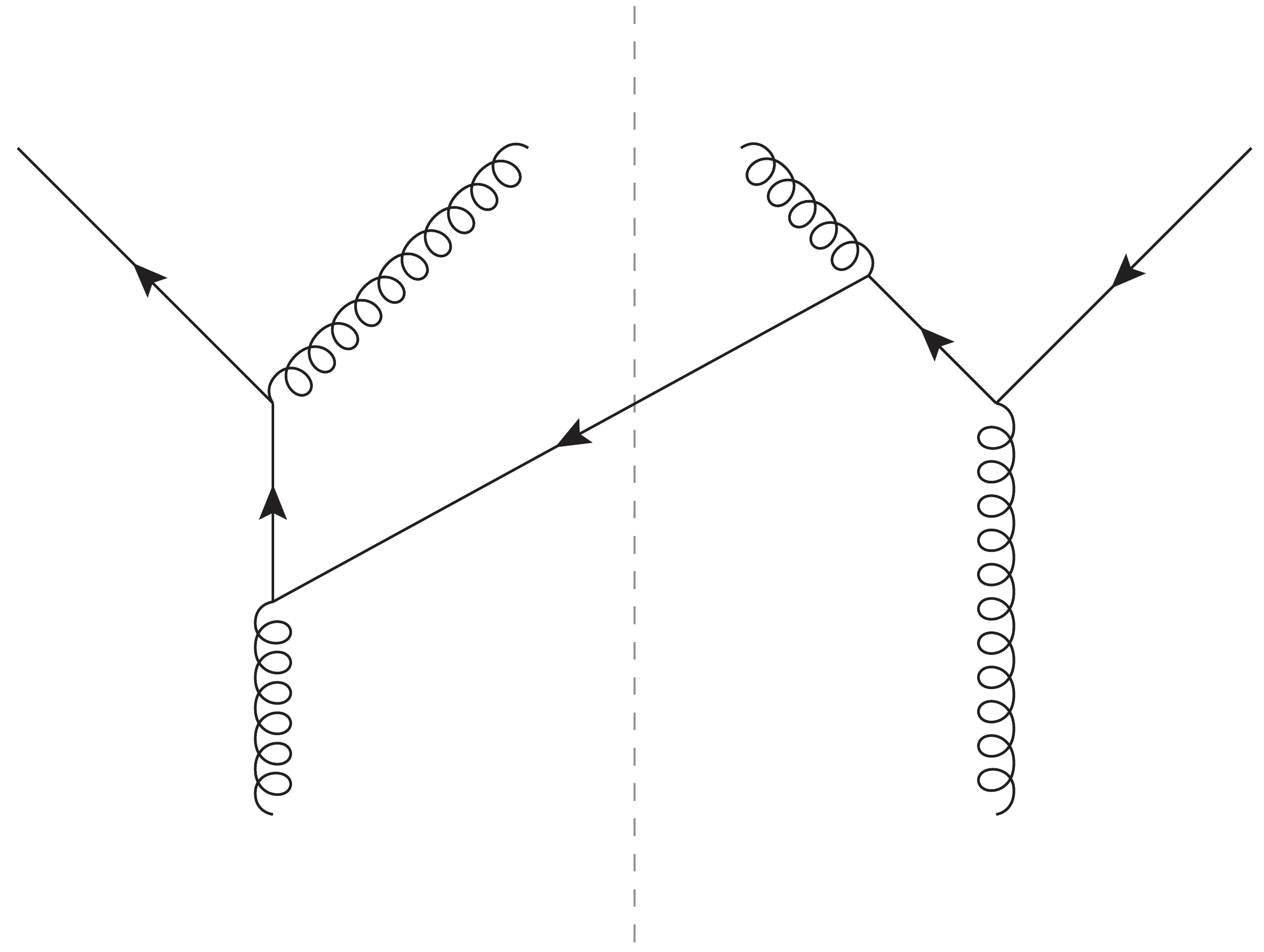}}
\\[2em]
    \subfigure[\label{fig:qgg-LD}]{\includegraphics[width=0.23\textwidth]{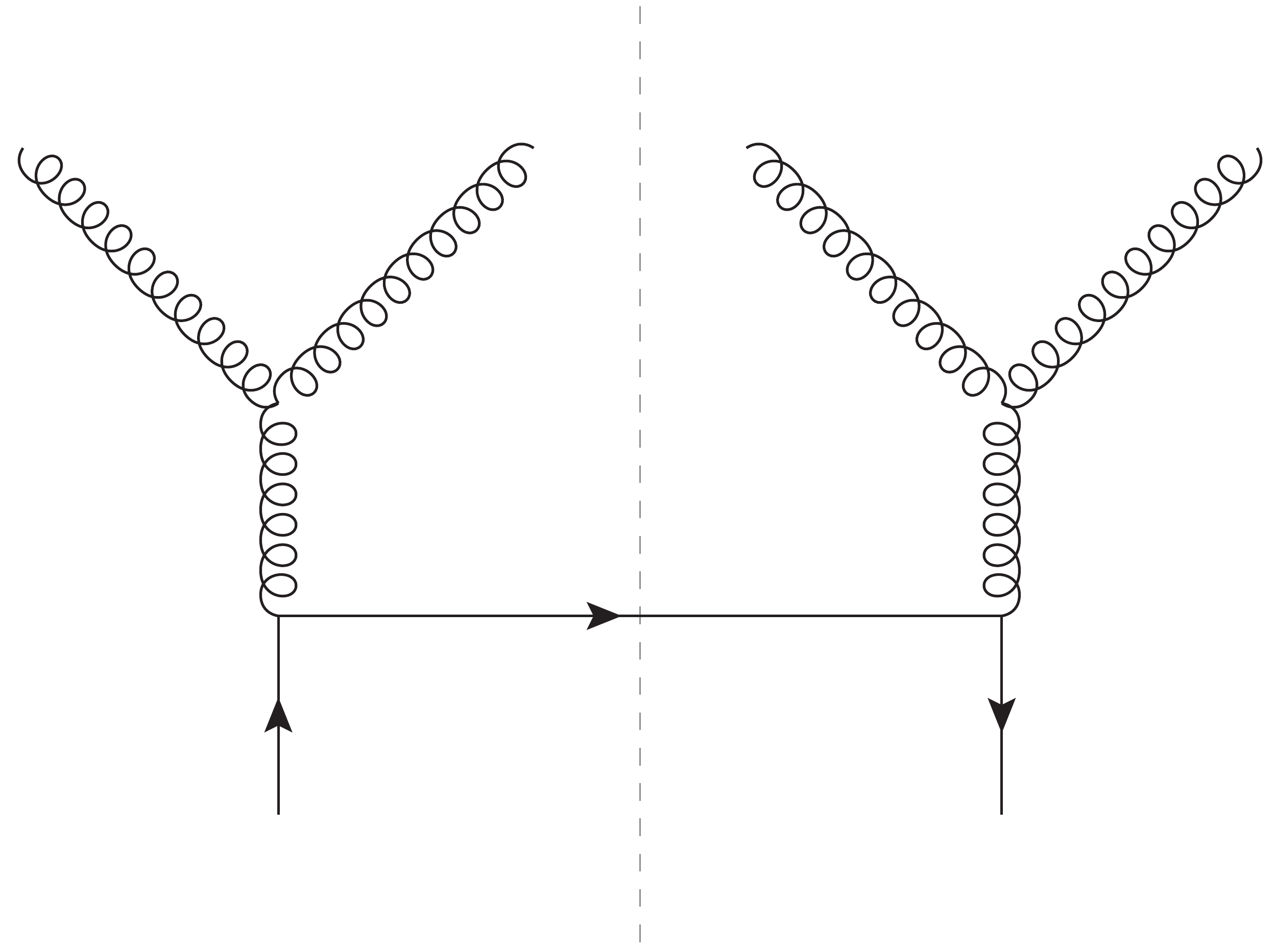}}
    \hspace{0.3em}
    \subfigure[\label{fig:qgg-UD}]{\includegraphics[width=0.23\textwidth]{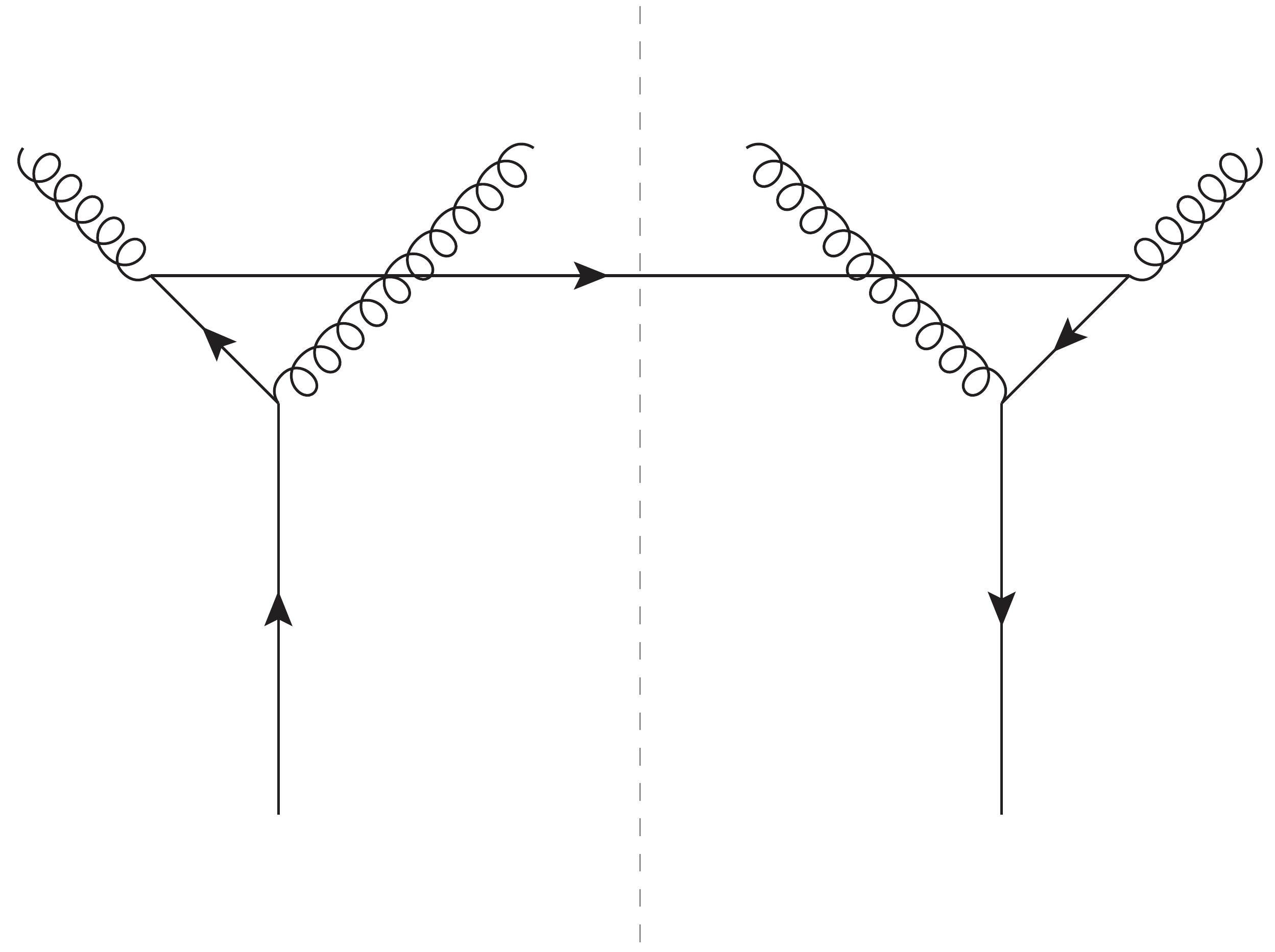}}
    \hspace{0.3em}
    \subfigure[\label{fig:qgg-UND}]{\includegraphics[width=0.23\textwidth]{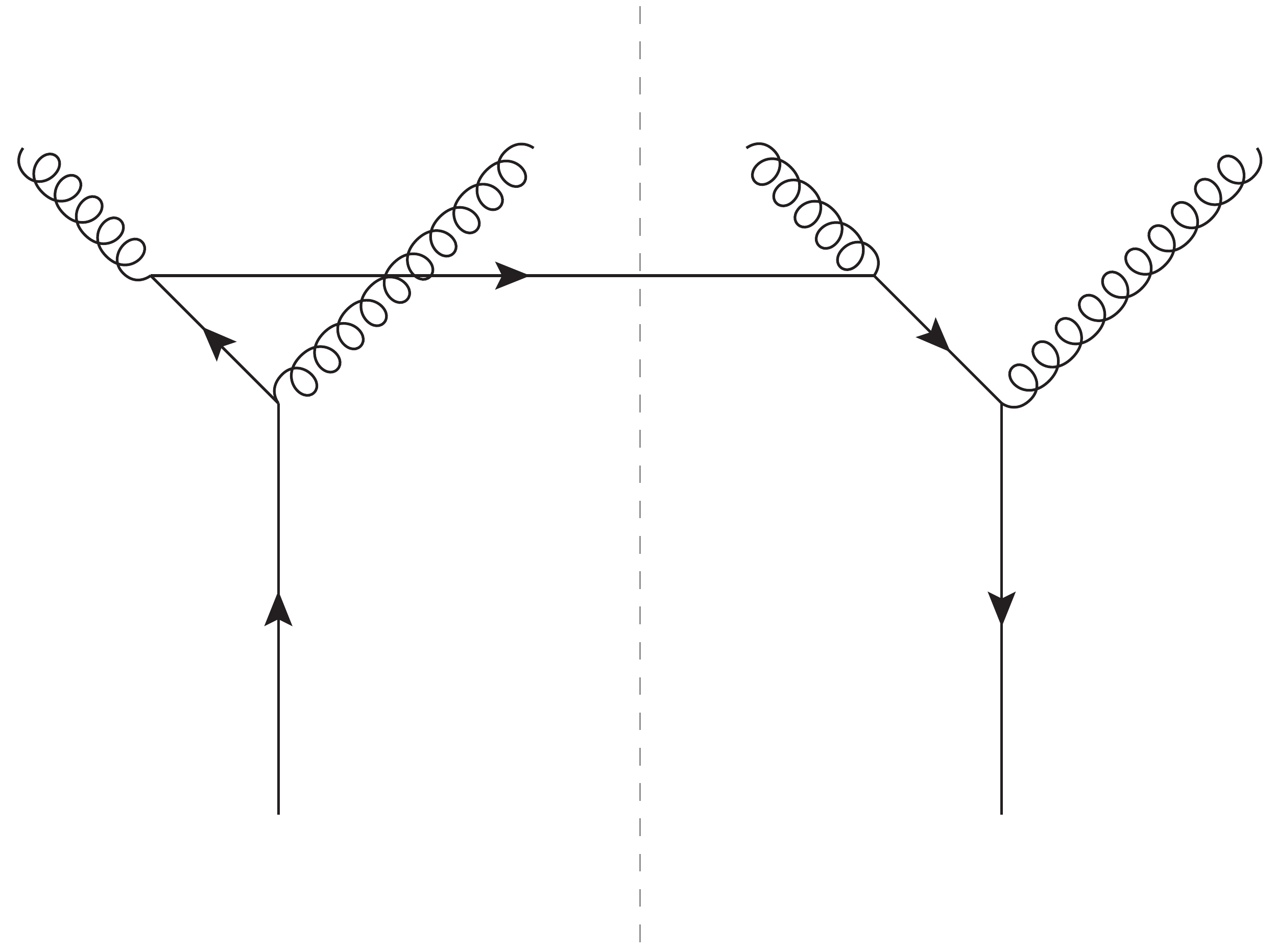}}
    \hspace{0.3em}
    \subfigure[\label{fig:qgg-T2B}]{\includegraphics[width=0.23\textwidth]{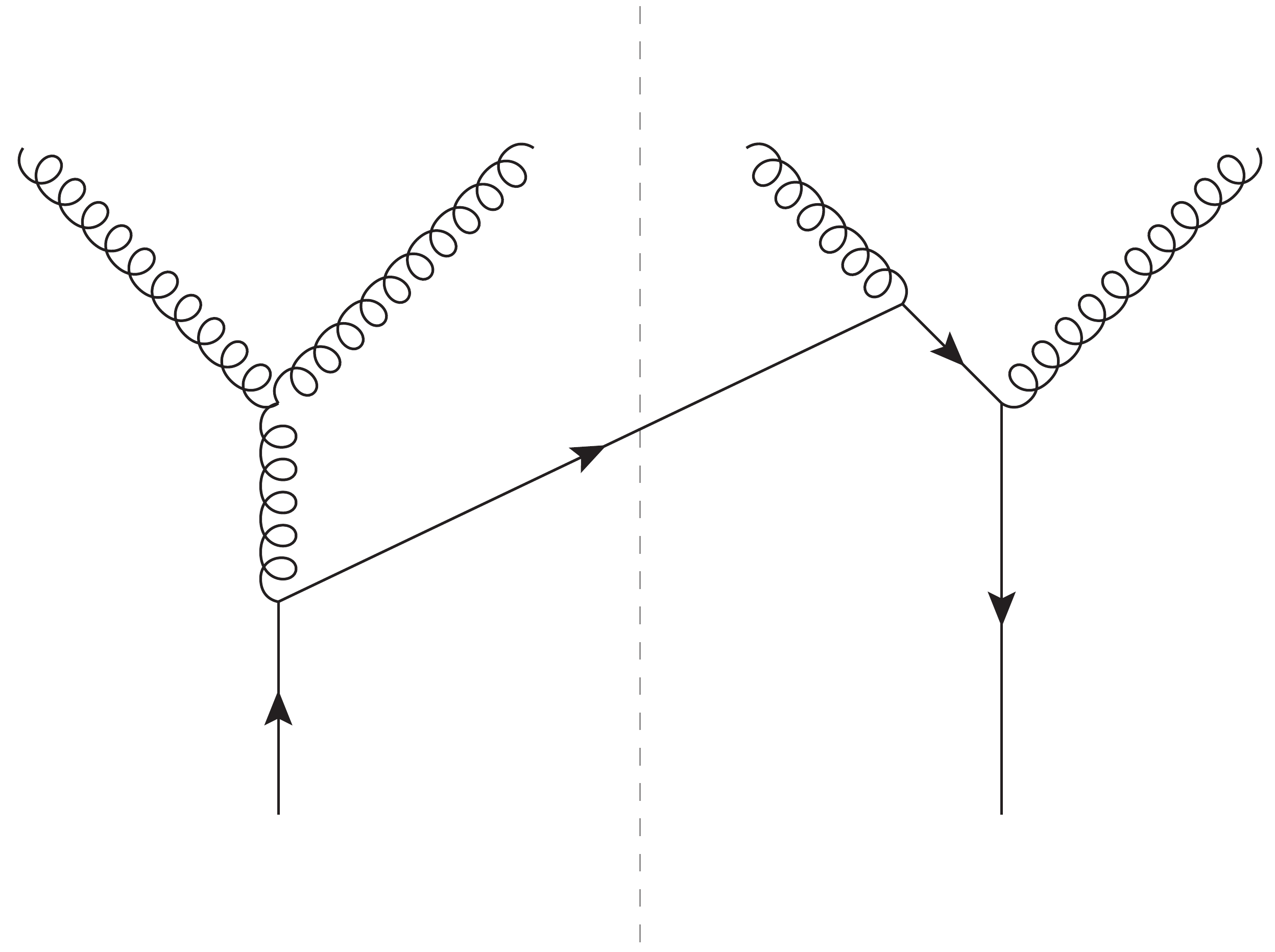}}
\\[2em]
    \subfigure[\label{fig:qqbar-1} $V_{q'\! \bar{q}'\!, q}$]{\includegraphics[width=0.23\textwidth]{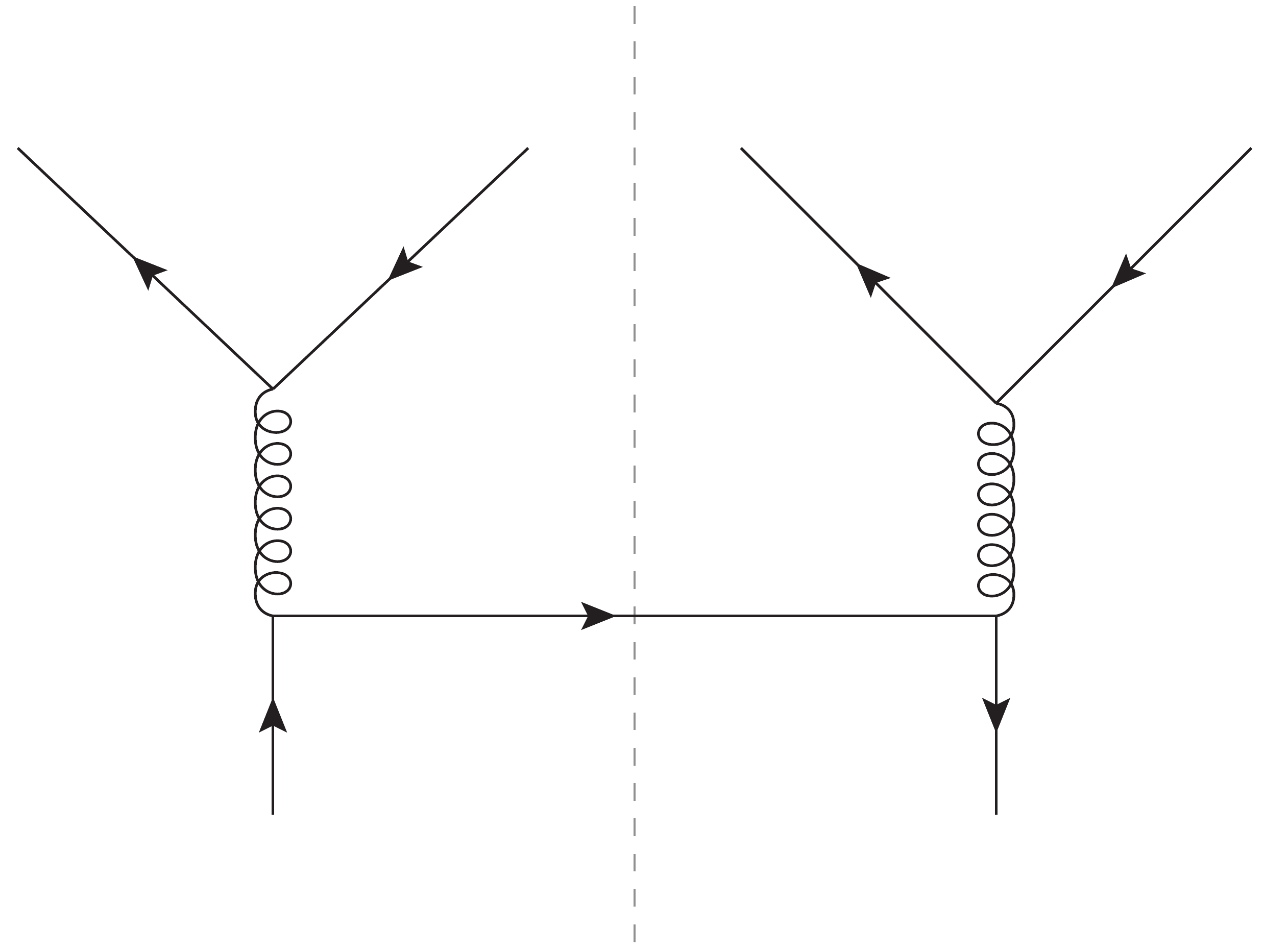}}
    \hspace{0.3em}
    \subfigure[\label{fig:qqbar-2} $V_{q \bar{q}'\!,q}$]{\includegraphics[width=0.23\textwidth]{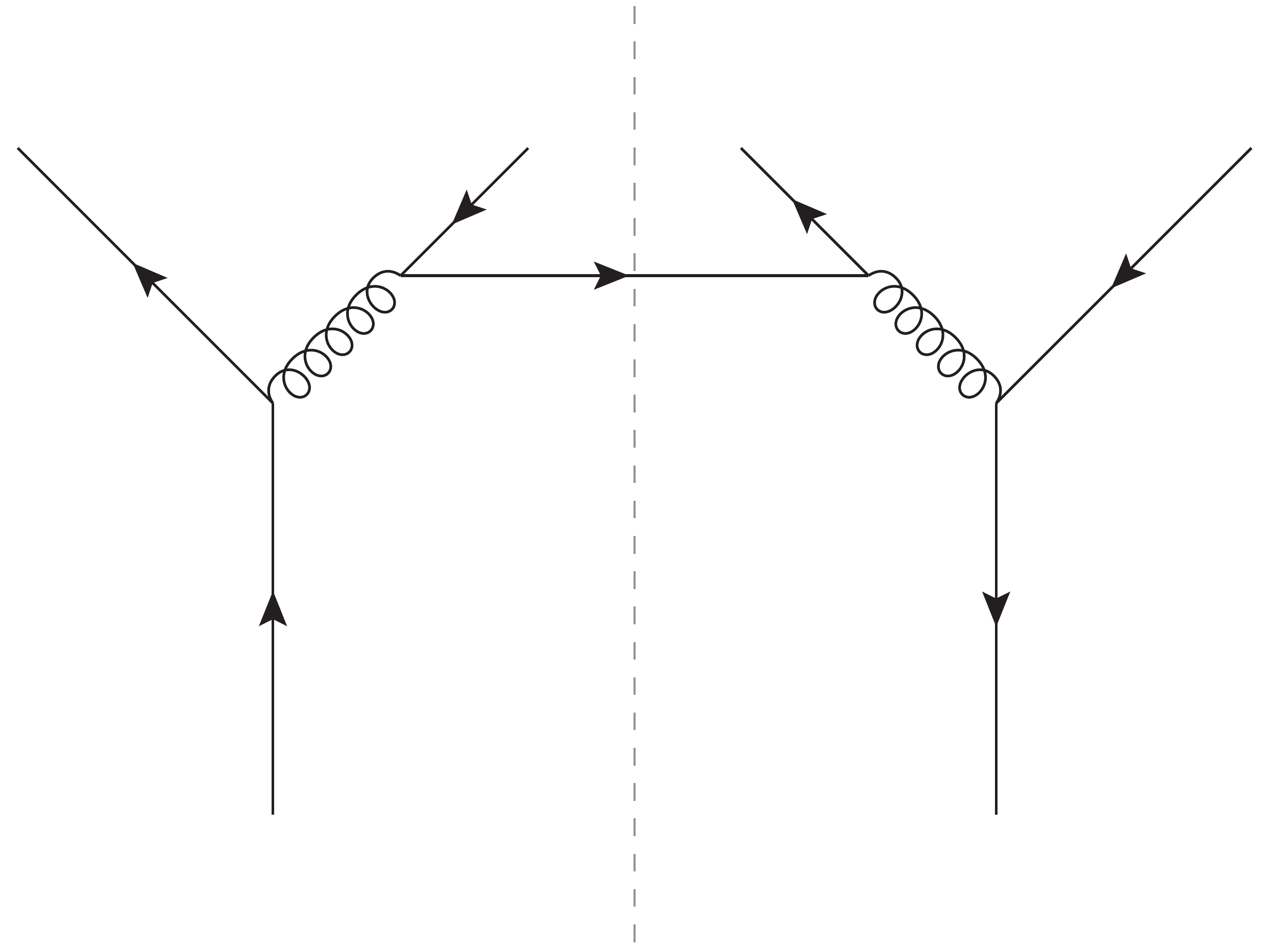}}
    \hspace{0.3em}
    \subfigure[\label{fig:qqbar-3} $V_{q \bar{q},q}^v$]{\includegraphics[width=0.23\textwidth]{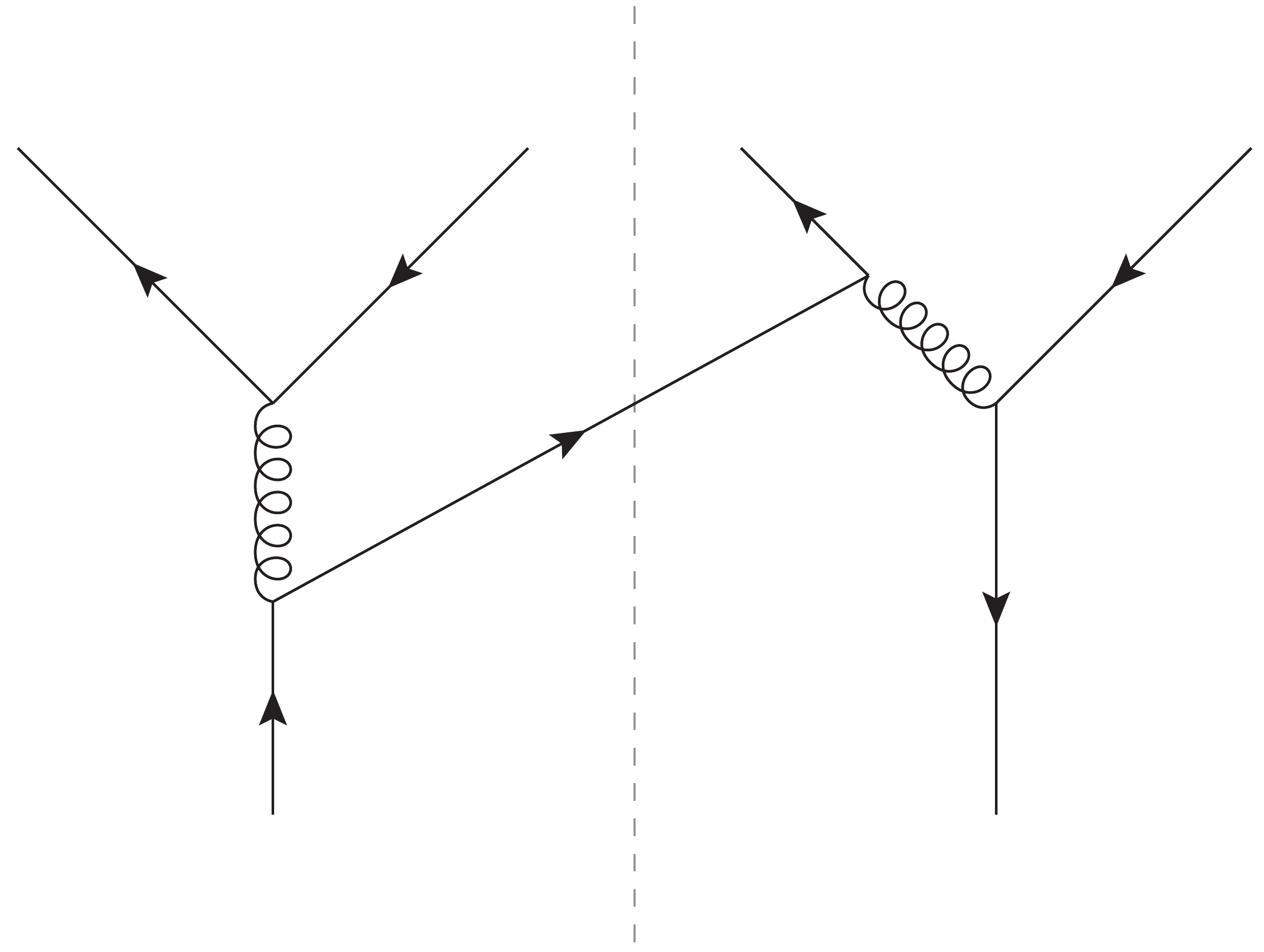}}
\\[2em]
    \subfigure[\label{fig:qq-1} $V_{q q'\!,q}$]{\includegraphics[width=0.23\textwidth]{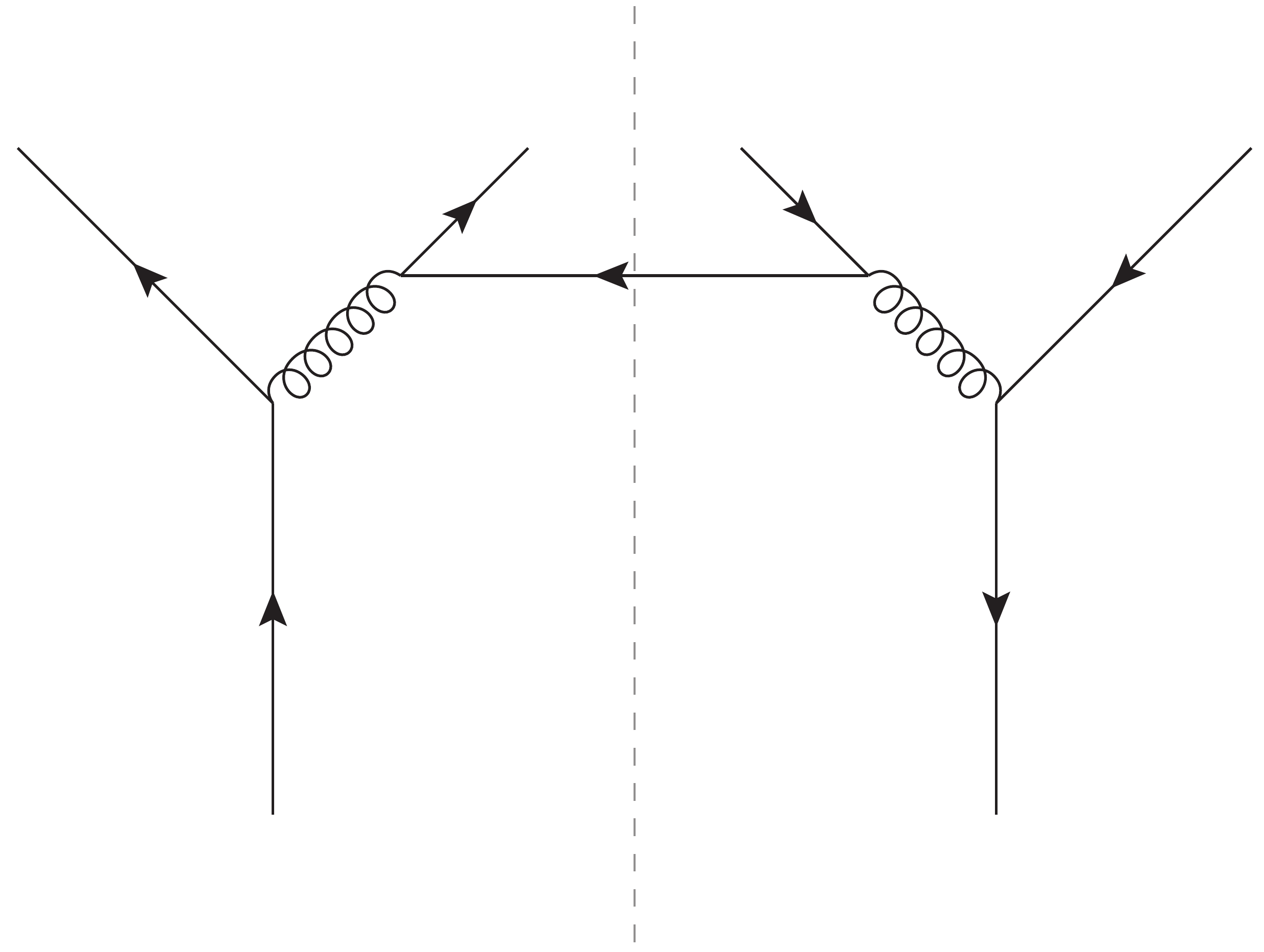}}
    \hspace{0.3em}
    \subfigure[\label{fig:qq-2} $V_{q q,q}^v$]{\includegraphics[width=0.23\textwidth]{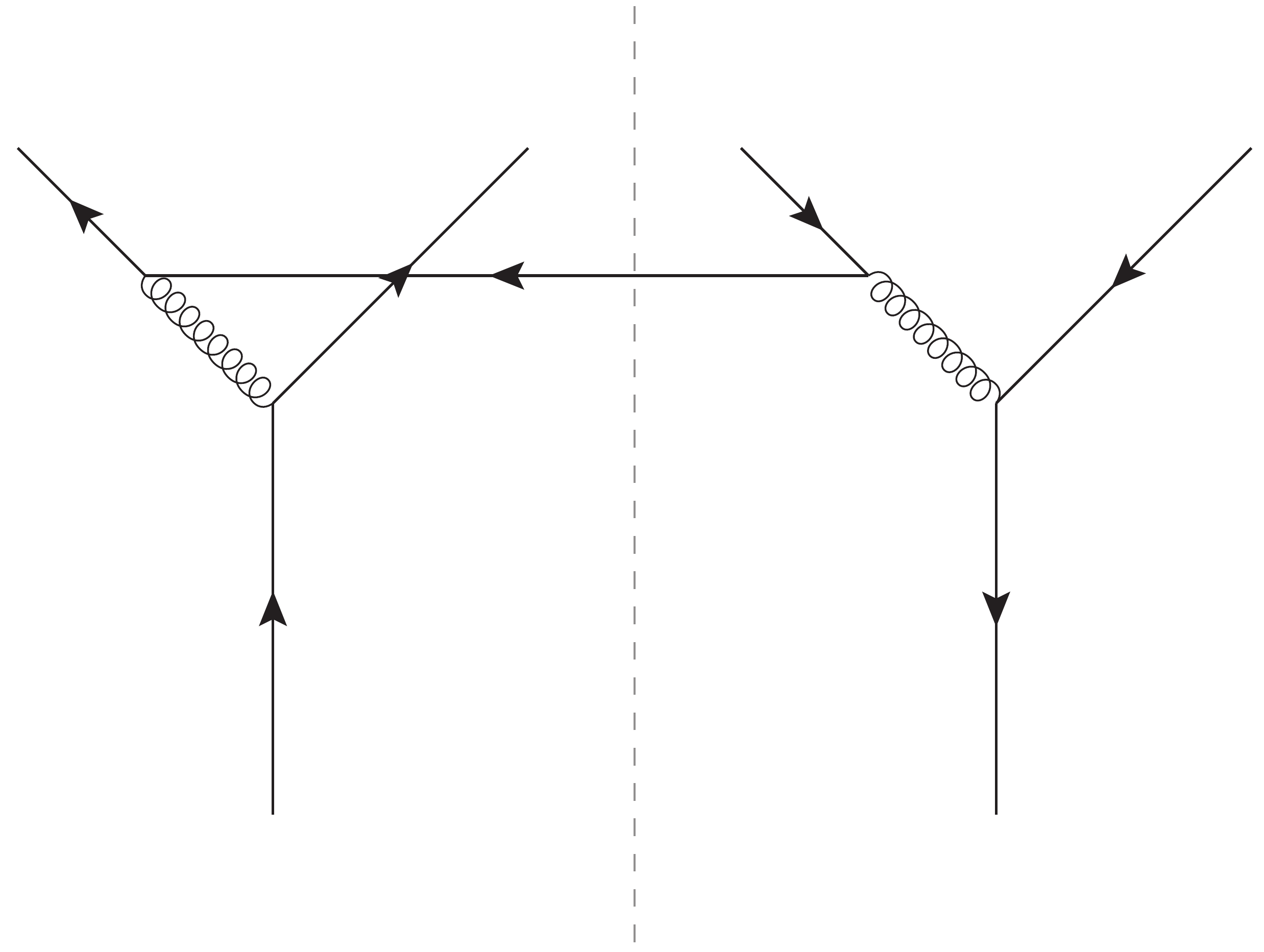}}
    \caption{\label{fig:real-NLO} Real graphs for NLO channels.  Additional graphs are obtained by interchanging $a_1 \leftrightarrow a_2$ and by reflection w.r.t.\ the final state cut.  The association of parton lines is as in figure~\protect\ref{fig:real-LO}.}
  \end{center}
\end{figure}

\begin{figure}
  \begin{center}
    \subfigure[\label{fig:ggg_vertex}]{\includegraphics[width=0.12\textwidth]{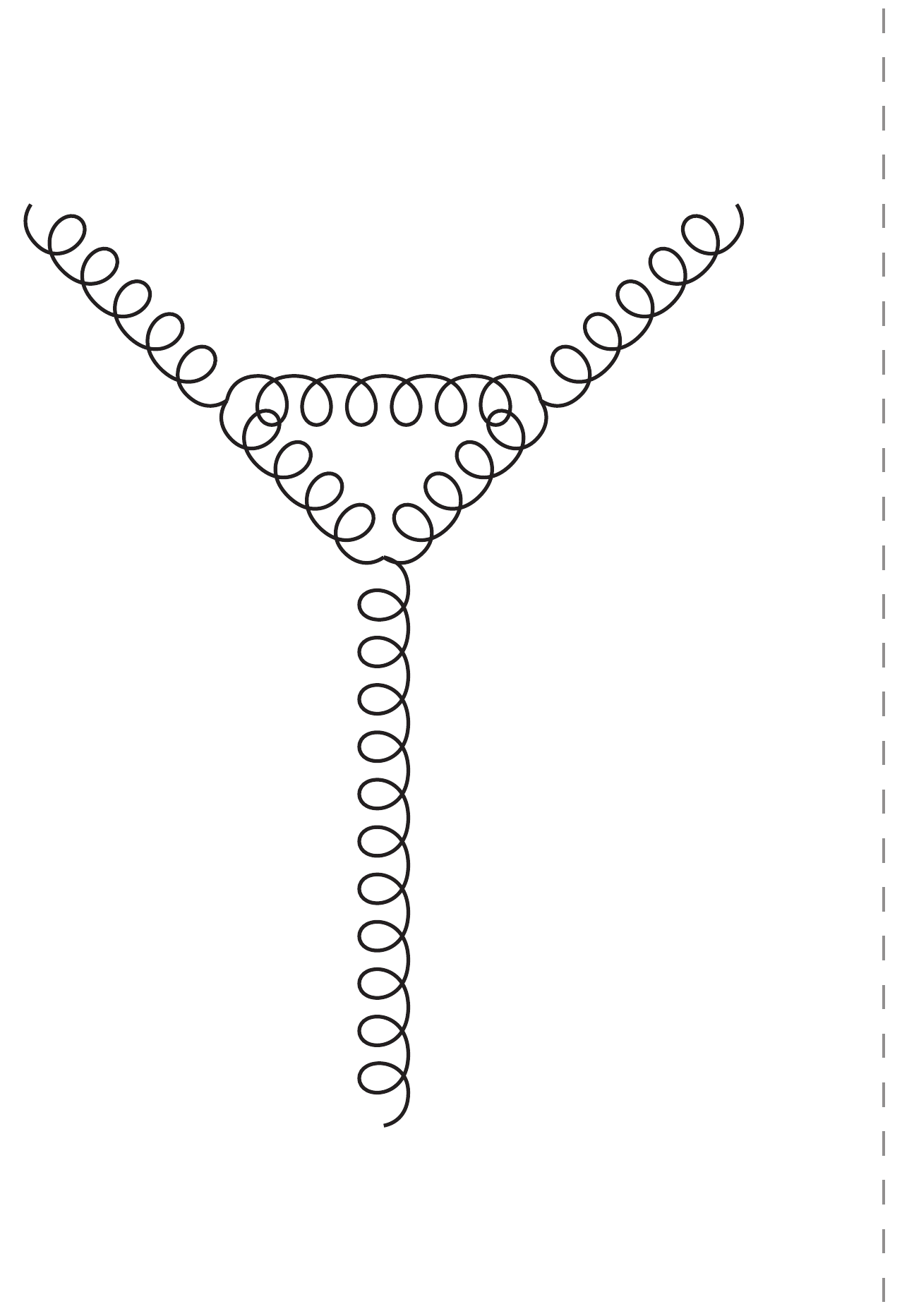}}
    \hspace{2.5em}
    \subfigure[]{\includegraphics[width=0.12\textwidth]{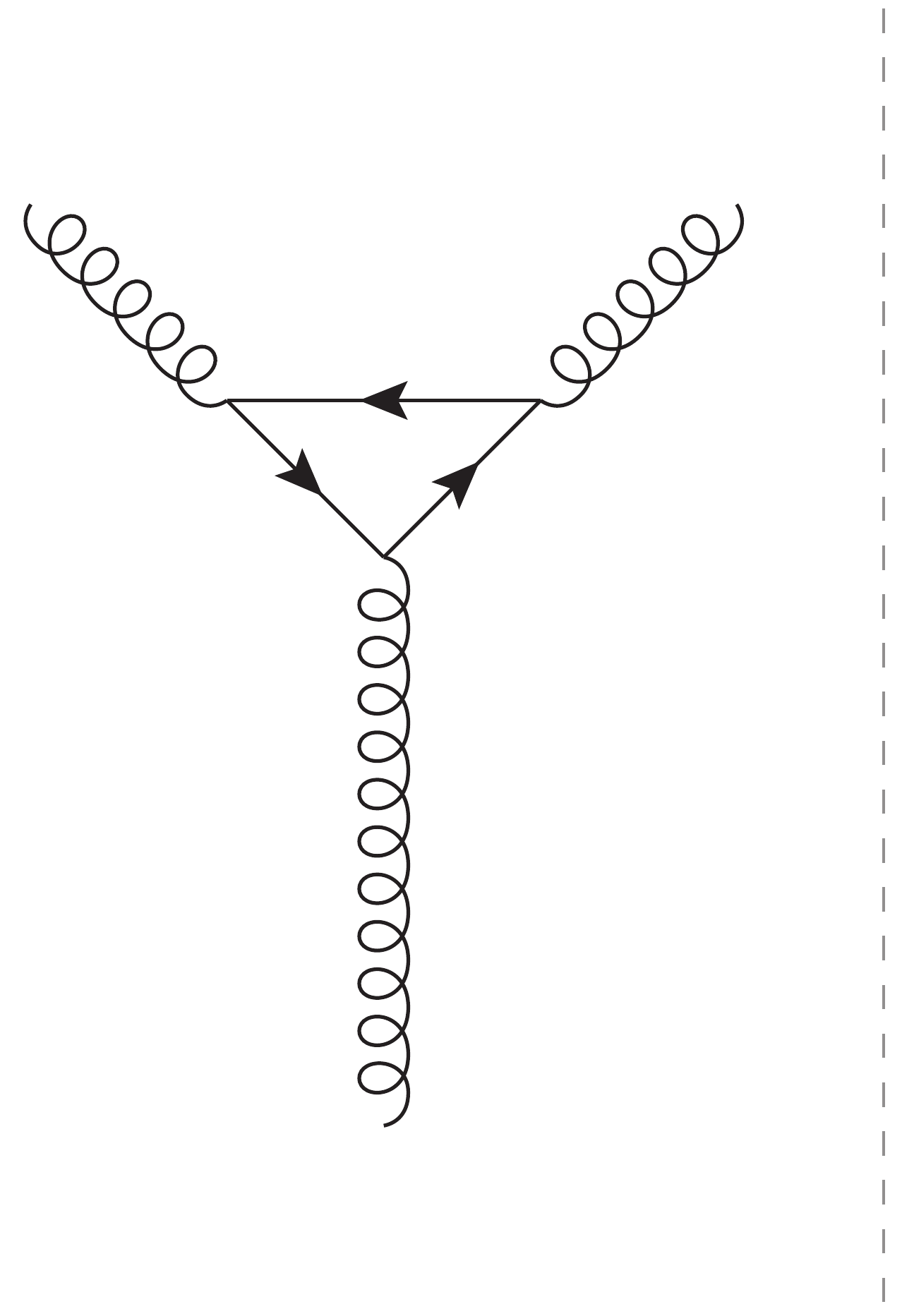}}
    \hspace{2.5em}
    \subfigure[]{\includegraphics[width=0.12\textwidth]{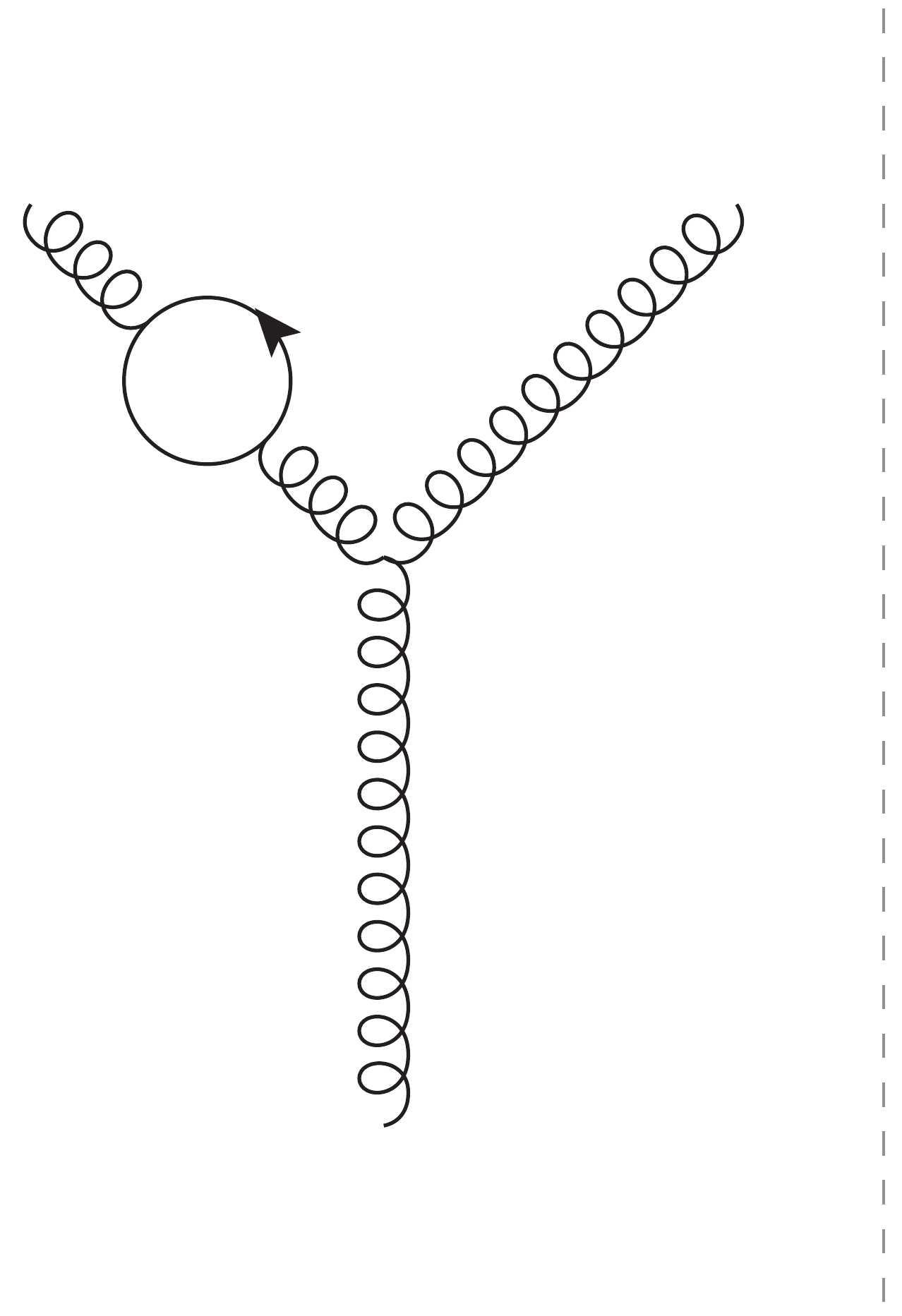}}
    \hspace{2.5em}
    \subfigure[\label{fig:gg_prop}]{\includegraphics[width=0.12\textwidth]{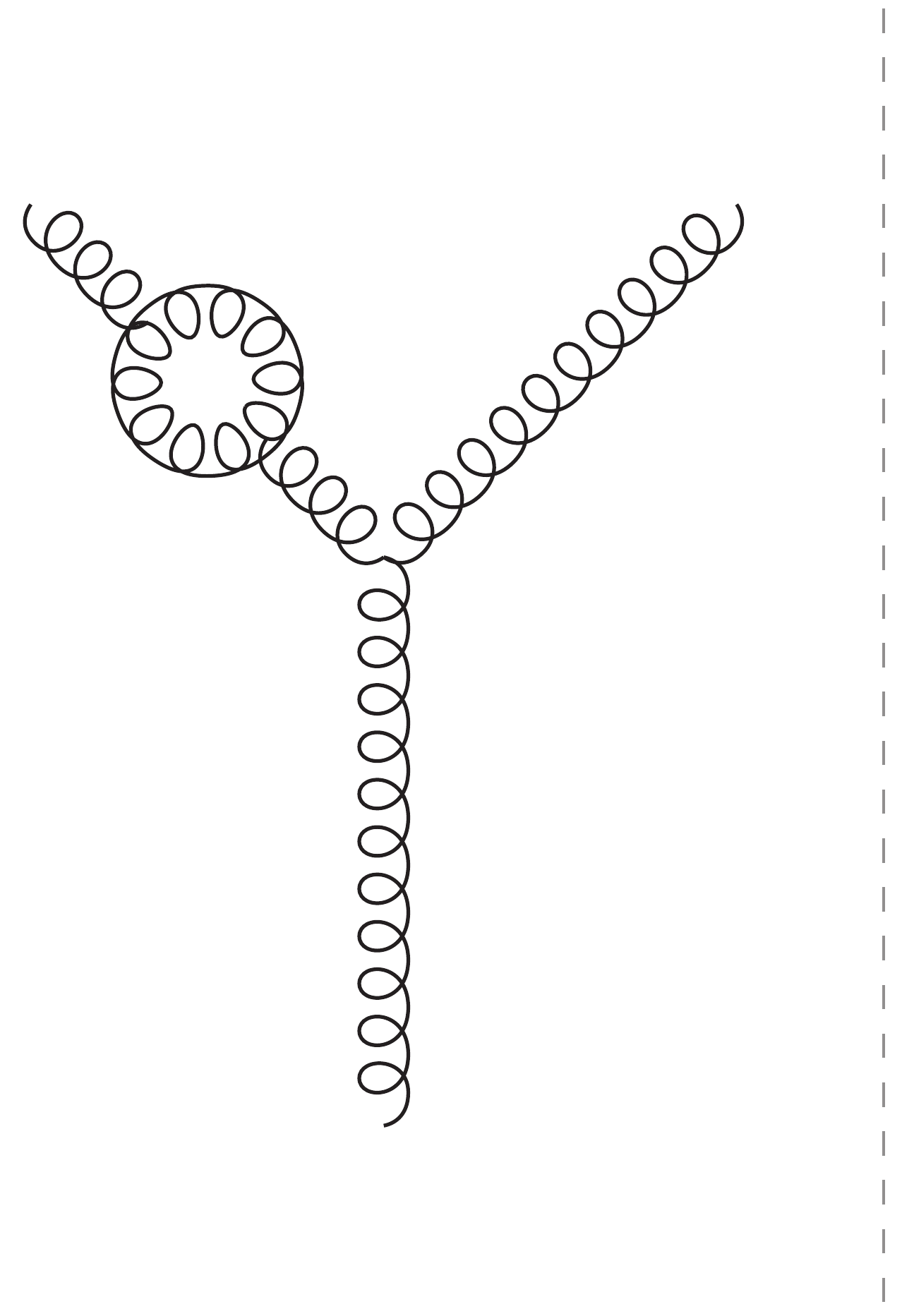}}
    \hspace{2.5em}
    \subfigure[]{\includegraphics[width=0.12\textwidth]{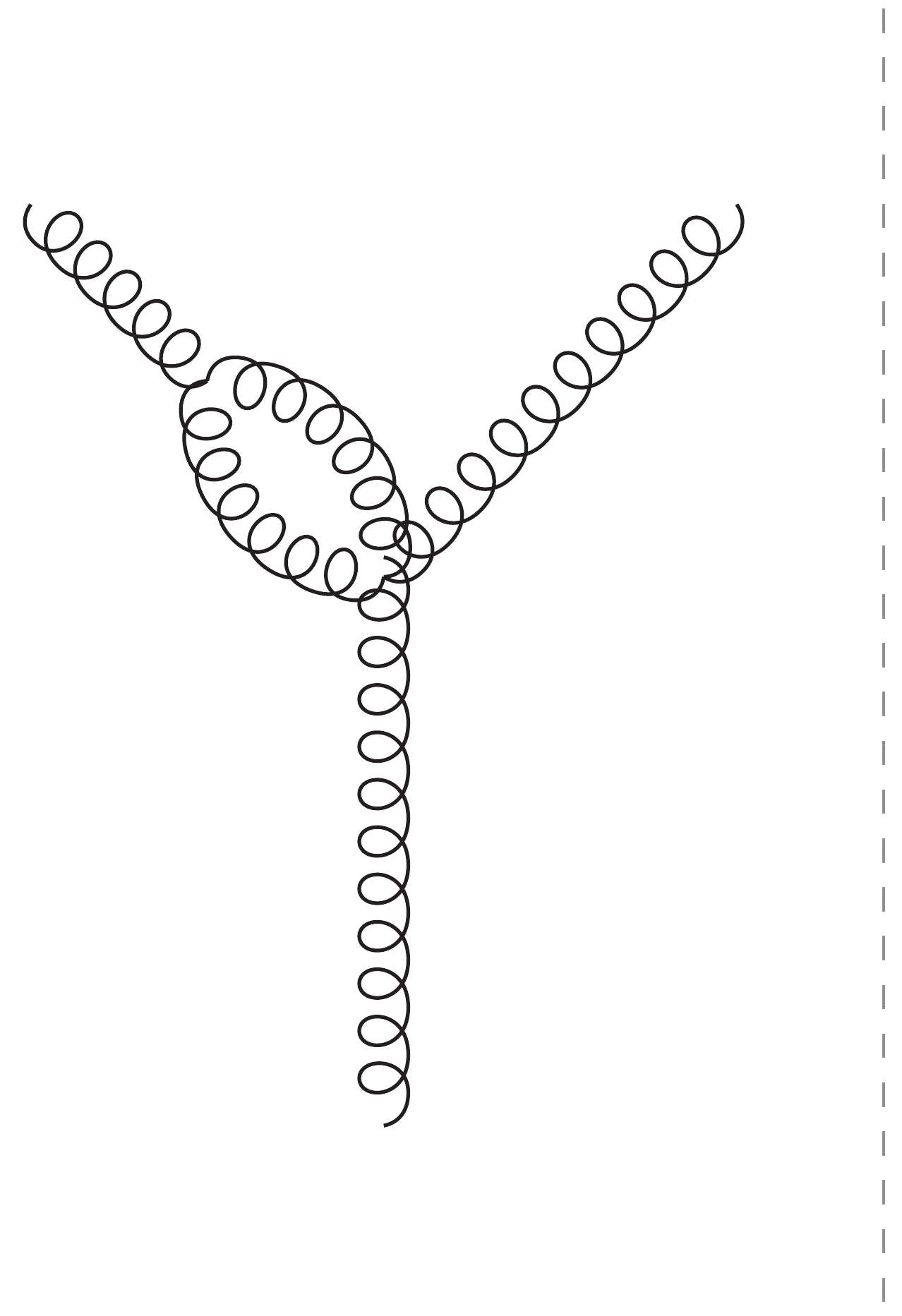}}
\\[2em]
    \subfigure[\label{fig:gqqbar_vertex}]{\includegraphics[width=0.12\textwidth]{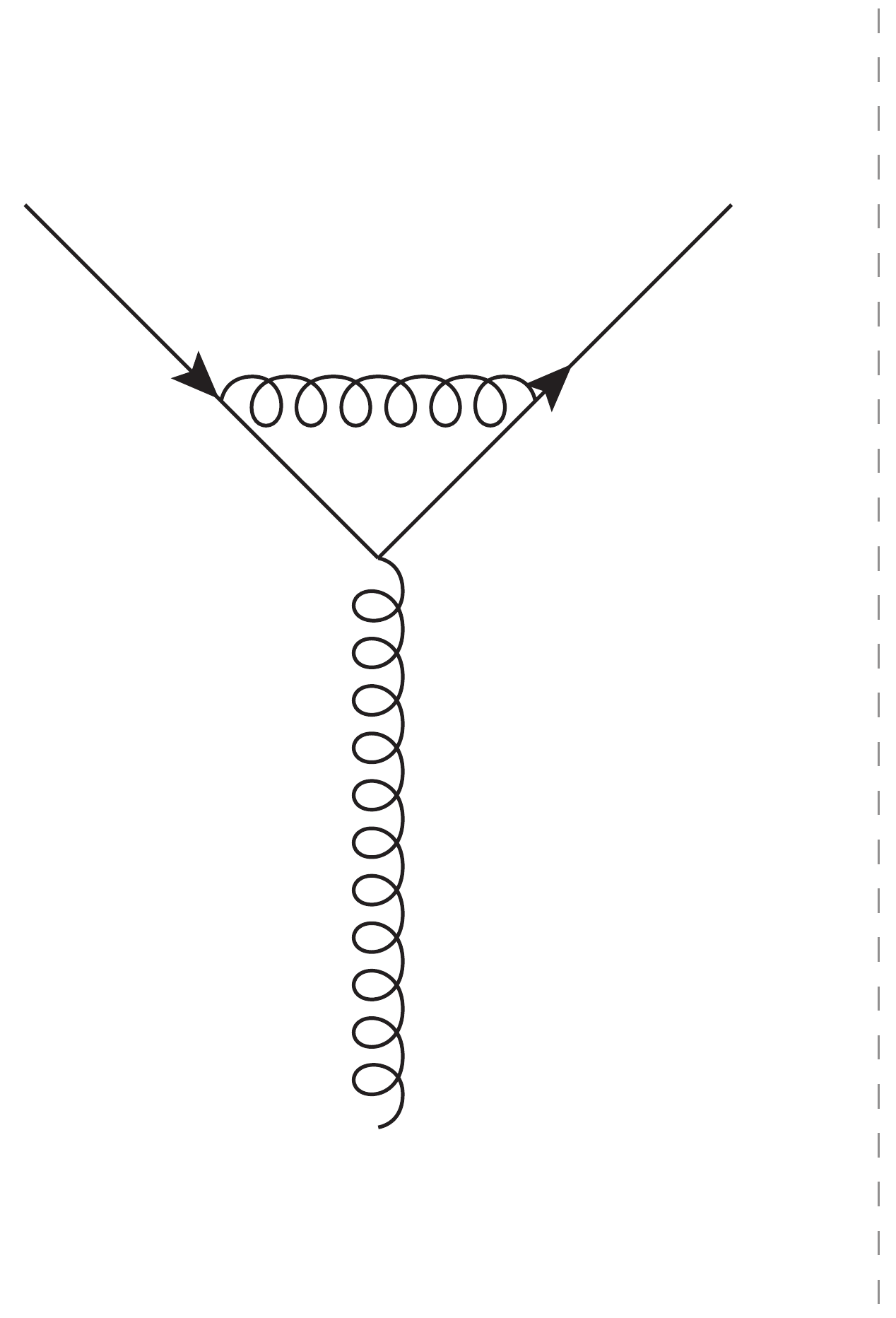}}
    \hspace{2.5em}
    \subfigure[\label{fig:gqqbar_vertex_2}]{\includegraphics[width=0.12\textwidth]{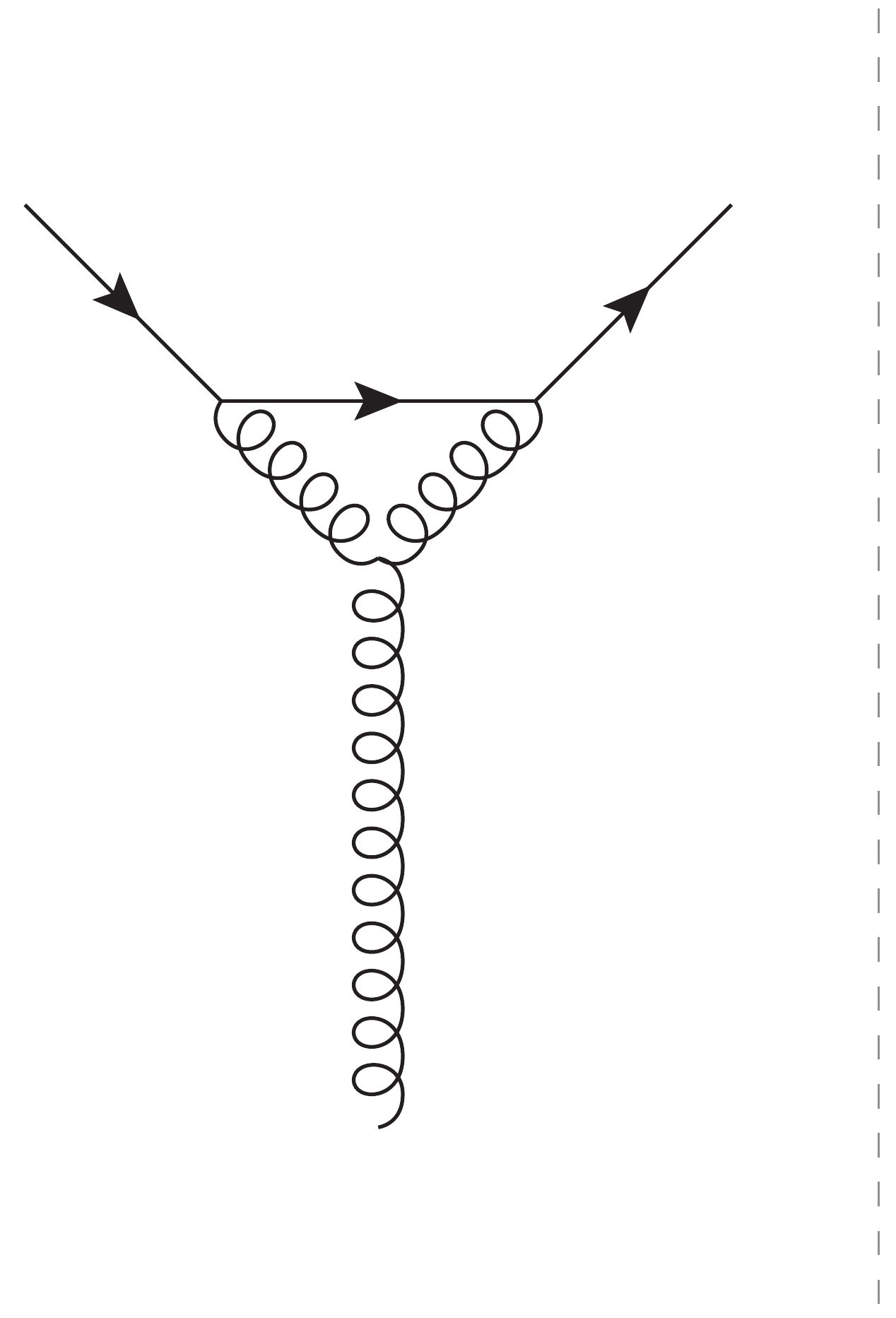}}
    \hspace{2.5em}
    \subfigure[]{\includegraphics[width=0.12\textwidth]{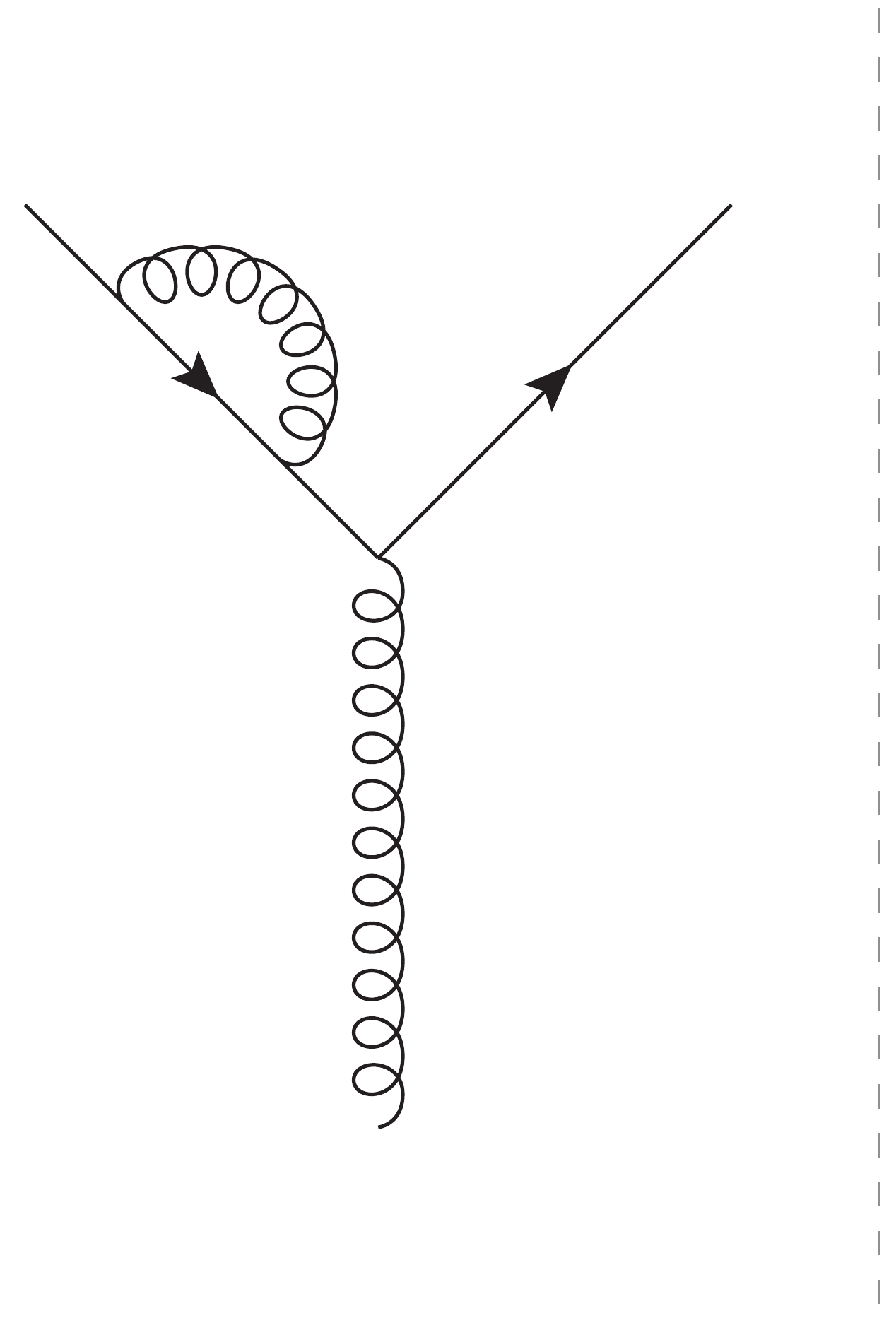}}
\\[2em]
    \subfigure[]{\includegraphics[width=0.12\textwidth]{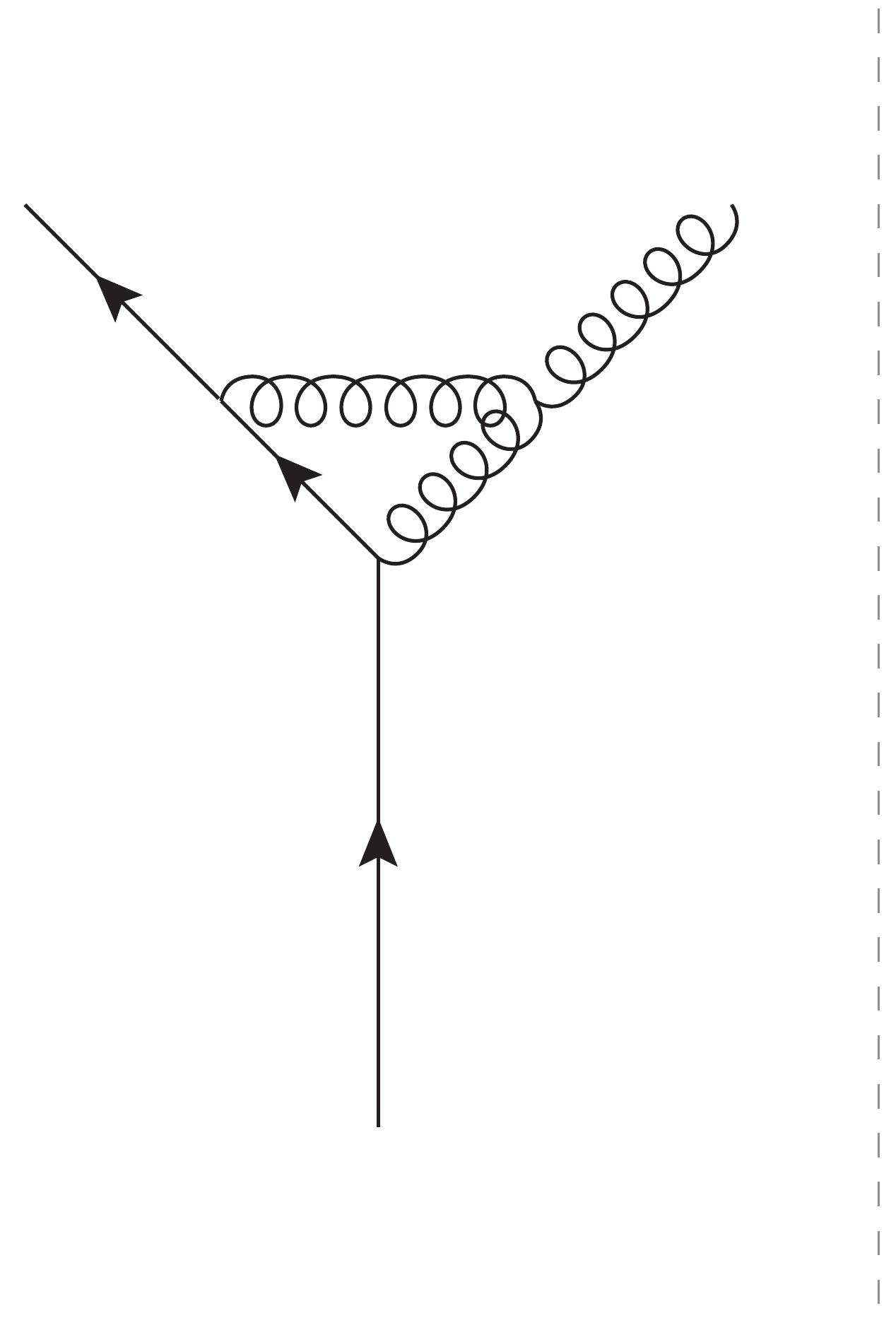}}
    \hspace{2.5em}
    \subfigure[]{\includegraphics[width=0.12\textwidth]{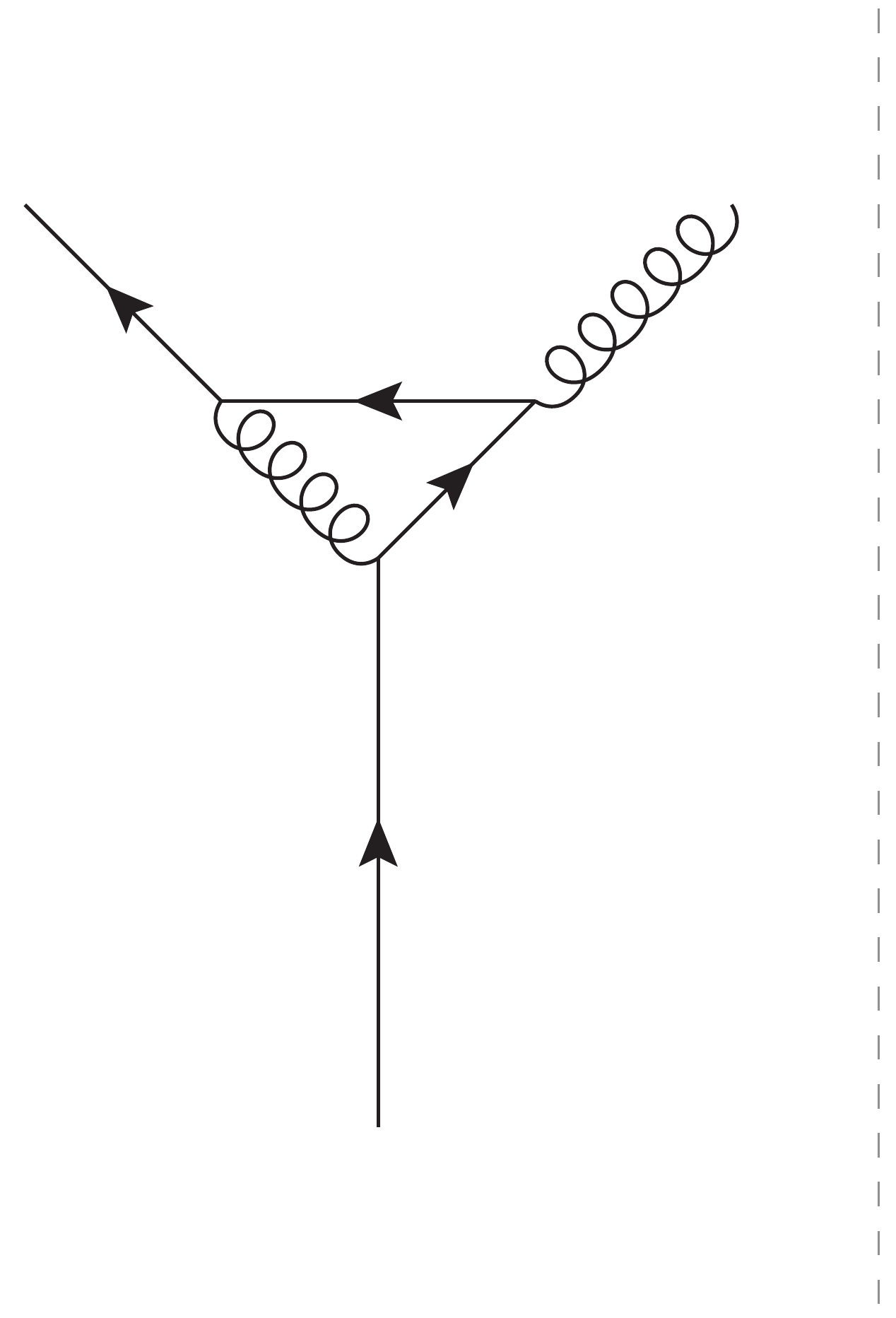}}
    \hspace{2.5em}
    \subfigure[]{\includegraphics[width=0.12\textwidth]{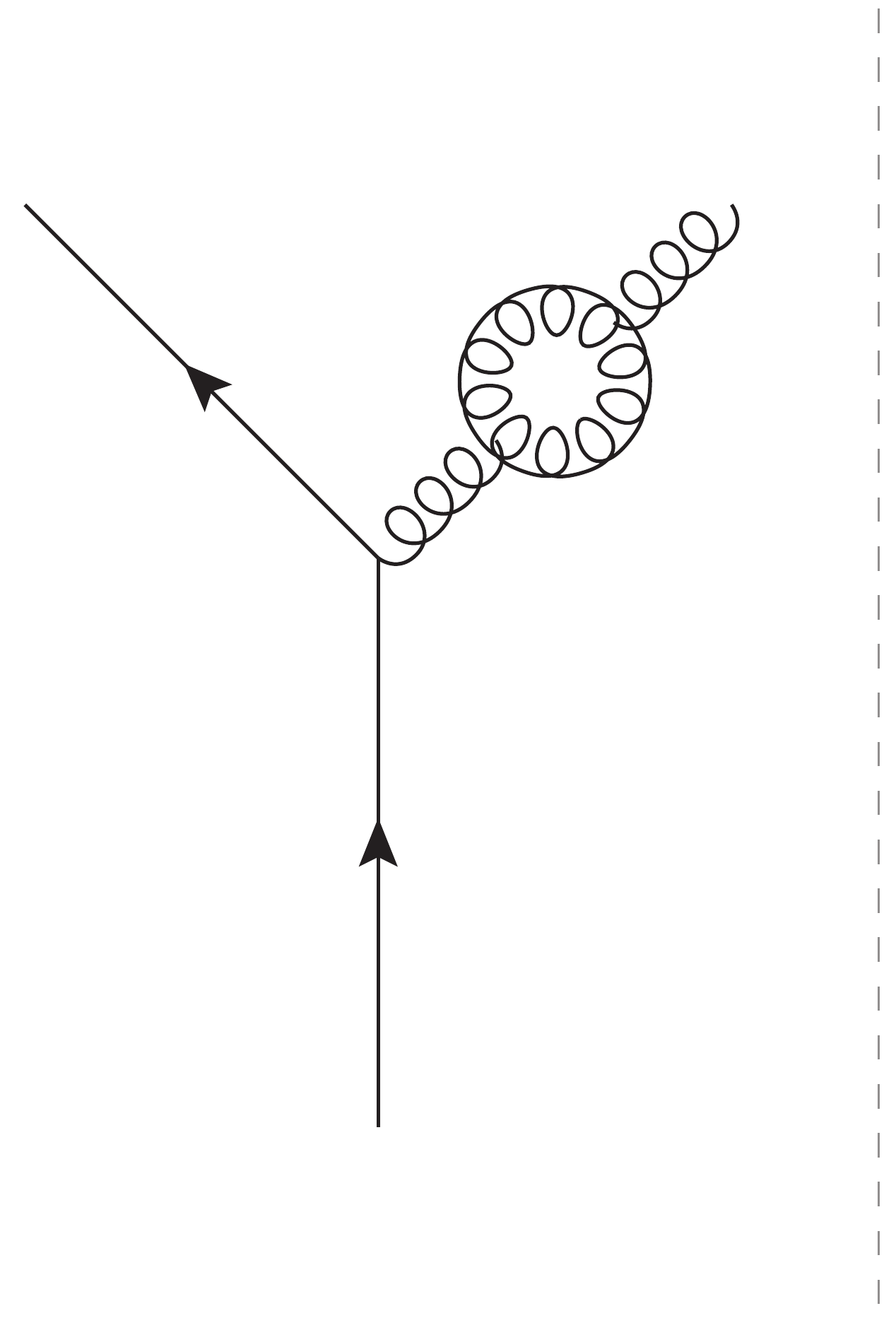}}
    \hspace{2.5em}
    \subfigure[]{\includegraphics[width=0.12\textwidth]{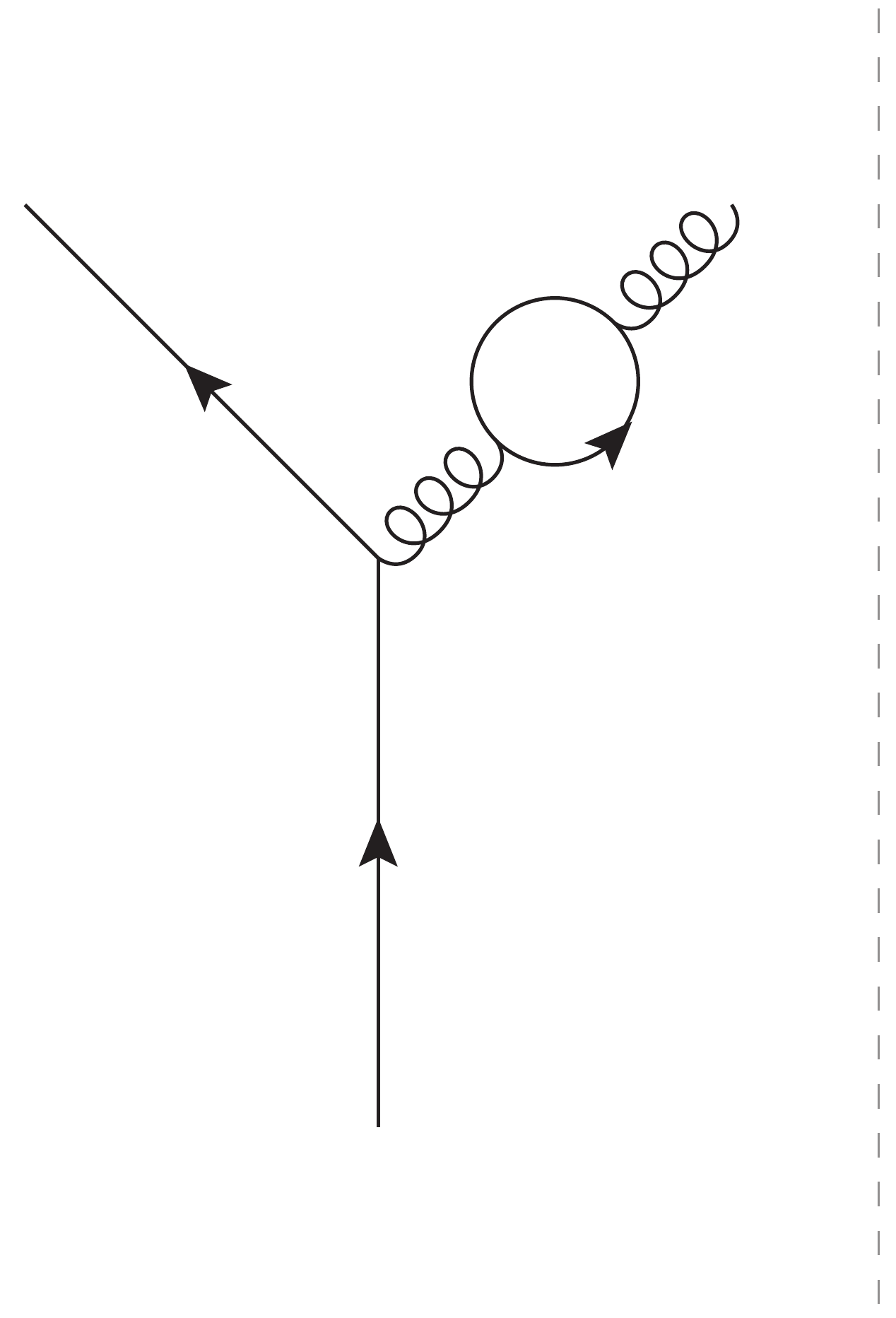}}
    \hspace{2.5em}
    \subfigure[]{\includegraphics[width=0.12\textwidth]{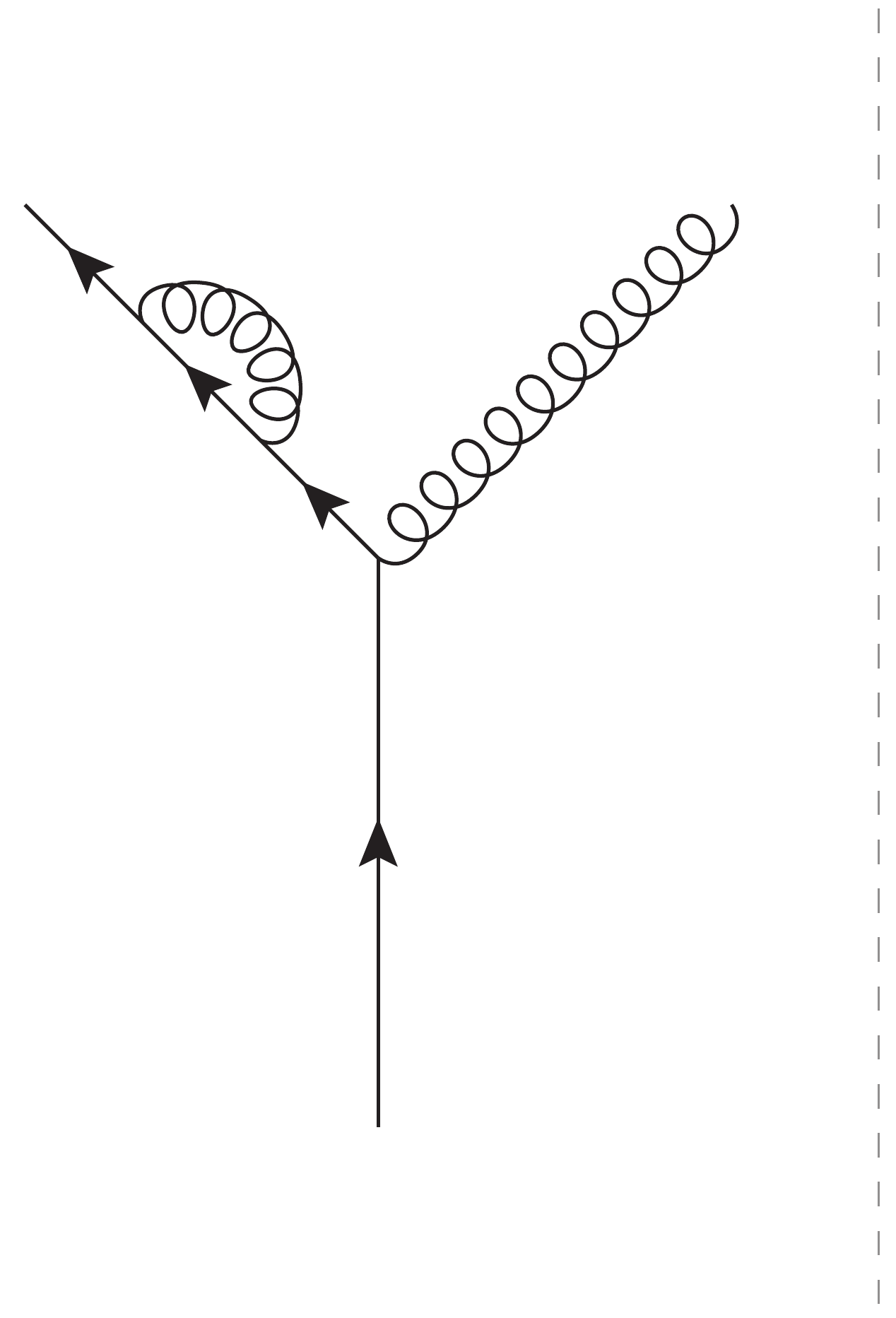}}
    \caption{\label{fig:virt-LO} Virtual graphs for the amplitude in LO channels.  The tree graphs in the complex conjugate amplitude are not shown.  In Feynman gauge, there is an additional graph with Faddeev-Popov ghosts for each graph with a closed gluon loop.}
  \end{center}
\end{figure}

At two-loop order, we have real graphs with one unobserved parton in the final state, as shown in \figs{\ref{fig:real-LO}} and \ref{fig:real-NLO}.  In addition, there are virtual graphs with a vertex or propagator correction on one side of the final state-cut, see \fig{\ref{fig:virt-LO}}.  In \fig{\ref{fig:real-LO}}, we have indicated those graphs that have topology UND (upper non-diagonal).  They play a special role and will be discussed  below.

\begin{figure}
\begin{center}
   \subfigure[\label{fig:WL_gg_1}]{\includegraphics[width=0.29\textwidth]{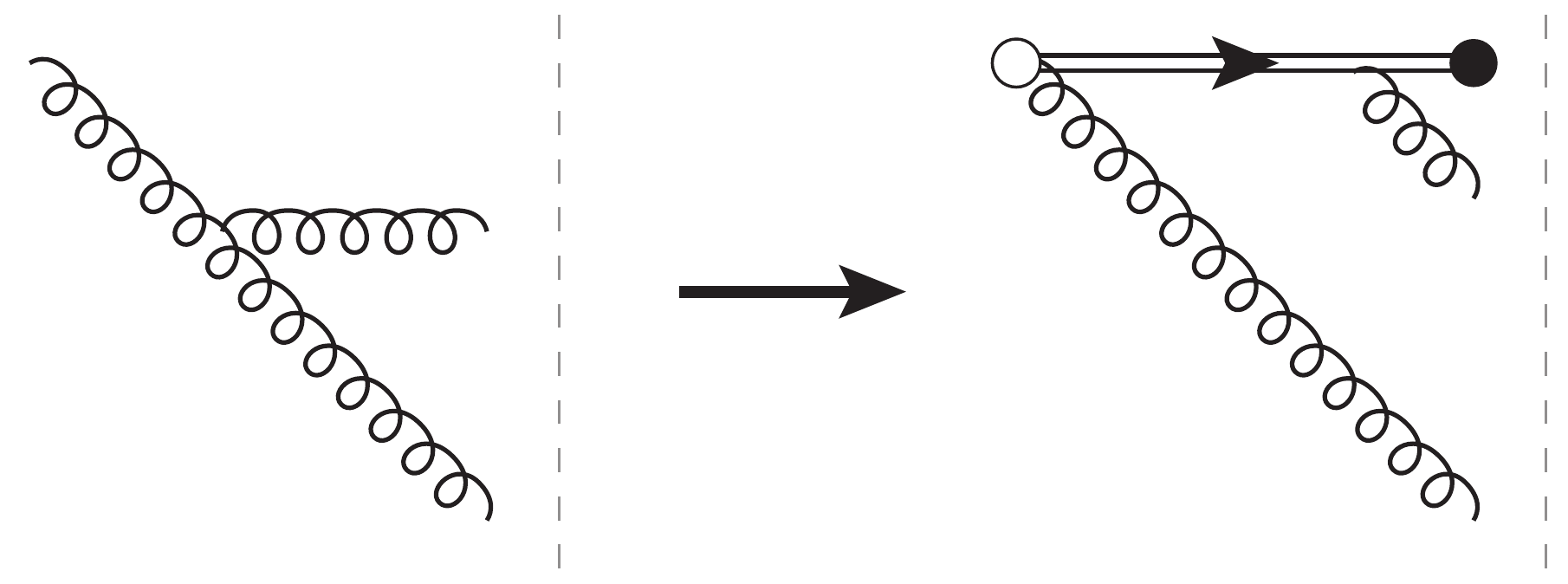}}
   \hspace{5em}{}
   \subfigure[\label{fig:WL_gg_2}]{\includegraphics[width=0.29\textwidth]{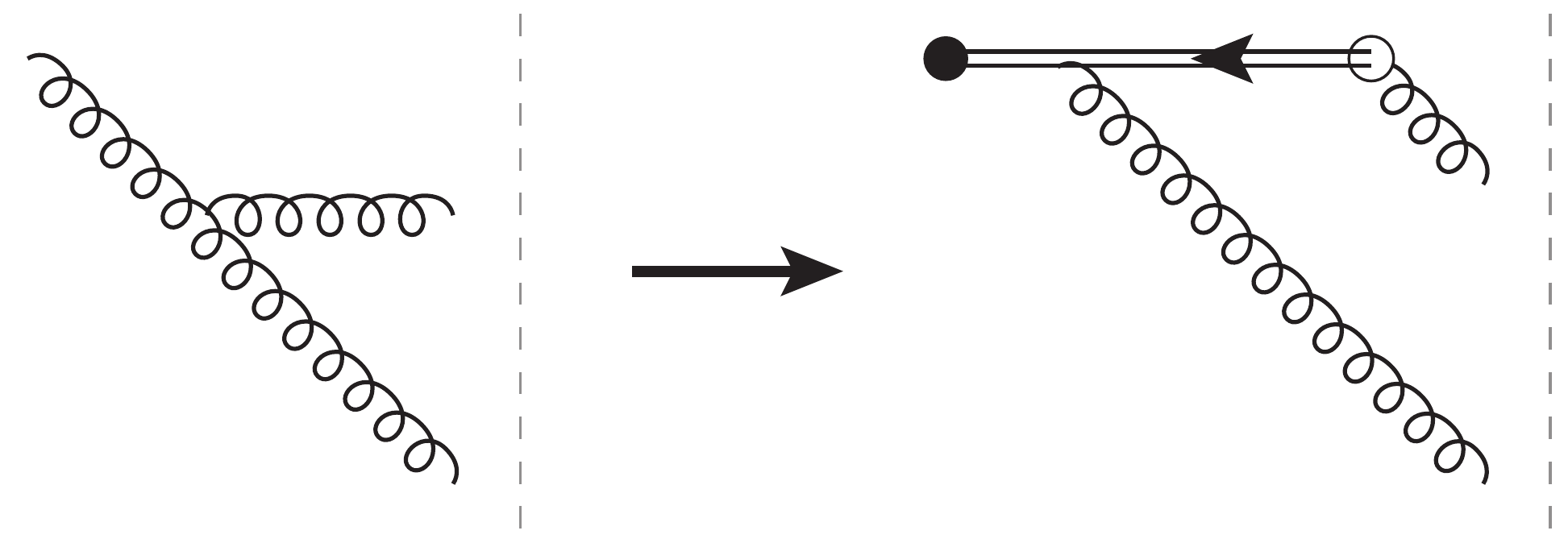}}
\\[2em]
   \subfigure[\label{fig:WL_gq}]{\includegraphics[width=0.29\textwidth]{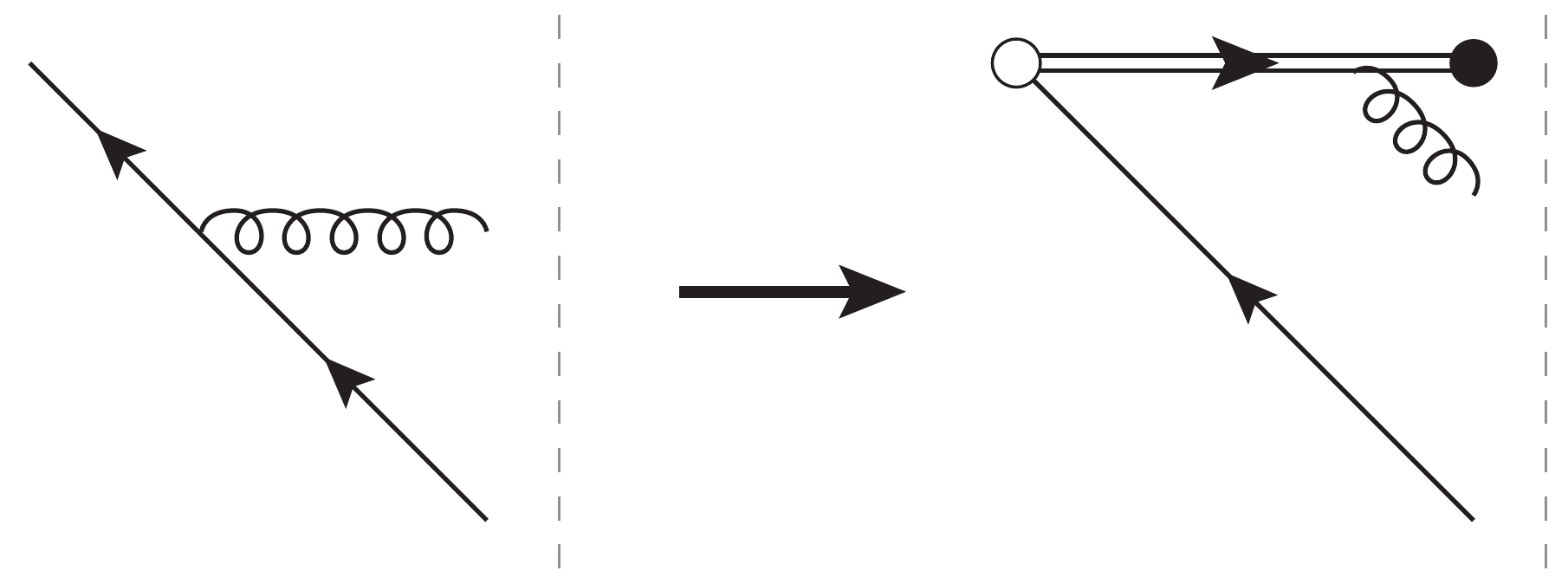}}
   \hspace{5em}{}
   \subfigure[\label{fig:WL_qg}]{\includegraphics[width=0.29\textwidth]{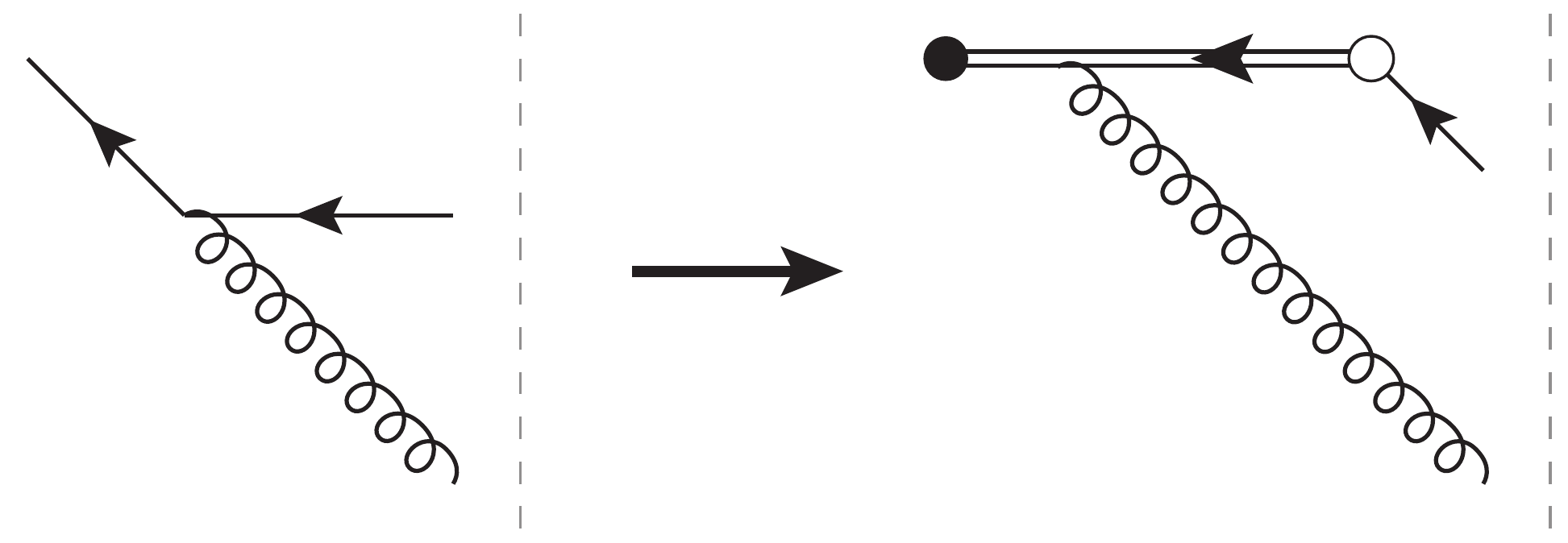}}
   \caption{\label{fig:eik-recipe} Rules for obtaining graphs with eikonal lines in Feynman gauge from the graphs in \figs{\ref{fig:real-LO}} to \ref{fig:virt-LO}.  The the top left line in each panel is an observed parton, $a_1$ or $a_2$.  Analogous rules hold on the right of the final state cut.}
\end{center}
\end{figure}

In Feynman gauge, one has additional graphs with an eikonal line.  These graphs can be obtained from the ones without eikonal lines by applying the graphical rules given in \fig{\ref{fig:eik-recipe}}.  Graphs obtained by applying these rules to graphs with UND topology will be referred to as UND graphs as well.

The flavour structure of the channels without external gluons can be decomposed as
\begin{align}
\label{quark-flavour-decomp}
V_{q_j \bar{q}_k, q_i}
  &= \delta_{j k}\ms V_{q'\bs \bar{q}'\!,q}
     + \delta_{i j}\ms V_{q \bar{q}'\!,q}
     + \delta_{i j}\ms \delta_{j k}\, V_{q\bar{q},q}^v \,,
\nonumber \\
V_{q_j q_k, q_j} &= \delta_{j k}\ms V_{q'\bs q,q}
     + \phantom{\delta_{i j}\, } V_{q q'\!,q}
     + \phantom{\delta_{i j}\, } \delta_{j k}\ms V_{q q,q}^v \,,
\end{align}
where $V_{q'\bs q,q}(z_1,z_2) = V_{q q'\!,q}(z_2,z_1)$.  The graphs for the kernels on the r.h.s.\ are shown in \figs{\ref{fig:qqbar-1}} to \ref{fig:qq-2}.

\paragraph{Colour structure of graphs.}
Let us now take a closer look how the different graphs behave as a function of the colour representations $R_1$ and $R_2$.  It is simplest to discuss this in light-cone gauge $A^+ = 0$, where graphs with eikonal lines are absent.

At LO, one has a single  graph without eikonal lines for each splitting process.  As a consequence, a kernel for the representations $R_1$ and $R_2$ can be obtained from the colour singlet one by a simple rescaling:
\begin{align}
\label{LO-rescaling}
\prn{\RR}{V}^{(1)}_{a_1 a_2, a_0} &=
   c_{a_1 a_2, a_0}^{}(R_1 R_2)\; \prn{11}{V}^{(1)}_{a_1 a_2, a_0}
\end{align}
with factors $c_{a_1 a_2, a_0}(R_1 R_2)$ given in \eqref{rescaling-factors}.
We have exactly the same rescaling for virtual correction graphs at NLO.  This is because for a given channel, all virtual subgraphs in \fig{\ref{fig:virt-LO}} depend on the colour indices of the external legs via $t^a_{i j}$ or $f^{a b c}$, as does the tree-level graph in the same channel.

For real emission graphs, the situation is less simple.  We find that graphs that have the same colour factors in the singlet channel do not necessarily have the same colour factors for other representations.  Examples are graphs \ref{fig:ggg-LD}, \ref{fig:ggg-UD}, \ref{fig:ggg-UND} and graphs \ref{fig:gqg-LD}, \ref{fig:gqg-UD}.  Interestingly, several graphs with UND topology have a zero colour factor for some representations.  Examples are graph \ref{fig:ggg-UND} for $R_1 R_2 = AA, SS$, graphs \ref{fig:qgq-UND} and \ref{fig:gqg-UND} for $R_1 R_2 = 8A$ and $8S$, and graph \ref{fig:qgg-UND} for $R_1 R_2 = AA$.

Despite these complications, we do find a number of simple patterns in the colour factors for the real graphs.  Together with the simple scaling rule for virtual graphs, we find the following.
\begin{itemize}
\item Denoting the sum of real and virtual two-loop graphs for a channel by $\prn{\RR}{\ms\Gamma}_{a_1 a_2, a_0}$, we find that the two octet channels for gluons are related as
\begin{align}
\label{relation-AS}
\frac{\prn{AA}{\ms\Gamma}_{g g, g}}{\prn{SS}{\ms\Gamma}_{g g, g}}
   &= \frac{c_{g g, g}(AA)}{c_{g g, g}(SS)}
    = - 1 \,,
\nonumber \\
\frac{\prn{8A}{\ms\Gamma}_{q g, q}}{\prn{8S}{\ms\Gamma}_{q g, q}}
    = \frac{\prn{8A}{\ms\Gamma}_{q g, g}}{\prn{8S}{\ms\Gamma}_{q g, g}}
   &= \frac{c_{q g, q}(8A)}{c_{q g, q}(8S)}
    = - \frac{N}{\sqrt{N^2 - 4}} \,.
\end{align}
However, there is no such relation in the $q \to g g$ channel.
\item With the same notation, we find
\begin{align}
\label{relation-D}
\prn{\tentenbar}{\ms\Gamma}_{g g, a}
&= \prn{\tenbarten}{\ms\Gamma}_{g g, a} = 0
& \text{for } a = g, q.
\end{align}
This result is valid for general $N$, as well as the equality $c_{g g, g}(\tentenbartext) = c_{g g, g}(\tenbartentext) = 0$.  As a consequence, all splitting kernels are zero in the decuplet sector, both at LO and at NLO.
\item For the sum $\prn{\RR}{\ms\Gamma}^{\text{real}}_{a_1 a_2, a_0}$ of real two-loop graphs, we find a structure
\begin{align}
\prn{11}{\ms\Gamma}^{\text{real}}_{q\bar{q}, g} &=
  C_A\, \Gamma^{A}_{q\bar{q}, g}
  + C_F\, \Gamma^{F}_{q\bar{q}, g} \,,
\nonumber \\
\prn{88}{\ms\Gamma}^{\text{real}}_{q\bar{q}, g} &=
  c_{q\bar{q}, g}(88)\, \biggl[ C_A\, \Gamma^{A}_{q\bar{q}, g}
  - \frac{1}{N^2 - 1}\, C_F\, \Gamma^{F}_{q\bar{q}, g} \biggr]
\end{align}
for $g \to q\bar{q}$, and
\begin{align}
\label{colour-relations-q-gg}
\prn{11}{\ms\Gamma}^{\text{real}}_{g g, q} &=
  C_A\ms C_F\, \Gamma^{A}_{g g, q}
  + C_F^2\, \Gamma^{F}_{g g, q} \,,
  \phantom{\frac{N^2}{2 \sqrt{N^2 - 1}}}
\nonumber \\
\prn{AA}{\ms\Gamma}^{\text{real}}_{g g, q} &=
  c_{g g, g}(AA)\, \biggl[ C_A\ms C_F\, \Gamma^{A}_{g g, q}
  + \frac{N^2}{N^2 - 1}\, C_F^2\, \Gamma^{F}_{g g, q} \biggr] \,,
\nonumber \\
\prn{SS}{\ms\Gamma}^{\text{real}}_{g g, q} &=
  c_{g g, g}(SS)\, \biggl[ C_A\ms C_F\, \Gamma^{A}_{g g, q}
  + \frac{N^2 - 4}{N^2 - 1}\, C_F^2\, \Gamma^{F}_{g g, q} \biggr] \,,
\nonumber \\
\prn{27\, 27}{\ms\Gamma}^{\text{real}}_{g g, q} &=
  c_{g g, g}(27\, 27)\, C_A\ms C_F\, \Gamma^{A}_{g g, q}
  \phantom{\frac{N^2}{2 \sqrt{N^2 - 1}}}
\end{align}
for $q \to g g$, where $N=3$ for $R_1 R_2 = 27\,27$ as usual.  The functions $\Gamma^F$ and $\Gamma^A$ are colour independent.  We observe that the terms going with $\Gamma^A$ have the same scaling with $R_1 R_2$ as the LO and the virtual NLO graphs, whereas the terms going with $\Gamma^F$ have a different scaling.  We note that the colour factors of single graphs for $R_1 R_2 = 11$ may be linear combinations of $C_A$ and $C_F$ or of $C_A\ms C_F$ and $C_F^2$.  For instance, the colour factor of graph \ref{fig:gqqbar-UND} in the singlet channel is $C_A - 2 C_F$, and the one of graph \ref{fig:qgg-UND} is $C_A\ms C_F - 2 C_F^2$.

Using $- C_F / (N^2-1) = C_F - C_A/2$, we can rewrite
\begin{align}
\label{colour-relation-g-qqbar}
\prn{88}{\ms\Gamma}^{\text{real}}_{q\bar{q}, g} &=
  c_{q\bar{q}, g}(88)\, C_A\,
  \bigl[\ms \Gamma^{A}_{q\bar{q}, g}
      - \Gamma^{F}_{q\bar{q}, g} \big/2 \ms\bigr]
  + c_{q\bar{q}, g}(88)\, C_F\, \Gamma^{F}_{q\bar{q}, g} \,,
\end{align}
which is the form we will use in the presentation of our results.
\item For $q\to g g$, we obtain nonzero results also in the mixed octet channels.  They satisfy
\begin{align}
\label{extra-channels-q-gg}
\prn{AS}{\ms\Gamma}^{\text{real}}_{g g, q}
&= -\; \prn{S\bs A}{\ms\Gamma}^{\text{real}}_{g g, q}
\end{align}
and cannot be written in terms of $\Gamma^{A}_{g g, q}$ and $\Gamma^{F}_{g g, q}$.
\end{itemize}

Apart from the contribution from real and virtual graphs to ${V}_{\bus}$, the full two-loop kernel ${V}^{(2)}$ receives contributions from the ultraviolet counterterms \eqref{Vct-def}.  These are single or double poles in $1/\epsilon$ that cancel the corresponding poles in ${V}_{\bus}$, so that their colour structure must match that of the ultraviolet divergent terms in the sum of real and virtual two-loop graphs.  Of course, virtual two-loop graphs are only present in LO channels.  The explicit colour dependence of the one-loop splitting kernels $\prb{R \Rp}{\hat{P}}$ appearing in the counterterms is given in \eqn{(7.91)} of \cite{Buffing:2017mqm}.

In LO channels, we finally need to include the contribution going with the one-loop soft factor in \eqref{bare-V}.  According to \eqref{SB-collins} and \eqref{SB-delta}, $\pr{R}{S}_B^{(1)}$ depends on the representation $R$ via a global factor $\prn{R}{\gamma}_J^{(0)}$, so that for its contribution to $\prn{\RR}{V}^{(2)}$ we have
\begin{align}
- \frac{1}{2}\; \pr{R_1}{S}_B^{(1)}\;
   \prn{\RR}{V}_{B,\ms a_1 a_2, a_0}^{(1)}
& \,\propto\,
   \prn{R_1}{\gamma}_J^{(0)} \; \prn{\RR}{V}_{B,\ms a_1 a_2, a_0}^{(1)} \,.
\end{align}
This is the colour structure of the counterterm $V_{\text{ct\,2}}$ and of the terms with double logarithms in the expression \eqref{two-loop-final} of ${V}^{(2)}$.  It gives zero for $R_1 R_2 = 11$, where rapidity divergences cancel between real and virtual graphs.

The scaling relations \eqref{relation-AS} and \eqref{relation-D} hold for the sum of  real and virtual two-loop graphs, and hence also for all counterterms.  Therefore, they hold for the full two-loop kernels ${V}^{(2)}$ in the channels $g\to g g$, $g\to q g$, and $q\to q g$.  Since they also hold for the LO kernels, they hold for the full splitting DPDs up to order $\as^2$.

For channels without external gluons, the kernels on the r.h.s.\ of \eqref{quark-flavour-decomp} receive contributions from exactly one graph, see \fig{\ref{fig:qqbar-1}} to \ref{fig:qq-2}.  For each of them, one therefore has a unique colour factor at given $R_1 R_2$, which gives the colour factor of the full two-loop kernel and hence for the full splitting DPDs up to order $\as^2$.  These factors will be given in \eqref{eq:RVqqq2-scaling}.

For $q\to g g$, the relations in \eqref{colour-relations-q-gg} and \eqref{extra-channels-q-gg} extend to the full two-loop kernel because only real graphs and their counterterms contribute in that case.  The colour structure in the channel $g\to q\bar{q}$ is more complicated than the relation \eqref{colour-relation-g-qqbar} for real graphs and will be given in \eqref{eq:RVqqbarg2} and \eqref{eq:RVqqbarg2-scaling}.


\subsection{Computation of the graphs and handling of rapidity dependence}
\label{sec:compute}

\begin{figure}
  \begin{center}
    \subfigure[real graph\label{fig:real-kin}]{\includegraphics[height=10em,trim=0 25 0 25,clip]{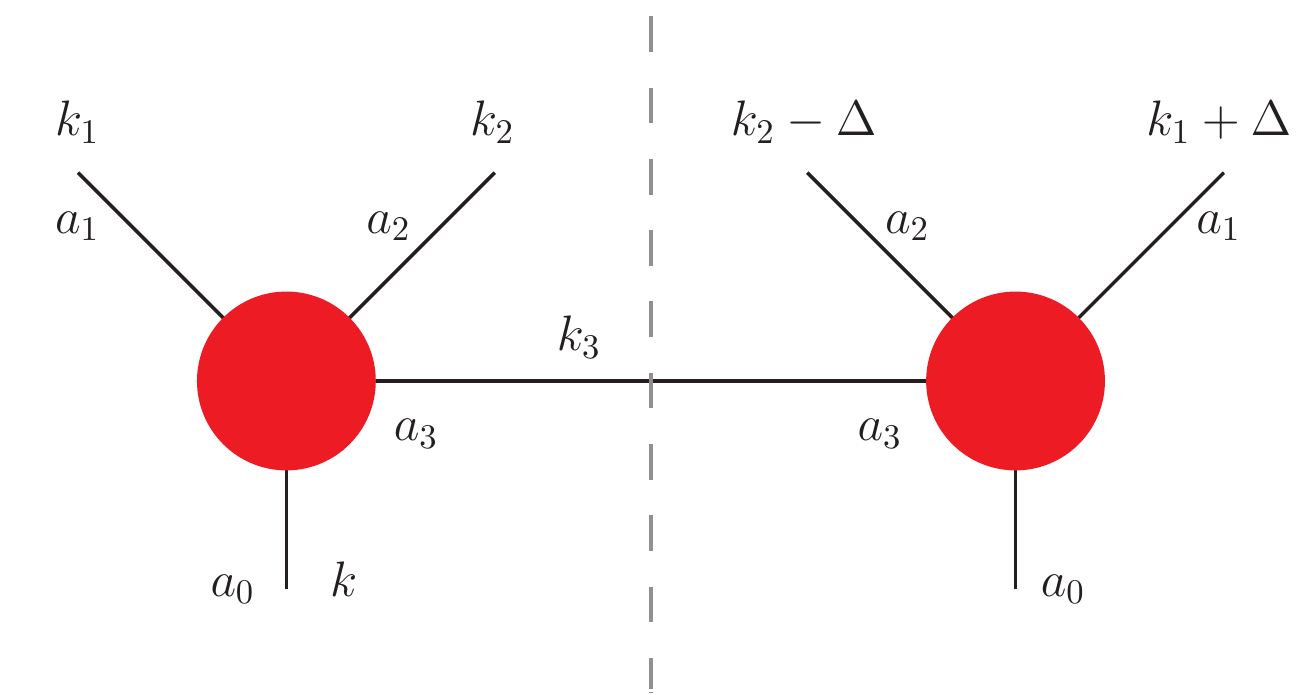}}
    \hspace{3em}
    \subfigure[vertex correction\label{fig:virt-kin-vertex}]{\includegraphics[height=10em]{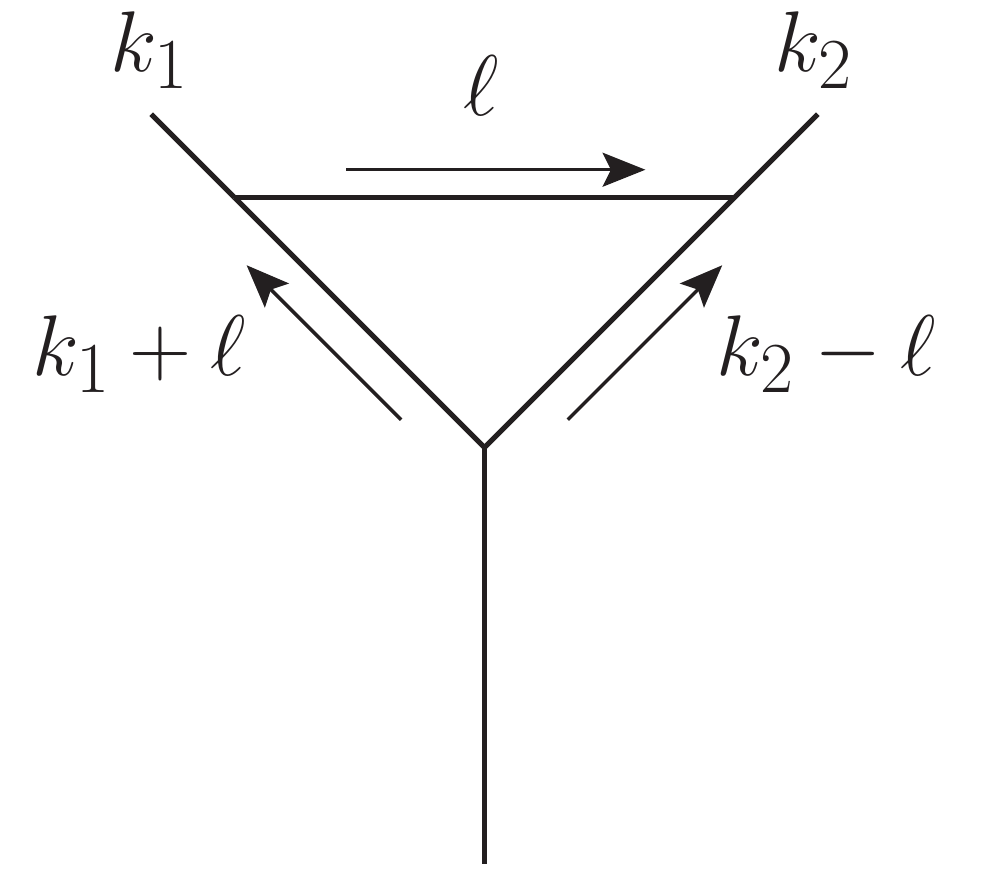}}
    \caption{\label{fig:kinematics} Left: Assignment of momenta in real emission graphs.  Plus momenta are fixed to $k_1^+ = z_1\ms k^+$, $k_2^+ = z_2\ms k^+$, and $\Delta^+ = 0$.  Compared with the symmetric assignment in \protect\cite{Diehl:2017kgu} we have shifted $k_1$ and $k_2$ such that $\Delta$ appears only on the r.h.s.\ of the cut.  Right: Assignment of momenta in vertex correction graphs to the left of the cut.  An analogous assignment holds for loops to the right of the cut, with $k_1$ and $k_2$ shifted by $+ \Delta$ and $- \Delta$, respectively.}
  \end{center}
\end{figure}

Let us briefly specify the kinematics of the two-loop graphs.  We work in a frame where the momentum of the incoming on-shell parton $a_0$ has a plus-component $k^+$ and zero minus- and transverse components.  Momenta are assigned as shown in \fig{\ref{fig:real-kin}} for real graphs; for virtual graphs one has to remove the line $a_3$ and set $k_3 = 0$.  The minus- and transverse components of $k_1$ and $k_2$ must be integrated over, because the fields in the operators \eqref{op-defs} have a light-like distance from each other.  In addition, one must integrate over $\Delta^-$, given that $y^+ = 0$ in the matrix element~\eqref{dpd-def}.

We write $z_1 = k_1^+ / k^+$ and $z_2 = k_2^+ / k^+$ for the plus-momentum fractions of the observed partons, which corresponds to writing the splitting kernel as $V(z_1, z_2)$.  In the convolution \eqref{conv-12-def} of this kernel with a PDF, one integrates over $z$ with $z_1 = u z$ and $z_2 = \bar{u} z$, with $u$ and $\bar{u}$ given in \eqref{x-u-def}.

\paragraph{Real graphs.}
One of the minus-momentum integrals in real graphs can be carried out trivially by using the on-shell condition of the emitted parton $a_3$.  The two remaining minus-momentum integrals are performed by closing the integration contour in the complex plane, picking up exactly one propagator pole per integral.  Details are given in \sect{4.2.1} of \cite{Diehl:2017kgu}.  After these steps, the remaining integrals are over the transverse momenta $\tvec{k}_1$ and $\tvec{k}_2$ in $2 - 2\epsilon$ dimensions.

We find it useful to write
\begin{align}
z_3 &= 1 - z_1 - z_2 \,,
&
\tvec{k}_3 = - \tvec{k}_1 - \tvec{k}_2
\end{align}
for the plus-momentum fraction and transverse momentum of the emitted parton $a_3$.  The on-shell condition then reads
\begin{align}
\label{k3-on-shell}
k_3^- &= \frac{\tvec{k}_3^2}{2 z_3^{} \ms k^+} \,,
\end{align}
and the convolution \eqref{conv-12-def} of the splitting kernel with a PDF then can be written as an integration over $z_3$ from $0$ to $1-x$ with
\begin{align}
\label{conv-12-z3}
z_1 &= u (1-z_3) \,,
&
z_2 &= \ub (1-z_3) \,.
\end{align}

Let us first assume that we take lightlike Wilson lines without any rapidity regulator in the operators \eqref{op-defs}.  For nonzero $z_3$ this is sufficient for obtaining well-defined results.  Instead of Feynman gauge, one can then also use the light-cone gauge $A^+ = 0$ to compute the graphs.  We computed in both gauges and found agreement.

At the point $z_3=0$, we encounter two types of singularities in the real graphs.  They arise if parton $a_3$ is a gluon, which must couple to an eikonal line if one computes in Feynman gauge.
\begin{enumerate}
\item \emph{Soft singularities} are due to the region where both $\tvec{k}_3$ and $z_3$ go to zero at fixed $\tvec{k}_3^2 / z_3$.  The minus-momentum $k_3^-$ of the final state parton $a_3$ remains finite in this limit and thus does not lead to a suppression due to other lines in the graph going far off shell.

We regulate these singularities by working in $4 - 2\epsilon$ dimensions.  The volume of the transverse momentum integration $d^{2-2\epsilon} \tvec{k}_3$ then contributes a fractional power $z_3^{1-\epsilon}$ in the relevant momentum region.  With one factor $1/z_3$ from the invariant phase space of $a_3$ and another factor $1/z_3$ from the eikonal propagator, one obtains a behaviour $z_3^{-1 -\epsilon}$ for the graph, unless there are additional factors of $z_3$ in the numerator.  The convolution integral in $z_3$ is well defined, provided that one takes $\epsilon < 0$, as is appropriate for the regularisation of divergences in the infrared.
\item \emph{Rapidity singularities} are due to the region in which $z_3$ becomes small but $\tvec{k}_3$ does not.  With $k_3^-$ scaling like $1/z_3$ in this limit, at least one line in the graph goes far off shell, and its propagator denominator gives a factor $z_3$.  With $1/z_3$ from the phase space of $a_3$ and $1/z_3$ from the unregulated eikonal propagator, one obtains a singular $z_3^{-1}$ behaviour of the kernel if there are no additional powers of $z_3$ from additional off-shell lines or from the numerator of the graph.

We find that the only graphs with rapidity divergences are those that have UND topology (see \fig{\ref{fig:real-LO}} and \fig{\ref{fig:eik-recipe}}).  These graphs must be evaluated with a rapidity regulator.  All other graphs can be evaluated without such a regulator, with a caveat that will be discussed in \app{\ref{sec:collins-integrals}}.
\end{enumerate}

Let us note in passing that the two kinematic regions just discussed resemble the ultrasoft and soft modes in soft-collinear effective theory (SCET) \cite{Bauer:2001yt, Beneke:2002ph}.  Ultrasoft gluons have sufficiently small momentum components to couple to collinear lines without carrying them far off-shell.  By contrast, the minus-momentum of soft gluons is large enough to carry right-moving collinear lines far off shell.

We now show how the two different rapidity regulators work for the two-loop graphs under consideration.  For this part of the calculation, we work in Feynman gauge.  If we use the $\delta$ regulator, the eikonal propagators with momentum $k_3$ have a denominator $k_3^+ \pm i \delta^+$.  In the sum over complex conjugate graphs, we find that only the real part
\begin{align}
\frac{1}{k_3^+ + i \delta^+} + \text{c.c.}
  &= \frac{2}{k^+} \, \frac{z_3}{z_3^2 + z_1^{} z_2^{} /\rho}
\end{align}
remains, where c.c.\ denotes the complex conjugate and $\rho = k_1^+ k_2^+ / (\delta^+)^2$.  Note that this is consistent with the definition \eqref{rho-def-delta} for $\rho$, because in the DPD we have $k_1^+ = x_1^{} \ms p^+$ and $k_2^+ = x_2^{} \ms p^+$ for the observed partons.  If we use the Collins regulator, we have
\begin{align}
\label{eikonal-collins}
\underset{\varepsilon \to 0}{\lim} \;
   \frac{1}{v_L^- k_3^+ + v_L^+ k_3^- + i \varepsilon} + \text{c.c.}
&= \frac{2}{v_L^- \ms k^+} \,
   \operatorname{PV}
   \frac{z_3}{z_3^2 - \tvec{k}_3^2 \ms z_1^{} z_2^{} \big/ \rho}
\end{align}
in the sum over complex conjugate graphs, where $\operatorname{PV}$ denotes the principal value prescription and $\rho = 2 k_1^+ k_2^+\, v_L^- \big/ |v_L^+|$ in accordance with \eqref{rho-def-collins}.  Here we have used the on-shell condition \eqref{k3-on-shell} for the emitted gluon.
Notice that for the $\delta$ regulator, the eikonal propagators do not affect the loop integrals over transverse momenta, whereas for the Collins regulator they do.  We explain in \app{\ref{sec:collins-integrals}} that it is not convenient to carry out the transverse-momentum integration with eikonal propagators of the form \eqref{eikonal-collins}.  Instead, we first make use of simplifications that arise in the limit $\rho \to \infty$.

In order to combine the rapidity singularities in the real graphs with those from virtual graphs and from the soft factor, we use the distributional identities
\begin{align}
\label{distr-delta}
\underset{\rho \to \infty}{\lim} \;
   \frac{z_3}{z_3^2 + z_1^{} z_2^{} /\rho} &=
   \mathcal{L}_0(z_3) + \frac{1}{2}\ms \delta(z_3)
      \Bigl[ \log\rho - \log (z_1 z_2) \Bigr]
\intertext{and}
\label{distr-collins}
\underset{\rho \to \infty}{\lim} \;
\operatorname{PV}
\frac{z_3}{z_3^2 - \tvec{k}_3^2 \ms z_1^{} z_2^{} \big/ \rho} &=
  \mathcal{L}_0(z_3)
  + \frac{1}{2}\ms \delta(z_3) \, \biggl[ \log \frac{\rho}{\tvec{\Delta^2}}
     - \log (z_1 z_2)
     - \log \frac{\tvec{k}_3^2}{\tvec{\Delta^2}} \biggr] \,,
\end{align}
where following \cite{Ligeti:2008ac} we use the notation
\begin{align}
\label{distr-L}
\mathcal{L}_n(z_3) &= \biggl[ \frac{\ln^n(z_3)}{z_3} \biggr]_+
  = \frac{\ln^n(z_3)}{z_3}
    - \delta(z_3) \int_0^1 dz'\, \frac{\ln^n(z')}{z'}
\end{align}
for plus distributions.  It is understood that the functions multiplying the distributions in  \eqref{distr-delta} to \eqref{distr-L} are sufficiently smooth and vanish outside the interval $0\le z_3 \le 1$.  A derivation of \eqref{distr-delta} and \eqref{distr-collins} is given in \app{\ref{sec:distrib}}.  Note that $z_1$ and $z_2$ depend on $z_3$ as specified in \eqref{conv-12-z3}.

Comparing \eqref{distr-delta} with \eqref{distr-collins}, we see that the term with $\log\rho$ involves the same loop integral for both regulators.  This implies that the contribution from real graphs to the coefficient $V_{\bus}^{(2,1)}$ in \eqref{V-log-delta} and \eqref{V-log-collins} is identical for both regulators.  By contrast, for the Collins regulator the coefficient $V_{\bus}^{(2,0)}$ receives an additional contribution from the last term in the square brackets of \eqref{distr-collins}.

The loop integrals for the $\log\rho$ terms \eqref{distr-delta} or \eqref{distr-collins} can be carried out with elementary methods, once one has taken the limit $z_3 \to 0$ in the integrands.  The same holds for the integrals involving the last term in the square brackets of \eqref{distr-collins} if we write
\begin{align}
\log \bigl( \tvec{k}_3^2 \ms\big/ \tvec{\Delta}^2 \bigr)
  &= \biggl[ \frac{\partial}{\partial \alpha} \,
     \bigl( \tvec{k}_3^2 \ms\big/ \tvec{\Delta}^2 \bigr)^\alpha
     \bigg]_{\alpha=0} \,,
\end{align}
perform the integrals for fixed $\alpha$, and then take the derivative at the point $\alpha=0$.
To carry out the loop integrals going with $\mathcal{L}_0(z_3)$ in \eqref{distr-delta} or \eqref{distr-collins}, as well as the loop integrals for graphs without rapidity divergences, we proceed as described in \cite{Diehl:2019rdh}.  We use integration-by-part reduction \cite{Chetyrkin:1981qh,Tkachov:1981wb} as implemented in \texttt{LiteRed} \cite{Lee:2013mka}.  The resulting master integrals are evaluated using the method of differential equations \cite{Kotikov:1990kg,Bern:1993kr,Remiddi:1997ny,Gehrmann:1999as}.  For the latter, we perform in particular a transformation to the canonical basis \cite{Henn:2013pwa} using \texttt{Fuchsia} \cite{Gituliar:2017vzm}.  The initial conditions for the differential equations are taken at the point $z_3\to 0$.  To compute the loop integrals in this limit, we use the method of regions  \cite{Beneke:1997zp}.

After carrying out the loop integrations, we have terms in which $\mathcal{L}_0(z_3)$ is multiplied by a function going like $z_3^0$ or like $z_3^{-\epsilon}$ in the limit $z_3\to 0$, which respectively corresponds to a rapidity or to a soft singularity.  In the former case, the plus-prescription regulates the $1/z_3$ behaviour in convolutions over $z_3$.  In the latter case, we can rewrite $\mathcal{L}_0(z_3^{}) \, z_3^{-\epsilon} = z_3^{-1-\epsilon}$ because $\epsilon$ is to be taken negative for soft singularities.  In a final step, we expand our results around $\epsilon = 0$ and use the distributional identity \cite{Gaunt:2014xga}
\begin{align}
\label{distr-eps}
z_3^{-1-\epsilon}
 = -\frac{1}{\epsilon} \, \delta(z_3) + \mathcal{L}_0(z_3) - \epsilon \ms \mathcal{L}_1(z_3) + \dfrac{\epsilon^2}{2} \, \mathcal{L}_2(z_3) + \mathcal{O}(\epsilon^3) \,.
\end{align}
We find that in the sum over all real graphs, the terms explicitly given on the r.h.s.\ are multiplied with a factor $z_3$.  As a result, the $\delta(z_3)$ contribution disappears, whereas the plus-distribution terms simplify to $z_3 \, \mathcal{L}_n(z_3) = \log^n z_3$ with $n=0,1,2$.


\paragraph{Virtual graphs.}
For virtual graphs, we have the kinematic constraints $k_2^- = - k_1^-$ and $\tvec{k}_2 = - \tvec{k}_1$, and we take $k_1^-$ and $\tvec{k}_1$ as independent integration variables.  In addition, we must integrate over a loop momentum $\ell$.  For vertex correction graphs, the momentum routing is shown in \fig{\ref{fig:virt-kin-vertex}}.  The integrals over $\Delta^-$, $k_1^-$, and $\ell^-$ are performed by complex contour integration in a similar way as for the real graphs.  This leaves us with integrals over $\tvec{k}_1$, $\tvec{\ell}$, and $\ell^+$.  The $\ell^+$ integral runs over a finite interval: outside this interval, one of the minus-momentum integrals gives zero because all poles are on the same side of the real axis.

Writing $\ell^+ = z_\ell \, k^+$, we find endpoint singularities at $z_\ell =0$, which are either soft singularities or rapidity singularities.  Their discussion is fully analogous to the one for $z_3 = 0$ singularities in the real graphs.  Rapidity divergences are made explicit by using the analogues of the distributional identities \eqref{distr-delta} and \eqref{distr-collins} with $z_\ell$ instead of $z_3$.  Note that in this case $z_1$ and $z_2$ are  fixed during the integration over $z_\ell$, which does however not change the r.h.s.\ of the identities.  Soft singularities are made explicit with the analogue of \eqref{distr-eps} after the transverse-momentum integrals have been performed.

The virtual graphs that have rapidity divergences are those obtained by the rules in \fig{\ref{fig:eik-recipe}} from vertex correction graphs in which the momentum $\ell$ is carried by a gluon, such as in \fig{\ref{fig:gqqbar_vertex}}.


\paragraph{Combining all graphs.}
After the steps described so far, we can add the contributions from real and virtual graphs, using that the latter are multiplied with $\delta(z_3)$ due to kinematics.  Since the loop integrals accompanied by $\log\rho$ are elementary, we can compute $V_{\bus}^{(2,1)}$ to all orders in $\epsilon$ and thus verify the condition \eqref{rap-relation}, which ensures that all rapidity divergences cancel when combining the unsubtracted kernels with the soft factor.

For the $\delta$ regulator, we find that the single poles in $V_{\bus}^{(2,0)}$ satisfy the consistency relation \eqref{single-pole-delta}.  For the Collins regulator, we denote by $\delta V_{\bus}^{(2,0)}$ the sum of contributions from the last term in \eqref{distr-collins} and from its analogue for $z_\ell$.  We find that
\begin{align}
\Bigl[ \delta V_{\bus}^{(2,0)} \Bigr]_{-2 + k}
&= \frac{3}{2} \, \frac{\gamma_J^{(0)}}{2}\,
   \Bigl[ R_{\epsilon}^{-1} \, V_{B}^{(1)} \Bigr]_{k}
\qquad \text{for $k=0,1,2$.}
\end{align}
This is exactly what is required to fulfil the consistency relations \eqref{double-pole-collins} and \eqref{single-pole-collins} for double and single poles with the Collins regulator and to obtain the same result for the finite part $V_{\text{fin}}^{(2)}$ with both regulators (see \eqref{V_fin-delta} and \eqref{V_fin-collins}).

Our calculation thus establishes the cancellation of rapidity divergences at all orders in $\epsilon$, as well as the cancellation of all poles in $\epsilon$ for two rapidity regulators, and we obtain the same finite result for $V^{(2)}$ in both cases.  We regard this is as a strong cross check of our results.

At this point, we must point out an issue with the definition of the $\delta$ regulator.  In \cite{Echevarria:2016scs} it is stated that in the twist-two operator defining an unsubtracted TMD, the parameter $\delta$ should be rescaled by the parton momentum fraction $x$; see \eqn{(3.9)} in that paper.  A corresponding rescaling in our case would lead to some modification of the term $\log\rho \; V_{\bus}^{(2,1)}(\epsilon)$ in \eqref{VB2-start-delta}.  The second term with $\log \rho$ in the same equation comes from the soft factor and would not be modified.  One would then obtain a different result for the final kernel $V^{(2)}$ and thus lose agreement with the result for the Collins regulator.  Moreover, the extra term from the modification would be multiplied by $V_{\bus}^{(2,1)}(\epsilon)$, which has a pole in $1/\epsilon$ according to \eqref{rap-relation}.  Since the counterterms in our calculation are completely fixed, the modification would not give a finite result for $\epsilon = 0$.  Clearly, our calculation does not admit any rescaling of $\delta$ in the unsubtracted DPD.

The rescaling prescription for $\delta$ in \cite{Echevarria:2016scs} is a consequence of the fact that the $\zeta$ parameter appearing in the $Z$ factor for TMDs was defined with respect to $p^+$ rather than $x p^+$ in that work.  As we explained in \sect{\ref{sec:renormalisation}}, the rapidity parameter in a $Z$ factor must refer to the parton momentum in the renormalised operator.  If this is the case, no rescaling prescription for $\delta$ should be applied.\footnote{We thank A.~Vladimirov for discussions about this issue.}

\section{Results for the two-loop kernels}
\label{sec:results}

In this section, we present several general features of our results for the DPD splitting kernels, with a focus on their colour structure and various kinematic limits.  Numerical illustrations will be given in \sect{\ref{sec:numerics}}.  The full analytic expressions of the kernels are given in the ancillary files associated with this paper on \href{https://arxiv.org/abs/2105.08425}{arXiv}.

According to \eqref{two-loop-final}, the two-loop kernels have the form
\begin{align}
\label{V-NLO-form}
\prn{\RR}{V}^{(2)}_{a_1 a_2, a_0}(z_1, z_2, y, \mu, \zeta)
&= \prn{\RR}{V}^{[2,0]}_{a_1 a_2, a_0}(z_1, z_2)
   + L \; \prn{\RR}{V}^{[2,1]}_{a_1 a_2, a_0}(z_1, z_2)
\nonumber \\[0.2em]
& \quad
   + \biggl( L \log\frac{\mu^2}{\zeta} - \frac{L^2}{2}
    + c_{\msbars} \biggr) \,
    \frac{\prn{R_1}{\gamma}_J^{(0)}}{2} \;
    \prn{\RR}{V}^{(1)}_{a_1 a_2, a_0}(z_1, z_2)
\end{align}
with $L$ defined in \eqref{L-b0-def}, $c_{\msbars}$ given in \eqref{MSbar-c}, and $\pr{R}{\gamma}_J^{(0)}$ in \eqref{gammaJ-LO}.  The full splitting DPD at two-loop accuracy is obtained from \eqref{V-expansion}.  In the following, we call ${V}^{[2,0]}$ the non-logarithmic part and ${V}^{[2,1]}$ the logarithmic part of the two-loop kernel.  The last line in \eqref{V-NLO-form}, including the constant $c_{\msbars}$, will be referred to as the double-logarithmic part.

Let us recall that the one-loop splitting kernels have the form
\begin{align}
\prn{\RR}{V}^{(1)}(z_1, z_2)
&= \delta(1 - z_1 - z_2) \, \prn{\RR}{V}^{(1)}(z_1) \,,
\end{align}
so that the convolution \eqref{conv-12-def} simplifies to
\begin{align}
\Bigl[ \prn{\RR}{V}^{(1)} \conv{12} f \,\Bigr](x_1, x_2)
&= \prn{\RR}{V}^{(1)}(u) \, f(x) \big/ x
\end{align}
with $u = x_1 / x$ and $x = x_1 + x_2$.  These kernels were already computed in \cite{Diehl:2011yj} and are given by
\begin{align}
\label{LO-rescaling-2}
\prn{\RR}{V}^{(1)}_{a_1 a_2, a_0} &=
   c_{a_1 a_2, a_0}^{}(R_1 R_2)\; \prn{11}{V}^{(1)}_{a_1 a_2, a_0}
\end{align}
with scaling factors
\begin{align}
\label{rescaling-factors}
c_{g g, g}(AA) & = - \frac{\sqrt{N^2-1}}{2} \,,
&
c_{g g, g}(SS) & = - c_{g g, g}(AA) \,,
\nonumber \\[0.9em]
c_{g g, g}(AS) &= c_{g g, g}(S\bs A) = 0
&
c_{g g, g}(\tentenbartext) &= c_{g g, g}(\tenbartentext) = 0 \,,
&
c_{g g, g}(27\, 27) & \underset{N=3}{=}\bs - \sqrt{3} \,,
\nonumber \\[0.6em]
c_{q\bar{q}, g}(88) & = -\frac{1}{\sqrt{N^2-1}}\,,
\nonumber \\
c_{q g, q}(8A) & = - \frac{N}{\sqrt{2}}\,,
&
c_{q g, q}(8S) & = \sqrt{\frac{N^2-4}{2}} \,.
\end{align}
By definition, one has $c_{a_1 a_2, a_0}(11) = 1$ for all channels.  The colour-singlet kernels $\prn{11}{V}^{(1)}(u)$ are equal to the familiar DGLAP splitting functions without the plus prescriptions and delta functions:
\begin{align}
\label{one-loop-singlet-V}
\prn{11}{V}^{(1)}_{g g, g}(u)
&= 2 C_A \, p_{g g}(u)  \,,
&
\prn{11}{V}^{(1)}_{q \bar{q}, g}(u)
&= T_F \, p_{q g}(u) \,,
&
\prn{11}{V}^{(1)}_{q g, q}(u)
&= C_F\, p_{q q}(u) \,,
\intertext{where}
  \label{eq:buildingblocks}
p_{g g}(u)
   & = \frac{\ub}{u} + \frac{u}{\ub} + u \ub \,,
&
p_{q g}(u)
   & = u^2 + \ub^2
&
p_{q q}(u)
  & = \frac{1 + u^2}{\ub} \,.
\end{align}
We recall that $\ub = 1-u$.
Setting $N=3$ in \eqref{rescaling-factors}, we obtain numerical values
\begin{align}
\label{rescaling-factors-num}
c_{g g, g}(AA) & \approx -1.4 \,,
&
c_{g g, g}(27\, 27) & \approx -1.7 \,,
&
c_{q \bar{q}, g}(88) & \approx - 0.35 \,,
\nonumber \\
c_{q g, q}(8A) & \approx - 2.1 \,,
&
c_{q g, g}(8S) & \approx 1.6 \,,
\end{align}
which shows that colour correlations are quite large for the splitting into $g g$ and $q g$.

The dependence of the two-loop kernels on the momentum fractions $z_1$ and $z_2$ can be cast into the form
\begin{align}
\label{V-gen-struct}
{V}^{[2,0]}(u z,\ub z)
&= {V}^{[2,0]}_{\text{reg}}(u z, \ub z)
   + \delta(1 - z)\, {V}_{\delta}^{[2,0]}(u) \,,
\nonumber \\
{V}^{[2,1]}(u z,\ub z)
&= {V}^{[2,1]}_{\text{reg}}(u z, \ub z)
   + \frac{1}{[ 1 - z ]_+}\, {V}^{[2,1]}_p(u)
   + \delta(1 - z)\, {V}^{[2,1]}_{\delta}(u)
\end{align}
with regular terms ${V}^{[2,k]}_{\text{reg}}(u z, \ub z)$ that are finite or have an integrable singularity at $z=1$ for $0 < u < 1$.  Such a singularity is at most a power of $\log (1-z)$.  Note that the values $u=0$ and $u=1$ are outside the region relevant for DPDs in cross section formulae, where both $x_1$ and $x_2$ must be strictly positive.  The behaviour of
the kernels in the limit $u\to 0$ or $u\to 1$ is discussed in \sect{\ref{sec:small_u_or_ubar}}.

Terms with $1/[1-z]_+$ and $\delta(1-z)$ appear only in LO channels.  It is noteworthy that ${V}^{[2,0]}$ does not have any term going with the plus distribution $1/[1-z]_+$.  All such terms appearing in two-loop graphs cancel when the one-loop counterterms are added.  We have no explanation for this finding.

We note that a more general form of the kernel is given in \eqn{(110)} of \cite{Diehl:2019rdh}, where the function multiplying $1/[1-z]_+$ depends on both $u$ and $z$.  This can be brought into the form \eqref{V-gen-struct} by expanding that function around $z=1$ and moving all terms with powers of $(1-z)$ into the regular term.

\paragraph{Symmetries.}
Let us now discuss the properties of splitting kernels that follow from charge conjugation invariance.  Channels in which all external partons are quarks or antiquarks will be called ``pure quark channels'' in the following.  The kernels for pure quark channels that are related by charge conjugation are equal.  For kernels with external gluons, we have
\begin{align}
   \label{C-rel-1}
\prn{R R}{V}_{q \bar{q}, g}(z_1,z_2)
   &= \prn{R R}{V}_{\bar{q} q, g}(z_1,z_2)\,,
\nonumber \\[0.1em]
\prn{\RR}{V}_{q g, q}(z_1,z_2)
   &= \eta(R_2)\, \prn{\RR}{V}_{\bar{q} g, \bar{q}}(z_1,z_2) \,,
\nonumber \\[0.1em]
\prn{\RR}{V}_{q g, g}(z_1,z_2)
   &= \eta(R_2)\, \prn{\RR}{V}_{\bar{q} g, g}(z_1,z_2) \,,
\nonumber \\
\prn{\RR}{V}_{g g, q}(z_1,z_2)
   &= \eta(R_1)\ms \eta(R_2)\,
      \prn{\Rbar_1\bs \Rbar_2}{V}_{g g, \bar{q}}(z_1,z_2) \,,
\nonumber \\
\prn{\RR}{V}_{g g, g}(z_1,z_2)
   &= \eta(R_1)\ms \eta(R_2)\,
      \prn{\Rbar_1\bs \Rbar_2}{V}_{g g, g}(z_1,z_2) \,,
\end{align}
where the sign factor $\eta(R)$ is $-1$ for $R=A$ and $+1$ otherwise.
The relation in the last line implies
\begin{align}
\prn{AS}{V}_{g g, g} &= \prn{S\bs A}{V}_{g g, g} = 0 \,,
&
\prn{\tentenbar}{V}_{g g, g} &= \prn{\tenbarten}{V}_{g g, g} \,.
\end{align}

A detailed proof of these relations is somewhat lengthy and shall not be given here.  In a first step, one can prove relations for the unsubtracted distributions $F_{\us, a_1 a_2}$, using the charge conjugation properties for the colour projected twist-two operators given in \eqref{C-parity-ops}.  At this stage, a sign factor $\eta(A) = -1$ appears for each gluon pair coupled to an antisymmetric octet.  Extending these relations to the subtracted distributions $F_{a_1 a_2}$ is easy, because the soft factor appearing in that step depends only on the multiplicity of the colour representations and does not distinguish between quarks and antiquarks.  Corresponding relations for the splitting kernels are then obtained from the splitting formula \eqref{split-master} by using the charge conjugation relations for PDFs.

The preceding relations hold at all orders in the coupling.  In addition, explicit calculation yields
\begin{align}
\label{C-rel-2}
\prn{R R}{V}^{(2)}_{q'\bs \bar{q}'\!, q}(z_1,z_2)
   &= \prn{R R}{V}^{(2)}_{\bar{q}'\bs q'\!, q}(z_1,z_2)
\\
\label{C-rel-3}
\prn{AS}{V}_{g g, q}(z_1,z_2)
   &= -\; \prn{S\bs A}{V}_{g g, q}(z_1,z_2)
\end{align}
and
\begin{align}
\label{C-rel-4}
\prn{\tentenbar}{V}_{g g, g}(z_1,z_2)
&= \prn{\tentenbar}{V}_{g g, q}(z_1,z_2)
 = \prn{\tenbarten}{V}_{g g, q}(z_1,z_2)
 = 0
\end{align}
at two-loop accuracy.

Together with the trivial symmetry relation
\begin{align}
\label{V-triv-symm}
\prn{\RR}{V}_{a_1 a_2, a_0}(z_1,z_2)
   &= \prn{R_2\bs R_1}{V}_{a_2\ms a_1, a_0}(z_2,z_1) \,,
\end{align}
we find that the kernels
\begin{align}
\prn{R R}{V}^{(2)}_{q'\bs \bar{q}'\!, q}(z_1,z_2) \,,
\qquad
\prn{R R}{V}_{q \bar{q}, g}(z_1,z_2) \,,
\qquad
\prn{\RR}{V}_{g g, g}(z_1,z_2)
\end{align}
are even under the exchange $z_1 \leftrightarrow z_2$, whereas $\prn{\RR}{V}_{g g, q}(z_1,z_2)$ is odd for the combinations $\RR = AS$ and $S\bs A$ and even for all others.


\subsection{Colour dependence}
\label{sec:colour-results}

We now specify the colour dependence of the two-loop kernels for each partonic channel.  In the following, $R(u z, \ub z)$, $D(u)$, and $S(u)$ respectively denote contributions to the terms $V_{\text{reg}}(u z, \ub z)$, $\delta(1-z)\, V_{\delta}(u)$, and $V_{p}(u) \, \big/ [1-z]_+$ in \eqref{V-gen-struct}.  The functions $R$, $D$, and $S$ are independent of the number of colours $N$.  They may or may not depend on the representations $R_1 R_2$, as indicated by the presence or absence of a corresponding superscript.  The coefficient $\beta_0$, defined in \eqref{beta0-def}, is the only place where a dependence on the number $n_F$ of active quarks appears in the kernels.

From now on, we set $T_F = 1/2$ for simplicity.  We also recall that all results for $R=27$ are valid for $N=3$ only.

\paragraph{$\boldsymbol{g \to g g} \ms$:}
Using the scaling factors in \eqref{rescaling-factors} one may write the non-logarithmic and logarithmic parts of the $g \to g g$ kernel as
\begin{align}
  \label{eq:RVggg2}
    \prn{\RR}{V}_{g g, g}^{[2,0]}
  & = c_{g g, g}(R_1 R_2)
    \left[\,
      C_A^2
      \left(
        \pr{\RR}{R}_{g g, g}^{A \, [2,0]} + D_{g g, g}^{A \, [2,0]}
      \right)
      + C_A\ms \beta_0 \, D_{g g, g}^{\beta \, [2,0]}
    \,\right] \,,
  \nonumber\\[1em]
    \prn{\RR}{V}_{g g, g}^{[2,1]}
  & = c_{g g, g}(R_1 R_2)
    \left[\,
      C_A^2
      \left(
        \pr{\RR}{R}_{g g, g}^{A \, [2,1]} + \pr{\RR}{S}_{g g, g}^{A \, [2,1]}
        + \pr{\RR}{D}_{g g, g}^{A \, [2,1]}
      \right) \right.
  \nonumber\\[0.5em]
  & \phantom{{}= c_{g g, g}(R_1 R_2) \Bigl[}
    \left.
      {}+ C_A\ms \beta_0 \, D_{g g, g}^{\beta \, [2,1]}
    \,\right] \,.
\end{align}
For some of the functions in \eqref{eq:RVggg2} one finds simple relations between different colour channels:
\begin{itemize}
  \item The $\delta$ distribution terms $D_{g g, g}^{A \, [2,0]}$, $D_{g g, g}^{\beta \, [2,0]}$, and $D_{g g, g}^{\beta \, [2,1]}$ are independent of $R$.   They originate from contributions with simple LO scaling, such as virtual graphs and associated counterterms.  In particular one finds that
  \begin{align}
    \label{eq:Dggg21beta}
   C_A \ms D_{g g, g}^{\beta \, [2,1]}
    & = \prn{11}{V}^{(1)}_{g g, g} \,.
  \end{align}
  \item As a consequence of \eqref{relation-AS}, one has
  \begin{align}
    \label{eq:RVggg2-octets}
      \prn{AA}{V}_{g g, g}^{(2)}
    & = {}- \prn{SS}{V}_{g g, g}^{(2)}
  \end{align}
  for the full two-loop kernels.  The functions $R$, $S$, and $D$ in \eqref{eq:RVggg2} are hence equal for $R_1 R_2 = AA$ and $R_1 R_2 = SS$.
  \item Since $c_{g g, g}(\tentenbartext) = 0$, one has $\prn{\tentenbar}{V}^{(1)}_{g g, g} = \prn{\tentenbar}{V}^{(2)}_{g g, g} = 0$.  The same holds for the conju\-gate channel $R_1 R_2 = \tenbartentext$.
  \item Due to charge conjugation invariance, the kernels for $R_1 R_2 = AS$ and $S\bs A$ are zero at all orders.
  \item An interesting feature is that the plus distribution terms in the octet channels vanish,
  \begin{align}
    \label{eq:RVggg2-octets-L0}
      \prn{AA}{S}_{g g, g}^{[2,1]}
    & = \prn{SS}{S}_{g g, g}^{[2,1]}
      = 0\,.
  \end{align}
  This is because these channels have a vanishing colour factor for the eikonal line graphs with UND topology, which are the only two-loop graphs giving rise to plus distribution terms.
\end{itemize}
%
%
\paragraph{$\boldsymbol{g \to q \bar{q}} \ms$:}
For the $g \to q \bar{q}$ kernel, one finds
\begin{align}
  \label{eq:RVqqbarg2}
    \prn{RR}{V}_{q \bar{q}, g}^{[2,0]}
  & = c_{q \bar{q}, g}(RR)
    \left[
      C_A
      \left(
        \pr{RR}{R}_{q \bar{q}, g}^{A \, [2,0]} + D_{q \bar{q}, g}^{A \, [2,0]}
      \right)
      + C_F
      \left(
        R_{q \bar{q}, g}^{F \, [2,0]} + D_{q \bar{q}, g}^{F \, [2,0]}
      \right)
    \right] \,,
  \nonumber\\[1em]
    \prn{RR}{V}_{q \bar{q}, g}^{[2,1]}
  & = c_{q \bar{q}, g}(RR)
    \left[\ms
      C_A
      \left(
        \pr{RR}{R}_{q \bar{q}, g}^{A \, [2,1]} + \pr{RR}{S}_{q \bar{q}, g}^{A \, [2,1]}
        + \pr{RR}{D}_{q \bar{q}, g}^{A \, [2,1]}
      \right)
    \right.
  \nonumber\\[0.5em]
  & \phantom{{} = c_{q \bar{q}, g}(RR)\Bigl[}
    \left.
      {}+ C_F
      \left(
        R_{q \bar{q}, g}^{F \, [2,1]} + S_{q \bar{q}, g}^{F \, [2,1]}
        + D_{q \bar{q}, g}^{F \, [2,1]}
      \right)
    \right] \,.
\end{align}
Except for $\pr{RR}{D}_{q \bar{q}, g}^{A \, [2,1]}$, all terms have simple relations between the octet and the singlet:
\begin{itemize}
  \item The $\delta$ distribution terms in the non-logarithmic part exhibit LO scaling, such that $D_{q \bar{q}, g}^{A \, [2,0]}$ and $D_{q \bar{q}, g}^{F \, [2,0]}$ are independent of $R$.
  \item All terms going with $C_F$ scale like the LO kernels.
  \item The relations \eqref{colour-relation-g-qqbar} for the sum of real graphs lead to the following simple relations for the regular and plus-distribution terms with colour factor $C_A\ms$:
  \begin{align}
    \label{eq:RVqqbarg2-scaling}
      \prb{88}{R}_{q \bar{q}, g}^{A \, [2,k]}
    & = \prn{11}{R}_{q \bar{q}, g}^{A \, [2,k]}
      - \frac{1}{2} \, R_{q \bar{q}, g}^{F \, [2,k]}
    &&\text{for } k = 0, 1 \,,
    \nonumber\\[0.3em]
      \pr{88}{S}_{q \bar{q}, g}^{A \, [2,1]}
    & = \prn{11}{S}_{q \bar{q}, g}^{A \, [2,1]}
      - \frac{1}{2} \, S_{q \bar{q}, g}^{F \, [2,1]} \,.
  \end{align}
\end{itemize}
%
%
\paragraph{$\boldsymbol{q \to q g} \ms$:}
The colour structure for this channel has the form
\begin{align}
  \label{eq:RVqgq2}
    \prn{\RR}{V}_{q g, q}^{[2,0]}
  & = c_{q g, q}(R_1 R_2)
    \left[\,
      C_A\ms C_F
      \left(
        \pr{\RR}{R}_{q g, q}^{A \, [2,0]} + D_{q g, q}^{A \, [2,0]}
      \right) \right.
  \nonumber\\[0.5em]
  & \phantom{{} = c_{q g, q}(R_1 R_2)\Bigl[}
    \left.
      {}+ C_F^2
      \left(
        R_{q g, q}^{F \, [2,0]} + D_{q g, q}^{F \, [2,0]}
      \right)
      + C_F \beta_0 \, D_{q g, q}^{\beta \, [2,0]}
    \,\right] \,,
  \nonumber\\[1em]
    \prn{\RR}{V}_{q g, q}^{[2,1]}
  & = c_{q g, q}(R_1 R_2)
    \left[\,
      C_A\ms C_F
      \left(
        \pr{\RR}{R}_{q g, q}^{A \, [2,1]} + \pr{\RR}{S}_{q g, q}^{A \, [2,1]}
        + \pr{\RR}{D}_{q g, q}^{A \, [2,1]}
      \right)
    \right.
  \nonumber\\[0.5em]
  & \phantom{{} = c_{q g, q}(R_1 R_2)\Bigl[}
    \left.
      {}+ C_F^2
      \left(
        R_{q g, q}^{F \, [2,1]} + S_{q g, q}^{F \, [2,1]}
        + D_{q g, q}^{F \, [2,1]}
      \right)
      + C_F \beta_0 \, D_{q g, q}^{\beta \, [2,1]}
    \,\right] \,.
\end{align}
Here, some but not all functions obey simple relations between different colour representations:
\begin{itemize}
  \item As in the previous channels, the $\delta$ distribution terms in the non-logarithmic part exhibit LO scaling, such that $D_{q g, q}^{A \, [2,0]}$, $D_{q g, q}^{F \, [2,0]}$, and $D_{q g, q}^{\beta \, [2,0]}$ are independent of $R$.  The same holds for the term going with $\beta_0$ in the logarithmic part, which is given by
  \begin{align}
    \label{eq:Dqgq21beta}
    C_F\ms  D_{q g, q}^{\beta \, [2,1]}
    & = \prn{11}{V}^{(1)}_{q g, q} \,
  \end{align}
  in analogy to \eqref{eq:Dggg21beta}.
  \item All terms with $C_F^2$ scale like the LO kernels.
  \item As a consequence of \eqref{relation-AS}, there is a simple scaling relation for the full kernels in the two octet representations:
  \begin{align}
    \label{eq:RVqgq2-octets}
    \prn{8A}{V}_{q g, q}^{(2)} \,\big/\, \prn{8S}{V}_{q g, q}^{(2)}
    & =  c_{q g, q}(8A) \big/\ms c_{q g, q}(8S)
      = - N \big/ \sqrt{N^2 - 4} \,.
  \end{align}
  The functions on the r.h.s.\ of \eqref{eq:RVqgq2} are hence equal for $R_1 R_2 = 8A$ and $R_1 R_2 = 8S$.
  \item As in the $g\to g g$ channel, one finds that the plus-distribution terms vanish for the octet representations,
  \begin{align}
    \label{eq:RVqqg2-octets-L0}
      \prn{8A}{S}_{q g, q}^{[2,1]}
    & = \prn{8S}{S}_{q g, q}^{[2,1]}
      = 0 \,.
  \end{align}
  This can be again explained by a vanishing colour factor of the eikonal line graphs with UND topology.
\end{itemize}
%
%
\vspace{\baselineskip}

We now turn to the kernels for the NLO channels, which contain regular terms but no $\delta$ or plus distributions.  The double logarithmic part of ${V}^{(2)}$ is zero in these channels.

\paragraph{$\boldsymbol{g \to q g} \ms$:}
The kernel for $g \to q g$ has the form
\begin{align}
  \label{eq:RVqgg2}
    \prn{\RR}{V}_{q g, g}^{[2,k]}
  & = c_{q g, q}(R_1 R_2)
    \left[\,
      C_A \, \pr{\RR}{R}_{q g, g}^{A \, [2,k]} + C_F \, R_{q g, g}^{F \, [2,k]}
    \,\right]
  &&\text{for }
    k = 0,1 \,.
\end{align}
Noteworthy features of this channel are:
\begin{itemize}
  \item The terms with $C_F$ scale like the LO kernels.
  \item The full kernels for the two octets are related by
  \begin{align}
    \label{eq:RVqgg2-octets}
    \prn{8A}{V}_{q g, g}^{(2)} \,\big/\, \prn{8S}{V}_{q g, g}^{(2)}
    & =  c_{q g, q}(8A) \big/\ms c_{q g, q}(8S)
      = - N \big/ \sqrt{N^2 - 4} \,,
  \end{align}
  as a consequence of \eqref{relation-AS}.  The functions for $R_1 R_2 = 8A$ and $R_1 R_2 = 8S$ on the r.h.s.\ of \eqref{eq:RVqgg2} are hence equal.
\end{itemize}
%
%
\paragraph{$\boldsymbol{q \to g g} \ms$:}
As a consequence of the relations in \eqref{colour-relations-q-gg}, one can write
\begin{align}
  \label{eq:RVggq2}
    \prn{\RR}{V}_{g g, q}^{[2,k]}
  & = c_{g g, g}(R_1 R_2)
    \left[\,
      C_A\ms C_F\, R_{g g, q}^{A \, [2,k]}
      + C_F^2 \; \tilde{c}(R_1 R_2) \, R_{g g, q}^{F \, [2,k]}
    \,\right]
  &&\text{for }
    k = 0,1
\end{align}
in all colour channels except for the mixed octets.
\begin{itemize}
\item The terms multiplying $C_A\ms C_F$ in the square brackets are colour independent, whereas the terms multiplying $C_F^2$ have an additional scaling factor
\begin{align}
\label{eq:ctilde}
\tilde{c}(11) &= 1 \,,
&
\tilde{c}(AA) &= \frac{N^2}{N^2 - 1} \,,
&
\tilde{c}(SS) &= \frac{N^2 - 4}{N^2 - 1} \,,
&
\tilde{c}(27\, 27) &= 0 \,.
\end{align}
\item Since $c_{g g, g}(\tentenbartext) = 0$, one has $\prn{\tentenbar}{V}^{(2)}_{g g, q} = 0$, and the same holds for $R_1 R_2 = \tenbartentext$.
\end{itemize}
In the mixed octet channels, one has
\begin{align}
  \label{eq:RVggq2-new}
\prn{AS}{V}_{g g, q}^{[2,k]}
&= -\,\, \prn{S\bs A}{V}_{g g, q}^{[2,k]}
 = \tilde{c}(AS) \, \pr{AS}{R}_{g g, q}^{[2,k]}
&&\text{for }
    k = 0,1
\end{align}
with a global colour factor
\begin{align}
\label{eq:ctilde-new}
\tilde{c}(AS) &= C_F\ms \sqrt{N^2 - 1}\, \sqrt{N^2 - 4} \,.
\end{align}
%
%
\paragraph{Pure quark channels.}
Using the decomposition \eqref{quark-flavour-decomp} and the symmetry relations \eqref{C-rel-2} and \eqref{V-triv-symm}, one can deduce results for all pure quark channels from the five kernels
\begin{align}
&& \prn{RR}{V}_{q q, q}^{v\ms (2)} \,,
&& \prn{RR}{V}_{q \bar{q}, q}^{v\ms (2)} \,,
&& \prn{RR}{V}_{q q'\!, q}^{(2)} \,,
&& \prn{RR}{V}_{q \bar{q}'\!, q}^{(2)} \,,
&& \prn{RR}{V}_{q'\bs \bar{q}'\!, q}^{(2)} \,.
\end{align}
For $R = 1$, the first two have a colour factor $C_F \ms (C_A - 2 C_F)$, whereas the last three have a colour factor $C_F$.  One finds
\begin{align}
\label{Rpure-quark-C}
\prn{11}{V}_{q q'\!, q}^{(2)} &= \prn{11}{V}_{q \bar{q}'\!, q}^{(2)} \,.
\end{align}
Because each kernel corresponds to a single two-loop graph, one finds simple scaling relations between octet and singlet:
\begin{align}
  \label{eq:RVqqq2-scaling}
\prn{88}{V}_{q q, q}^{v\ms (2)}
\big/ \, \prn{11}{V}_{q q, q}^{v\ms (2)}
  & = (N^2 + 1) \, c_{q \bar{q}, g}(88) \,,
&
\prn{88}{V}_{q \bar{q}, q}^{v\ms (2)}
\big/\, \prn{11}{V}_{q \bar{q}, q}^{v\ms (2)}
  & = c_{q \bar{q}, g}(88) \,,
\nonumber\\[1em]
\prn{88}{V}_{q q'\!, q}^{(2)}
\,\big/\, \prn{11}{V}_{q q'\!, q}^{(2)}
  & = 2 \ms c_{q \bar{q}, g}(88) \,,
&
\prn{88}{V}_{q \bar{q}'\!, q}^{(2)}
\,\big/\, \prn{11}{V}_{q \bar{q}'\!, q}^{(2)}
  & = - (N^2 - 2) \, c_{q \bar{q}, g}(88) \,,
\nonumber\\[1em]
\prn{88}{V}_{q' \bs \bar{q}'\!, q}^{(2)}
\,\big/\, \prn{11}{V}_{q' \bs \bar{q}'\!, q}^{(2)}
  & = c_{q \bar{q}, g}(88) \,.
\end{align}
%
%
\paragraph{Relations between DPDs at small $y$.}
According to \eqref{eq:RVqgq2-octets} and \eqref{eq:RVqgg2-octets}, the two-loop kernels for both $q\to q g$ and $g\to q g$ follow the same scaling relation between the two octets as the $q\to q g$ splitting at one loop.  We therefore have the relation
\begin{align}
\label{Fqg-AS}
\pr{8A}{F}_{q g} \ms\big/\ms \pr{8S}{F}_{q g}
&= - N \big/ \sqrt{N^2 - 4}
\end{align}
for the DPDs at small $y$.  Corrections to this and the following three relations are of order $\mathcal{O}(\as^3)$ and $\mathcal{O}(y^2 \Lambda^2)$.

The kernels for the octet combinations $AA$ and $SS$ are opposite to each other in the $g\to g g$ channel, but they are not for $q\to g g$ because of the terms with $C_F^2$ in \eqref{eq:RVggq2}.  Using the values of the rescaling factors $c$ and $\tilde{c}$, we obtain the relation
\begin{align}
\label{Fgg-AS}
\pr{AA}{F}_{g g} + \pr{SS}{F}_{g g}
&= - \frac{\as^2}{\pi y^2} \, \frac{2\ms C_F^2}{\sqrt{N^2 - 1}}\;
     \Bigl[ R_{g g, q}^{F \, [2,0]} + L \ms R_{g g, q}^{F \, [2,1]} \Bigr]
     \,\conv{12}\, \sum_{q} \bigl( f_q + f_{\bar{q}} \bigr) \,.
\end{align}
We expect the relative difference between $\pr{AA}{F}_{g g}^{\phantom{[]}}$ and $- \,\pr{SS}{F}_{g g}$ to grow with $x_1 + x_2$, as the quark singlet distribution then becomes more important than the gluon distribution in the splitting DPDs.  This is confirmed by the numerical examples in \sect{\ref{sec:numerics}}.

The mixed octet distribution is driven by the valence combination $f_q - f_{\bar{q}}$,
\begin{align}
\label{Fgg-new}
\pr{AS}{F}_{g g}
& \underset{\phantom{N=3}}{=}
   \frac{\as^2}{\pi y^2} \, C_F\ms \sqrt{N^2 - 1}\, \sqrt{N^2 - 4}\;
     \Bigl[ \pr{AS}{R}_{g g, q}^{[2,0]}
            + L \, \pr{AS}{R}_{g g, q}^{[2,1]} \ms\Bigr]
     \,\conv{12}\, \sum_{q} \bigl( f_q - f_{\bar{q}} \bigr) \,,
\end{align}
with the minus sign between quark and antiquark distributions resulting from the fourth relation in \eqref{C-rel-1}.  Unlike all other two-gluon distributions, $\prn{AS}{F}_{g g}(x_1, x_2) = -\,\prn{S\bs A}{F}_{g g}(x_1, x_2)$ is odd under the exchange $x_1 \leftrightarrow x_2$.

For two gluons in the decuplet representation, we find
\begin{align}
\label{Fgg-dec}
\prn{\tentenbar}{F}_{g g} &= \prn{\tenbarten}{F}_{g g} = 0
\end{align}
at two-loop accuracy.  This result holds for all $N$.  It is an interesting question whether these distributions remain small when $y$ is in the non-perturbative region.

For pure quark channels, we get simple relations between the singlet and octet distributions if the quark flavors differ, because in this case only the kernel ${V}_{q q'\!, q}$ or ${V}_{q \bar{q}'\!, q}$ contributes.  As a consequence, one has
\begin{align}
\label{Fqq-8}
\pr{88}{F}_{q q'} \big/\ms \pr{11}{F}_{q q'}
  &= - 2 \big/ \sqrt{N^2 - 1} \,,
&
\pr{88}{F}_{q \bar{q}'} \big/\ms \pr{11}{F}_{q \bar{q}'}
  &= (N^2 - 2) \big/ \sqrt{N^2 - 1}
\end{align}
for $q \neq q'$.  For equal quark flavors, no simple relation is obtained at two-loop accuracy.

We emphasise that the relations \eqref{Fqg-AS} to \eqref{Fqq-8} should be evaluated for $\mu \sim \sqrt{x_1 x_2\ms \zeta_p} \sim 1/y$ in order to avoid logarithmically enhanced contributions from higher orders.  An exception is the relation \eqref{Fgg-dec}, which is stable under evolution.

In the following subsections, we discuss the behaviour of the two-loop kernels and their convolution with PDFs in various kinematics limits.  We closely follow the presentation for the colour singlet kernels in \cite{Diehl:2019rdh} where appropriate.  In the channels $g\to g g$, $q\to q g$, and $g\to q g$, we will not give results for kernels for $R_1 R_2 = SS$, which are proportional to the kernels for $R_1 R_2 = AA$ according to \eqref{eq:RVggg2-octets}, \eqref{eq:RVqgq2-octets}, and \eqref{eq:RVqgg2-octets}.  For the mixed octet combinations, we can limit our attention to $\prn{AS}{V}_{g g, q}$.  Finally, we will no longer discuss the kernels in the decuplet sector, which are all zero.


\subsection{Threshold limit: large \texorpdfstring{$x_1 + x_2$}{x1+x2}}
\label{sec:threshold_limit}
Let us start with the threshold limit $x \to 1$.  From \eqref{conv-12-def}, \eqref{V-NLO-form} and \eqref{V-gen-struct}, we deduce that the convolution of kernels with PDFs gives\footnote{
We note that the sign of the plus distribution term in \eqn{(149)} of \protect\cite{Diehl:2019rdh} is incorrect.}
\begin{align}
  \label{eq:leading-conv-threshold-limit}
& \prn{\RR}{V}^{(2)} \underset{12}{\otimes} f
\underset{x \to 1}{=}
    L\, \prn{\RR}{V}^{[2,1]}_p(u) \, \log (1 - x) \, f(x)
\nonumber \\[0.1em]
& \quad
  {}+ \Biggl[
         \prn{\RR}{V}^{[2,0]}_\delta(u) + L\, \prn{\RR}{V}^{[2,1]}_\delta(u)
         + \biggl( L \log\frac{\mu^2}{x_1 x_2\ms \zeta_p} - \frac{L^2}{2}
             + c_{\msbars} \biggr)
          \frac{\prn{R_1}{\gamma}_J^{(0)}}{2} \, \prn{\RR}{V}^{(1)}(u)
      \ms\Biggr] \ms f(x)
\nonumber \\[0.3em]
& \quad
  {}+ \mathcal{O}\bigl[ (1-x) \, \log^n(1-x) \bigl] \, f(x) \,.
\end{align}
The power for the subleading contributions in the last line is $n=0$ or $n=1$ and follows from the behaviour of ${V}^{[2,k]}_{\text{reg}}(u z, \ub z)$ for $z \to 1$.

The contribution enhanced by $\log(1-x)$ is due to plus distribution terms.  Such terms are absent in $V^{(1)}$.  They appear in $V^{(2)}$ only for LO channels, where they are accompanied with a logarithm $L$ and hence vanish for $\mu = y/b_0$.  The relevant coefficients read
\begin{align}
  \label{eq:plus-dist-terms}
    \prn{11}{V}_{g g, g}^{[2,1]} \ms \bigr|_{\ms p}
  & = 4 C_A^2 \; p_{g g}(u) \,,
  & \prn{27\, 27}{V}_{g g, g}^{[2,1]} \ms \bigr|_{\ms p}
  & = {}- c_{g g, g}(27\, 27) \; 60 \, p_{g g}(u) \,,
\nonumber \\[0.2em]
    \prn{11}{V}_{q \bar{q}, g}^{[2,1]} \ms \bigr|_{\ms p}
  & = {}- (C_A - 2 C_F) \; p_{q g}(u) \,,
  & \prn{88}{V}_{q \bar{q}, g}^{[2,1]} \ms \bigr|_{\ms p}
  & = {}- c_{q \bar{q}, g}(88) \; 2\ms (C_A - C_F) \; p_{q g}(u) \,,
\nonumber \\[0.2em]
    \prn{11}{V}_{q g, q}^{[2,1]} \ms \bigr|_{\ms p}
  & = 2 C_A\ms C_F \, p_{q q}(u)
\end{align}
with the functions $p(u)$ given in \eqref{eq:buildingblocks}.  Here we write $V_{\,\ldots\phantom{p}}^{[2,1]} \ms|_{\ms p}^{}$ instead of $V_{\,\ldots,\, p}^{[2,1]}$ for better readability.  The plus distribution terms in the $g\to g g$ and $q\to q g$ channels are zero for $R_1 R_2 = AA$ and $R_1 R_2 = SS$.
%
%
\subsection{Small \texorpdfstring{$x_1 + x_2$}{x1+x2}}
\label{sec:small_x}
To understand how the $x\to 0$ limit is analysed, let us first consider the ordinary Mellin convolution \eqref{conv-1-def} of two functions $P(z)$ and $f(z)$.  If both functions approximately behave like $1/z$ at small $z$, then the convolution integral approximately goes like $\int \! dz /z$ over a wide range $x \ll z \ll 1$ in which the arguments of both $P(z)$ and $f(x/z)$ are small.  If this range is wide enough, its contribution dominates the result.  The same observation applies to the convolution \eqref{conv-12-def} if $V(u z, \ub z) \sim 1/z^2$ for small $z$.

To be more specific, one finds that for $V(u z, \ub z) = w(u) / z^2$
\begin{align}
\Bigl[\ms V \conv{12} f \ms\Bigr](x)
   = \frac{w(u)}{x} \biggl[\ms \frac{1}{z} \otimes f \,\biggr](x)
  &= w(u)\, E(x)\, \frac{f(x)}{x}
\end{align}
with an enhancement factor
\begin{align}
   \label{enhancement-factor}
E(x) &=
\begin{cases}
k^{-1} \ms \log (1/x)
& \text{for }\ms x f(x) = c\, \log^{k-1} (1/x) \,,
\\[0.3em]
(1 - x^{\alpha}) / \alpha
& \text{for }\ms x f(x) = c\ms x^{-\alpha} \,,
\end{cases}
\end{align}
where $k$ is a positive integer in the upper line of \eqref{enhancement-factor}.  The form of $f(x)$ in the lower line provides a reasonable first approximation for the small $x$ behaviour of most PDFs, with $\alpha$ not too far away from zero.

Based on this discussion, we now extract the leading $1/z^2$ power behaviour of the two-loop kernels for the different channels.
%
\paragraph{$\boldsymbol{g \to g g} \ms:$}
\begin{align}
  \label{eq:small_x_RVggg2}
    \prn{11}{V}_{g g, g}^{[2,0]}
  & \approx \frac{2 C_{A}^{2}}{z^2}
    \biggl[
      1 - 6 u \bar{u}
      + \frac{2 u^3 - 2 u^2 + 4 u - 1}{ u } \log \bar{u}
      + \frac{2 \ub^3 - 2 \ub^2 + 4 \ub - 1}{ \bar{u} } \log u
    \biggr] \,,
\nonumber\\[0.5em]
    \prn{AA}{V}_{g g, g}^{[2,0]}
  & \approx c_{g g, g}(AA) \, \prn{11}{V}_{g g, g}^{[2,0]} \,,
    \phantom{\frac{1}{1}}
\nonumber\\[0.5em]
    \prn{27\, 27}{V}_{g g, g}^{[2,0]}
  & \approx c_{g g, g}(27\, 27) \, \prn{11}{V}_{g g, g}^{[2,0]} \,,
    \phantom{\frac{1}{1}}
\nonumber\\[1em]
    \prn{11}{V}_{g g, g}^{[2,1]}
  & \approx \frac{4 C_{A}^{2}}{z^2} \,
    \frac{1 + 2u \bar{u} - (u \bar{u})^2}{u \bar{u}}\,,
\nonumber\\[0.5em]
    \prn{AA}{V}_{g g, g}^{[2,1]}
  & \approx c_{g g, g}(AA) \, \frac{4 C_{A}^{2}}{z^2} \,
    \bigl( 2 - u \bar{u} \bigr)\,,
\nonumber\\[0.5em]
    \prn{27\, 27}{V}_{g g, g}^{[2,1]}
  & \approx {}- c_{g g, g}(27\, 27) \, \frac{12}{z^2} \;
    \frac{5 - 6 u \bar{u} + 3 (u \bar{u})^2}{u \bar{u}}\,.
\end{align}
%
\paragraph{$\boldsymbol{g \to q \bar{q}} \ms:$}
\begin{align}
  \label{eq:small_x_RVqqbarg2}
    \prn{11}{V}_{q \bar{q}, g}^{[2,0]}
  & \approx \frac{C_{A}}{z^2}
    \Bigl[
      2 - p_{q g}(u)
      \bigl(
        \log (u \bar{u}) + 3
      \bigr)
    \Bigr] \,,
  & \prn{11}{V}_{q \bar{q}, g}^{[2,1]}
  & \approx {}- \frac{C_{A}}{z^2} \, p_{q g}(u) \,,
  \nonumber\\[0.5em]
    \prn{88}{V}_{q \bar{q}, g}^{[2,0]}
  & \approx c_{q \bar{q}, g}(88) \, \prn{11}{V}_{q \bar{q}, g}^{[2,0]} \,,
  & \prn{88}{V}_{q \bar{q}, g}^{[2,1]}
  & \approx c_{q \bar{q}, g}(88) \, \prn{11}{V}_{q \bar{q}, g}^{[2,1]} \,.
\end{align}
%
\paragraph{$\boldsymbol{q \to g g} \ms:$}
In several (but not all) cases, we observe Casimir scaling by a factor $C_F / C_A$ between the $q\to g g$ and $g\to g g$ kernels:
\begin{align}
  \label{eq:small_x_RVggq2}
\prn{\RR}{V}_{g g, q}^{[2,0]}
  & \approx \frac{C_F}{C_A} \; \prn{\RR}{V}_{g g, g}^{[2,0]}
    \qquad \text{for $R_1 R_2 = 11, AA, SS, 27\, 27$} \,,
\nonumber\\[0.5em]
\prn{11}{V}_{g g, q}^{[2,1]}
  & \approx \frac{4}{z^2} \ms
    \left[
      2C_F^2\, \frac{1}{u \bar{u}} - C_A\ms C_F \, p_{g g}(u)
    \right]\,,
\nonumber \\[0.5em]
\prn{AA}{V}_{g g, q}^{[2,1]}
  & \approx \frac{C_F}{C_A} \; \prn{AA}{V}_{g g, g}^{[2,1]} \,,
\nonumber \\[0.5em]
\prn{SS}{V}_{g g, q}^{[2,1]}
  & \approx c_{g g, g}(SS) \, \frac{4}{z^2} \,
    \left[
       \frac{N^2 - 4}{N}\, C_F\, \frac{1}{u \ub} - C_A\ms C_F \, p_{g g}(u)
    \right]\,,
\nonumber \\[0.5em]
\prn{27\, 27}{V}_{g g, q}^{[2,1]}
  & \approx {}- c_{g g, g}(27\, 27) \, \frac{16}{z^2} \; p_{g g}(u) \,.
    \hspace{18em}
\end{align}
The kernel $\prn{AS}{V}_{g g, q}^{[2,0]}(u z, \ub z)$ goes like $1/z^2$ at small $z$, but the region $z \ll 1$ does not dominate the small $x$ limit of its convolution with PDFs in \eqref{Fgg-new}.  This is because the valence combination $f_q(z) - f_{\bar{q}}(z)$ has a small $z$ behaviour far away from the $1/z$ law that underlies the argument at the beginning of this subsection.
%
\paragraph{Remaining channels:}
For the splitting $q\to q' \bar{q}'$, we also find Casimir scaling:\footnote{
The result for $W^{[2,2]}_{q' \bs \bar{q}'\!, q}$ in \eqn{(161)} of \protect\cite{Diehl:2019rdh} is wrong and should read $W^{[2,2]}_{q' \bs \bar{q}'\!, q} = (C_F / C_A)\, W^{[2,2]}_{q' \bs \bar{q}'\!, g}\ms$.}
\begin{align}
  \label{eq:small_x_RVqprimeqprimebarq2}
    \prn{RR}{V}_{q' \bs \bar{q}'\!, q}^{[2,k]}
  & \approx \frac{C_F}{C_A} \; \prn{RR}{V}_{q \bar{q}, g}^{[2,k]}
  &&\text{for all $R$ and } k = 0,1\,.
\end{align}
The kernels for the other pure quark channels, as well as those for $q \to q g$ and $g\to q g$ have no leading $1/z^2$ behaviour at small $z$.  They behave like $z^n$ with integer $n \ge -1$, and their convolution with a PDF is not dominated by the region $x \ll z \ll 1$.

We observe that in the channels with a leading $1/z^2$ behaviour, the non-logarithmic part $\prn{\RR}{V}^{[2,0]}$ of the kernel scales with $c(R_1 R_2)$ like the LO terms, whereas the logarithmic part $\prn{\RR}{V}^{[2,1]}$ in general does not.  (An exception to this rule is the case $R_1 R_2 = AS$, which is special as we just explained.)

%
\subsection{Small \texorpdfstring{$x_1$}{x1} or \texorpdfstring{$x_2$}{x2}}
\label{sec:small_u_or_ubar}
We now analyse the limit in which $x_1$ or $x_2$ becomes small compared with $x_1 + x_2$, i.e.\ the limit $u\to 0$ or $u\to 1$.  The second limit can be reduced to the first by swapping the parton indices $a_1$ and $a_2$, and we will always take $u\to 0$ for definiteness.

The limit $u\to 0$ is subtle because it gives rise to singularities in the two-loop kernels that are absent for finite $u$.  We isolate these singularities by writing
\begin{align}
  \label{eq:V2-decomposed}
V^{[2,k]}_{\text{reg}}(u z, \ub z)
&= V^{[2,k]}_{r}(u z, \ub z) + V^{[2,k]}_{s}(u z, \ub z)
& \text{for } k=0,1 \,,
\end{align}
where the first term is integrable over $z$ in the limit $u \to 0$, whereas the second one develops a $1/ (1-z)$ singularity at $z=1$.  This singularity arises from graphs in which the initial parton $a_0$ splits into $a_2$ and another parton.  That parton has a momentum fraction $1 - z_2 = 1 - z (1-u)$, which appears in the denominator of its propagator.

For $V_{r}$ and for the terms with a plus or $\delta$ distribution in $V^{(2)}$, one can readily take the limit $u\to 0$.  One obtains $1/u$ for the leading behaviour.  For $V_{s}$, one must take the $u\to 0$ limit at the level of the convolution $V_{s} \otimes f$ and not at the level of the kernel.  This part of the kernel has the general form
\begin{align}
  \label{eq:V22-structure}
z\ms V_s(u z, \ub z)
  & = \sum_{n = 2,3,4}
    \frac{\zb^{n - 2} \, h_{n}(\zb, u)}{(\zb + u - \zb u)^n}
    + \frac{\log u - \log(\zb + u - \zb u)}{\zb}\,
      \frac{h_{1, a}}{\zb + u - \zb u}
  \nonumber\\
  & \quad
    + \frac{\log \zb - \log(\zb + u - \zb u)}{u}\,
      \frac{h_{1, b}}{\zb + u - \zb u}
    + \frac{h_{1, c}(\zb, u)}{\zb + u - \zb u} \,,
\end{align}
where on the r.h.s.\ we have changed variables from $z$ to $\zb = 1 - z$.  For ease of writing, we omit the superscript $[2,k]$ on $V_s$ and on the coefficients $h$ henceforth.  $h_{1, a}$ and $h_{1, b}$ are numerical constants, whilst $h_{n}(\zb, u)$ and $h_{1, c}(\zb, u)$ are functions which in the limits of small $\zb$ or small $u$ are either finite or have a singularity of order $\log \zb$ or $\log u$.

Notice that no explicit factor $1/u$ appears when taking the $u\to 0$ limit of $V_s$.  We hence erroneously discarded the contribution from $V_s$ when discussing the limit of small $x_1$ in our previous work \cite{Diehl:2019rdh}.  However, all terms except for the last one in \eqref{eq:V22-structure} give rise to a $1/u$ behaviour of the convolution
\begin{align}
  \label{eq:V22-conv-1}
    V_s\underset{12}{\otimes} f
  & = \frac{1}{x} \, \int_0^{1 - x} \! d \zb \;
    V_s(u z, \ub z) \, f \left(\frac{x}{1 - \zb}\right) \,.
\end{align}
This is revealed by power counting (with $\log u$ counting as order $1$).  In the region $\zb \sim u$, one finds $V_s \sim 1/u^2$ and thus obtains a contribution of order $1/u$ to the integral.  By contrast, $V_s \sim 1$ for $\zb \sim 1$, so that this region contributes to the integral only at subleading level.  In the term proportional to $h_{1,b} \ms$, one must use $\log \zb - \log(\zb + u + \zb u) \sim u$ for $z \sim 1$ to obtain this result.

Using the method of regions \cite{Beneke:1997zp}, we can compute the leading behaviour of the integral by expanding the integrand for small $\zb$ at fixed $\zb / u$.  To implement this, we rescale $u \to \varepsilon u$ and $\zb \to \varepsilon \zb$, approximate the integrand for $\varepsilon \to 0$, and finally set $\varepsilon = 1$.  This gives
\begin{align}
  \label{eq:V22-conv-2}
    V_s \underset{12}{\otimes} f
  = \frac{1}{x} \, \int_0^{1 - x} \! d \zb \;
   & \Biggl[\,
      \sum_{n = 2,3,4}
    \zb^{n - 2} \; \frac{h_{n,0} + h_{n,1}\ms \log \zb}{(\zb + u)^n}
    + \frac{\log u - \log(\zb + u )}{\zb}\, \frac{h_{1, a}}{\zb + u}
  \nonumber\\
  & \,
   + \frac{\log \zb - \log(\zb + u)}{u}\, \frac{h_{1, b}}{\zb + u}
   \,\Biggr] \, f(x)
   + \ldots \,,
\end{align}
where we used the leading behaviour $h_{n}(\zb, u) \sim h_{n,0} + h_{n,1}\, \log \zb$ in the limit just specified.
The ellipsis denotes subleading terms in the $\varepsilon$ expansion and in particular contains the full contribution from the term with $h_{1,c}(\zb, u)$ in \eqref{eq:V22-structure}.  One can readily carry out the integral over $\zb$ for the terms given in \eqref{eq:V22-conv-2} and then take the limit $u=0$.  The result goes like $1/u$ and is enhanced by $\log u$ for the terms with nonzero coefficients $h_{n,1}$.

With these preliminaries worked out, we can determine the small $u$ limit of the two-loop kernels for each channel in turn.
\paragraph{$\boldsymbol{g \to g g} \ms$:}
\begin{align}
  \label{eq:Vggg2_small_u}
    \prn{11}{V}_{g g, g}^{[2,0]} \underset{12}{\otimes} f
  & \approx \frac{1}{u \ms x}
    \left(
      \frac{10}{3} \ms C_A\ms \beta_0 - \frac{110}{9} \ms C_A^2
    \right) f(x) \,,
\nonumber\\[0.5em]
    \prn{AA}{V}_{g g, g}^{[2,0]} \underset{12}{\otimes} f
  & \approx \frac{c_{g g, g}(AA)}{u \ms x}
    \left(
      \frac{10}{3} C_A\ms \beta_0 - C_A^2 \, \frac{43 + 3 \pi^2}{9}
    \right) f(x) \,,
\nonumber\\[0.5em]
    \prn{27\, 27}{V}_{g g, g}^{[2,0]} \underset{12}{\otimes}f
  & \approx \frac{c_{g g, g}(27\, 27)}{u \ms x}
    \left(
      10 \beta_0 + \frac{206 - 24 \pi^2}{3}
    \right) f(x) \,,
\nonumber\\[1em]
    \prn{11}{V}_{g g, g}^{[2,1]} \underset{12}{\otimes}f
  & \approx \frac{1}{u \ms x}
    \left[
      \left(
        2 C_A\ms \beta_0 - C_A^2 \, \frac{22 + 12 \log u}{3}
      \right) f(x)
      + 4 C_A^2 \,
      \left( \ms
        \frac{z}{[\zb]_+} + \frac{\zb}{z} + z \zb
      \right) \otimes f
    \,\right]\,,
\nonumber \\[0.5em]
    \prn{AA}{V}_{g g, g}^{[2,1]} \underset{12}{\otimes}f
  & \approx \frac{c_{g g, g}(AA)}{u \ms x}
    \left(
      2 C_A\ms \beta_0 - C_A^2 \, \frac{11 + 6 \log u}{3}
    \right) f(x) \,,
\nonumber \\[0.5em]
    \prn{27\, 27}{V}_{g g, g}^{[2,1]} \underset{12}{\otimes}f
  & \approx \frac{c_{g g, g}(27\, 27)}{u \ms x}
    \left[
      \bigl(
        6 \beta_0 + 22 + 12 \log u
      \bigr) \, f(x)
      - 60
      \left( \ms
        \frac{z}{[\zb]_+} + \frac{\zb}{z} + z \zb
      \right) \otimes f
    \,\right]\,.
\end{align}
The terms with an explicit factor $f(x)$ in these expressions are due to $V^{[2,k]}_s$ and $V^{[2,k]}_{\delta}$, whereas the Mellin convolutions
\begin{align}
\label{phi-conv}
\varphi(z) \otimes f
 &= \int_x^1 \frac{dz}{z} \, \varphi(z) \, f\biggl( \frac{x}{z} \biggr)
\end{align}
originate from $V^{[2,k]}_r$ and $V^{[2,1]}_p \big/\ms [\zb]_+$.  For the sake of legibility, we do not indicate the argument $x$ of these convolutions in \eqref{eq:Vggg2_small_u} and the following equations.  We note that the double logarithmic part of the $V^{(2)}$ also contributes to the small~$u$ limit of $V^{(2)} \smash{\conv{12}}\, f$.  Its contribution is readily obtained by extracting the leading term of ${V}^{(1)}(u)$ for $u \to 0$.
%
$\phantom{V^{[2,k]}_\delta}$
%
\paragraph{$\boldsymbol{q \to g q} \ms$:}
\begin{align}
  \label{eq:Vgqq2_small_u}
    \prn{11}{V}_{g q, q}^{[2,0]} \underset{12}{\otimes} f
  & \approx \frac{1}{u \ms x}
    \left(
      \frac{10}{3}\ms C_F \beta_0 - \frac{110}{9}\ms C_A\ms C_F
    \right) f(x) \,,
\nonumber\\[0.5em]
    \prn{A8}{V}_{g q, q}^{[2,0]} \underset{12}{\otimes} f
  & \approx \frac{c_{q g,q}(8A)}{u \ms x}
    \left[
      \left(
        \frac{10}{3} C_F \beta_0 - C_A\ms C_F \, \frac{43 + 3 \pi^2}{9}
      \right) f(x)
      + C_A\ms C_F \, \zb \otimes f
    \,\right]\,,
\nonumber\\[1em]
    \prn{11}{V}_{g q, q}^{[2,1]} \underset{12}{\otimes}f
  & \approx \frac{1}{u \ms x}
    \left[
      \left(
        2 C_F \beta_0 - C_A\ms C_F \, \frac{22 + 12 \log u}{3}
      \right) f(x)
      + 2 C_A\ms C_F\,
        \frac{1 + z^2}{[\zb]_+}
      \otimes f
    \,\right]\,,
\nonumber \\[0.5em]
    \prn{A8}{V}_{g q, q}^{[2,1]} \underset{12}{\otimes}f
  & \approx \frac{c_{q g,q}(8A)}{u \ms x}
    \left(
      2 C_F \beta_0 - C_A\ms C_F \, \frac{11 + 6 \log u}{3}
    \right) f(x) \,.
\end{align}
We remark that the only cases for which the $1/u$ behaviour is enhanced by $\log u$ are the convolutions of ${V}_{g g, g}^{[2,1]}$ and ${V}_{g q, q}^{[2,1]}$ with $f$.  The $\log u$ factor is thus always accompanied by a logarithm $L$ of the scale $\mu$.
%
$\phantom{V^{[]}_g}$
%
\paragraph{$\boldsymbol{g \to q g}$ and $\boldsymbol{g \to g q} \ms$:}
\begin{align}
  \label{eq:Vqgg2_small_u}
    \prn{11}{V}_{q g, g}^{[2,0]} \underset{12}{\otimes} f
  & \approx \frac{1}{u \ms x} \, \frac{10}{9} \ms C_A \, f(x) \,,
  & \prn{11}{V}_{q g, g}^{[2,1]} \underset{12}{\otimes}f
  & \approx \frac{1}{u \ms x} \, \frac{2}{3} \ms C_A \, f(x) \,,
\nonumber \\[1em]
    \prn{8A}{V}_{q g, g}^{[2,0]} \underset{12}{\otimes} f
  & \approx \frac{1}{2}\, c_{q g,q}(8A) \; \prn{11}{V}_{q g, g}^{[2,0]} \,,
  & \prn{8A}{V}_{q g, g}^{[2,1]} \underset{12}{\otimes}f
  & \approx \frac{1}{2}\, c_{q g,q}(8A) \; \prn{11}{V}_{q g, g}^{[2,1]} \,,
\\[1.5em]
  \label{eq:Vgqg2_small_u}
    \prn{11}{V}_{g q, g}^{[2,0]} \underset{12}{\otimes} f
  & \approx - \frac{2}{u \ms x} \, (C_A - C_F) \; (z \zb)
    \otimes f\,,
  &
    \prn{11}{V}_{g q, g}^{[2,1]} \underset{12}{\otimes}f
  & \approx \frac{1}{u \ms x} \, C_A \, (z^2 + \zb^2)
    \otimes f\,,
\nonumber \\[1em]
    \prn{A8}{V}_{g q, g}^{[2,0]} \underset{12}{\otimes} f
  & \approx - \frac{c_{q g,q}(8A)}{u \ms x} \, (C_A - 2C_F) \; (z \zb)
    \otimes f\,,
  &
    \prn{A8}{V}_{g q, g}^{[2,1]} \underset{12}{\otimes}f
  & \approx \mathcal{O}(u^0)\,.
\end{align}
%
\paragraph{$\boldsymbol{q \to g g} \ms$:}
\begin{align}
  \label{eq:Vggq2_small_u}
    \prn{11}{V}_{g g, q}^{[2,0]} \underset{12}{\otimes} f
  & \approx \frac{2}{u \ms x} \, C_F \ms (C_A - C_F) \; z
    \otimes f \,,
&
   \prn{11}{V}_{g g, q}^{[2,1]} \underset{12}{\otimes}f
  & \approx - \frac{2}{u \ms x} \, C_F (C_A - 2 C_F) \, \frac{1 + \zb^2}{z}
    \otimes f \,,
\nonumber \\[1em]
    \prn{AA}{V}_{g g, q}^{[2,0]} \underset{12}{\otimes} f
  & \approx \frac{c_{g g, g}(AA)}{u \ms x} \, C_A\ms C_F \, z
    \otimes f\,,
&
    \prn{AA}{V}_{g g, q}^{[2,1]} \underset{12}{\otimes} f
  & \approx \mathcal{O}(u^0) \,,
\nonumber \\[1em]
   \prn{SS}{V}_{g g, q}^{[2,0]} \underset{12}{\otimes} f
  & \approx \frac{c_{g g, g}(SS)}{u \ms x} \,
    \frac{N^2 + 4}{N}\, C_F\, z \otimes f \,,
    \hspace{-1.1em}
&
   \prn{SS}{V}_{g g, q}^{[2,1]} \underset{12}{\otimes} f
  & \approx 4 \ms c_{g g, g}(SS) \;
      \prn{11}{V}_{g g, q}^{[2,1]} \underset{12}{\otimes}f \,,
\nonumber \\[1em]
    \prn{27\, 27}{V}_{g g, q}^{[2,0]} \underset{12}{\otimes}f
  & \approx \frac{c_{g g, g}(27\, 27)}{u \ms x} \; 8 z
    \otimes f \,,
&
    \prn{27\, 27}{V}_{g g, q}^{[2,1]} \underset{12}{\otimes}f
  & \approx - \frac{c_{g g, g}(27\, 27)}{u \ms x} \; 8 \ms \frac{1 + \zb^2}{z}
    \otimes f \,,
\nonumber \\[1em]
   \prn{AS}{V}_{g g, q}^{[2,0]} \underset{12}{\otimes} f
  & \approx \frac{\tilde{c}(AS)}{u \ms x} \, \frac{z}{2} \otimes f \,,
& \prn{AS}{V}_{g g, q}^{[2,1]} \underset{12}{\otimes} f
  & \approx \mathcal{O}(u^0) \,,
\end{align}
where $\tilde{c}(AS)$ is given in \eqref{eq:ctilde-new}.
%
\paragraph{Remaining channels:}
For $q\to q' q$ in the colour singlet representation, we have
\begin{align}
  \label{eq:Vqprimeqq2_small_u}
    \prn{11}{V}_{q' \bs q, q}^{[2,0]} \underset{12}{\otimes} f
  & \approx \frac{1}{u \ms x} \, C_F \, \frac{10}{9} \, f(x) \,,
  & \prn{11}{V}_{q' \bs q, q}^{[2,1]} \underset{12}{\otimes}f
  & \approx \frac{1}{u \ms x} \,C_F \, \frac{2}{3} \, f(x) \,.
\end{align}
Corresponding expressions for $\prn{88}{V}_{q' \bs q, q}$, $\prn{11}{V}_{\bar{q}' \bs q, q}$ and $\prn{88}{V}_{\bar{q}' \bs q, q}$ follow from \eqref{Rpure-quark-C} and \eqref{eq:RVqqq2-scaling} together with \eqref{V-triv-symm}.
The convolutions for $g\to q \bar{q}$, $q\to q g$, and for the other pure quark channels do not have a leading $1/u$ behaviour.

Notice that in $V_{q g, g}$ and $V_{q' \bs q, q}$ we have a $1/u$ enhancement for a slow \emph{quark} rather than a slow gluon.  This enhancement is due to graphs in which the observed slow quark originates from a slow gluon, see \figs{\ref{fig:gqg-UD-2}} and \ref{fig:qq-1}.

%
\subsection{Triple Regge limit}
\label{sec:triple-Regge}

We finally consider the ``triple Regge limit'' $x_1 \ll x_1 + x_2 \ll 1$ or $x_2 \ll x_1 + x_2 \ll 1$.  The kernels with a leading behaviour turn out the be symmetric under the interchange of $a_1$ and $a_2$, so that we can restrict  ourselves to the first of the two cases.

The limit can be approached in two ways.  One can either start from the small $x$ results in \sect{\ref{sec:small_x}} and take their limit for $u\to 0$, or one can start from the small $u$ results in \sect{\ref{sec:small_u_or_ubar}} and collect their leading terms for $x\to 0$.  Let us discuss the two procedures in turn.


Taking the limit $u\to 0$ of the kernels expanded for small $z$ in \eqref{eq:small_x_RVggg2} to \eqref{eq:small_x_RVqprimeqprimebarq2} is straightforward and yields the following expressions.
\paragraph{$\boldsymbol{g \to g g} \ms$:}
\begin{align}
  \label{eq:small_u_small_x_RVggg2}
    \prn{11}{V}_{g g, g}^{[2,0]}
  & \approx \frac{2 C_{A}^{2}}{z^2} \,
    \bigl( 2 + 3 \log u \bigr) \,,
  & \prn{11}{V}_{g g, g}^{[2,1]}
  & \approx \frac{4 C_{A}^{2}}{z^2} \, \frac{1}{u} \,,
  \nonumber\\[0.5em]
    \prn{AA}{V}_{g g, g}^{[2,0]}
  & \approx c_{g g, g}(AA) \, \prn{11}{V}_{g g, g}^{[2,0]} \,,
  & \prn{AA}{V}_{g g, g}^{[2,1]}
  & \approx c_{g g, g}(AA) \, \frac{8 C_{A}^{2}}{z^2} \,,
  \nonumber\\[0.5em]
    \prn{27\, 27}{V}_{g g, g}^{[2,0]}
  & \approx c_{g g, g}(27\, 27) \, \prn{11}{V}_{g g, g}^{[2,0]} \,,
  &  \prn{27\, 27}{V}_{g g, g}^{[2,1]}
  & \approx -c_{g g, g}(27\, 27) \, \frac{60}{z^2} \ms \frac{1}{u} \,.
\end{align}
%
\paragraph{$\boldsymbol{g \to q \bar{q}} \ms$:}
\begin{align}
  \label{eq:small_u_small_x_RVqqbarg2}
    \prn{11}{V}_{q \bar{q}, g}^{[2,0]}
  & \approx - \frac{C_{A}}{z^2} \,
  \bigl( 1 + \log u \bigr) \,,
  & \prn{11}{V}_{q \bar{q}, g}^{[2,1]}
  & \approx - \frac{C_{A}}{z^2} \,,
  \nonumber\\[0.5em]
    \prn{88}{V}_{q \bar{q}, g}^{[2,0]}
  & \approx c_{q \bar{q}, g}(88) \, \prn{11}{V}_{q \bar{q}, g}^{[2,0]} \,,
  & \prn{88}{V}_{q \bar{q}, g}^{[2,1]}
  & \approx c_{q \bar{q}, g}(88) \, \prn{11}{V}_{q \bar{q}, g}^{[2,1]} \,.
\end{align}
%
\paragraph{$\boldsymbol{q \to g g} \ms$:}
\begin{align}
  \label{eq:small_u_small_x_RVggq2}
    \prn{\RR}{V}_{g g, q}^{[2,0]}
  & \approx \frac{C_F}{C_A} \; \prn{\RR}{V}_{g g, g}^{[2,0]}
  \qquad \text{for $R_1 R_2 = 11, AA, SS, 27\, 27$} \,,
\nonumber \\[0.5em]
  \prn{11}{V}_{g g, q}^{[2,1]}
  & \approx - \frac{4}{z^2} \ms \frac{C_F \ms (C_A - 2 C_F)}{u} \,,
\nonumber \\[0.5em]
    \prn{AA}{V}_{g g, q}^{[2,1]}
  & \approx \frac{C_F}{C_A} \; \prn{AA}{V}_{g g, g}^{[2,1]} \,,
\nonumber \\[0.5em]
\prn{SS}{V}_{g g, q}^{[2,1]}
  & \approx - c_{g g, g}(SS) \, \frac{16}{z^2} \,
      \frac{C_F \ms (C_A - 2 C_F)}{u} \,,
\nonumber \\[0.5em]
\prn{27\, 27}{V}_{g g, q}^{[2,1]}
  & \approx - c_{g g, g}(27\, 27) \, \frac{16}{z^2} \, \frac{1}{u} \,,
    \hspace{16em}
\end{align}
where we omit the case $R_1 R_2 = AS$ for the reason given after \eqref{eq:small_x_RVggq2}.
%
\paragraph{Remaining channels:}
\begin{align}
  \label{eq:small_u_small_x_RVqqbarq2}
    \prn{RR}{V}_{q' \bar{q}', q}^{[2,k]}
  & \approx \frac{C_F}{C_A} \; \prn{RR}{V}_{q \bar{q}, g}^{[2,k]} \,,
  &&\text{for all $R$ and } k = 0,1 \,.
\end{align}
The kernels for all other channels have no leading $1/z^2$ behaviour.


Taking the small $x$ limit of the small $u$ results in \sect{\ref{sec:small_u_or_ubar}} means to select the convolution terms $(1/z) \otimes f$, since this will give a small $x$ enhancement as explained in \sect{\ref{sec:small_x}}.  Such a behaviour is only found in very few cases, namely in
\begin{align}
  \label{eq:small_u_small_x_RVggg2-revisited}
\prn{11}{V}_{g g, g}^{[2,1]}
  & \approx \frac{4 C_{A}^{2}}{z^2\ms u} \,,
&
\prn{27\, 27}{V}_{g g, g}^{[2,1]}
  & \approx {}- c_{g g, g}(27\, 27) \, \frac{60}{z^2\ms u} \,,
\intertext{and}
  \label{eq:small_u_small_x_RVggq2-revisited}
\prn{11}{V}_{g g, q}^{[2,1]}
  & \approx {}- \frac{4 C_F \ms (C_A - 2C_F)}{z^2\ms u} \,,
\nonumber \\
\prn{SS}{V}_{g g, q}^{[2,1]}
  & \approx 4\ms c_{g g, g}(SS) \, \prn{11}{V}_{g g, q}^{[2,1]} \,,
&
\prn{27\, 27}{V}_{g g, q}^{[2,1]}
  & \approx {}- c_{g g, g}(27\, 27) \, \frac{16}{z^2\ms u} \,.
\end{align}
These expressions are in agreement with the ones in \eqref{eq:small_u_small_x_RVggg2} and \eqref{eq:small_u_small_x_RVggq2}, which were obtained by taking the opposite order of the two limits.
In fact, the only cases in which the two orders of limits give the same result are those in which the kernels have a leading behaviour for small $z$ \emph{and} small $u$.

As an example in which the two limits do not commute, let us take the kernel $\prn{AA}{V}_{g g, g}^{[2,1]}$.  Taking first the small $x$ limit and then $u\to 0$, we obtain
\begin{align}
  \label{eq:AVggg21-small-x}
    \prn{AA}{V}_{g g, g}^{[2,1]}\underset{12}{\otimes} f
  & \approx c_{g g, g}(AA) \; \frac{1}{x}\,
       \biggl( \frac{8 C_{A}^{2}}{z} \otimes f \biggr)
    \approx c_{g g, g}(AA) \; 8 C_{A}^{2} \; E(x) \, \frac{f(x)}{x} \,,
\end{align}
where in the last step we have performed the convolution for specific choices of the PDF, as detailed in \eqref{enhancement-factor}.
If we first take $u\to 0$ and then the small $x$ limit, we have instead
\begin{align}
  \label{eq:AVggg21-small-u}
    \prn{AA}{V}_{g g, g}^{[2,1]} \underset{12}{\otimes} f
  & \approx \frac{c_{g g, g}(AA)}{u}
    \left(
      2 C_A\ms \beta_0 - C_A^2 \, \frac{11 + 6 \log u}{3}
    \right) \frac{f(x)}{x}
\end{align}
with a leading $u^{-1}\, \log u$ behaviour but no small $x$ enhancement.  Clearly, the approach to the triple Regge limit strongly depends on the direction in the plane of $x_1$ and $x_2$ in this case.
%

\section{Numerical illustration}
\label{sec:numerics}

In this section, we illustrate the impact of NLO corrections on the splitting DPDs.  For easier comparison, we use the same settings as in \sect{5.4} of our study \cite{Diehl:2019rdh} for the colour singlet case.  We take $y = 0.022 \fm$ for the inter-parton distance.  This corresponds to a scale
\begin{align}
   \label{y-choice}
\mu_y &= b_0 / y = 10 \gev
\end{align}
at which the logarithm $L$ in \eqref{L-b0-def} is zero.

For the PDFs in the splitting formula, we take the NLO set of CT14 \cite{Dulat:2015mca}, along with the coupling $\alpha_s$ used in that set.\footnote{We use the set \texttt{CT14nlo} obtained from LHAPDF \cite{Buckley:2014ana} via the \texttt{ManeParse} interface \cite{Clark:2016jgm}.}
This gives
\begin{align}
\alpha_s(10 \gev) &= 0.178 \,,
&
\alpha_s(5 \gev) &= 0.212 \,,
&
\alpha_s(20 \gev) &= 0.153
\end{align}
at the scales \rev{on which we focus in the following study}.  We work with $n_F = 5$ active quark flavours.

The splitting DPD at two-loop accuracy is given by the sum of the LO and NLO terms $F^{(1)}$ and $F^{(2)}$ defined in \eqref{F-expansion}.  Note that we evaluate the one-loop expression $F^{(1)}$ with NLO PDFs here.  Thus, the comparison of $F^{(2)}$ with $F^{(1)}$ directly reveals the impact of the two-loop splitting kernels.  As we saw in \cite{Diehl:2019rdh}, the difference between evaluating $F^{(1)}$ with LO or NLO PDFs is typically quite large, at least in the $g g$ channel.  This simply reflects that PDFs at moderately large scales depend quite strongly on the order at which they are determined and evolved.


\subsection{Two-gluon distributions}

\begin{figure}[p]
\begin{center}
\subfigure[$x_2 = x_1$]{
   \includegraphics[height=5cm,trim=0 0 0 20,clip]{
   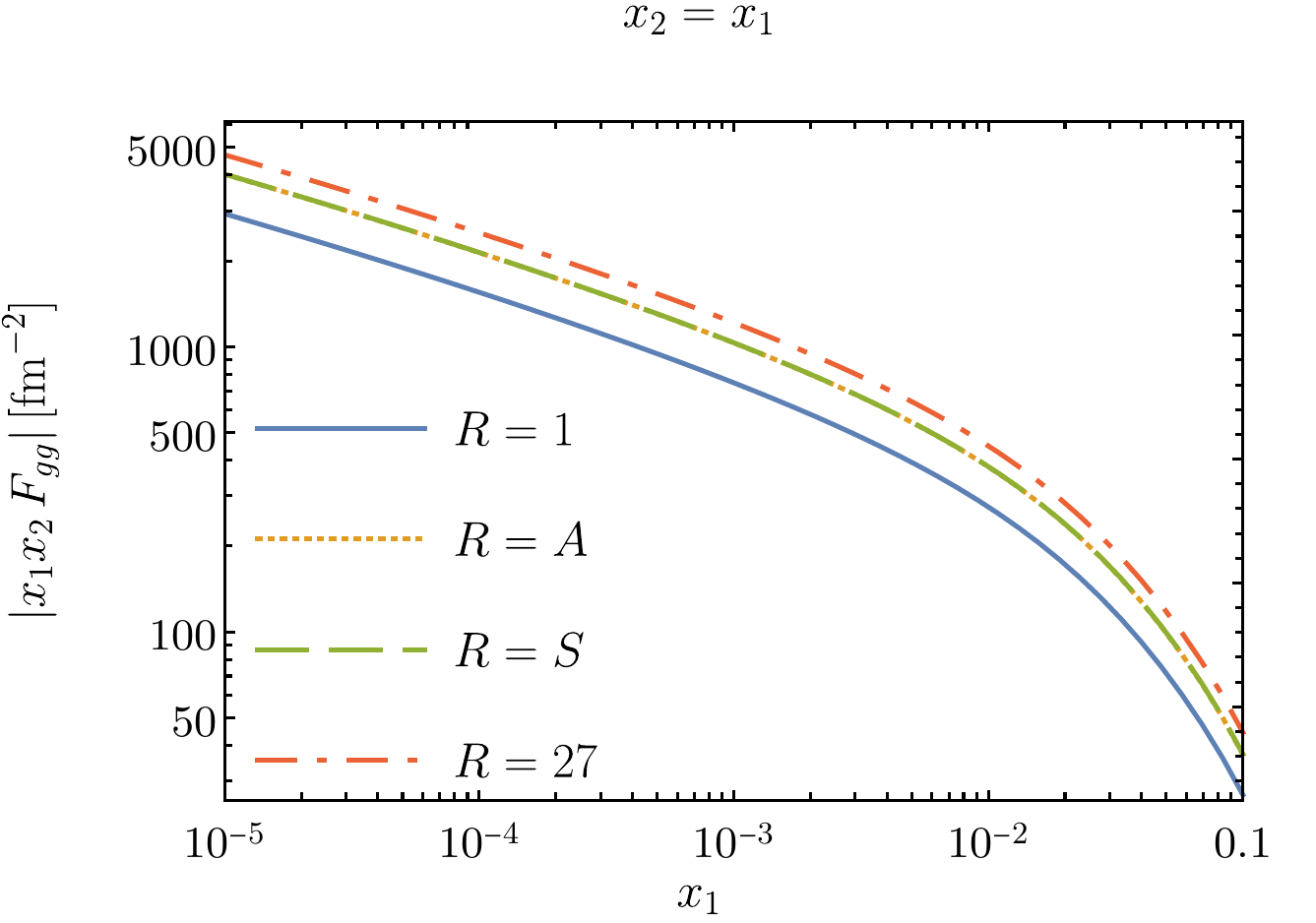}
\hspace{0.5em}
   \includegraphics[height=5cm,trim=0 0 0 20,clip]{
   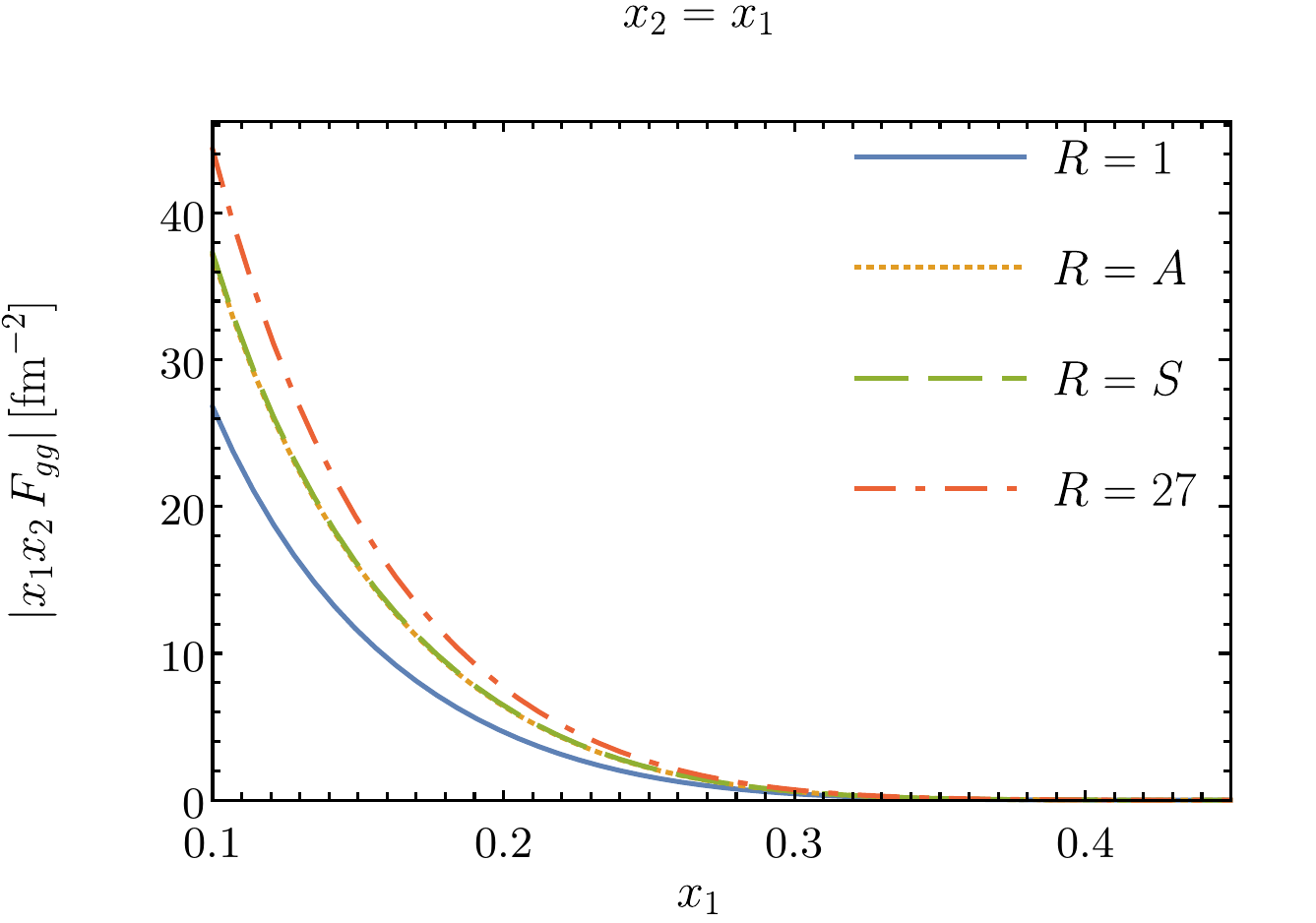}
} \\[1em]
\subfigure[$x_2 = 0.1$]{
   \includegraphics[height=5cm,trim=0 0 0 20,clip]{
   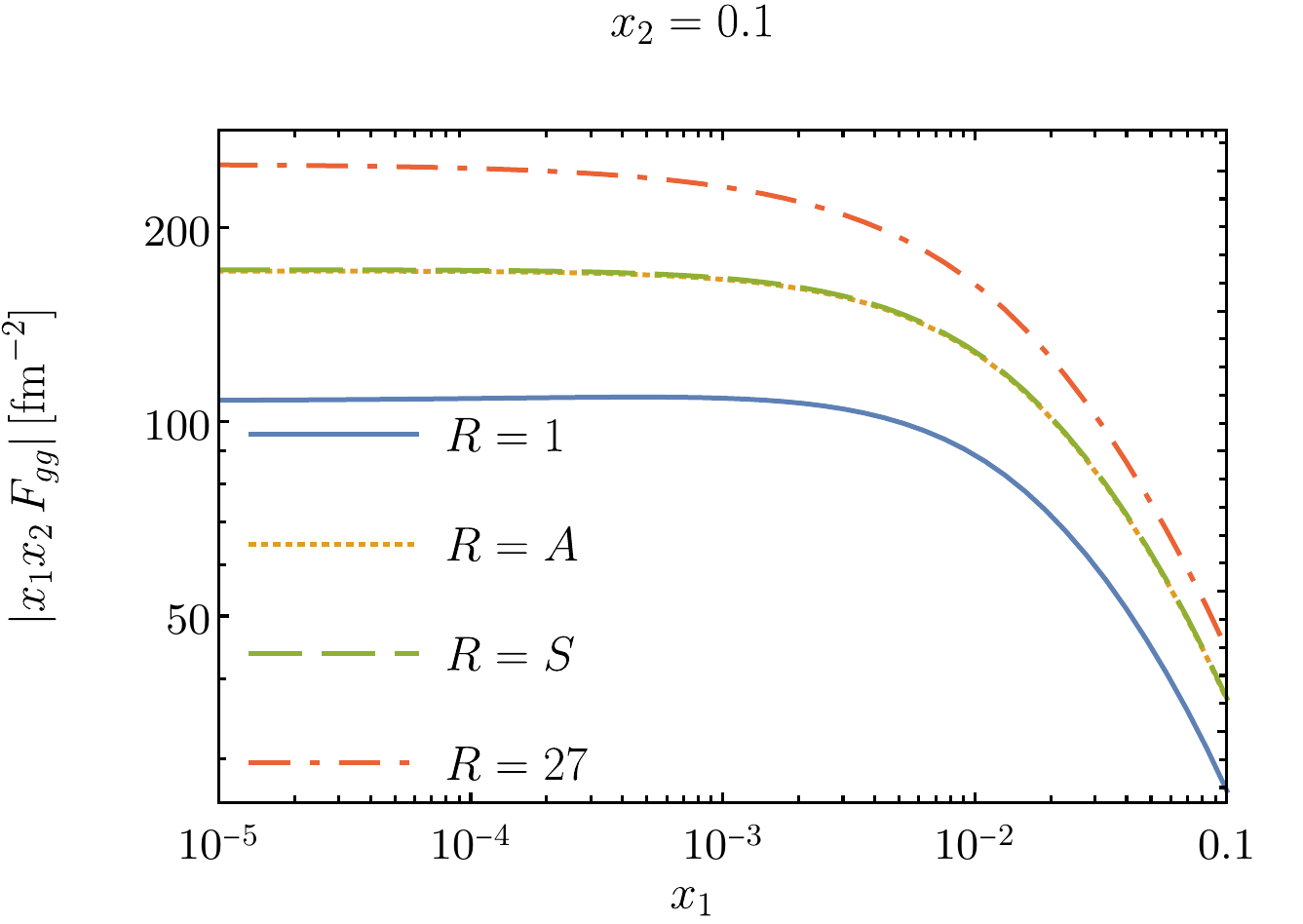}
\hspace{0.5em}
   \includegraphics[height=5cm,trim=0 0 0 20,clip]{
   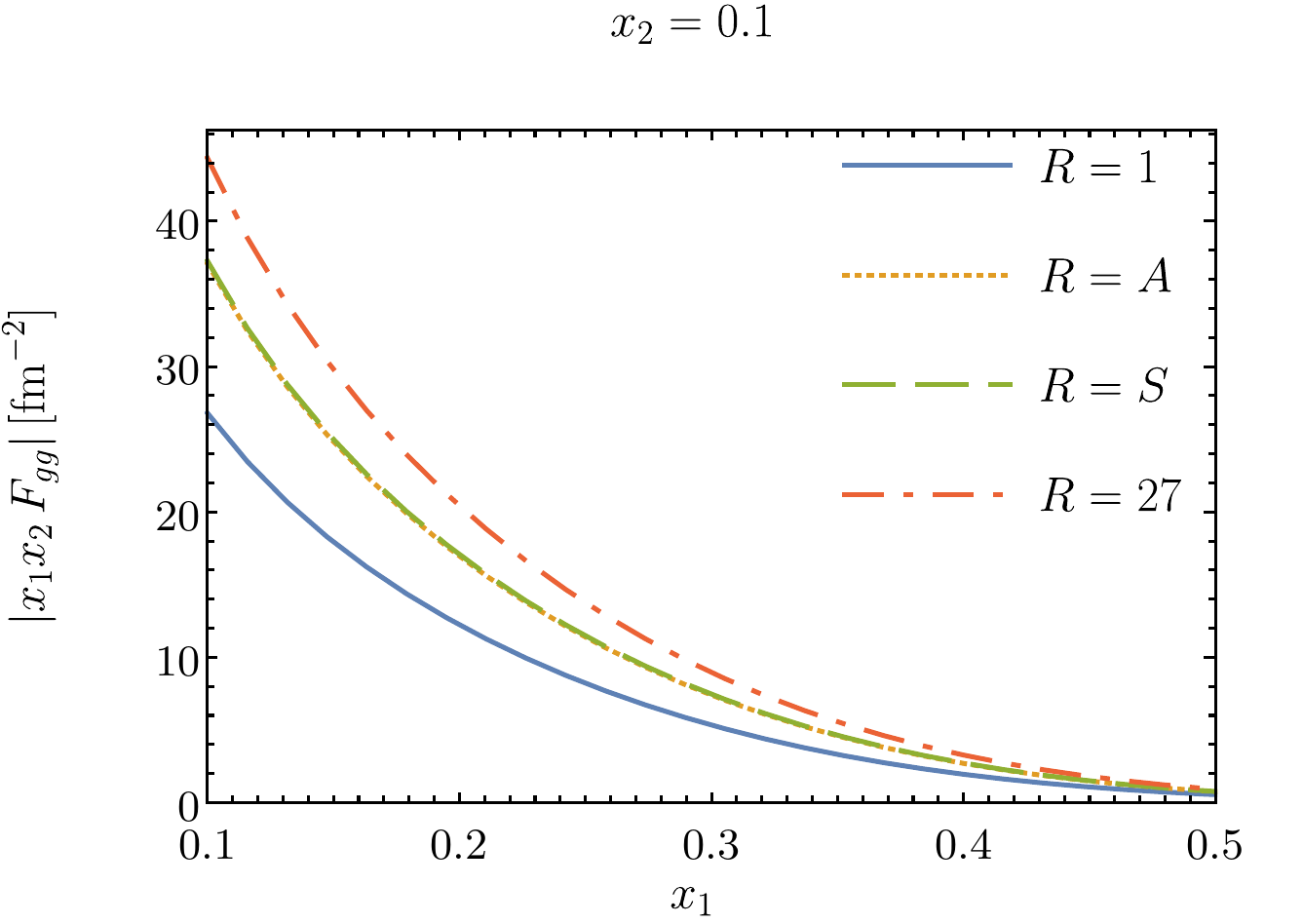}
} \\[1em]
\subfigure[$x_2 = 10^{-3}$]{
   \includegraphics[height=5cm,trim=0 0 0 20,clip]{
   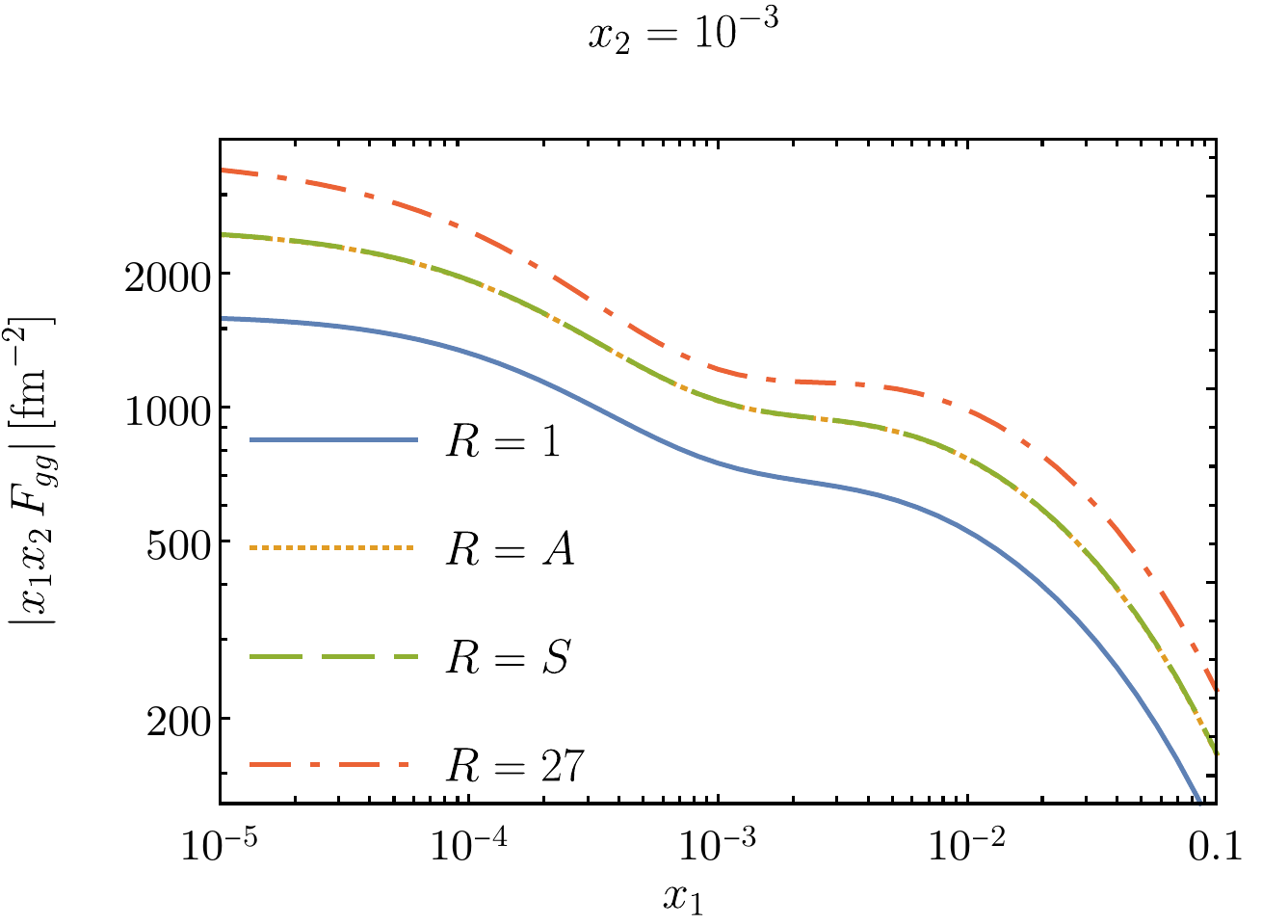}
\hspace{0.5em}
   \includegraphics[height=5cm,trim=0 0 0 20,clip]{
   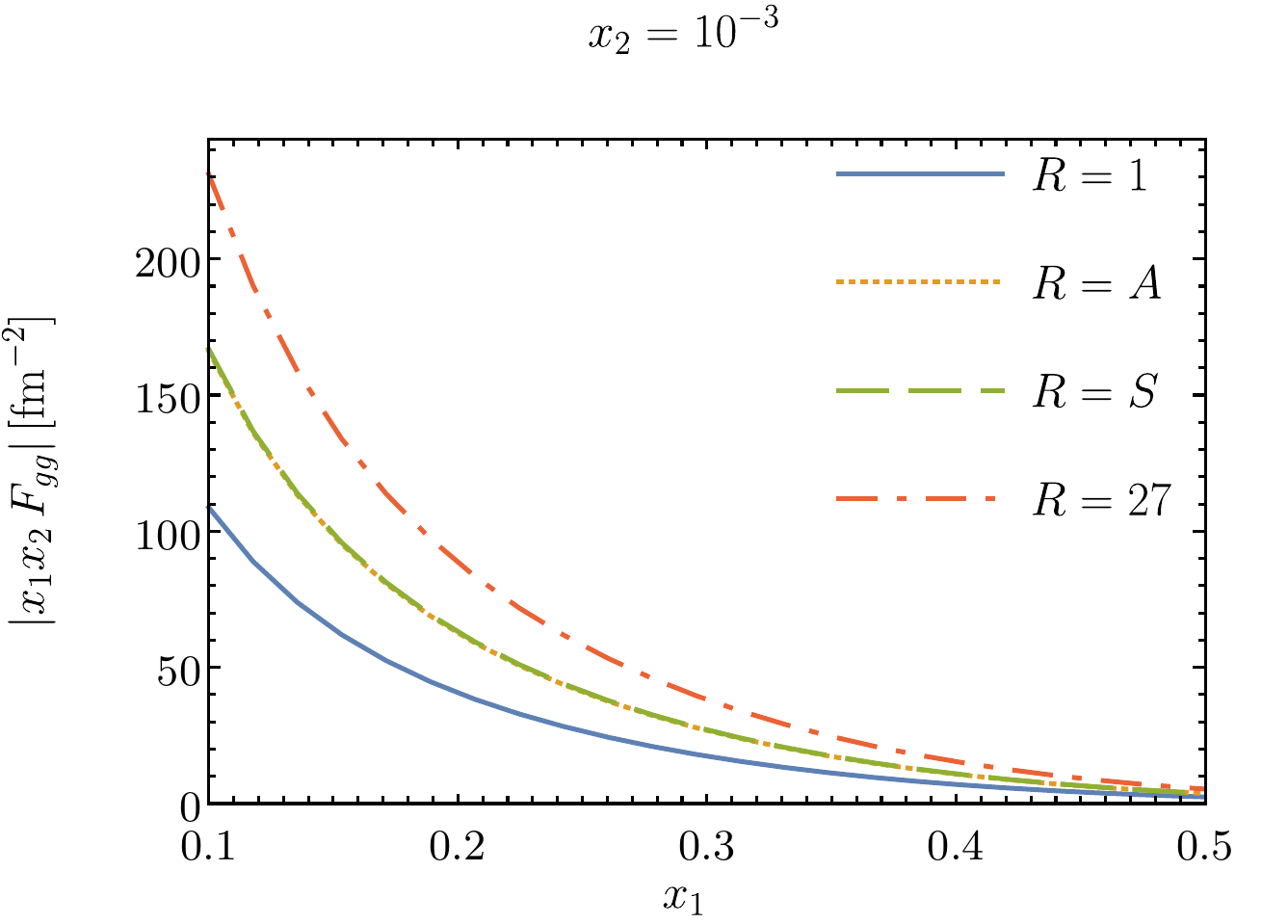}
}
\caption{\label{fig:F-values} Absolute value of the splitting DPD $\pr{RR}{F}_{g g}^{(1)} + \pr{RR}{F}_{g g}^{(2)}$ evaluated at two-loop accuracy with central values $\mu = \sqrt{x_1 x_2\ms \zeta_p} = \mu_y$ of the scales.  The curves for $R=A$ and $R=S$ cannot be distinguished at the scale of the plots.  Unless mentioned otherwise, all plots in this section are for CT14 NLO PDFs and the standard definition of the \msbar scheme (i.e.\ for $c_{\msbars} = - \zeta_2/2$).}
\end{center}
\end{figure}

In \fig{\ref{fig:F-values}} we show the full splitting DPD $F_{g g} = F^{(1)}_{g g} + F^{(2)}_{g g}$ as a function of $x_1$ for different values of $x_2$.  We take scales $\mu = \sqrt{x_1 x_2\ms \zeta_p} = \mu_y$ and use the standard definition of the \msbar scheme, so that according to \eqref{MSbar-c} the scheme parameter in the two-loop kernels \eqref{V-NLO-form} is $c_{\msbars} = - \zeta_2/2$.
The sign of the distributions (not visible in the logarithmic plots) is the same as at LO, i.e.\ $\pr{11}{F}_{g g}$ and $\pr{SS}{F}_{g g}$ are positive and $\pr{AA}{F}_{g g}$ and $\pr{27\, 27}{F}_{g g}$ are negative.

\begin{figure}[p]
\begin{center}
\subfigure[$x_2 = x_1$]{
   \includegraphics[height=6.3cm,trim=0 0 0 20,clip]{
   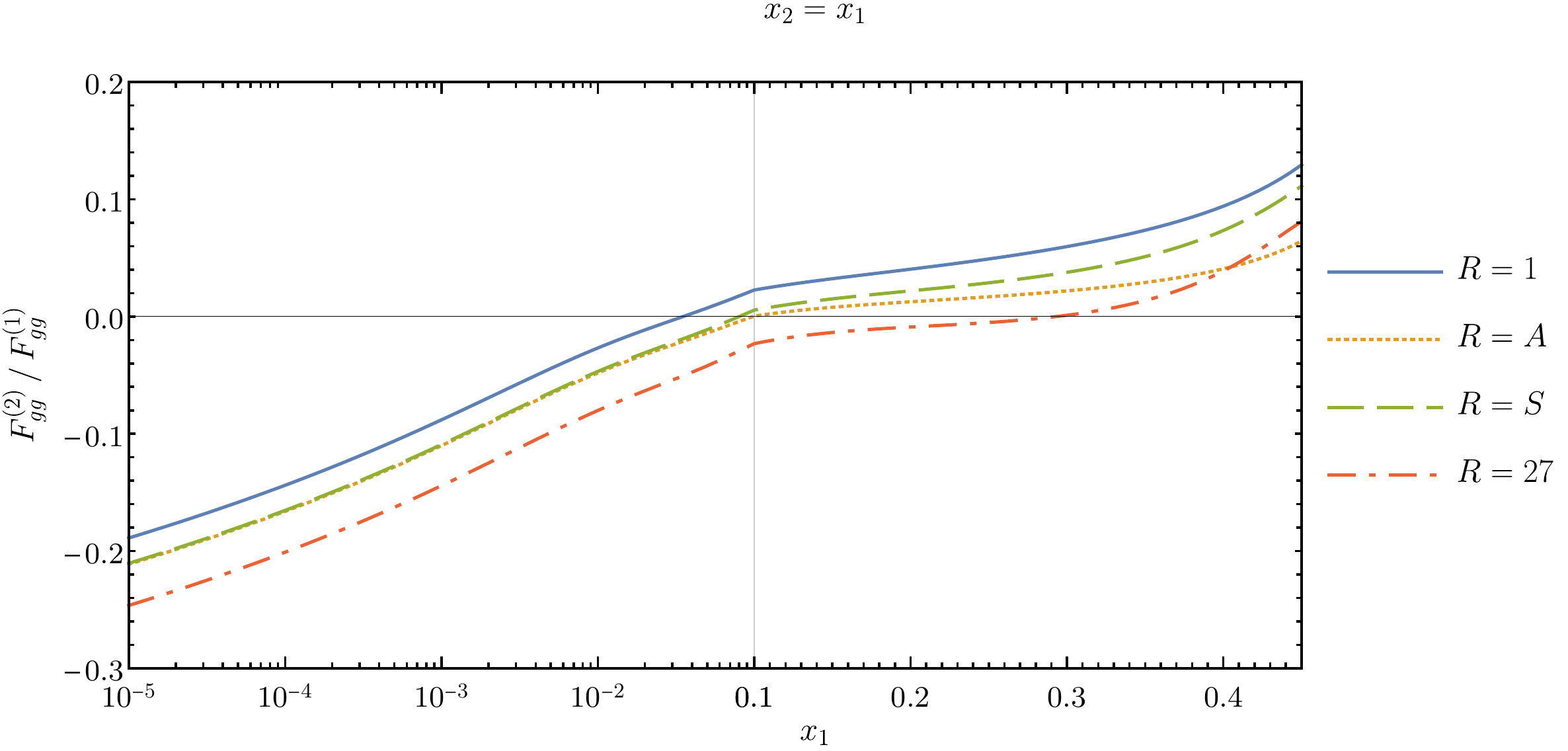}
} \\[1em]
\subfigure[$x_2 = 0.1$]{
   \includegraphics[height=6.3cm,trim=0 0 0 20,clip]{
   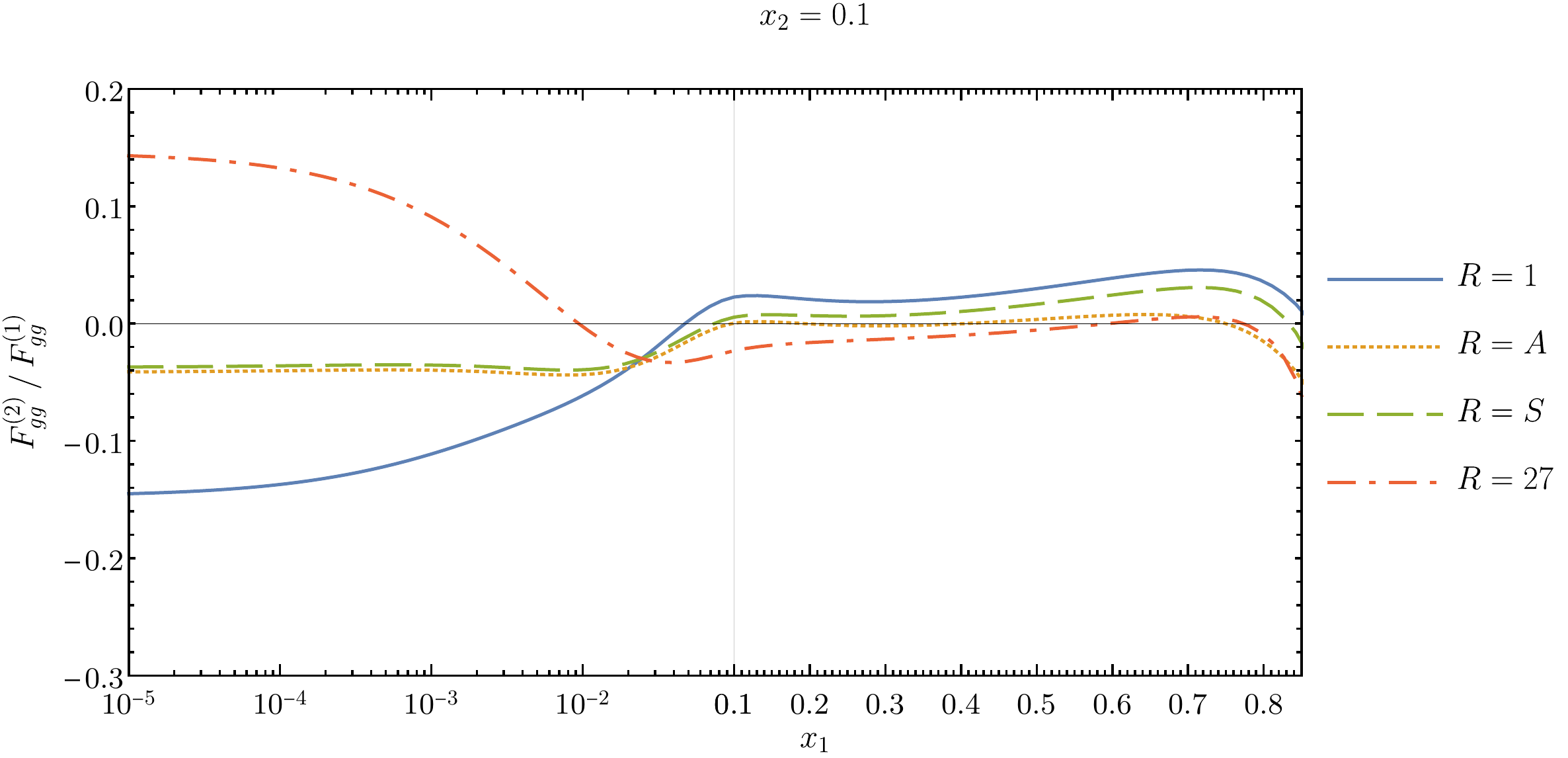}
} \\[1em]
\subfigure[\label{fig:F-ratios-central-small-x1}$x_2 = 10^{-3}$]{
   \includegraphics[height=6.3cm,trim=0 0 0 20,clip]{
   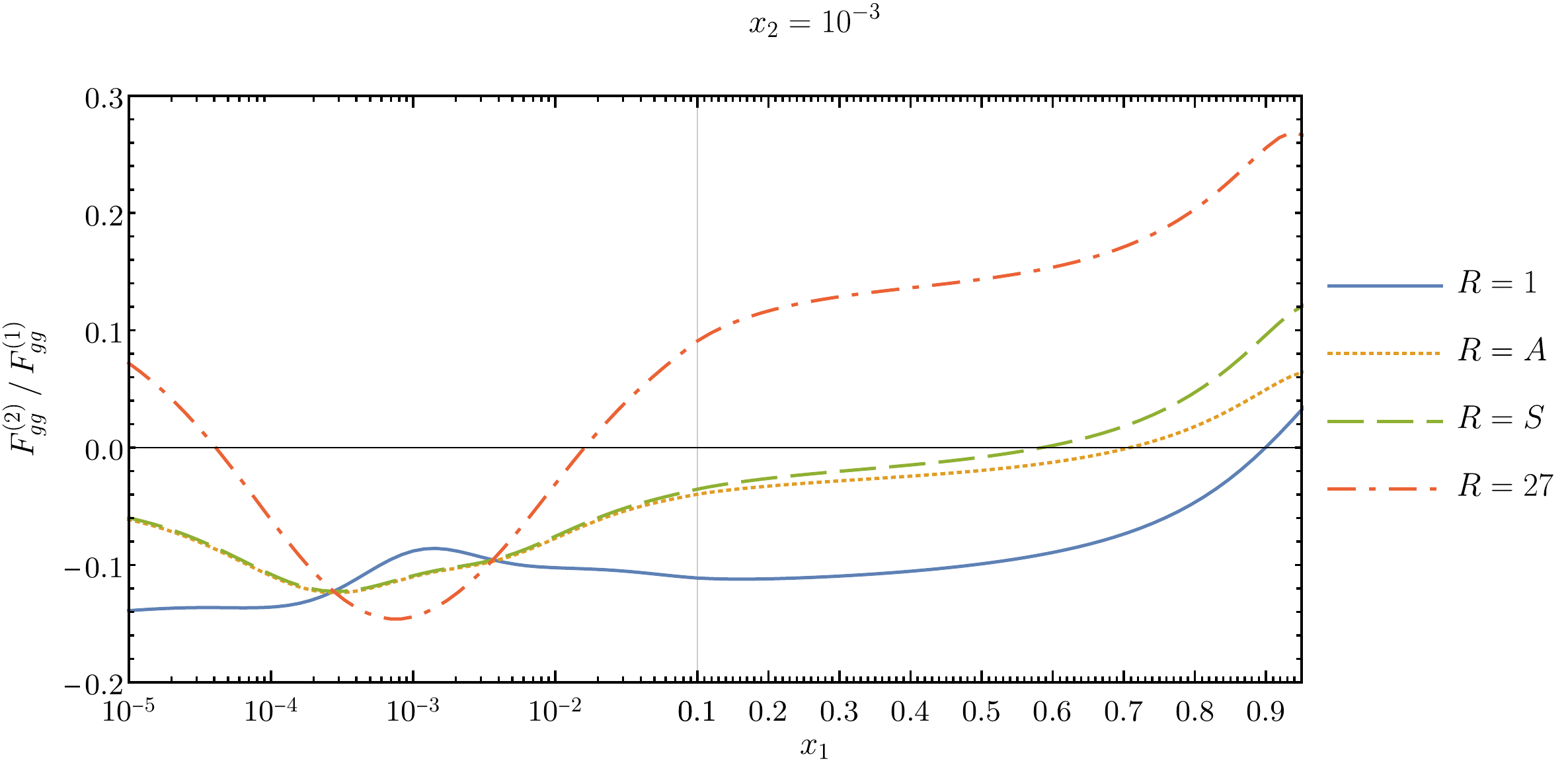}
}
\caption{\label{fig:F-ratios-central} The ratio $\pr{RR}{F}^{(2)}_{g g} \big/ \pr{RR}{F}^{(1)}_{g g}$, evaluated at central values of the scales as in \fig{\protect\ref{fig:F-values}}.}
\end{center}
\end{figure}

The relative size of $\pr{RR}{F}_{g g}$ for different $R$ reflects the scaling factors of the LO kernels given in \eqref{rescaling-factors} and \eqref{rescaling-factors-num}.  In some panels of \fig{\ref{fig:F-values}}, one can see slight differences in the shape of the distributions, which are due to the two-loop terms.  To see this more clearly, we plot the ratio ${F}^{(2)} \big/ {F}^{(1)}$ in \fig{\ref{fig:F-ratios-central}}.  We find that the $\as^2$ corrections exhibit a quite varied structure as a function of the momentum fractions $x_1$ and $x_2$.  Not exceeding $30 \%$, they are of moderate size in this setting.  Note that $\pr{RR}{F}^{(2)} \big/ \pr{RR}{F}^{(1)}$ is nearly equal for $R=A$ and $R=S$ at small $x_1 + x_2$, whilst for large values of $x_1 + x_2$ a clear difference between the two representations is observed.  The growing difference with increasing $x_1 + x_2$ is what we anticipated when discussing our analytic result~\eqref{Fgg-AS}.


\begin{figure}[t]
\begin{center}
\vspace{0.5em}   
   \includegraphics[height=6.3cm,trim=0 0 0 20,clip]{
   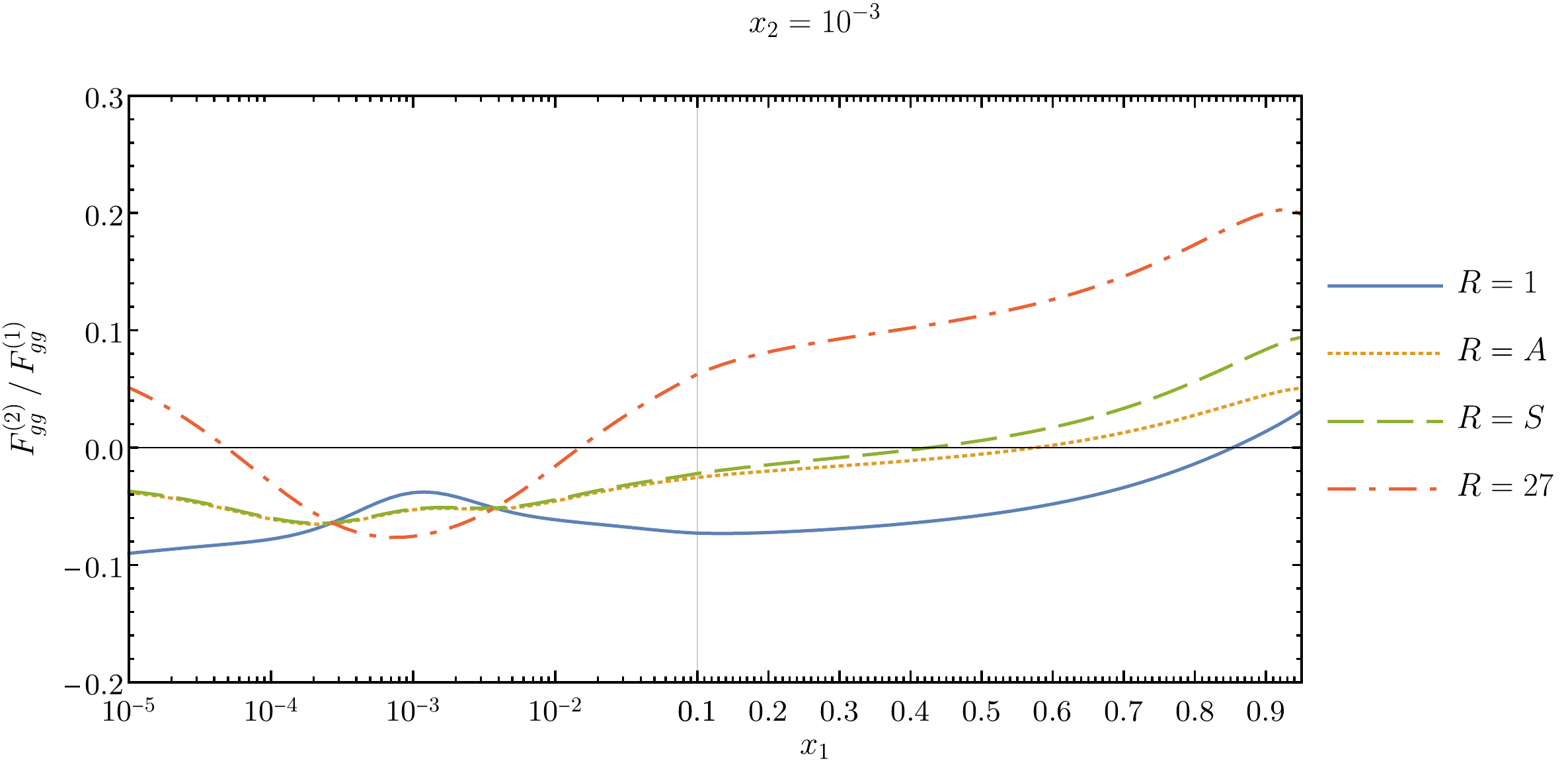}
\caption{\label{fig:F-ratios-y} \rev{As \fig{\protect\ref{fig:F-ratios-central-small-x1}} but for $\mu_y = 80 \gev$ and $\alpha_s(80 \gev) = 0.120$.  The corresponding curves for $\mu_y = 5 \gev$ are similar in shape but differ in overall size, with absolute values up to $0.3$ for $R=27$ and up to $0.2$ for the octets and the singlet.}}
\end{center}
\end{figure}

\rev{We also evaluated ${F}^{(2)} \big/ {F}^{(1)}$ for $\mu_y = 5 \gev$ and $\mu_y = 80 \gev$, always with central values $\mu = \sqrt{x_1 x_2\ms \zeta_p} = \mu_y$ of the scales.  Compared with the choice in \eqref{y-choice}, the overall size of the ratio changes due to the changed value of $\alpha_s(\mu_y)$.  There is also a mild change in the shape of the curves, due to the scale dependence of the PDFs in the splitting formula.  An example is shown in \fig\ref{fig:F-ratios-y}.  We continue our study with the value $\mu_y = 10 \gev$ in~\eqref{y-choice}.}

\begin{figure}[p]
\begin{center}
\subfigure[MSHT20 NLO]{
   \includegraphics[height=6.3cm,trim=0 0 0 20,clip]{
   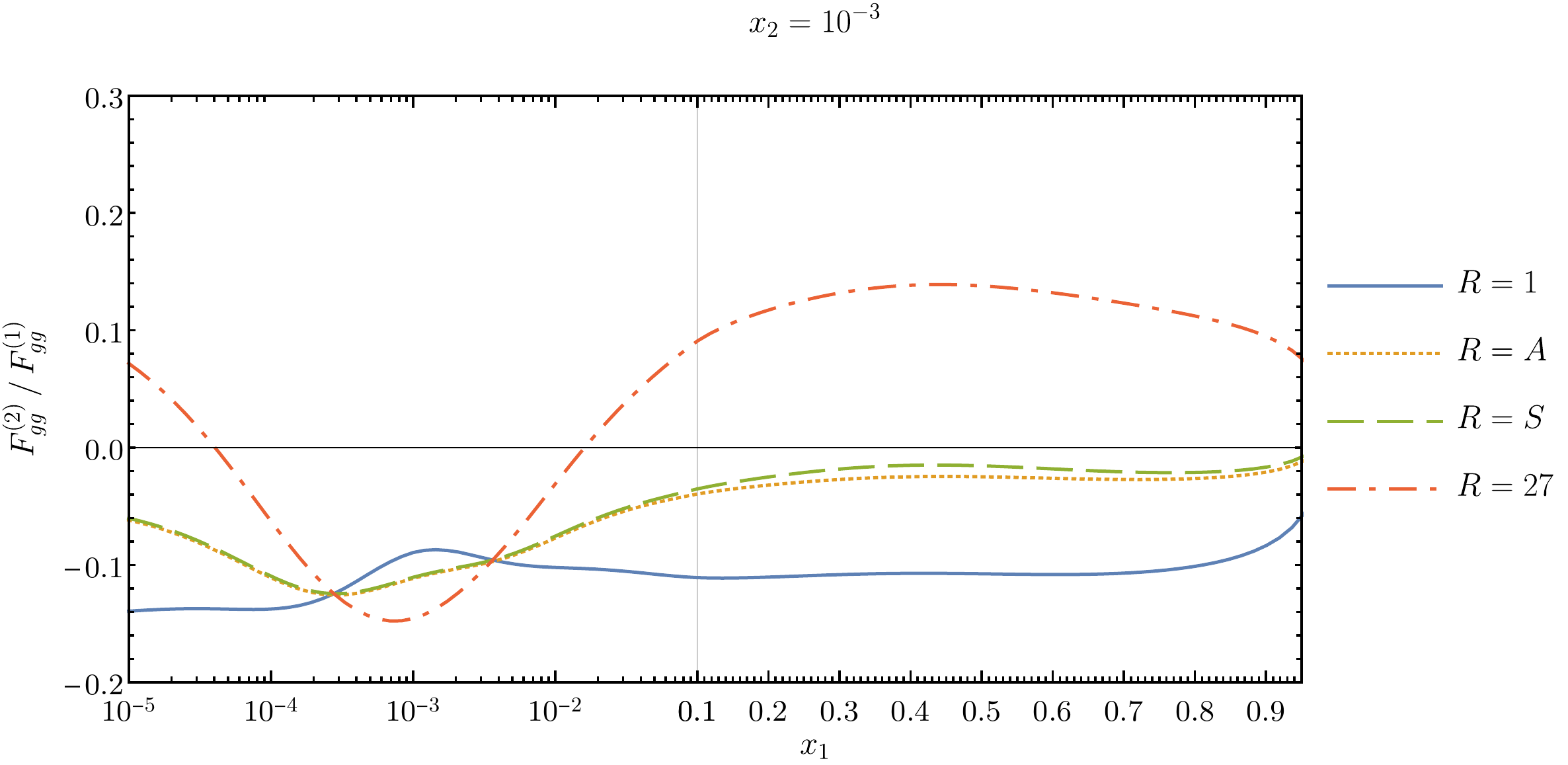}
}
\subfigure[ABMP16 NLO]{
   \includegraphics[height=6.3cm,trim=0 0 0 20,clip]{
   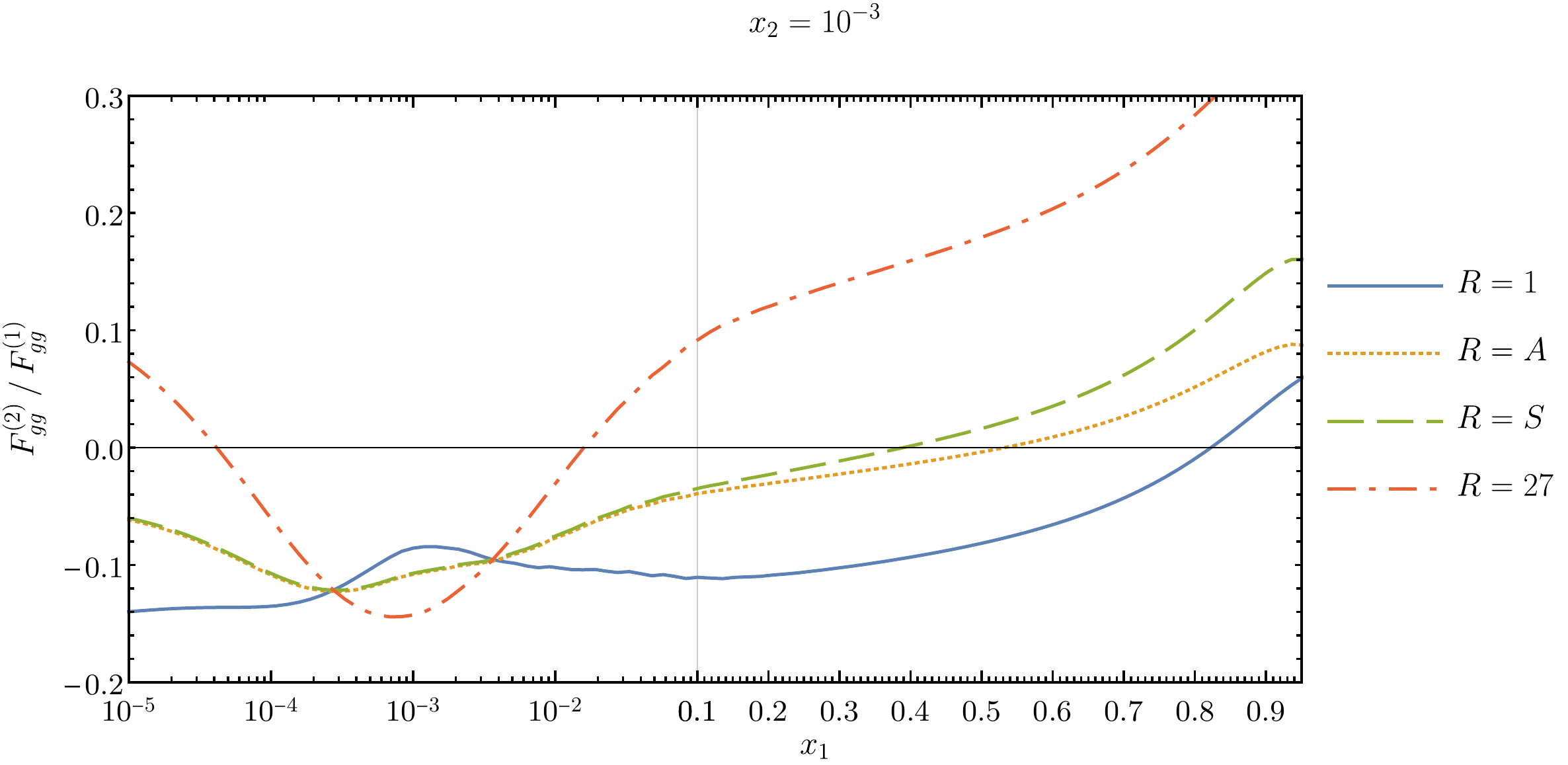}
}
\caption{\label{fig:F-ratios-PDFs} The ratio $\pr{RR}{F}^{(2)}_{g g} \big/ \pr{RR}{F}^{(1)}_{g g}$ for $x_2 = 10^{-3}$, evaluated at central scales for different PDFs.  The corresponding plot for the CT14 NLO set is in \fig{\protect\ref{fig:F-ratios-central-small-x1}}.}
\end{center}
\end{figure}

To gauge the dependence of our results on the PDFs used in the splitting formula, we produced plots for the NLO sets of ABMP16 \cite{Alekhin:2017kpj}, NNPDF3.1 \cite{Ball:2017nwa}, and MSHT20 \cite{Bailey:2020ooq}.\footnote{The corresponding LHAPDF identifiers are \texttt{ABMP16\_5\_nlo}, \texttt{NNPDF31\_nlo\_as\_0118}, and \texttt{MSHT20nlo\_as118}.}
As illustrated in \fig{\ref{fig:F-ratios-PDFs}}, the curves for ABMP16 and MSHT20 are quite similar to those for CT14 up to about $x_1 + x_2 \sim 0.4$, whereas differences become more pronounced for larger $x_1 + x_2$.  The same holds for NNPDF3.1.  In this case, the gluon distribution at $\mu = 10 \gev$ has a zero for $x \approx 0.7$, which leads to a corresponding zero in $\pr{RR}{F}^{(1)}_{g g}$.

\begin{figure}[p]
\begin{center}
   \includegraphics[height=6.3cm,trim=0 0 0 20,clip]{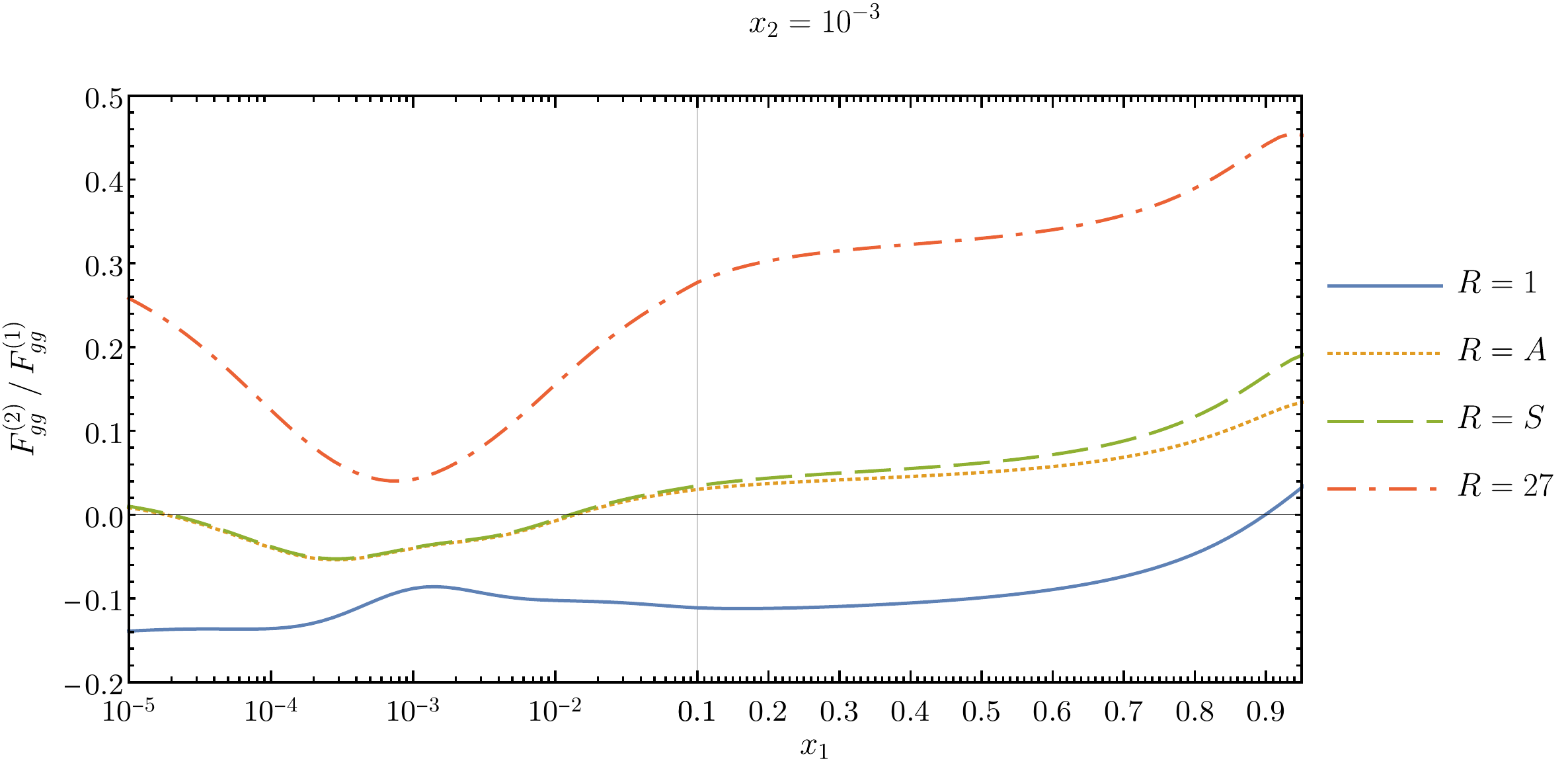}
\caption{\label{fig:F-ratios-MSbar} As \fig{\protect\ref{fig:F-ratios-central-small-x1}}, but with the \msbar scheme defined in \protect\cite{Collins:2011zzd} (i.e.\ with $c_{\msbars} = 0$).}
\end{center}
\end{figure}

At order $\as^2$, the splitting DPDs have a dependence on the \msbar scheme definition, which cancels at the level of DPS cross sections as explained at the end of \sect{\ref{sec:two-loop-kernels}}.  According to \eqref{V-NLO-form}, a change in the scheme parameter $c_{\msbars}$ results in a shift in the ratio $\pr{RR}{F}^{(2)} \big/ \pr{RR}{F}^{(1)}$ that is proportional to $\as \prn{R}{\gamma}_J^{(0)}$ and independent of $x_1$ and $x_2$.  Comparing \fig{\ref{fig:F-ratios-MSbar}} with \fig{\ref{fig:F-ratios-central-small-x1}}, we see that this shift is moderate for two-gluon distributions in the octet channels but quite large for $R=27$.


\begin{figure}
\begin{center}
\subfigure[$\mu = \mu_y$]{
   \includegraphics[height=6.3cm,trim=0 0 0 20,clip]{
   colors_x1eq0p001_ratios_central_scale.pdf}
} \\[0.8em]
\subfigure[$\mu = \mu_y / 2$]{
   \includegraphics[height=6.3cm,trim=0 0 0 20,clip]{
   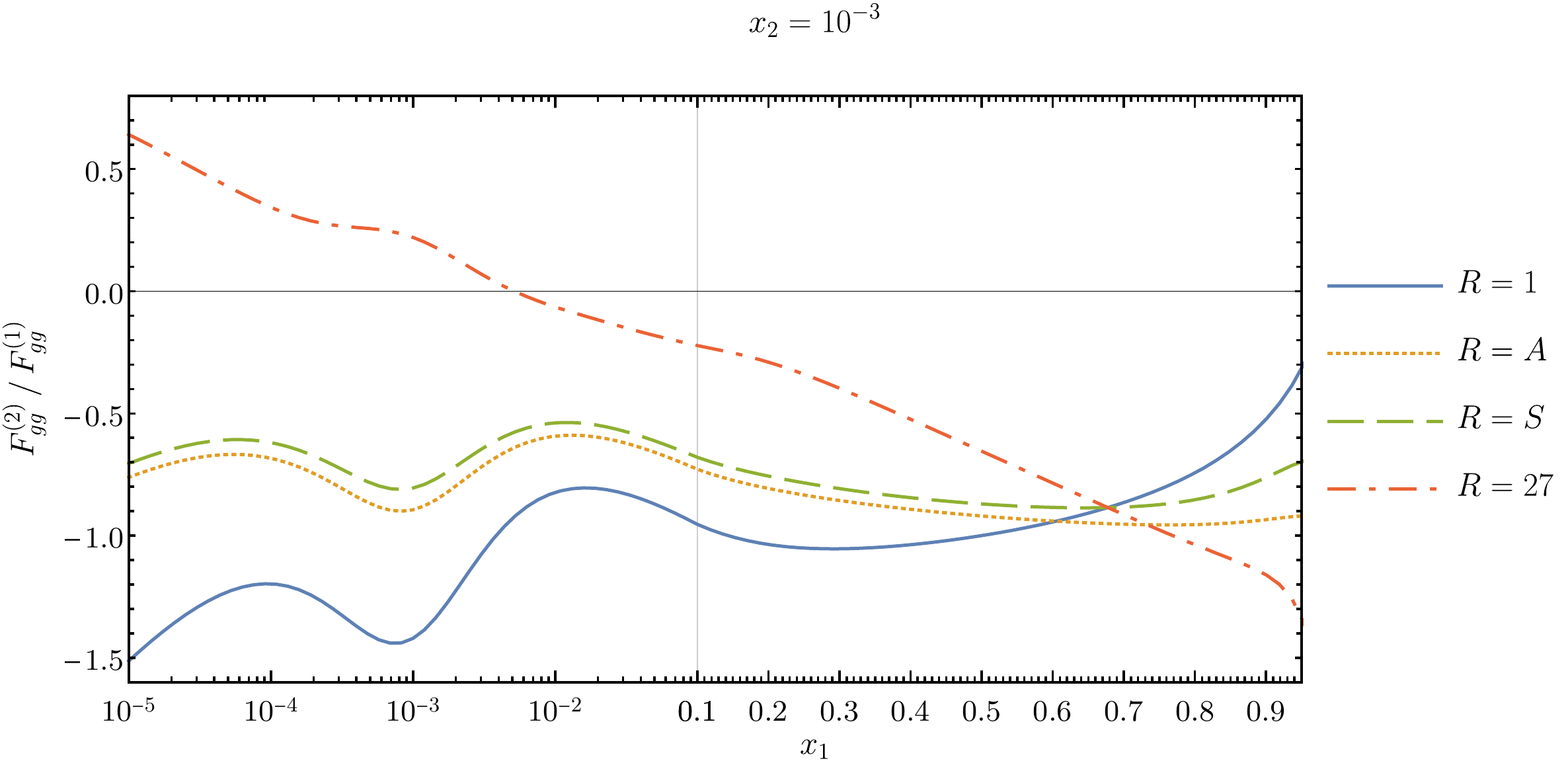}
} \\[0.8em]
\subfigure[$\mu = 2 \mu_y$]{
   \includegraphics[height=6.3cm,trim=0 0 0 20,clip]{
   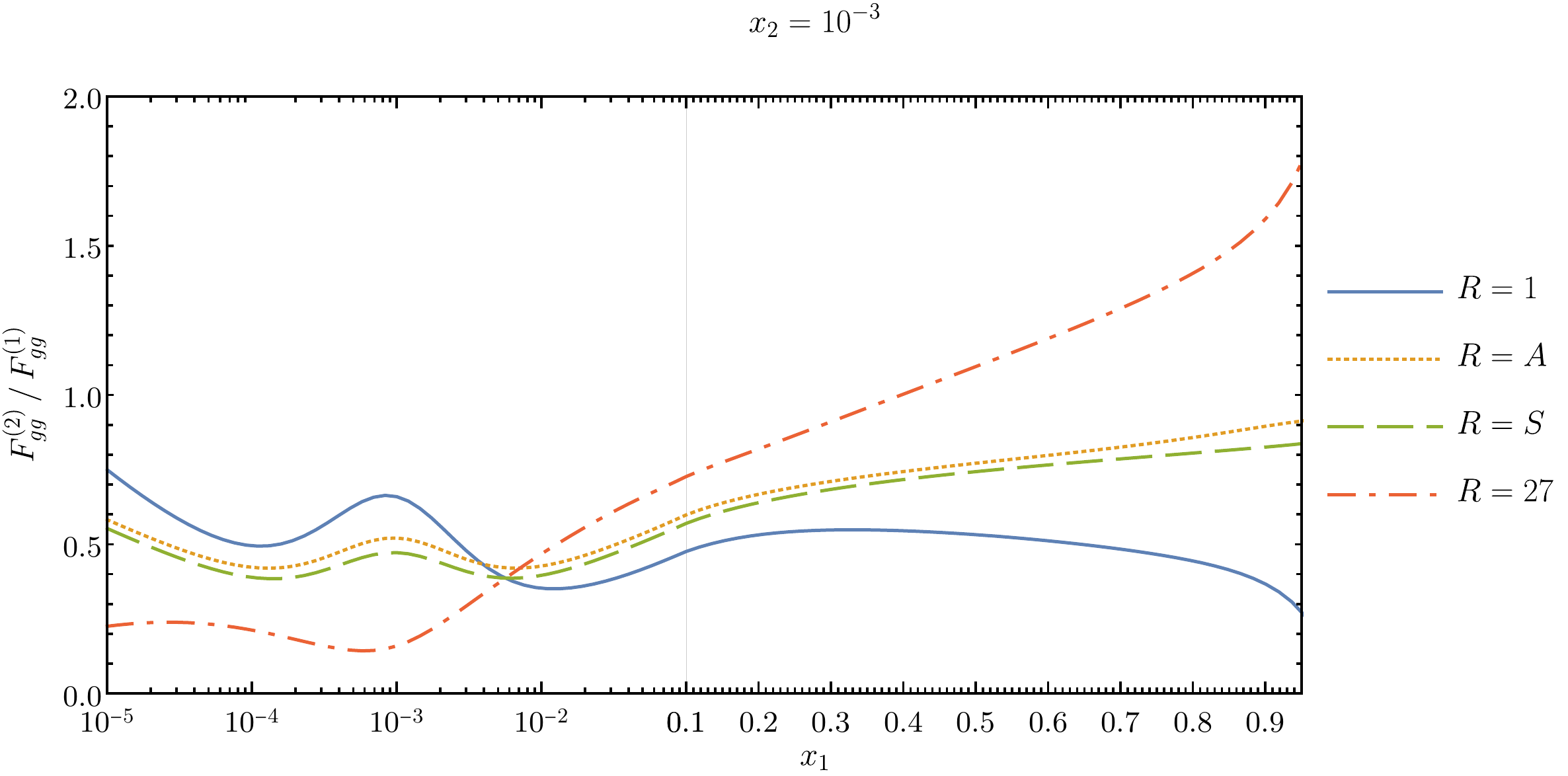}
}
\caption{\label{fig:F-ratios-mu_var} The ratio $\pr{RR}{F}^{(2)}_{g g} \big/ \pr{RR}{F}^{(1)}_{g g}$ for $x_2 = 10^{-3}$, evaluated at different values of the scale $\mu$ at $\sqrt{x_1 x_2\ms \zeta_p} = \mu_y$.  The plot in panel (a) is the same as in \fig{\protect\ref{fig:F-ratios-central-small-x1}} and shown again for easier comparison.}
\end{center}
\end{figure}

We now investigate how the importance of $\as^2$ contribution changes if we evaluate the DPDs at different scales.  In \fig{\ref{fig:F-ratios-mu_var}}, we show the ratio $F^{(2)}_{g g} \big/ F^{(1)}_{g g}$ for three different values of $\mu$, taking $\sqrt{x_1 x_2\ms \zeta_p} = \mu_y$ in all three cases.  We see that for both $\mu = \mu_y/2$ and $\mu = 2 \mu_y$, the impact of the two-loop kernels becomes substantial at least in some region of $x_1$ and $x_2$.  This holds for all colour representations and is in stark contrast to the case $\mu = \mu_y$.


\begin{figure}
\begin{center}
\subfigure[$\mu = \sqrt{x_1 x_2\ms \zeta_p} = \mu_y / 2$]{
   \includegraphics[height=6.3cm,trim=0 0 0 20,clip]{
   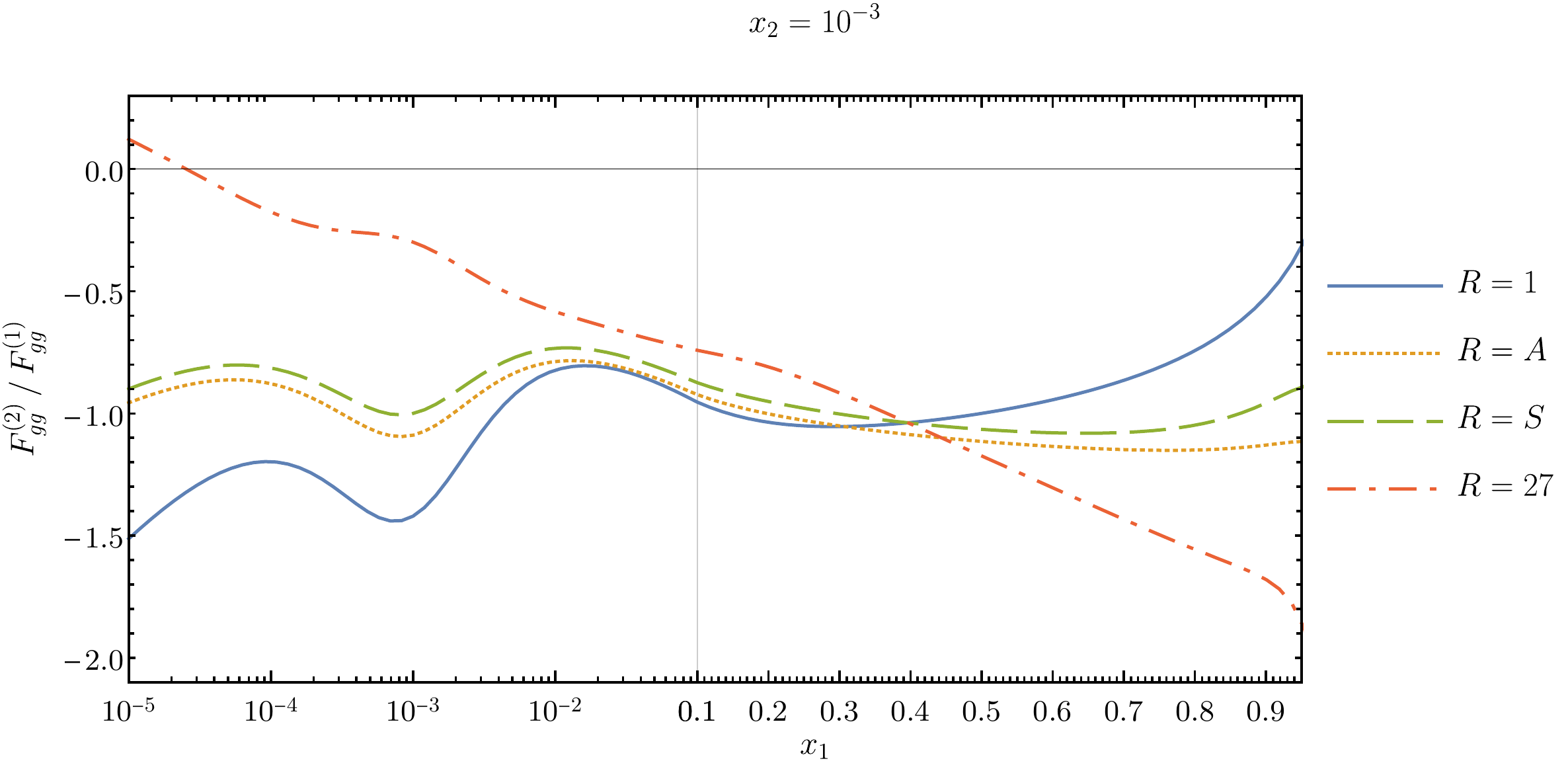}
} \\[1em]
\subfigure[$\mu = \sqrt{x_1 x_2\ms \zeta_p} = 2 \mu_y$]{
   \includegraphics[height=6.3cm,trim=0 0 0 20,clip]{
   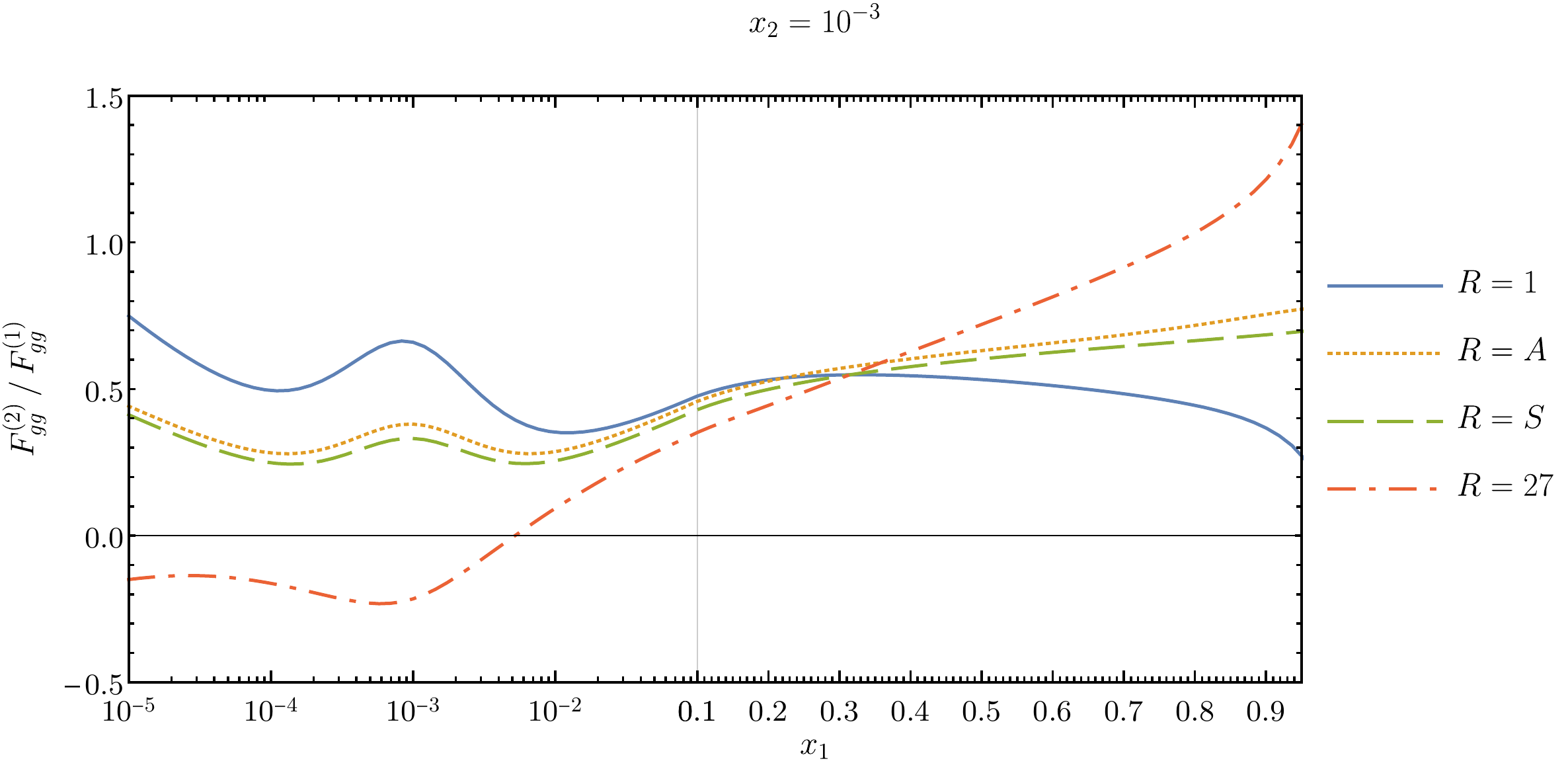}
}
\caption{\label{fig:F-ratios-zeta_var} The ratio $F^{(2)}_{g g} \big/ F^{(1)}_{g g}$ for $x_2 = 10^{-3}$, evaluated at different values of $\mu = \sqrt{x_1 x_2\ms \zeta_p}$.}
\end{center}
\end{figure}

A change of the rapidity parameter $\zeta_p$ at given $\mu$ affects only
the double logarithmic part of the kernel \eqref{V-NLO-form}.  For the ratio
$\pr{RR}{F}^{(2)}_{g g} \big/ \pr{RR}{F}^{(1)}_{g g}$, this induces a shift proportional to $\as \prn{R}{\gamma}_J^{(0)}$ and independent of $x_1$ and $x_2$ (as does the change of \msbar scheme definition).  We show the ratio for $\sqrt{x_1 x_2\ms \zeta_p} = \mu$ with $\mu = \mu_y / 2$ and $\mu = 2 \mu_y$ in \fig{\ref{fig:F-ratios-zeta_var}} and see that also in this case the impact of the two-loop terms is in general large.  The large NLO corrections are mainly due to the single logarithmic terms $L\, V^{[2,1]}$ in the two-loop kernel.

We conclude that, at least for two-gluon DPDs, it is most favourable to use the fixed-order splitting formula for $\mu$ and $\sqrt{x_1 x_2\ms \zeta_p}$ very close to $\mu_y$, so as to avoid large radiative corrections from higher orders.  The DPD at any other scale can then be obtained by using the evolution equations \eqref{CS-F} and \eqref{DGLAP-F}.


\begin{figure}
\begin{center}
\subfigure[$x_2 = 0.1$]{
   \includegraphics[height=5cm,trim=0 0 0 20,clip]{
   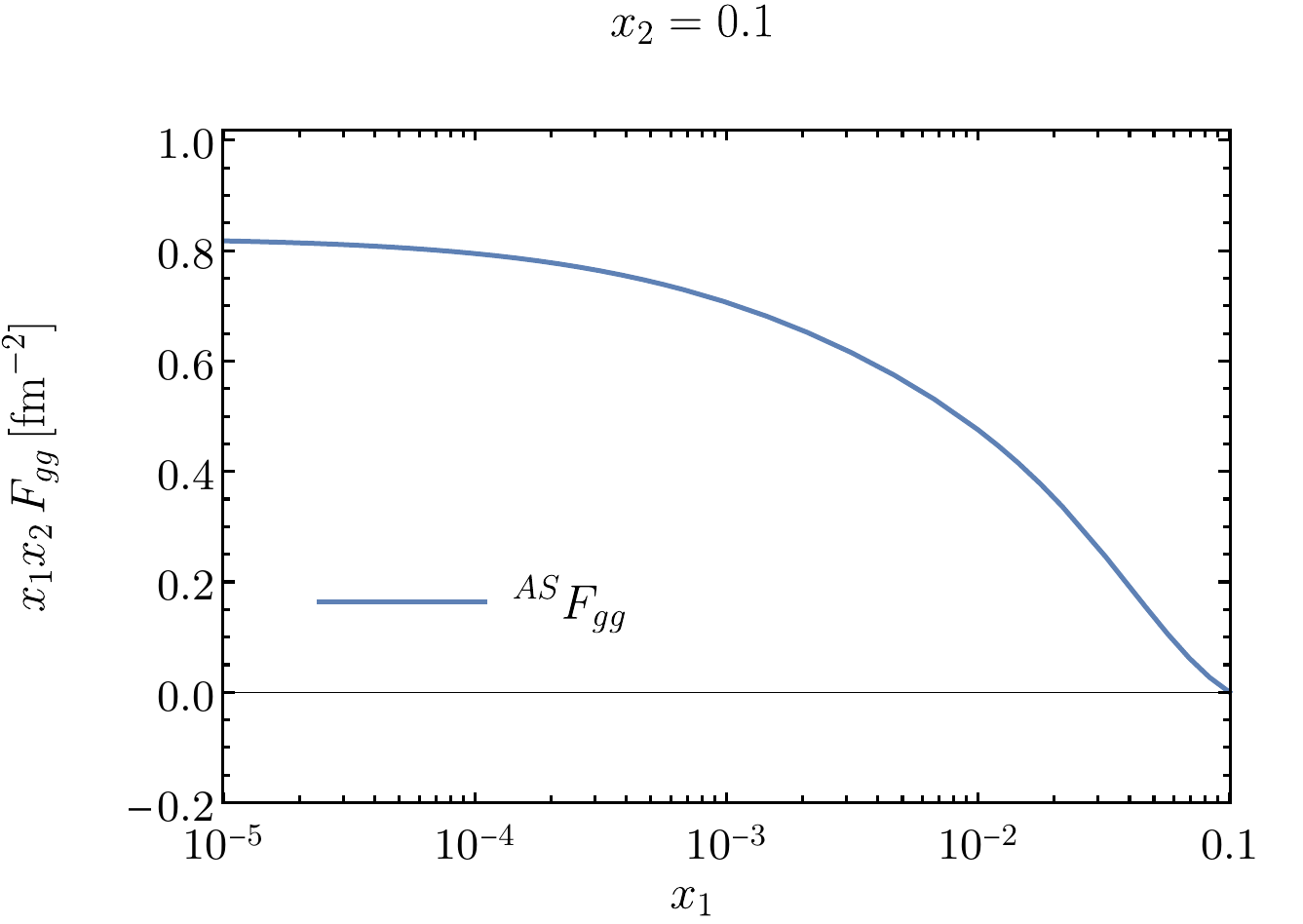}
\hspace{0.5em}
   \includegraphics[height=5cm,trim=0 0 0 20,clip]{
   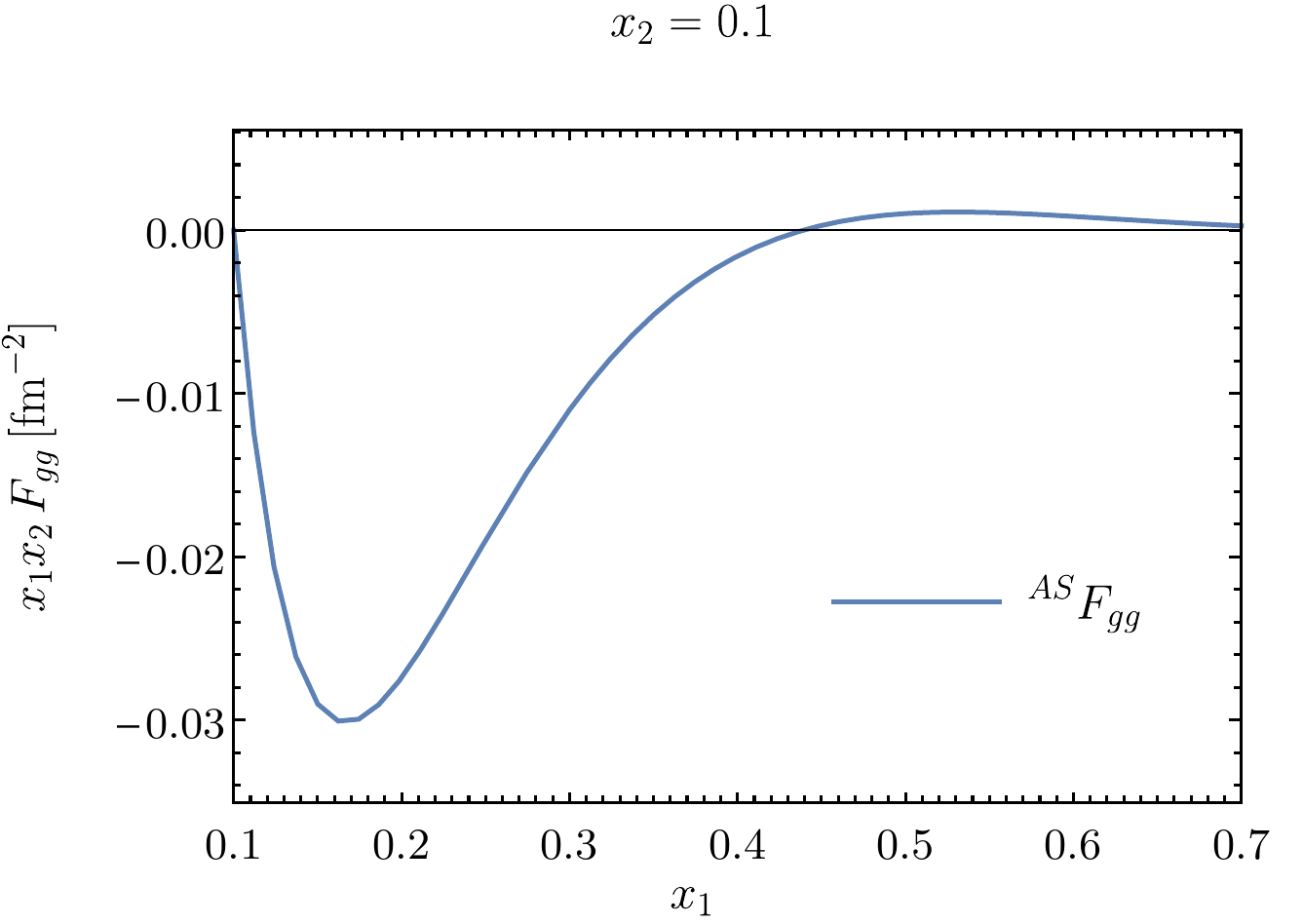}
} \\[1em]
\subfigure[$x_2 = 10^{-3}$]{
   \includegraphics[height=5cm,trim=0 0 0 20,clip]{
   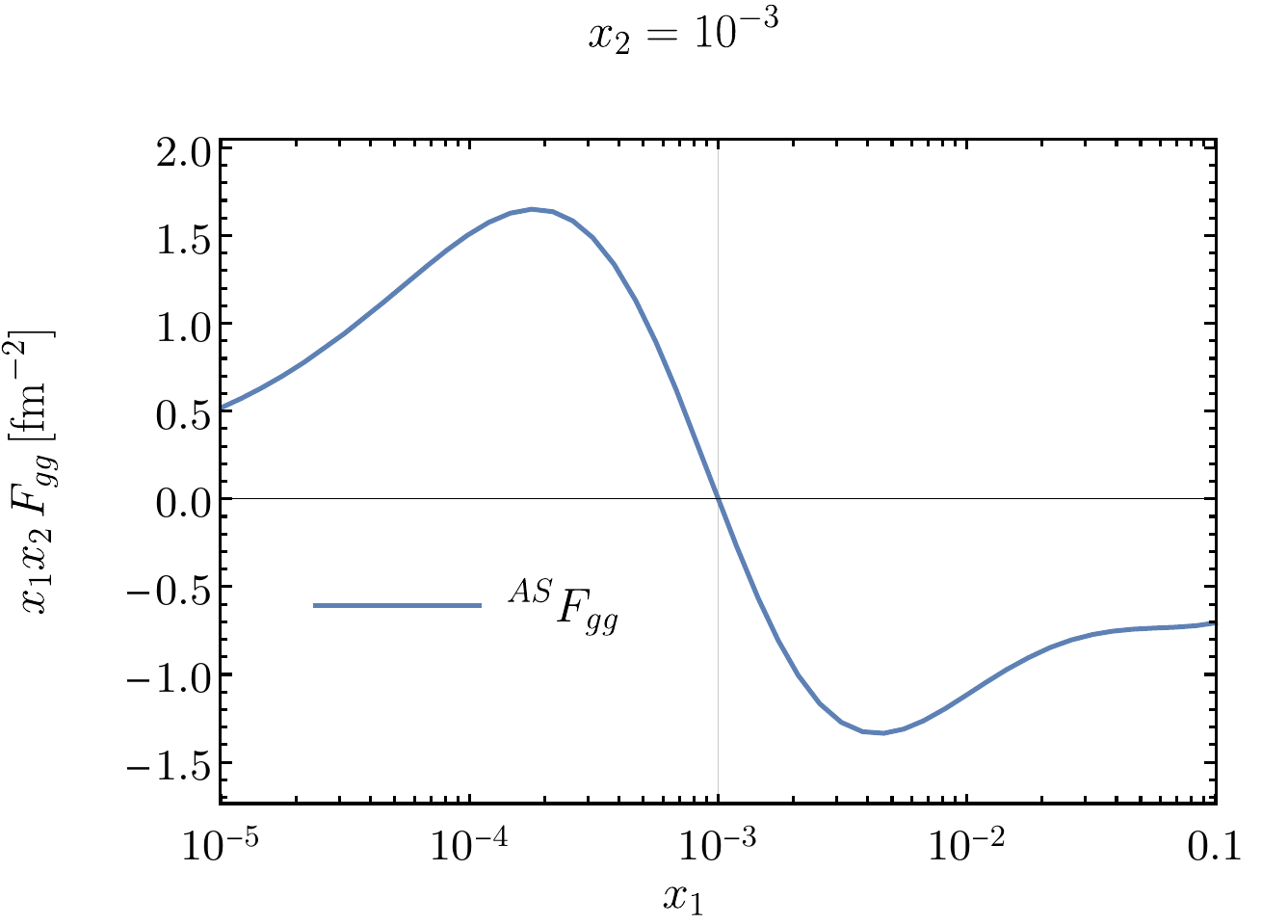}
\hspace{0.5em}
   \includegraphics[height=5cm,trim=0 0 0 20,clip]{
   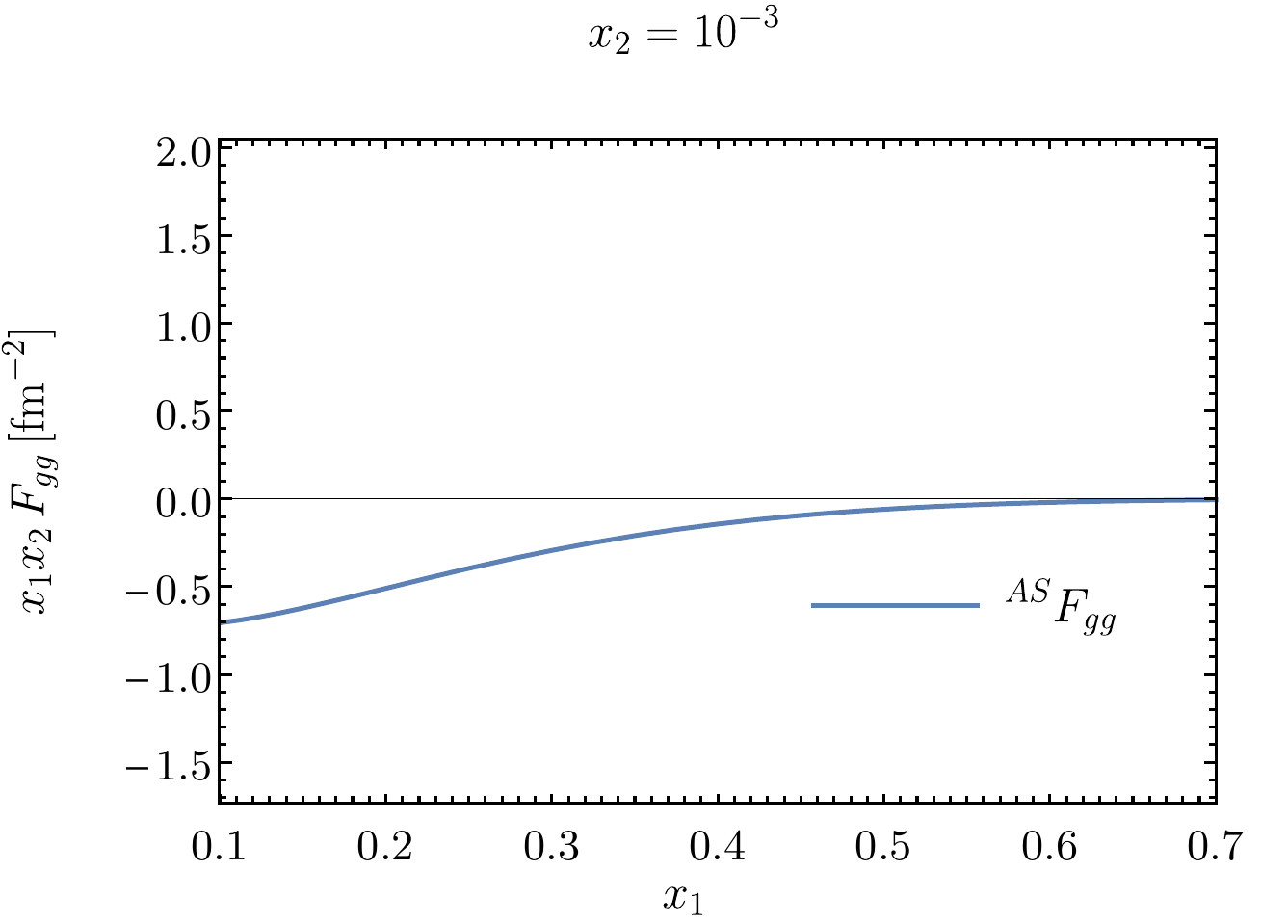}
}
\caption{\label{fig:F-mixed-octet} The distribution $\pr{AS}{F}_{g g}$, evaluated at central values of the scales as in \fig{\protect\ref{fig:F-values}}.}
\end{center}
\end{figure}

Let us finally take a look at the mixed octet distribution $\pr{AS}{F}_{g g}$, which starts at order $\as^2$ and is driven by the quark valence distributions.  Comparing the plots in \fig{\ref{fig:F-mixed-octet}} with those in \fig{\ref{fig:F-values}}, we see that $\pr{AS}{F}_{g g}$ is tiny compared with the two-gluon distributions in the channels that receive a contribution from $g\to g g$ splitting.  The zero of $\pr{AS}{F}_{g g}$ at $x_1 = x_2$ is a consequence of the asymmetry of this distribution under the exchange $x_1 \leftrightarrow x_2$.

\subsection{Quark-gluon distributions}

\begin{figure}
\begin{center}
\subfigure[$R_1 R_2 = 11$]{
   \includegraphics[height=5.3cm,trim=0 0 0 20,clip]{
   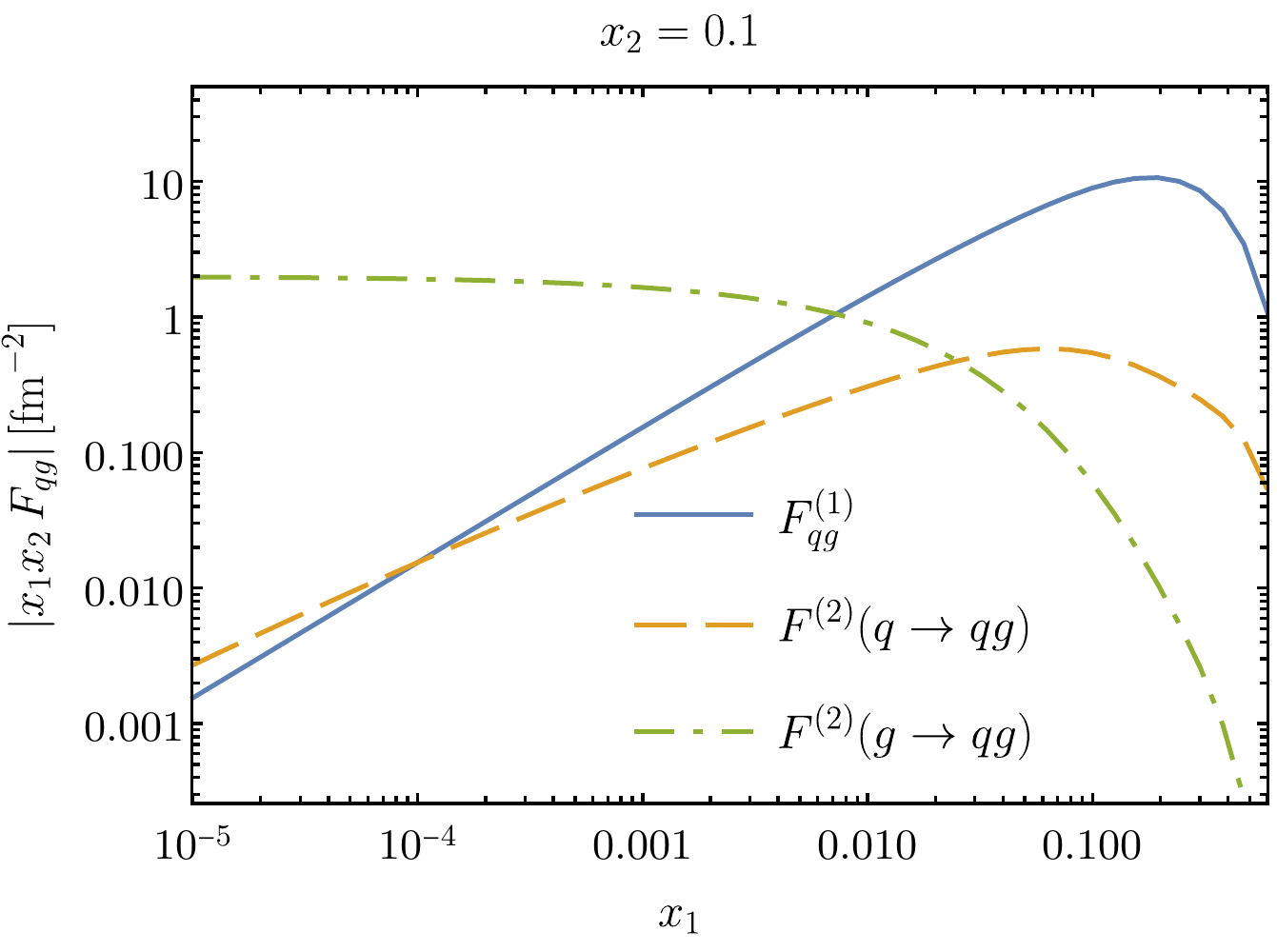}
}
\subfigure[$R_1 R_2 = 8A$]{
   \includegraphics[height=5.3cm,trim=22 0 0 20,clip]{
   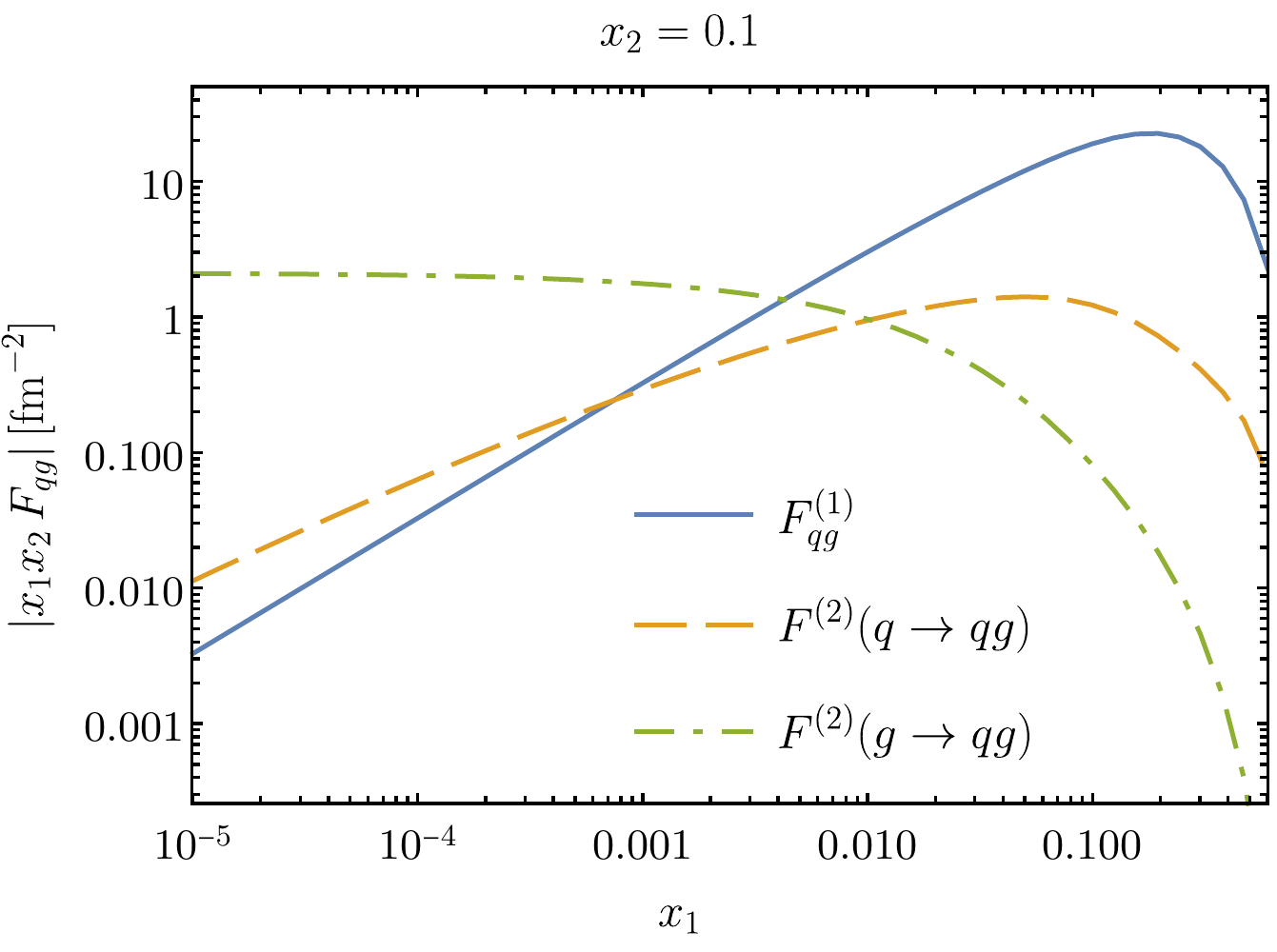}
}
\caption{\label{fig:F_qg} Absolute values of the different contributions to $\pr{\RR}{F}_{u g}$ at $x_2 = 0.1$, evaluated at central scales with the CT14 PDFs.  The case $R_1 R_2 = 8S$ is not shown here, because according to \protect\eqref{LO-rescaling-2} and \protect\ref{eq:RVqgq2-octets} one has $\pr{8S}{F}_{u g} = {}- \sqrt{5/9}\,\, \pr{8A}{F}_{u g}$ for each separate contribution.}
\end{center}
\end{figure}

We now turn our attention to the quark-gluon channel.  Here, a new phenomenon occurs compared with the two-gluon case.  This is seen in \fig{\ref{fig:F_qg}}, where we show the term $F^{(1)}$ and the separate contributions from the subprocesses $q\to q g$ and $g\to q g$ to $F^{(2)}$.  For $x_1 \sim x_2$ and for $x_1 > x_2$ , the $\as^2$ contributions are small compared with the $\as$ term, but when $x_1$ becomes much smaller than $x_2$, the contribution from $g\to q g$ quickly overtakes the $\as$ term and completely dominates the sum of terms.

This can be understood from our results for the limit of small $u = x_1 / (x_1 + x_2)$ in \sect{\ref{sec:small_u_or_ubar}}.  According to \eqref{eq:Vqgg2_small_u}, the two-loop contribution $F^{(2)}(g\to q g)$ behaves like $1/u$ in that limit, whereas $F^{(1)}$ and $F^{(2)}(q\to q g)$ do not have a $1/u$ enhancement.  Since in \fig{\ref{fig:F_qg}} we plot contributions to the scaled DPD $x_1 \ms x_2 \ms F$ at fixed $x_2$, the contribution from $g\to q g$ tends to a constant for small $x_1$ and the contributions from $q\to q g$ go to zero.  Closer inspection reveals that $F^{(2)}(q\to q g)$ has a \rev{$\log^2 u$} enhancement compared with $F^{(1)}$ for small $u$, which explains why the $\as^2$ term for $q\to q g$ slowly overtakes the $\as$ term in that limit.

As explained in \sect{\ref{sec:small_u_or_ubar}}, the $1/u$ enhancement of $F^{(2)}(g\to q g)$ originates from the graph in \fig{\ref{fig:gqg-UD-2}}, where a gluon with momentum fraction of order $u$ is emitted from a gluon with much larger momentum fraction.  This is not possible in the $q\to q g$ channel at order $\as^2$.  As is well known in small $x$ physics, higher orders will enhance the $1/u$ behaviour by factors of $\log^k u$ but not lead to higher powers like $1/u^2$.  Hence, one does not expect a change as drastic as the step from one to two loops in \fig{\ref{fig:F_qg}} when going to three loops or higher.
This situation is reminiscent of hard-scattering cross sections, where the step from LO to NLO accuracy can be huge if qualitatively new contributions first appear at NLO.

The graph in \fig{\ref{fig:gqg-UD-2}} is not only responsible for the $1/u$ behaviour of $F^{(2)}(g\to q g)$, but it also contributes to the LO splitting $g\ \to g g$ followed by one step of LO evolution with the splitting of a gluon into $q \bar{q}$.  At small $u$, one has $| {F}_{g g}^{(1)} | \gg | {F}_{q g}^{(1)} |$ in all relevant colour channels, so that $| {F}_{q g}^{(1)} |$ will quickly increase in size when evolved to higher scales.

\section{Conclusions}
\label{sec:conclude}

DPDs at small inter-parton distance $y$ are dominated by the splitting of one parton into two.  We computed the kernels that describe this splitting at order $\as^2$ for all colour configurations of the observed partons.  All partons are taken to be unpolarised.  The colour dependence of the DPDs is classified by a pair of irreducible representations $R_1$ and $R_2$ of the colour group.  For representations different from the singlet, Sudakov double logarithms in the renormalisation scale $\mu$ and the rapidity parameter $\zeta_p$ appear in the splitting kernels.  These logarithms go along with rapidity divergences in the intermediate stages of the calculation, which need to be regularised appropriately.  We computed the kernels independently with the Collins regulator, which uses spacelike Wilson lines, and with the $\delta$ regulator of Echevarria et al.  Intermediate results differ in the two cases, for instance by the order of ultraviolet poles in DPD matrix elements and the soft factor, but in both cases we find complete cancellation of ultraviolet and rapidity divergences when all building blocks are assembled.  The resulting splitting kernels are identical for both regulators.

In the context of multi-loop calculations, the Collins regulator is more involved because eikonal propagators depend not only on plus-momenta.  However, in the limit of large Wilson line rapidities, these propagators take a simple distributional form.  With this form, loop integrals are no more difficult than those to be computed with the $\delta$ regulator, where eikonal propagators depend only on plus-momenta.  Whilst in our two-loop calculation, there is at most one eikonal propagator per graph, we see no impediment to using the same procedure at higher orders.

At order $\as$, the splitting DPDs for a given parton combination are proportional to each other for the different colour channels $R_1 R_2$.  This no longer holds at order $\as^2$, where the colour dependence of the splitting kernels presents a rather varied picture.  Nevertheless, we identified a number of simple patterns.
\begin{enumerate}
\item At two-loop accuracy, the splitting DPD for $g g$ in decuplet representations is zero, and the distributions $\pr{8A}{F}_{q g}$ and $\pr{8S}{F}_{q g}$ for the gluon in the antisymmetric and symmetric octet are proportional to each other.
\item At two loops, terms containing Sudakov double logarithms are proportional to the one-loop splitting result times the one-loop anomalous dimension \eqref{gammaJ-LO} of the Collins-Soper kernel for the rapidity dependence of the DPD.
\item A threshold logarithm $\log(1 - x_1 -x_2)$ appears in the splitting DPDs at two loops and is accompanied by a renormalisation group logarithm $L = 2 \log(y \mu /b_0)$.  Its coefficient is nonzero for the distributions $\pr{11}{F}_{g g}$, $\pr{27\, 27}{F}_{g g}$, $\pr{11}{F}_{q \bar{q}}$, $\pr{88}{F}_{q \bar{q}}$, and $\pr{11}{F}_{q g}$, but vanishes for $g g$ and $q g$ distributions in the octet channels.
\item When the sum $z = z_1+z_2$ of parton momentum fractions in the two-loop kernel becomes small, a leading $1/z^2$ behaviour is found for $g\to g g$, $q\to g g$, and $g\to q \bar{q}$.  This gives rise to a high-energy logarithm $\log(x_1 + x_2)$ when convoluted with PDFs of a suitable shape, as specified in \eqref{enhancement-factor}.  If $\mu$ equals  $\mu_y = b_0/y$, the leading $1/z^2$ terms for different channels $R_1 R_2$ scale exactly like the one-loop kernels, but for other values of $\mu$ they do not.
\item There is no simple pattern for the colour dependence of splitting DPDs in the limit of small $u = x_1 / (x_1+x_2)$.  The strongest enhancement in this limit is by a factor $u^{-1} \log u$.  It appears in the $g\to g g$ and $q\to g q$ channels and goes along with a renormalisation group logarithm $L$.
\item In the triple Regge limit $x_1 \ll x_1 + x_2 \ll 1$, a double enhancement like $u^{-1} \log(x_1 + x_2)$ appears in $\pr{11}{F}_{g g}$, $\pr{SS}{F}_{g g}$, and $\pr{27\, 27}{F}_{g g}$ for PDFs of a suitable shape.  It originates from a $1 /(u z^2)$ behaviour of the splitting kernels and is accompanied by a factor $L$.  In the case of $\pr{SS}{F}_{g g}$, the subprocess $q\to g g$ is enhanced in this way, whereas $g\to g g$ is not.  The behaviour of other DPDs depends on the order in which the limits $x_1 \ll x_1+x_2$ and $x_1+x_2 \ll 1$ are taken.
\item Because of charge conjugation invariance, the distributions $\pr{AS}{F}_{g g}$ or $\pr{S\bs A}{F}_{g g}$ receive no contribution from $g\to g g$, whilst $q\to g g$ and $\bar{q}\to g g$ contribute with opposite sign.  The splitting formula for these channels hence involves the valence distributions $f_q - f_{\bar{q}}$.  At two-loop accuracy we find that $\pr{AS}{F}_{g g} = {}- \pr{S\bs A}{F}_{g g}$.
\end{enumerate}
Full analytic expressions of the kernels are given in the ancillary files associated with this paper on \href{https://arxiv.org/abs/2105.08425}{arXiv}.

A detailed numerical study of two-loop splitting effects in DPDs is beyond the scope of this work.  However, the plots in \sect{\ref{sec:numerics}} show a few examples in selected kinematics.  We find that for $g g$ distributions, the $\as^2$ kernels provide corrections not exceeding a few $10\%$ relative to the $\as$ terms if we evaluate the splitting formula at $\mu = \sqrt{x_1 x_2\ms \zeta_p} = \mu_y$.  However, when we change $\mu$ by a factor of $2$ in either direction, the $\as^2$ corrections become of order $100\%$ and larger in parts of the $x_1$--\,$x_2$ plane.  To obtain reliable perturbative results for scales different from $\mu_y$, it is therefore better to evaluate the splitting formula at $\mu_y$ and then use the appropriate evolution equations.  The distribution $\pr{AS}{F}_{g g}$ is found to be tiny compared with distributions that receive a contribution from $g\to g g$ splitting.

For $q g$ distributions at $\mu = \sqrt{x_1 x_2\ms \zeta_p} = \mu_y$, we find small to moderate $\as^2$ corrections for $x_1 \sim x_2$ and $x_1 > x_2$.  For $x_1 \ll x_2$, however, the $\as^2$ contribution from $g\to q g$ splitting completely dominates ${F}_{q g}$, because it has a $1/u$ enhancement that is absent in the $q\to q g$ channel up to that order.

With the results of the present paper, we have added a further ingredient for an analysis of double parton scattering at NLO.  An extension to the case of polarised DPDs seems straightforward with the methods developed here and in  \cite{Diehl:2019rdh}.  Complications due to the use of $\gamma_5$ and $\epsilon^{\mu\nu\rho\sigma}$ in dimensional regularisation can be avoided using the procedure employed in \cite{Buffing:2017mqm}.  We intend to come back to this in future work.

\appendix
\section{Colour space matrices}
\label{sec:projectors}

In this appendix, we list the explicit matrices in colour space that appear in the relation \eqref{F-R-projection} between DPDs for definite colour representations $R_1, R_2$ and DPDs with open colour indices.  We have
\begin{align}
\label{projectors}
M_{11}^{i i'  \ms j j'} &= \delta_{i' i} \, \delta_{j'\bs j} \,,
&
M_{88}^{i i'  \ms j j'}
&= \frac{2 N}{\sqrt{N^2-1}}\, t^a_{i' i} \, t^a_{j'\bs j}
\end{align}
for quark-quark distributions $F_{q q}^{i i'  \ms j j'}$,
\begin{align}
M_{11}^{j j'  \ms a a'}
&= M_{11}^{a a'\ms j j'} = \delta_{j'\bs j} \, \delta^{a a'} \,,
\nonumber \\
M_{8A}^{j j'  \ms a a'}
&= M_{A8}^{a a'\ms j j'} = \sqrt{2}\, t^c_{j'\bs j}\, i f^{a a' c} \,,
&
M_{8S}^{j j'  \ms a a'}
&= M_{S8}^{a a'\ms j j'}
 = \sqrt\frac{2 N^2}{N^2 - 4}\, t^c_{j'\bs j}\, d^{a a' c}
\end{align}
for mixed quark-gluon distributions $F_{q g}^{i i'  \ms a a'}$ and $F_{g q}^{a a' \ms i i' }$, and
\begin{align}
M_{11}^{a a' \ms b b'} &= \delta^{a a'} \delta^{b b'} \,,
   \phantom{\frac{0}{0}}
\nonumber \\
M_{AA}^{a a' \ms b b'} &= - \frac{\sqrt{N^2 - 1}}{N}\, f^{a a' c} f^{b b' c} \,,
\qquad
M_{SS}^{a a' \ms b b'} = \frac{N\ms \sqrt{N^2 - 1}}{N^2-4}\,
   d^{a a' c} d^{b b' c} \,,
\nonumber \\
M_{AS}^{a a'\, b b'} &= \frac{\sqrt{N^2 - 1}}{\sqrt{N^2-4}}\,
   i f^{a a' c} d^{b b' c} \,,
\qquad\;\;
M_{S\bs A}^{a a'\, b b'} = \frac{\sqrt{N^2 - 1}}{\sqrt{N^2-4}}\;
   i\ms d^{a a' c} f^{b b' c} \,,
\nonumber \\
M_{\tentenbar}^{a a' \ms b b'} &= \frac{\sqrt{N^2 - 1}}{\sqrt{N^2-4}} \,
   \biggl[
      \frac{1}{2} \bigl( \delta^{a b} \delta^{a' b'}
         - \delta^{a b'} \delta^{a' b} \bigr)
      - \frac{1}{N}\ms f^{a a' c} f^{b b' c}
      + \frac{i}{2} \bigl( d^{a b c} f^{a' b' c}
         + f^{a b c} d^{a' b' c} \bigr)
   \biggr] \,,
\nonumber \\
M_{\tenbarten}^{a a' \ms b b'}
   &= \bigl[ M_{\tentenbar}^{a a' \ms b b'} \bigr]^{*} \,, \phantom{\frac{0}{0}}
\nonumber \\
M_{27\, 27}^{a a' \ms b b'} & \!\!\!\underset{N=3}{=} \frac{4}{\sqrt{27}} \,
   \biggl[
      \delta^{a b} \delta^{a' b'} + \delta^{a b'} \delta^{a' b}
      - \frac{1}{4}\ms \delta^{a a'} \delta^{b b'}
      - \frac{6}{5}\ms d^{a a' c} d^{b b' c}
   \biggr]
\end{align}
for two-gluon distributions $F_{g g}^{a a' \ms b b'}$.  These matrices are obtained from \eqn{(I.11)} of \cite{Buffing:2017mqm}.

The matrices for antiquarks are the same as for quarks.  Note that according to the relation between quark and antiquark operators given below \eqref{op-defs}, the index $i$ in $\mathcal{O}^{i i'}_q$ refers to the quark in the amplitude, whereas in $\mathcal{O}^{i i'}_{\bar{q}}$ it refers to the antiquark in the conjugate amplitude.

\paragraph{Charge conjugation.}
Let us see how the colour projected products of twist-two operators transform under charge conjugation.  One can use light-cone gauge $A^+ = 0$ for this purpose, which avoids the need to transform Wilson lines.  Denoting the charge conjugation operator in the space of fields by $U_c$, we have
\begin{align}
\label{ops-C-transform}
U_c \, \mathcal{O}^{i i'}_{q} \, U_c^\dagger
   &= \mathcal{O}^{i' i}_{\bar{q}} \,,
&
t^a_{i j}\, t^{a'}_{i'\bs j'}\, U_c \, \mathcal{O}^{a a'}_{g} \, U_c^\dagger
   &= t^a_{j i}\, t^{a'}_{j'\bs i'}\; \mathcal{O}^{a a'}_{g} \,,
\end{align}
where the transformation for $\mathcal{O}_g$ follows from the transformation  $t^a_{i j} \, U_c\, A^{\mu a} U_c^\dagger = - t^a_{j i} \, A^{\mu a}$ of the gluon potential.  The space-time arguments of fields do not change under charge conjugation.

We see that charge conjugation is tantamount to interchanging quark and antiquark labels and transposing the colour indices in the fundamental representation, provided that one contracts each adjoint index $a$ of gluonic operators with a generator $t^a$.
To make use of this, we rewrite the colour matrices $M_{\RR}$ as
\begin{align}
   \label{mixed-matrices}
M_{88}^{i i'  \ms j j'}
&= \frac{N}{\sqrt{N^2-1}}\,
   \biggl( \delta_{i' j} \, \delta_{j' i}
      - \frac{1}{N}\, \delta_{i' i} \, \delta_{j'\bs j} \biggr) \,,
\nonumber \\
M_{11}^{j j' \ms a a'} &= \delta_{j'\bs j} \tr \bigl\{ t^a, t^{a'} \bigr\} \,,
\qquad\qquad\qquad
M_{8A}^{j j' \ms a a'} = \sqrt{2}\, \bigl[ t^a, t^{a'} \bigr]{}_{j'\bs j} \,,
   \phantom{\frac{1}{1}}
\nonumber \\
M_{8S}^{j j'  \ms a a'} &= \sqrt\frac{2 N^2}{N^2 - 4}\;
   \biggl( \bigl\{ t^a, t^{a'} \bigr\}{}_{j'\bs j}
      - \frac{1}{N}\, \delta_{j'\bs j} \tr \bigl\{ t^a, t^{a'} \bigr\} \biggr)
\end{align}
in the quark and mixed quark-gluon sectors, whereas in the
pure gluon sector we replace all tensors $\delta^{a b}$, $f^{a b c}$ and $d^{a b c}$ using
\begin{align}
   \label{replace-adjoints}
\delta^{a a'} &= \tr \bigl\{ t^a, t^{a'} \bigr\} \,,
\nonumber \\
f^{a a' c} f^{b b' c}
   &= - 2 \tr \bigl[ t^a, t^{a'} \bigr] \bigl[ t^b, t^{b'} \bigr] \,,
\qquad\qquad
i f^{a a' c} d^{b b' c}
   = 2 \tr \bigl[ t^a, t^{a'} \bigr] \bigl\{ t^b, t^{b'} \bigr\} \,,
\nonumber \\
d^{a a' c} d^{b b' c}
   &= 2 \tr \bigl\{ t^a, t^{a'} \bigr\} \big\{ t^b, t^{b'} \bigr\}
      - 2 N^{-1}\, \tr \bigl\{ t^a, t^{a'} \bigr\}
                   \tr \bigl\{ t^b, t^{b'} \bigr\} \,.
\end{align}
After this, all generators $t^a$ in $M_{\RR}$ are associated with the adjoint index $a$ of a gluon operator.  The charge conjugation behaviour of the contraction
\begin{align}
\pr{\RR}{(\mathcal{O}_{a_1} \mathcal{O}_{a_2})}
  &= M_{\RR}^{r_1^{}\bs r_1' \, r_2^{} r_2'} \,
     \mathcal{O}^{r_1^{}\bs r_1'}_{a_1} \, \mathcal{O}^{r_2^{} r_2'}_{a_2}
\end{align}
is then readily worked out by transposing the fundamental matrices in \eqref{mixed-matrices} and \eqref{replace-adjoints}.  Using $[A^T, B^T] = - [A, B]^T$, $\{ A^T, B^T \} = \{ A, B \}^T$, and $\tr C^T D^T = \tr C D$, we find that
\begin{align}
   \label{C-parity-ops}
U_c \; \pr{\RR}{(\mathcal{O}_{a_1} \mathcal{O}_{a_2})} \, U_c^\dagger
&= \eta(R_1)\ms \eta(R_2)\;
   \pr{\Rbar_1\bs \Rbar_2}{(\mathcal{O}_{\bar{a}_1}
      \mathcal{O}_{\bar{a}_2})} \,.
\end{align}
The sign factor $\eta(R)$ is $-1$ for $R=A$ and $+1$ otherwise, and it understood that $\bar{g} = g$.  The minus sign for each antisymmetric octet reflects the fact that the tensor $f^{a a' c}$ leads to a commutator $[t^a, t^{a'}]$ in \eqref{mixed-matrices} and \eqref{replace-adjoints}.  Furthermore, we see that charge conjugation interchanges decuplet and antidecuplet.


\section{Fourier transforms}
\label{sec:FT}

The Fourier transforms in $2 - 2\epsilon$ transverse dimensions required in this work can be performed with the help of the integrals
\begin{align}
\label{FT_power}
\int \frac{d^{2-2\epsilon} \tvec{\Delta}}{(2 \pi)^{2-2\epsilon}} \,
    e^{-i \tvec{\Delta} \tvec{y}}\,
    \frac{1}{\bigl( \tvec{\Delta}^{2} \bigr)^{\alpha}}\,
& = \frac{1}{(\pi \tvec{y}^2)^{1 - \epsilon}} \,
    \left(\frac{\tvec{y}^2}{4}\right)^{\alpha} \,
    \frac{\Gamma(1-\epsilon-\alpha)}{\Gamma(\alpha)} \,,
\\
\label{FT_power_and_log}
\int \frac{d^{2-2\epsilon} \tvec{\Delta}}{(2 \pi)^{2-2\epsilon}} \,
    e^{-i \tvec{\Delta} \tvec{y}}\;
    \frac{\log \tvec{\Delta}^2}{\bigl( \tvec{\Delta}^{2} \bigr)^{\alpha}}\,
& = \frac{1}{(\pi \tvec{y}^2)^{1 - \epsilon}} \,
    \left(\frac{\tvec{y}^2}{4}\right)^{\alpha} \,
    \frac{\Gamma(1-\epsilon-\alpha)}{\Gamma(\alpha)} \,
\nonumber \\
& \quad \times
  \biggl[\ms \log \frac{4}{\tvec{y}^2}
  + \psi(\alpha) + \psi(1-\alpha-\epsilon) \ms\biggr] \,,
\end{align}
where $ \psi(x) = \frac{d}{dz} \log\Gamma(z)$ is the digamma function.  One
can obtain \eqref{FT_power_and_log} by differentiation of \eqref{FT_power} with respect to $\alpha$.  A useful relation is
\begin{align}
\psi(z) &= \psi(1+z) - 1/z \,,
\end{align}
which makes the pole at $z=0$ explicit.


\section{Loop integrals with the Collins regulator}
\label{sec:collins-integrals}

The use of spacelike Wilson lines leads to a non-trivial interplay between rapidity divergences and loop integrations, because eikonal propagators depend on both  plus- and minus-momenta.  After setting gluon momenta on shell, we have eikonal propagators depending on plus- and transverse momenta.  By contrast, eikonal propagators with the $\delta$ regulator depend only on plus-momenta.  In this appendix, we discuss some implications of the additional transverse-momentum dependence that appears with the Collins regulator.

Quite remarkably, there are loop integrals that give a $\delta(z_3)$ contribution for the Collins regulator but not for the $\delta$ regulator.  An example is
\begin{align}
\label{collins-integral}
\int d^{2-2\epsilon} \tvec{k}_1\, d^{2-2\epsilon} \tvec{k}_2\;
  \frac{\tvec{k}_3^2 / z_3^{}}{\tvec{k}_2^2 \,
      ( \tvec{k}_1 - \tvec{\Delta} )^2 \,
      \bigl( \tvec{k}_1^2 / z_1^{} + \tvec{k}_2^2 / z_2^{}
         + \tvec{k}_3^2 / z_3^{} \bigr)} \,
  \frac{z_3}{z_3^2 - \tvec{k}_3^2 \ms z_1^{} z_2^{} \big/ \rho} \,.
\end{align}
Using the distributional identity \eqref{distr-collins}, one finds that the terms going with $\delta(z_3)$ read
\begin{align}
& \frac{1}{2}\ms \delta(z_3)
\int d^{2-2\epsilon} \tvec{k}_1\, d^{2-2\epsilon} \tvec{k}_2\;
  \frac{1}{\tvec{k}_2^2 \,
      ( \tvec{k}_1^{} - \tvec{\Delta} )^2} \,
  \biggl[ \log \frac{\rho}{\tvec{\Delta^2}}
     - \log (z_1 z_2)
     - \log \frac{\tvec{k}_3^2}{\tvec{\Delta^2}} \biggr]
\nonumber \\
& \quad =
- \frac{1}{2}\ms \delta(z_3)
\int d^{2-2\epsilon} \tvec{k}_1\, d^{2-2\epsilon} \tvec{k}_2\;
  \frac{1}{\tvec{k}_2^2 \,
      ( \tvec{k}_1^{} - \tvec{\Delta} )^2} \,
  \log \frac{\tvec{k}_3^2}{\tvec{\Delta^2}} \,,
\end{align}
where we recall that $\tvec{k}_3 = - \tvec{k}_1 - \tvec{k}_2$.  The integrals going with the first two terms in the square brackets on the l.h.s.\ are scaleless and therefore give zero.  If one takes the $\delta$ regulator instead, this applies to all terms going with $\delta(z_3)$, because the corresponding identity \eqref{distr-delta} does not involve transverse loop momenta.

This example provides a caveat to our statement in \sect{\ref{sec:compute}} that graphs which give no rapidity divergence can simply be evaluated with the unregulated $1/k_3^+$ eikonal propagator.  This is not true for the Collins regulator when a would-be rapidity divergence is absent because an eikonal propagator is multiplied by a loop integral that is scaleless in the unregulated case but not scaleless in the presence of the regulator.  In other words, for integrals of the type \eqref{collins-integral} it matters whether one removes the rapidity regulator before or after performing the transverse loop integrals.
We find, however, that such integrals do not occur for the graphs in our calculation.


When using the Collins regulator to compute the real graphs, we take the $\rho \to \infty$ limit \eqref{distr-collins} of the eikonal propagators before evaluating the loop integrals.  One may wonder whether it is advantageous to calculate with the propagators \eqref{eikonal-collins} at finite $\rho$ instead, postponing the $\rho \to \infty$ limit to a later stage.

It is indeed possible to apply integration-by-part reduction to the integrals with finite $\rho$.  The set of master integrals is then augmented, because one has an additional denominator factor $(z_3^2 - \tvec{k}_3^2 \ms z_1^{} z_2^{} \big/ \rho)$.  After the reduction, each graph is a sum over different terms with a $z_3$ and $\rho$ dependence of the form
\begin{align}
\label{problem-int-1}
& \frac{z_3^{n+1}}{z_3^2 - \tvec{k}_3^2 \ms z_1^{} z_2^{} \big/ \rho}
& & \text{ with } n=-2,-1,0,1,2,3
\intertext{or}
\label{problem-int-2}
& \frac{\rho \ms z_3^{n+1}}{z_3^2 - \tvec{k}_3^2 \ms z_1^{} z_2^{} \big/ \rho}
& & \text{ with } n=0,1,2,3
\end{align}
times a function that is finite in the limit $z_3 \to 0$.  Evaluating the augmented set of master integrals with the differential equation technique would be a formidable task, so that it seems preferable to take $\rho \to \infty$ at this point.

The form \eqref{problem-int-1} with $n \ge 0$ has a simple $\rho \to \infty$ limit according to \eqref{distr-collins} and in this sense is unproblematic.  For $n<0$, however, one obtains a non-integrable singularity at $z_3 = 0$ when taking $\rho \to \infty$.  Moreover, terms of the form \eqref{problem-int-2} grow linearly with $\rho$ and thus have a stronger rapidity divergence than the full result, which we know to behave like $\log \rho$.  One thus finds that the individual terms after integration-by-part reduction are more singular than the graphs one needs to compute.  Combining individual terms so as to obtain less singular expressions would be at odds with the very idea of a reduction to master integrals.  We find this unattractive and prefer to take the limit $\rho \to \infty$ as early as possible in the calculation, as described in \sect{\ref{sec:compute}}.


\section{Distributional identities}
\label{sec:distrib}

In this appendix, we derive the distributional identity \eqref{distr-collins} required for the Collins regulator.   Its analogue \eqref{distr-delta} for the $\delta$ regulator can be established in a very similar manner, as discussed at the end.

To begin with, we cast \eqref{distr-collins} into a simpler form.  Replacing $z_1$ and $z_2$ in terms of $u$, $\ub$ and $z_3$ as specified in \eqref{conv-12-z3}, we obtain
\begin{align}
\label{distr-collins-2}
\underset{\sigma \to 0}{\lim} \;
\operatorname{PV} \frac{t}{t^2 - \sigma^2\, (1-t)^2} &=
  \mathcal{L}_0(t) - \frac{\log \sigma^2}{2}\ms \delta(t) \,,
\end{align}
where we write $t = z_3$ and $\sigma^2 = u \bar{u} \, \tvec{k}_3^2 \big/ \rho$.  It is understood that the identity \eqref{distr-collins-2} is to be integrated over $t \in [0,1]$ with a sufficiently regular test function $\varphi(t)$.

Let us first show that the term $\sigma^2\, (1-t^2)$ on the l.h.s.\ of \eqref{distr-collins-2} can be replaced with $\sigma^2$.  Integrating over a test function, we have for the difference of the two versions
\begin{align}
\label{Delta-I}
\Delta I &=
\underset{\sigma \to 0}{\lim} \;
\operatorname{PV} \int_0^1 d t\, \varphi(t) \biggl[
  \frac{t}{t^2 - \sigma^2} - \frac{t}{t^2 - \sigma^2\, (1-t)^2} \biggr]
\nonumber \\
&=
\underset{\sigma \to 0}{\lim} \; \sigma^2
\operatorname{PV} \int_0^{2\sigma} d t\; \varphi(t) \,
   \frac{2 - t}{t^2 - \sigma^2\, (1-t)^2} \;
   \frac{t^2}{t^2 - \sigma^2}
\nonumber \\
& \quad + \underset{\sigma \to 0}{\lim} \; \sigma^2
   \int_{2\sigma}^{1} d t\;
   \varphi(t) \, \frac{2 - t}{t^2 - \sigma^2\, (1-t)^2} \;
   \frac{t^2}{t^2 - \sigma^2} \,,
\end{align}
where in the second step we have split the integration into two parts.  The PV prescription applies to both singularities at $t = \sigma$ and $t = \sigma/(1+\sigma)$.  In the first integral on the r.h.s., we can replace $\varphi(t)$ with $\varphi(0)$, using that the test function is sufficiently smooth.  In the second integral, the factor multiplying $\varphi(t)$ is a monotonic function if  $\sigma$ is sufficiently small, so that we can use the mean value theorem for integrals and replace $\varphi(t)$ with $\varphi(t_\sigma)$ at some point $t_\sigma \in [2\sigma, 1]$.  Rescaling the integration variable as $t = \sigma \tau$, we obtain
\begin{align}
\label{rescaled-Delta-I}
\Delta I
&\quad = \underset{\sigma\to 0}{\lim} \; \sigma \varphi(0) \,
\operatorname{PV} \int_0^{2} d\tau\;
   \frac{2 - \sigma \tau}{\tau^2 - (1 - \sigma \tau)^2} \;
   \frac{\tau^2}{\tau^2 - 1}
\nonumber \\
&\qquad +
\underset{\sigma\to 0}{\lim} \; \sigma \varphi(t_\sigma)
   \int_2^{1/\sigma} d\tau\;
   \frac{2 - \sigma \tau}{\tau^2 - (1 - \sigma \tau)^2} \;
   \frac{\tau^2}{\tau^2 - 1} \,.
\end{align}
The two integrals over $\tau$ remain finite for $\sigma\to 0$, as can be checked by explicit calculation, so that $\Delta I = 0$ in that limit.

We now show \eqref{distr-collins-2} without the factor $(1-t)^2$.  To this end, we decompose the integral over a test function into three terms
\begin{align}
\underset{\sigma \to 0}{\lim} \;
\operatorname{PV} \int_0^1 d t\; \varphi(t) \, \frac{t}{t^2 - \sigma^2}
&= I_1 + I_2 + I_3
\end{align}
with
\begin{align}
I_1 &= \int_0^1 d t\; \frac{\varphi(t) - \varphi(0)}{t} \,,
&
I_2 &= \underset{\sigma \to 0}{\lim} \; \operatorname{PV} \int_0^1 d t\;
   \frac{\varphi(t) - \varphi(0)}{t}  \,
      \frac{\sigma^2}{t^2 - \sigma^2} \,,
\nonumber \\
I_3 &= \varphi(0) \;
   \underset{\sigma \to 0}{\lim} \; \operatorname{PV} \int_0^1 d t\;
   \frac{t}{t^2 - \sigma^2} \,. \hspace{-3em}
\end{align}
One readily sees that $I_1 = \int_0^1 d t\; \varphi(t)\, \mathcal{L}_0(t)$, and straightforward evaluation of the integral in $I_3$ yields
\begin{align}
I_3 &= - \frac{\log \sigma^2}{2} \int_0^1 d t\; \varphi(t) \, \delta(t) \,,
\end{align}
so that \eqref{distr-collins-2} without the factor $(1-t)^2$ is established by showing that $I_2 = 0$.

We now assume that $\varphi(t) - \varphi(0)$ is continuous and finite in the interval $t \in [0,1]$.  We also assume that it behaves like $c t^\alpha$ with $\alpha > 0$ for $t \to 0$.  Notice that this is less than requiring $\varphi(t)$ to be differentiable at $t=0$, because $\alpha$ need not be an integer.  The function
\begin{align}
\psi(t) &= \frac{\varphi(t) - \varphi(0)}{t^\beta}
&
\text{ with } \beta &= \min(\alpha, 1)
\end{align}
is then finite for $t \in [0,1]$, with $\psi(0) = c$ for $\alpha=\beta$ and $\psi(0) = 0$ otherwise.  Splitting the integration into two terms, we have
\begin{align}
\label{I2-split}
I_2 &= \underset{\sigma \to 0}{\lim} \;
   \operatorname{PV} \int_0^{2\sigma} d t\;
   \frac{\psi(t)}{t^{1-\beta}} \, \frac{\sigma^2}{t^2 - \sigma^2}
+ \underset{\sigma \to 0}{\lim} \; \int_{2 \sigma}^1 d t\;
  \frac{\psi(t)}{t^{1-\beta}}  \,
      \frac{\sigma^2}{t^2 - \sigma^2} \,.
\end{align}
For sufficiently small $\sigma$, we can replace $\psi(t)$ with $\psi(0)$ in the first term on the r.h.s., and for the second term we can use the mean value theorem for integrals.  This gives
\begin{align}
I_2 &= \psi(0) \, \underset{\sigma \to 0}{\lim} \;
   \operatorname{PV} \int_0^{2\sigma}
   \frac{d t}{t^{1-\beta}}  \, \frac{\sigma^2}{t^2 - \sigma^2}
+ \underset{\sigma \to 0}{\lim} \;
   \psi(t_\sigma) \int_{2 \sigma}^{1}
  \frac{d t}{t^{1-\beta}}  \, \frac{\sigma^2}{t^2 - \sigma^2}
\end{align}
with some value $t_\sigma \in [2\sigma, 1]$ different from the one that appears in \eqref{rescaled-Delta-I}.  Rescaling the integration variable as $t = \sigma \tau$, we finally obtain
\begin{align}
I_2 &= \psi(0) \, \underset{\sigma \to 0}{\lim} \; \sigma^\beta
  \operatorname{PV} \int_0^{2} \frac{d \tau}{\tau^{1-\beta}} \;
  \frac{1}{\tau^2 - 1}
+ \underset{\sigma \to 0}{\lim} \; \psi(t_\sigma) \, \sigma^\beta
  \int_{2}^{\infty} \frac{d \tau}{\tau^{1-\beta}} \; \frac{1}{\tau^2 - 1}
= 0 \,,
\end{align}
which concludes our proof.

The distributional identity \eqref{distr-delta} we use with the $\delta$ regulator has the same form as \eqref{distr-collins-2} with $t^2 + \sigma^2\, (1-t)^2$ instead of $t^2 - \sigma^2\, (1-t)^2$ on the l.h.s.  The PV prescription is not needed in that case.  The proof of this identity proceeds in analogy to what we have shown, with the simplification that one does not need to split integrals into two parts as we did in \eqref{Delta-I} and \eqref{I2-split}.


\section*{Acknowledgements}

We gratefully acknowledge discussions with Iain Stewart and Alexey Vladimirov on rapidity regulators.  We also thank Florian Fabry for feedback on the manuscript.

This work was in part supported by the Deutsche Forschungsgemeinschaft (DFG, German Research Foundation) -- Research Unit FOR 2926, grant number 409651613.  The work of JRG is supported by the Royal Society through Grant URF\textbackslash{}R1\textbackslash{}201500, and the work of PP in part by the BMBF project 05P2018 (ErUM-FSP T01).
MD and JRG would like to express their thanks to the Mainz Institute for Theoretical Physics (MITP) of the DFG Cluster of Excellence PRIMSA$^+$ (Project ID 39083149) for its hospitality and support.  The figures in this work were produced with JaxoDraw \cite{Binosi:2003yf,Binosi:2008ig}.


\phantomsection
\addcontentsline{toc}{section}{References}

\bibliographystyle{JHEP}
\bibliography{split.bib}

\end{document}